\documentclass{article}

\usepackage{arxiv}

\usepackage[utf8]{inputenc} 
\usepackage[T1]{fontenc}    
\usepackage{hyperref}       
\usepackage{url}            
\usepackage{booktabs}       
\usepackage{amsfonts}       
\usepackage{nicefrac}       
\usepackage{microtype}      
\usepackage{lipsum}		
\usepackage{graphicx}
\usepackage[numbers]{natbib}
\usepackage{doi}

\usepackage{caption}
\usepackage{subcaption}

\usepackage{chemformula}

\usepackage{siunitx}
\usepackage{xcolor}
\definecolor{darkred}{rgb}{0.65,0.0,0.0}
\definecolor{darkblue}{rgb}{0.0,0.0,0.5}

\hypersetup{                
    colorlinks=true,
    linkcolor=darkblue,
    filecolor=darkblue,
    urlcolor=darkblue,
    citecolor=darkblue
    }

\title{PMMA Pyrolysis Simulation -- from Micro- to Real-Scale}
\renewcommand{\shorttitle}{PMMA Pyrolysis Simulation -- from Micro- to Real-Scale}

\usepackage{authblk}

\author[a]{{\large Tristan Hehnen\,}}
\author[b,a]{{\large Lukas Arnold\,}}

\affil[a]{Chair of Computational Civil Engineering, University of Wuppertal, Pauluskirchstraße 7, 42285 Wuppertal, Germany}
\affil[b]{Institute for Advanced Simulation, Forschungszentrum  J\"ulich, Wilhelm-Johnen-Straße, 52428 J\"ulich, Germany\newline}

\affil[ ]{
Tristan Hehnen: \texttt{\href{mailto:hehnen@uni-wuppertal.de}{hehnen@uni-wuppertal.de}; \href{https://orcid.org/0000-0002-6123-261X}{ORCID: 0000-0002-6123-261X}}
}

\affil[ ]{
Lukas Arnold: \texttt{\href{mailto:l.arnold@fz-juelich.de}{l.arnold@fz-juelich.de}; \href{mailto:arnold@uni-wuppertal.de}{arnold@uni-wuppertal.de}; \href{https://orcid.org/0000-0002-5939-8995}{ORCID: 0000-0002-5939-8995}}
}

\hypersetup{
pdftitle={PMMA Pyrolysis Simulation -- from Micro- to Real-Scale},
pdfsubject={Fire Propagation Simulation},
pdfauthor={Tristan Hehnen, Lukas Arnold},
}

\begin{document}
\maketitle

\begin{abstract}
In fire spread simulations, heat transfer and pyrolysis are processes to describe the thermal degradation of solid material. 
In general, the necessary material parameters cannot be directly measured. They are implicitly deduced from micro- and bench-scale experiments, i.e. thermogravimetric analysis (TGA), micro-combustion (MCC) and cone calorimetry. Using a complex fire model, an inverse modelling process (IMP) is capable to find parameter sets, which are able to reproduce the experimental results.
In the real-scale, however, difficulties arise predicting the fire behaviour using the deduced parameter sets.
Here, we show an improved model to fit data of multiple small scale experiment types. Primarily, a gas mixture is used to model an average heat of combustion for the surrogate fuel. The pyrolysis scheme is using multiple reactions to match the mass loss (TGA), as well as the energy release (MCC). Additionally, a radiative heat flux map, based on higher resolution simulations, is used in the cone calorimeter setup.
With this method, polymethylmetacrylate (PMMA) micro-scale data can be reproduced well. For the bench-scale, IMP setups are used differing in cell size and targets, which all lead to similar and good results. Yet, they show significantly different performance in the real-scale parallel panel setup.
\end{abstract}


\section{Introduction}
\label{sec:intro}

The simulation of fire propagation is of great interest for the fire safety engineering community. It could lead to reduced costs for mitigation measures, since the fire scenario could be less over-predicting and fire protection measures could be better evaluated. It could even make certain types of assessments possible, for instance when the release of (radioactive) combustion products is to be determined, and not prescribed within a design fire. Much research is performed in this direction internationally~\cite{Matala2013PhdApplicationsOfPyrolysisModelling,Stoliarov2019PyrolysisModelDevelopment,ThermalDecompositionAndPyrolysisOfSolidFuels_Rogaume,LAUTENBERGER2006204,Alonso2019_LLDPEKineticPropertiesEstimation, lautenberger2011optimization,macfp_matl_git,AlexandraViitanen_CableTrays,Aleksi_WoodPyrolysis,Hehnen_fire3030033,HostikkaMatala2017PyrolysisBirchWood, Alonso2023_NumericalPredictionofCablesFireBehaviour}. 
This approach requires material parameter sets, which allow meaningful reaction to changed physical conditions near the fire. For example, reduced oxygen should lead to less energy release from the flame, which in turn reduces the heat transfer to the sample, impacts the release of combustible gas and ultimately leads to a smaller flame.
The performance of these parameter sets needs to be assessed over all involved length scales, not only in the micro- and bench-scale. Specifically, the transition from the bench- to the real-scale is important. Assessment of the parameter set's performance is only meaningful in the real-scale, i.e. in terms of validation. Thus, the real-scale should not be part of the estimation of the parameter set itself. 

PMMA is an often studied combustible solid in the fire safety engineering community. Research is performed on the pyrolysis reaction kinetics of the PMMA material, as well as on the gas phase combustion reaction, and the numerical simulation thereof, e.g.~\cite{Korobeinichev2019, Zeng_chemreac_burning_PMMA, FIOLA2021103083PMMA_Pyrolysis_Model}.
Fire propagation was investigated over PMMA slabs, oriented horizontally and vertically, e.g.~\cite{KARPOV2018937, KARPOV_UpwardFlameSpread2015, WU2003697, Rauwoens_UpwardFlameSpreadEnthalpy2010}. Simulation setups limited to two spatial dimensions allow the detailed study of interactions between flame and sample with high numerical resolution. These simulation setups focus on sample sizes of about 0.1~m up to 5.0~m in length. Material and pyrolysis reaction parameters have been determined from analytical methods and literature.
The upward flame spread over a vertical 5.0~m tall, 0.6~m wide and 2.5~cm thick PMMA panel was investigated by Kwon~\cite{KwonMasterThesisFDS4FlameSpreadPMMA}. They reproduced real-scale experiments with the Fire Dynamics Simulator (FDS, version 4), conducting full three-dimensional large eddy simulations. The employed setup incorporates the simulation of gas phase combustion and the associated heat feedback to the sample. The fluid cell size was chosen to be 2.5~cm. Material and pyrolysis kinetics parameters have been determined from cone calorimeter data. They identified issues in the coupling of the gas phase combustion and pyrolysis models. This lead to an overestimation of the flame height in the simulation. 
Recently, buoyancy driven real-scale flame spread was investigated in a corner configuration over PMMA slabs~\cite{CHAUDHARI2021109433PMMA_RoomCorner_Simulation}. The used material parameter set of black cast PMMA was determined in an earlier study~\cite{FIOLA2021103083PMMA_Pyrolysis_Model}. In this real-scale simulation setup, each panel was divided into~28 surface elements. Using ThermKin2Ds~\cite{StoliarovThermaKin}, each surface element was simulated independently. No gas phase combustion reactions are considered in the model, flame heat flux from experiments was imprinted as boundary condition to the individual elements directly.

In the literature, fire propagation over PMMA samples is often addressed with a focus on selected aspects. 
With this contribution we present a general strategy to estimate material parameter sets, solely based on small-scale experiment data. This creates a consistent material model. 
The procedure is built on existing approaches. These approaches are based on micro- and bench-scale tests, using cast black PMMA as an example. The parameter sets are applied for fire spread simulations with full three-dimensional large eddy simulations and gas phase combustion, using  FDS~6.7.6~\cite{fdsUserGuide676}.

In this work, for the parameter estimation, an optimisation algorithm is employed in an inverse modelling process (IMP). 
We deliberately assume no information on pyrolysis and combustion parameters of PMMA. With this, the modelled system gains many degrees of freedom to represent, e.g., the intermediate states and structural changes, which are in general not measurable from the virgin material. Additionally, these parameters may not be available in practical fire risk assessment scenarios, and the presented approach aims for a general applicability. 
As a general strategy, the process presented here is divided into three major steps, in an effort to reduce the otherwise significant computational demand. In the first step, reaction kinetics of the material decomposition (pyrolysis) and the energy release in the gas phase (combustion) are determined. This is based on micro-scale tests, thermogravimetric analysis (TGA) and micro-combustion calorimetry (MCC). A large number of parameters (33) is used to define the PMMA decomposition scheme. By using an extremely simplified micro-scale simulation setup the computational demand can be kept low -- in the order of days. In the second step, the thermophysical and optical parameters are determined in the bench-scale. The simplified cone calorimeter setup used in this step is computationally much more expensive. Looking at fewer parameters (15), the computational demand can be kept relatively low as well. Still, the needed computing time is in the order of months on a high performance computing cluster to complete a full IMP, including multiple sampling limit adjustments. Finally, the performance of the parameter sets is assessed in a real-scale simulation setup of a parallel panel test. This is considered here as validation step, since the goal is to determine parameter sets in the small-scale that lead to the appropriate behaviour to predict the fire development in the real-scale.

The here proposed method is built on state-of-the-art strategies of using FDS for pyrolysis simulation, e.g.~\cite{AlexandraViitanen_CableTrays,Aleksi_WoodPyrolysis,Hehnen_fire3030033,Moinuddin_PMMAConeModelling}. The primary changes are the use of multiple superimposed pyrolysis reactions, using a gas mixture as surrogate fuel and conducting simplified cone calorimeter simulations with higher fluid cell resolution during the inverse modelling. 
The surrogate fuel consists of combustible and non-combustible primitive species. During the IMP, their fractions are adjusted such that the average heat of combustion (HOC) of this mixture fits to the experiment data. With this, FDS can use the mass loss rate from the solid directly as input for the gas phase without the usual scaling. 

Furthermore, the radiative flux from the cone heater to the sample surface is not uniform~\cite{BOULET201253, HostikkaAxelssonRadiativeFeedbackFlamesConeCalorimeter}, 
which is taken into account here. A high resolution simulation with a conical heater geometry was conducted and the resulting radiative heat flux determined. This result is then baked into a flux map for a simplified cone calorimeter setup. This setup has a higher fluid cell resolution than is typically used. With the higher resolution and the inhomogeneous heat flux, it is possible to capture an uneven sample consumption.

In the presented theoretical study, we did not conduct any own experiments, but used publicly available experimental data for the micro- and bench-scale tests. These data sets are part of the MaCFP materials database~\cite{macfp_matl_git} (Git commit hash:~\texttt{7f89fd8}). The result of the inverse modelling is an effective material parameter set. Its performance is compared against real-scale experiment data of a parallel panel test, taken from the MaCFP database~\cite{Leventon2022ParallelPanel} (Git commit hash:~\texttt{25614bd}). 

This article is accompanied by a publicly available data repository on Zenodo~\cite{zenodo:ArticleDataset}, containing the simulation data and analysis scripts, as well as a video series on YouTube explaining how they are set up~\cite{firesimandcoding_playlist}.

\section{Methodology}
\label{sec:methodology}

\subsection{Numerical and Physical Models}
\label{subsec:num_phys_models}

Throughout this work, FDS (version \texttt{FDS6.7.6-810-ge59f90f-HEAD}) is used with default settings if not stated otherwise. All methods, models and default parameters are documented in the FDS user's guide~\cite{fdsUserGuide676} and technical reference guide~\cite{fdsTechGuide1_676}. As FDS is an open access model, including well documented sub-models, we will not make an attempt to summarise the publicly available documentation here. Yet, in the following sections, we highlight relevant modelling approaches, which may not be commonly used. Additionally, whenever useful, we explicitly state which FDS parameters or definitions we have used to enhance the comprehensibility.

\subsection{Inverse Modelling Process}
\label{subsec:IMP}

Throughout the presented work, different sets of material parameters are estimated by inverse modelling. The general idea in this is approach is, that a cost function is defined, which measures the deviation of the predicted to the target experimental data. Here, the root mean square error (RMSE) is used. Based on this function, an optimiser searches for its minimal value, i.e. the set of material parameters for which the prediction of the computational model matches the experimental data best. The search is in general an iterative process, where new parameter sets are determined and evaluated. In this work, the evaluation of a parameter set corresponds to the execution of an FDS simulation. 

The parameter estimation is conducted, by employing a shuffled complex evolutionary (SCE) algorithm~\cite{duan1993shuffled}. The SCE is designed to find the global optimum reliably in a multi-parameter space, without getting stuck at local optima. The algorithm combines deterministic and stochastic strategies, to efficiently sample the space of possible solutions. The generated parameter sets are further divided into complexes which are evolved competitively. Then the parameter sets are shuffled and if convergence was reached, the process is stopped. Each of these loops is called here a "generation". The number of assessed parameter sets per generation depends on the amount of parameters to be considered and on the chosen number of complexes. Here, the number of complexes $n_{\text{complex}}$ is chosen to equal the number of parameters $n_{\text{parameter}}$. The resulting generation size~$\Phi$ is determined from equation~\ref{formula:GenerationSize}. 

\begin{equation}
    \Phi = (2 \cdot n_{\text{parameter}} + 1) \cdot n_{\text{complex}}\quad \text{; with} \quad n_{\text{complex}} = n_{\text{parameter}}
    \label{formula:GenerationSize}
\end{equation}

The number of generations was chosen to be about 150 for the micro-scale simulations and about 100 generations for the simplified cone calorimeter (bench-scale). In general, this was a sufficient number of generations to reach convergence.

The initial sampling limits for the individual parameters are chosen as best guess. If during the IMP parameters get stuck at one of their limits, their sampling space is expanded in the respective direction.
Thus, the sampling space gets only larger and the same parameter combinations of the previous runs can still be reached. 
Typically, after a couple generations it is clear which parameter approaches a limit and a new IMP run can be set up. These changes in sampling limits are denoted with a capital "L" followed by a number, for example "L0" is the original sampling space, "L1" would be the first expansion and so on. See appendix~\ref{App:LimitAdjustmentExample} for an example. This leads to a staggering of the IMP runs.

The parameter estimation for the complete set is divided into two IMP steps. In the first IMP step, the reaction kinetics parameters for the material pyrolysis scheme are determined in the micro-scale. In the second step, the thermophysical and optical parameters are determined on the bench-scale, using the reaction kinetics of the first step. These individual IMP steps are described in more detail in section~\ref{subsec:methods_microscale} and section~\ref{subsec:methods_benchscale} respectively.

The splitting of the two scales aims to distribute the workload for the parameter estimation over two different simulation setups. 
Separating the reaction kinetics from the thermophysical parameters, is beneficial in two aspects. For one, it reduces the complexity of the inverse modelling itself ($n_{a+b}^2 > n_a^2 + n_b^2$). With increasing amount of parameters, the generation size grows more than quadratic, see equation~\ref{formula:GenerationSize}.
Furthermore, about two thirds of the parameters can be determined in a less costly setup. The computing time for the IMP massively depends on the fidelity of the employed simulation. Even though the number of simulations in a single IMP run at the micro-scale is about 6~times larger than the simplified cone calorimeter setup (bench-scale), the latter takes a good 30 times longer for the base case, as summarised in table~\ref{tab:ParametersIMP}. The simple cone setup with a coarse resolution (C2, as defined in \ref{subsec:methods_benchscale}) can be completed in about 2~weeks, while the parameters at the finest resolution (C5, as defined in \ref{subsec:methods_benchscale}) is estimated to take 10 months to over a year!

The inverse modelling is conducted with  PROPTI~\cite{propti_nancy, propti:ARNOLD2019102835, zenodo:PROPTI}, an in-house developed publicly available inverse modelling framework. New dependencies were implemented to determine the gas mixtures (Git commit hash:~\texttt{3a05366}), more recent versions of PROPTI should support this directly. In this work, PROPTI used the shuffled complex evolutionary algorithm from SPOTPY~\cite{houska2015spotting}, version 1.5.14.

\begin{table}[h]
    \caption{Overview of number of optimisation parameters during different steps of the IMP. The amount of used CPU cores equals the number of parameters. Time necessary in the bench-scale setup massively depends on the fluid cell number and size, listed here is the base case, as defined in \ref{subsec:methods_benchscale}. Counting of "IMP run time" begins with the start of L0 and ends with the stop of L3. 
    \label{tab:ParametersIMP}}
    \centering
    \begin{tabular}{l r r}
        \toprule
        & Micro-scale  & Bench-scale  \\
        \midrule
        Number of parameters    &   33         &  15        \\
        Generation size         & 2211         & 465        \\
        Number of generations   &  150         & 100        \\
        IMP run time (approx.)  & 3.5 days     & > 110 days \\
        \bottomrule
    \end{tabular}
\end{table}

\subsection{Pyrolysis Modelling Approach}
\label{subsec:pyrolysis_approach}

Fire spread on solid materials involves the transformation of the material into a combustible gas. This transformation, determined by the temperature of the solid, is called pyrolysis. In the case of PMMA, the long molecule chains of a polymer are split into smaller molecules. This is mostly its monomer methylmetacrylate~(MMA), more than 90~\%, and small amounts of carbon dioxide~\cite{Moinuddin_PMMAConeModelling,Korobeinichev2019,McNeill:ThermalDegradationMMA,Zeng_chemreac_burning_PMMA}. 
However, the MMA is not directly involved in the combustion~\cite{Zeng_chemreac_burning_PMMA}. It further decomposes into even smaller chemical compounds, among which are methane, acetylene, ethylene, carbon monoxide, carbon dioxide, and hydrogen~\cite{Zeng_chemreac_burning_PMMA}.
These smaller molecules are then taking part in the gas phase combustion (flame). The combustion reaction involves many intermediate species and reactions. Already for simple hydrocarbons like methane, reaction mechanisms are proposed~\cite{BowmanShockInitiatedMethaneOxidation, ZhukovCompactReactionMechanismMethane, ZhukovAutoignitionLeanPropaneAirMixtureHighPressures} that involve 30 to over 1200 intermediate reactions and over 200 intermediate species. Reaction mechanisms for longer carbon chains contain the reactions of the shorter molecules~\cite{WESTBROOKDetailedChemicalKineticReactionMechanismsHeptane}. Concentrations of the individual species also change across the combustion reaction zone~\cite{Wolfhard_1949}.  
It seems that the limiting factor to the fidelity of the reaction models is primarily the available computing  power~\cite{fdsUserGuide676,CURRANDevelopingDetailedChemicalKineticMechanisms}. Reaction schemes also differ by which PMMA decomposes, depending on the polymerisation method and molecular weight~\cite{McNeill:ThermalDegradationMMA,Zeng_chemreac_burning_PMMA,FIOLA2021103083PMMA_Pyrolysis_Model}, as well as with different test apparatus designs~\cite{KarenCorinna_PMMA}.

The material decomposition and combustion models within this work using FDS are strongly simplified, yet reflect common practise in scientific and engineering applications. 
These simplifications are justified in many practical fire risk assessments, due to outside restrictions. Fluid cell sizes are large when smoke spread is assessed during the design phase of buildings, typically in the order of 10~cm to~20~cm. Combustion reactions happen in length scales of millimetres, thus detailed chemical reaction schemes are not feasible. Secondly, combustible objects involved in a fire are typically only vaguely known. The material parameters are not known in the required amount of detail and determining them requires large effort as discussed within this work.
For the presented model, the solid sample material is transformed within a series of parallel elementary (single-step) pyrolysis reactions. From the original material, each elementary reaction produces a mass fraction of about 0.99 of a surrogate fuel (\texttt{NU\_SPEC}) and about 0.01 of an inert residue (\texttt{NU\_MATL}). This is summarised in equation~\ref{eq:PMMA_PrincipleDecomposition}. The pyrolysis is modelled by a first-order Arrhenius approach and consists of a set of parallel single-step decomposition reactions. The composition of the surrogate fuel mixture is determined by the first IMP step (micro-scale). It consists of the following mass fractions: 0.193008 methane, 0.315408 ethylene and 0.491585 carbon dioxide, see also equation~\ref{eq:PMMA_FuelMixture}. This mixture yields an average effective heat of combustion similar to the experiment. The fuel mixture skips over intermediate reaction steps of the PMMA and MMA decomposition~\cite{Zeng_chemreac_burning_PMMA}.
The released gas is directly involved in the combustion reaction, see section~\ref{subsec:methods_gas_phase_combustion}. Intermediate reaction steps of further decomposition of the MMA are neglected here. This is regarded as an intermediate approach, located between a single surrogate or many intermediate chemical reactions and species. It still maintains the benefit of the surrogate: the reduced computational cost, because fewer species and reactions need to be tracked.

\begin{equation}
    \ch{PMMA_i -> 0.99 Fuel~Mixture + 0.01 Residue}
    \label{eq:PMMA_PrincipleDecomposition}
\end{equation}

\begin{equation}
    \ch{Fuel~Mixture \,\, = \,\, 0.193008 CH_4 + 0.315408 C_2H_4 + 0.491585 CO_2}
    \label{eq:PMMA_FuelMixture}
\end{equation}

For the purpose of this work, peaks in the reaction rate plots of micro-scale experiments are interpreted as pyrolysis reactions. There is no detailed analysis conducted as to their chemical nature. Here, the decomposition reactions are used to approximate the development of the reaction rates determined from the micro-scale experiments.\\
The material definition in FDS (\texttt{MATL}) plays two roles important to the work presented here. In the first and obvious role, it defines the thermo-physical properties of the sample material, specifically the density, thermal conductivity and so on. In the second role, it contains information on the pyrolysis reaction, i.e. the parameters of the Arrhenius equation, the products of the decomposition reaction as well as the heat of reaction and/or combustion. The material definition needs to be realised as a representation of the sample, by connecting it to a boundary condition (\texttt{SURF}). In the boundary condition definition mass fractions are assigned to an individual material. With this it is defined how much of the sample material that can be consumed by a given reaction. Assume a fictional homogeneous sample, which exhibits distinct reaction rate peaks at two different temperatures. To replicate this behaviour two materials need to be defined. Both get the same thermo-physical parameters to satisfy the condition of the homogeneous sample material. However, they need different pyrolysis reaction definitions to reproduce the peaks at different temperatures.

\subsection{Gas Phase Combustion}
\label{subsec:methods_gas_phase_combustion}

In the here proposed method, the complete energy release is assumed to take place as a gas phase reaction in the flame -- no oxidation at the solid surface. Furthermore, it is assumed that the involved materials, the combustible gases and the polymer, are hydrocarbons. Thus, for different materials the strategy might need to be adjusted, but should be transferable in principle. The combustion reaction is a single step from the surrogate fuel mixture to the combustion products, see equation~\ref{eq:PMMA_MixtureCombustion}.

\begin{equation}
    \ch{Fuel~Mixture + 7.747198 Air -> 8.747198 ( 0.675303 N_2 + 0.221259 CO_2 + 0.002515 soot + 0.100923 H_2O ) }
    \label{eq:PMMA_MixtureCombustion}
\end{equation}

Since PMMA mostly decomposes into its monomer MMA when heated~\cite{Moinuddin_PMMAConeModelling,Korobeinichev2019,McNeill:ThermalDegradationMMA,Zeng_chemreac_burning_PMMA}, it could be considered as the gaseous fuel. This is also commonly done in practice, see for example the "NIST/NRC Parallel Panel Experiments" validation case for FDS~\cite{fdsValiGuide676}. Due to neglecting all intermediate reaction steps and species, this fuel is considered a surrogate, i.e. a surrogate fuel. In FDS, the surrogate is often chosen to be a pure, primitive species, like propane or the aforementioned MMA. This might lead to difficulties connecting it to the gas released in an experiment, of which the heat of combustion is likely different. FDS deals with this situation, by scaling the mass of combustible species introduced into the gas domain based on the energy release~\cite{fdsUserGuide676}. Appendix section~\ref{App:FlameHeight} provides an exemplary description of this concept. In the proposed method here, a simple gas mixture is used with the goal to get an average HOC so that the mass loss rates match for the solid and gas phase. With "simple" meaning only a few primitive components. It is built, using the lumped species concept in FDS. Here, components are chosen that are already implemented in FDS and are also part of the overall combustion reaction mechanism~\cite{Zeng_chemreac_burning_PMMA}. They differ in their respective molecular weight and heat of combustion. No specific emphasis is given to the radiative fraction (\texttt{RADIATIVE\_FRACTION}), thus it uses the default of~0.35 for unknown species~\cite{fdsUserGuide676}. The chosen species are: methane, ethylene and carbon dioxide, see section~\ref{subsec:pyrolysis_approach}. The fractions of methane and ethylene are directly adjusted during the IMP. Carbon dioxide is used to account the remaining difference. Using three adjustable components, the degrees of freedom for the mixture and therefore the computational demand is kept low. They are also part of a computationally inexpensive simulation setup, compare table~\ref{tab:ParametersIMP}, but still connect to the simulations with gas phase combustion. Since the combustible species are hydrocarbons, the "simple chemistry" approach of FDS~6.7.6 can be used, with a soot yield of 0.022~g/g taken from~\cite{Quintire:PrinciplesOfFireBehaviour}, table~8.1.

During the real-scale validation simulations, two different gas phase reaction definitions are used: the gas burner (propane) and the fuel mixture for PMMA. This requires the "complex chemistry" approach in FDS~6.7.6 and individual gas phase reactions. The stoichiometry of the gas mixture is extracted from the best parameter set of the first IMP step, for details see appendix~\ref{App:ComplexChem}. Two different gas phase reactions enable, for example, the assignment of different values for the radiative fraction and soot yield.

\section{Micro-Scale Setup  --- Pyrolysis Reaction Scheme}
\label{sec:pyrolysis_scheme}

\subsection{Methods}
\label{subsec:methods_microscale}

The focus of the micro-scale setups is to determine the temperature-dependent material decomposition reactions (pyrolysis). The experimental data, that is used as target during the parameter estimation, is taken from the open-access MaCFP git repository~\cite{macfp_matl_git}. Two data sets are used in two different IMP setups. First, the normalised residual mass data recorded by a TGA with a heating rate of 10~K/min and heat release rate data recorded by an MCC with a heating rate of 60~K/min. Both are provided by the National Institute of Standards and Technology (NIST). 
Secondly, normalised residual mass data by a TGA from Sandia National Laboratories (Sandia), recorded at a higher heating rate of 50~K/min. This is used in an effort to get close to matching heating rates for TGA and MCC.  
As of writing, no 60~K/min TGA data set or MCC experiments with different heating rates matching the existing TGA data are available from the MaCFP repository.
The TGA experiments by NIST were conducted in nitrogen atmosphere. For the MCC experiments the pyrolysis chamber was fed with nitrogen and oxygen was mixed into the effluents before entering the combustion chamber. The TGA experiments by Sandia were conducted in argon atmosphere.

\begin{figure}[h]%
    \centering

    \subfloat[\centering TGA normalised residual mass, for a  heating rate of 10~K/min in nitrogen atmosphere  (NIST).\label{fig:TGA_exp_NIST}]{{\includegraphics[width=0.45\columnwidth]{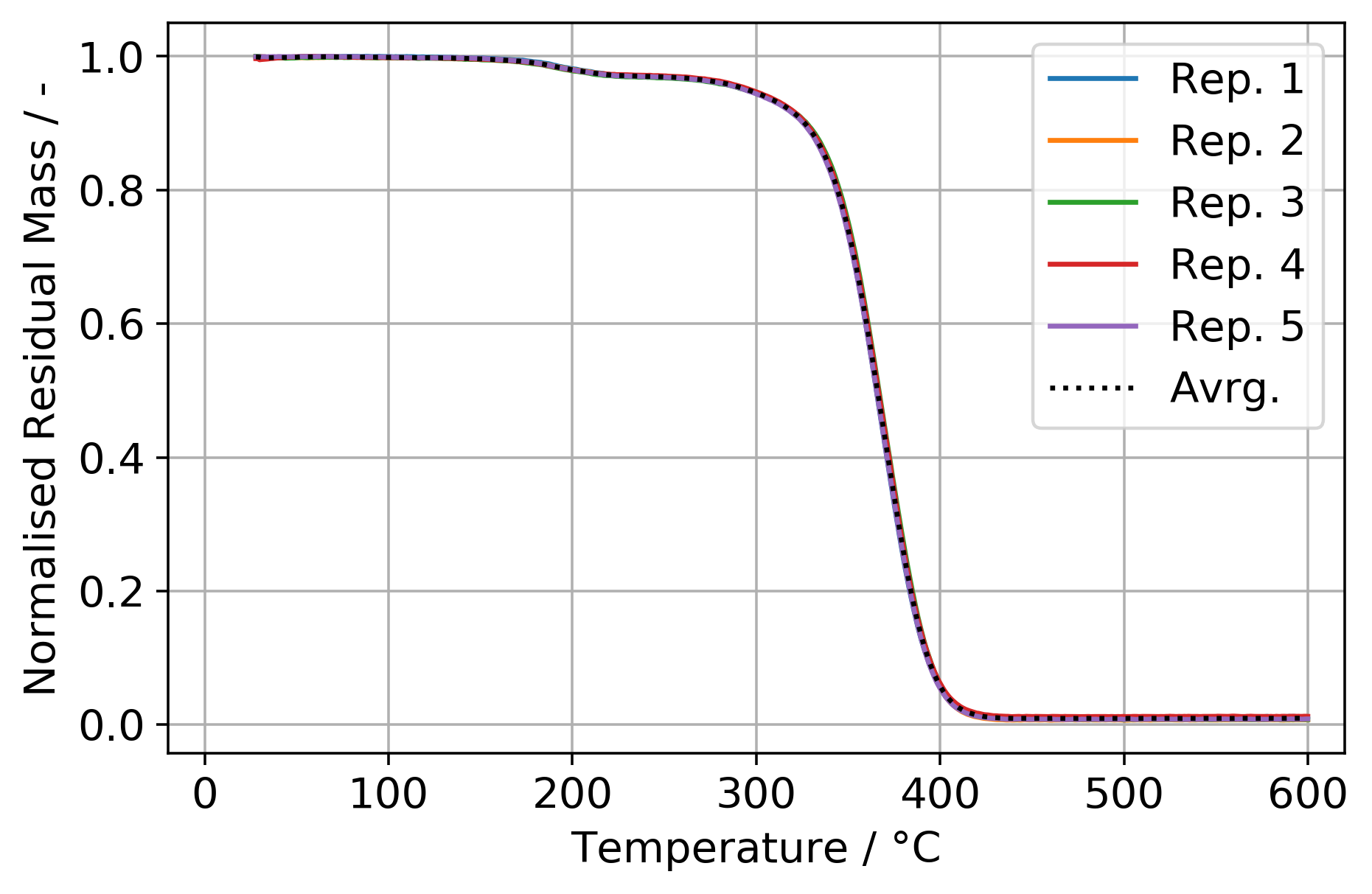} }}%
    \qquad
    \subfloat[\centering MCC heat release rate, for a  heating rate of 60~K/min in nitrogen atmosphere (NIST).\label{fig:MCC_exp_NIST}]{{\includegraphics[width=0.45\columnwidth]{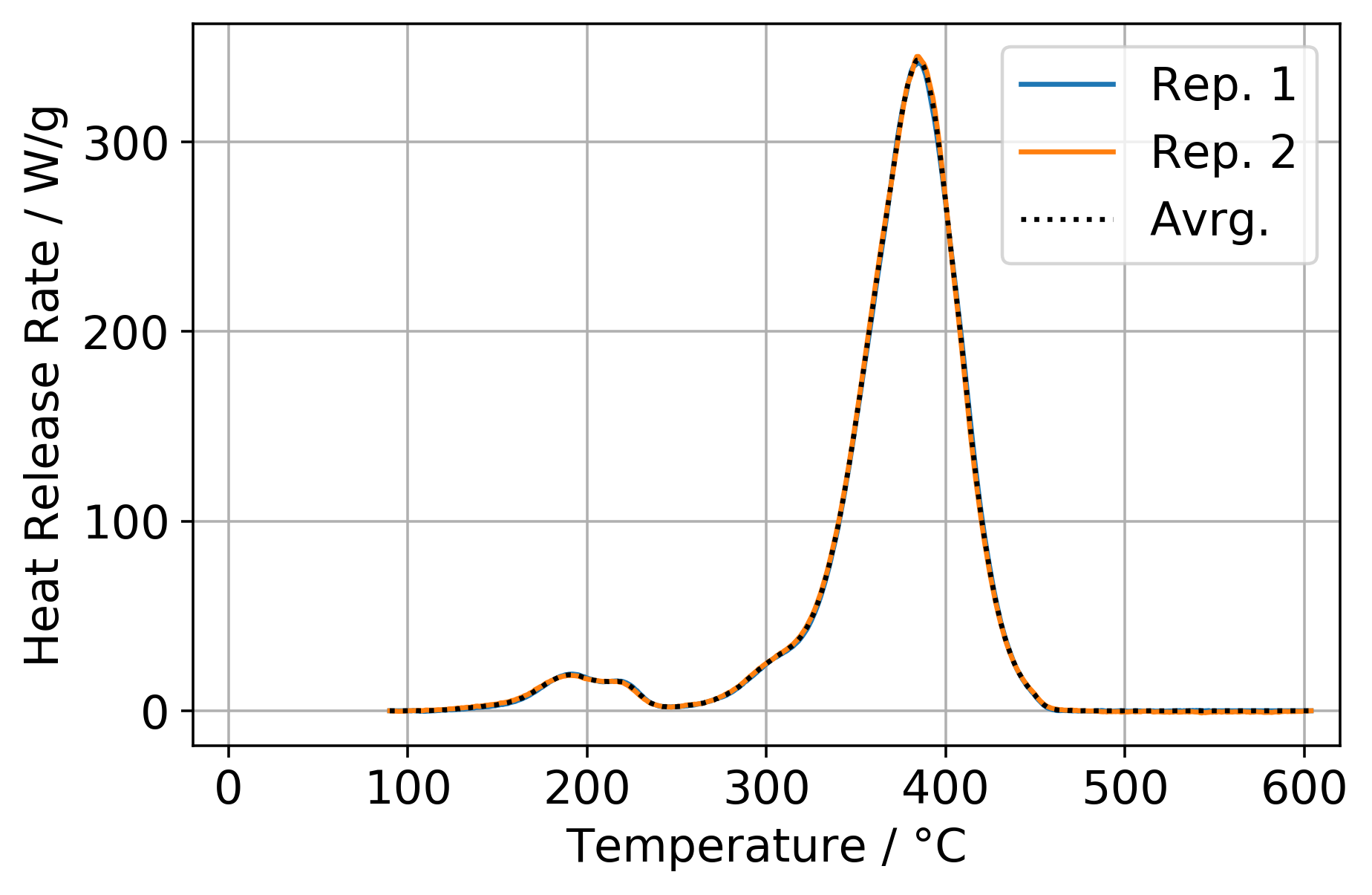} }}%
    \qquad
    \subfloat[\centering TGA normalised residual mass, for a  heating rate of 50~K/min in argon atmosphere (Sandia).\label{fig:TGA_exp_Sandia}]{{\includegraphics[width=0.45\columnwidth]{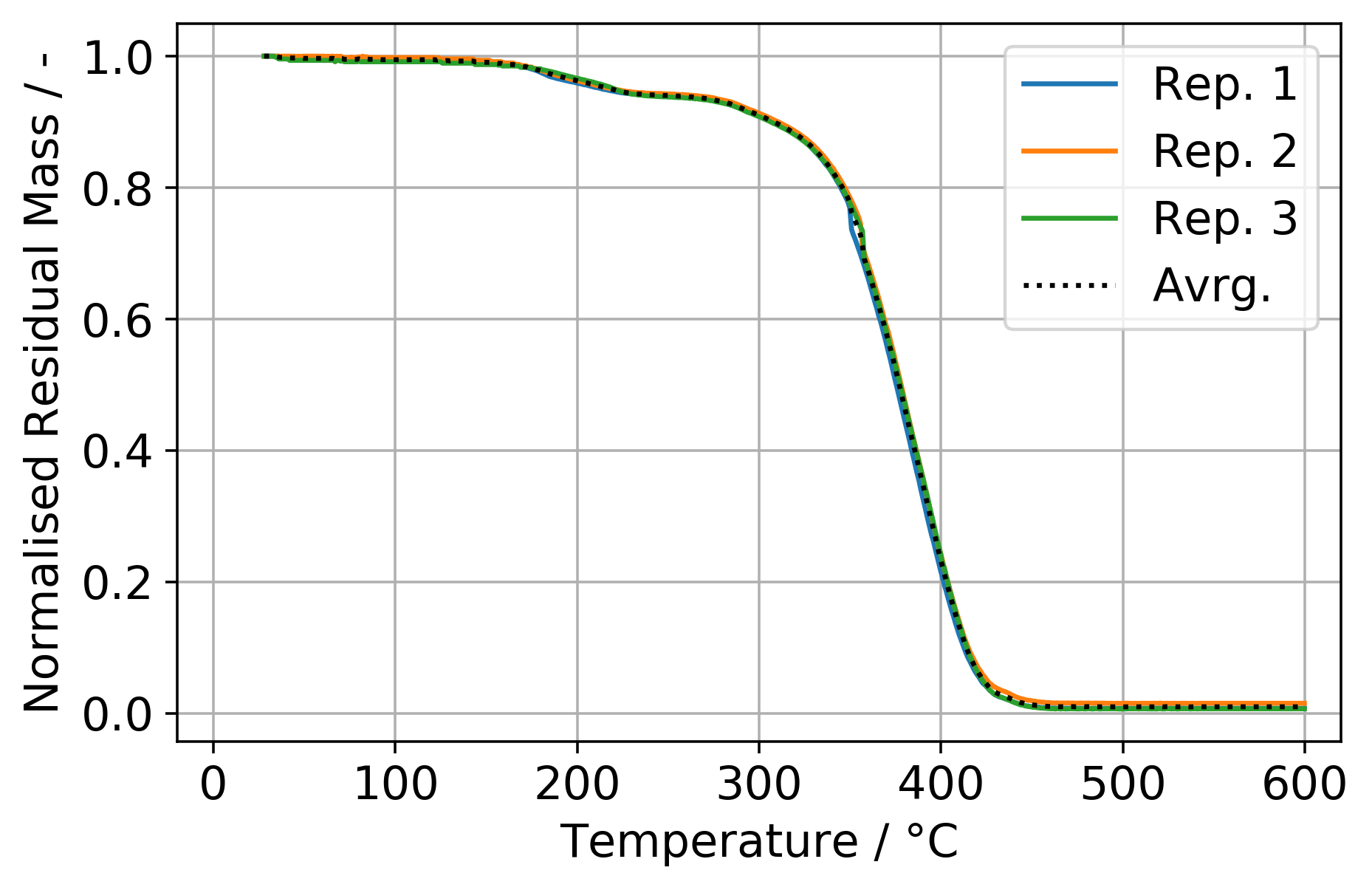} }}%

    \caption{Data from different repetitions (Rep.) of micro-scale experiments, MaCFP~\cite{macfp_matl_git}. Average (Avrg.) used as IMP target.}%
    \label{fig:MicroScaleEXP}%
\end{figure}

All the experimental data series are averaged within each group, to be used as target during the IMP, see figure~\ref{fig:MicroScaleEXP}. The residual mass data is normalised to get the normalised residual sample mass over temperature, and the final residue amount is determined (figure~\ref{fig:TGA_exp_NIST}). The amount of residue produced is very small, about one percent of the starting sample mass. 
This value is used directly as residue production for each decomposition reaction in FDS (\texttt{NU\_MATL}). 
From the processed heat release rate data the average effective heat of combustion is determined.

Two different IMP setups are designed, summarised in table~\ref{tab:MicroScaleIMP}. They assess each parameter sets performance in two different simulation setups simultaneously. With this two experiment setups with different heating rates can be taken into account. The first micro-scale IMP setup uses the normalised residual mass at a heating rate of 10~K/min (TGA, NIST) and the heat release rate at 60~K/min (MCC, NIST). The second micro-scale IMP setup uses the normalised residual mass at a heating rate of 50~K/min (TGA, Sandia) and the heat release rate at 60~K/min (MCC, NIST).
Thus, the first IMP setup covers a relatively large difference in the heating rate. The second setup aims to provide a heating rate in the TGA that closer matches the conditions during the MCC experiment, within the available data from MaCFP.

\begin{table}[h]
    \caption{Overview over the different micro-scale IMP setups, with two simultaneous simulations. 
    \label{tab:MicroScaleIMP}}
    \centering
    \begin{tabular}{l l l}
        \toprule
        IMP Setup  & IMP Target Details  &  \\
                   & Simulation 1 & Simulation 2 \\
        \midrule
        MCCTGA\_01 & normalised residual mass at 10~K/min (TGA) & heat release rate at 60~K/min (MCC) \\
        MCCTGA\_02 & normalised residual mass at 50~K/min (TGA) & heat release rate at 60~K/min (MCC) \\ 
        \bottomrule
    \end{tabular}
\end{table}

\begin{table*}[h]
    \caption{Overview over the parameters of the pyrolysis scheme that are adjusted in the micro-scale IMP step. 
    \label{tab:MicroScaleIMP_Paras}}
    \centering
    \begin{tabular}{l l}
        \toprule
        Component    & Parameter    \\
        \midrule
        PMMA 1 to 8  & Reference temperature \\
                     & Pyrolysis range \\ 
                     & Heat of reaction \\ 
                     & Mass fraction \\
        Fuel mixture & Carbon dioxide mass fraction \\
                & Ethylene mass fraction \\
                & Methane mass fraction \\
        \bottomrule
    \end{tabular}
\end{table*}

\begin{figure}[h]%
    \centering

    \subfloat[\centering Location of primary decomposition reactions.\label{fig:DecompositionScheme01}]{{\includegraphics[width=0.45\columnwidth]{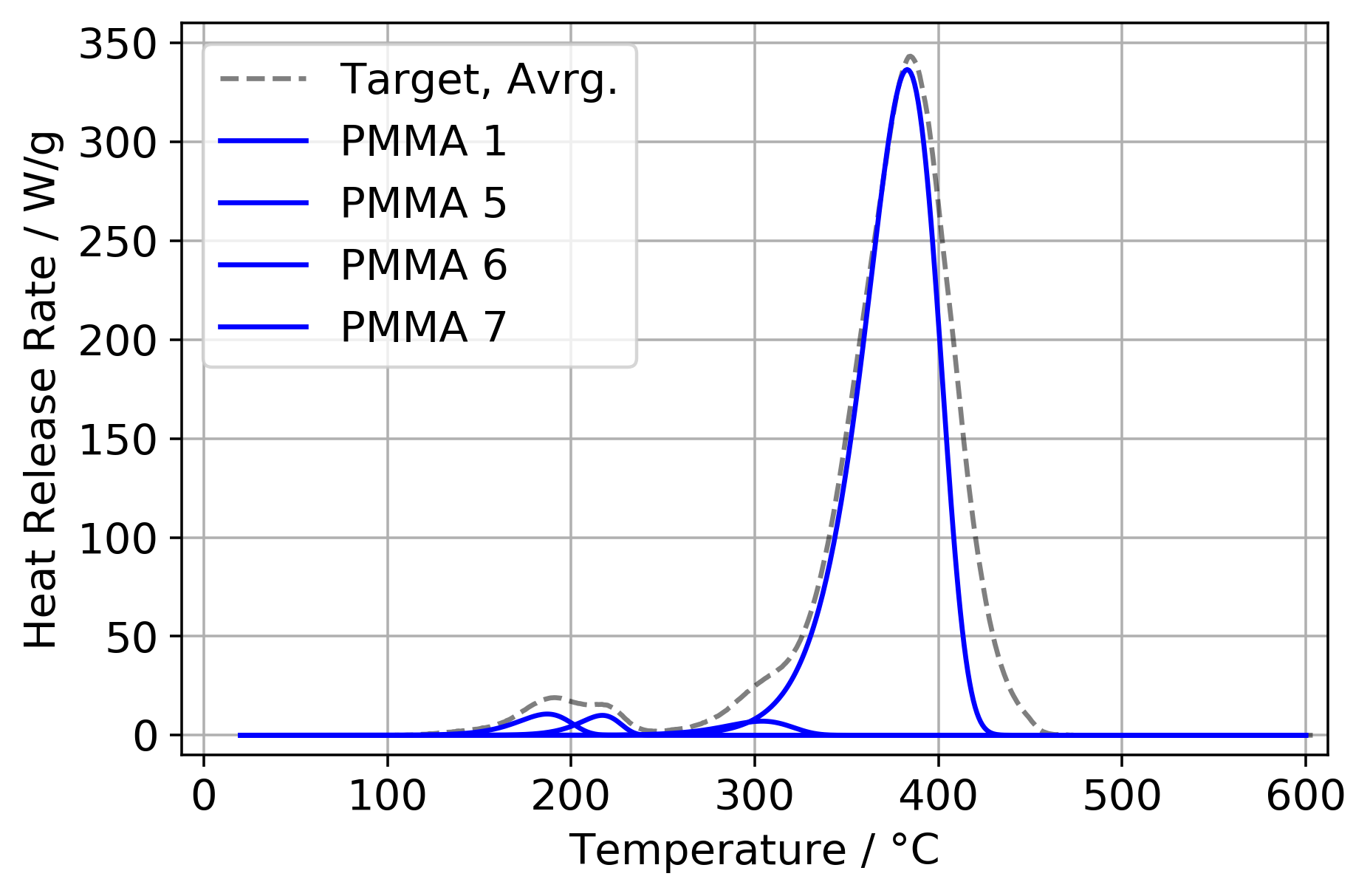} }}%
    \qquad
    \subfloat[\centering Location of secondary decomposition reactions.\label{fig:DecompositionScheme02}]{{\includegraphics[width=0.45\columnwidth]{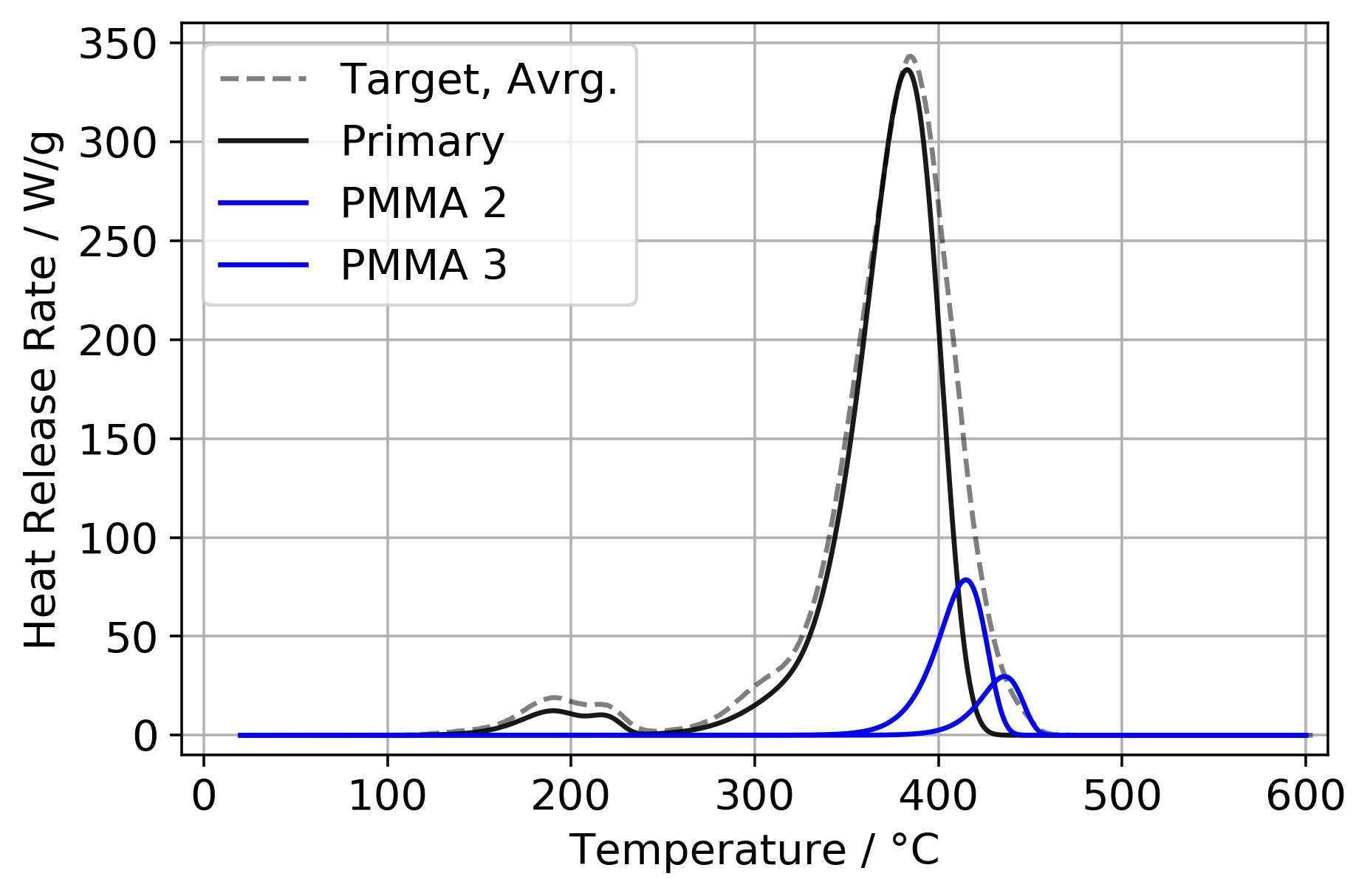} }}%
    \qquad
    \subfloat[\centering Location of tertiary decomposition reactions.\label{fig:DecompositionScheme03}]{{\includegraphics[width=0.45\columnwidth]{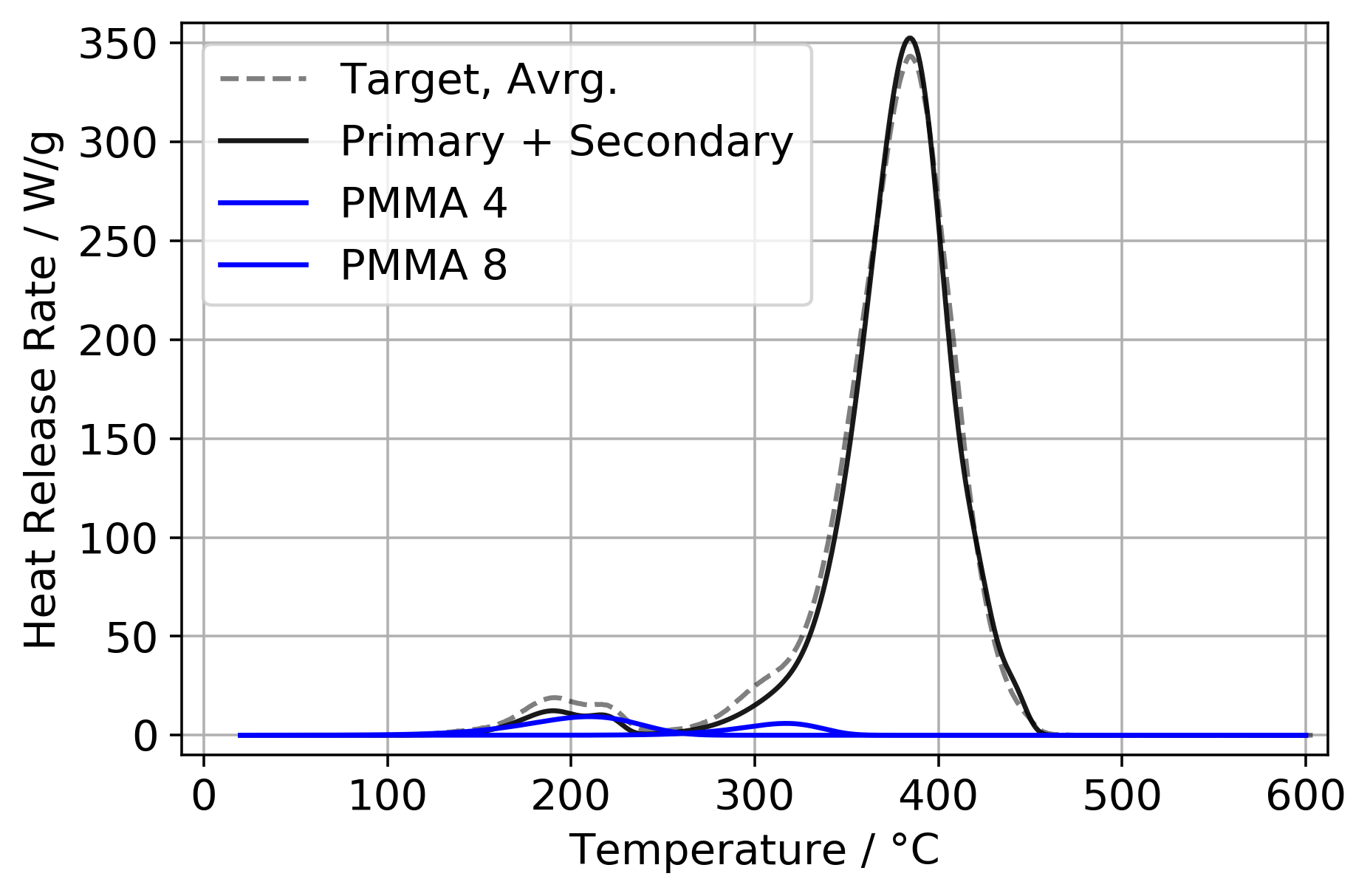} }}%
    \qquad
    \subfloat[\centering Initial guess with all decomposition reactions combined.\label{fig:DecompositionScheme04}]{{\includegraphics[width=0.45\columnwidth]{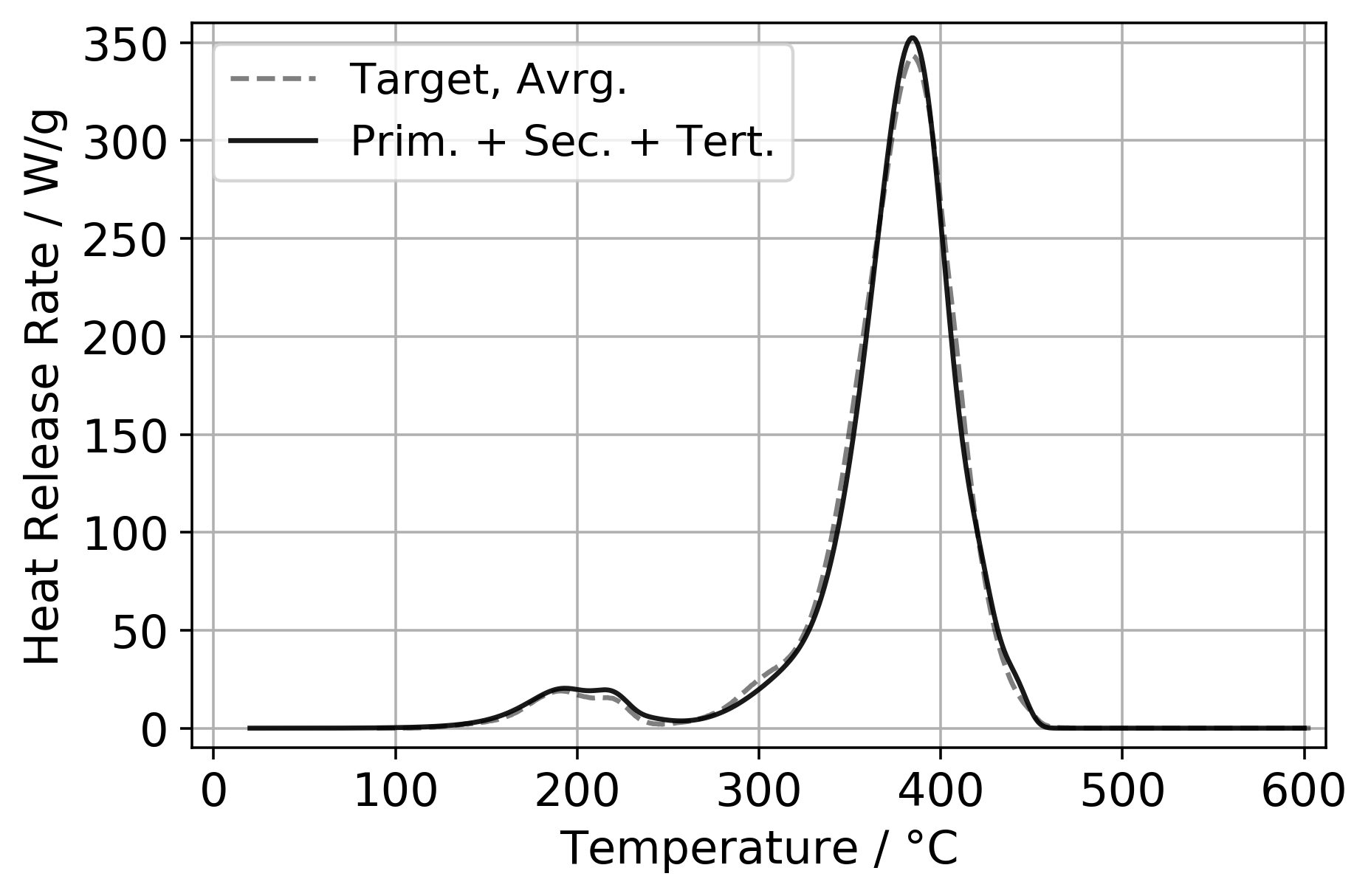} }}%

    \caption{Design of the PMMA pyrolysis scheme with eight parallel decomposition reactions. The initial guess is fine-tuned by the first IMP step. Target is MCC heat release rate data for a heating rate of 60~K/min (NIST)~\cite{macfp_matl_git}.}%
    \label{fig:DecompositionScheme}%
\end{figure}

Multiple parallel pyrolysis reactions are used to roughly approximate the experiment data. The goal is to use as few reactions as possible, yet approximate the experiment response to a high precision. The reactions are manually positioned, by defining the reference values by trail-and-error. The parameters of these reactions form the first guess. The positioning process is conducted as follows.
Individual decomposition reactions are identified from a plot of the MCC heat release rate experiment data, see figure~\ref{fig:MCC_exp_NIST}. Four peaks can be visually identified at about 187~°C, 217~°C, 304~°C and 383~°C, see figure~\ref{fig:DecompositionScheme01}. They form the primary set of decomposition reactions. A secondary set, figure~\ref{fig:DecompositionScheme02}, is used to capture the decay of the largest reaction peak. They are located at 415~°C and 436~°C. This is necessary, because to skew the peak, reaction orders larger than unity would be needed. The skewing behaviour needs therefore be reproduced by superposition of multiple reactions. Finally, a set of tertiary reactions is defined at about 210~°C and 317~°C, see figure~\ref{fig:DecompositionScheme03}. They are used to provide the optimiser with some flexibility to capture the reaction rate development. All these reaction sets combined form the initial guess of the decomposition scheme for the inverse modelling, see figure~\ref{fig:DecompositionScheme04}. Thus, these pyrolysis reactions are not necessarily directly chemical/physical in nature, but needed for the model.
In total, eight decomposition reactions are defined, labelled "PMMA 1" to "PMMA 8".
The fine-tuning of the parameters of these reactions, using an IMP, concludes the first step of the procedure.

The argument could be made, that eight pyrolysis reactions are too many. One might be able to exclude the tertiary reactions. However, the algorithm is not able to add more reactions if needed, but can only adjust predefined parameters. Since it is unclear in the beginning how many reactions are necessary more decomposition reactions are added to provide the algorithm with some flexibility.


The above decomposition scheme is implemented as follows. An FDS functionality (\texttt{TGA\_ANALYSIS}) is used to evaluate an extremely simplified micro-scale simulation setup. It linearly increases the temperature of a mass sample and tracks the pyrolysis.
The pyrolysis is simulated by employing Arrhenius equations for multiple decomposition reactions. The Arrhenius model is used as provided with FDS.

FDS offers some convenience functionalities where the Arrhenius parameters (pre-exponential factor~A, activation energy~E) can be deduced from the peak temperature of the decomposition reaction rate and the width of the peak. These reference values are entered by the user as \texttt{REFERENCE\_TEMPERATURE} and \texttt{PYROLYSIS\_RANGE}. In this procedure, the reaction order is assumed to be unity. The heating rate of the experiment that serves as starting point needs to be provided to FDS as well.

In FDS, the decomposition reactions are associated with the material definitions (\texttt{MATL}). Thus, the material definitions provide two functions: 1) define the thermophysical properties of a material and 2) define its pyrolysis reaction. While the pyrolysis reaction parameters are here determined during the first IMP step, the thermophysical parameters are determined in the second IMP step, see section~\ref{sec:benchscale}. The different materials (\texttt{MATL}) are realised into a sample, by combining them into a boundary condition (\texttt{SURF}). In the case here, the PMMA sample needs to combine eight different material definitions to account for the decomposition reactions. All the pyrolysis reactions are directly converting the PMMA material into a gas mixture and an inert residue, see section~\ref{subsec:pyrolysis_approach}. The individual contributions of the reactions are controlled by mass fractions of the materials, set in the PMMA sample definition. This means that each reaction has access to a predefined amount of sample mass. Furthermore, each material definition is assigned a heat of reaction (HOR), which is also determined during this first IMP step. In total, four parameters describe a single reaction and are determined with the IMP: the two reference values, the sample mass fraction per reaction and the respective HOR, see table~\ref{tab:MicroScaleIMP_Paras}. The overall fitness value is a combination of the performance in the MCC and in the TGA simulations.

The thermophysical parameters are identical for all eight material definitions, to realise a homogeneous PMMA sample, and are determined in the second IMP step, see section~\ref{sec:benchscale}.

\subsection{Results}
\label{subsec:results_microscale}

\begin{figure}[h]%
    \centering
    \includegraphics[width=0.45\columnwidth]{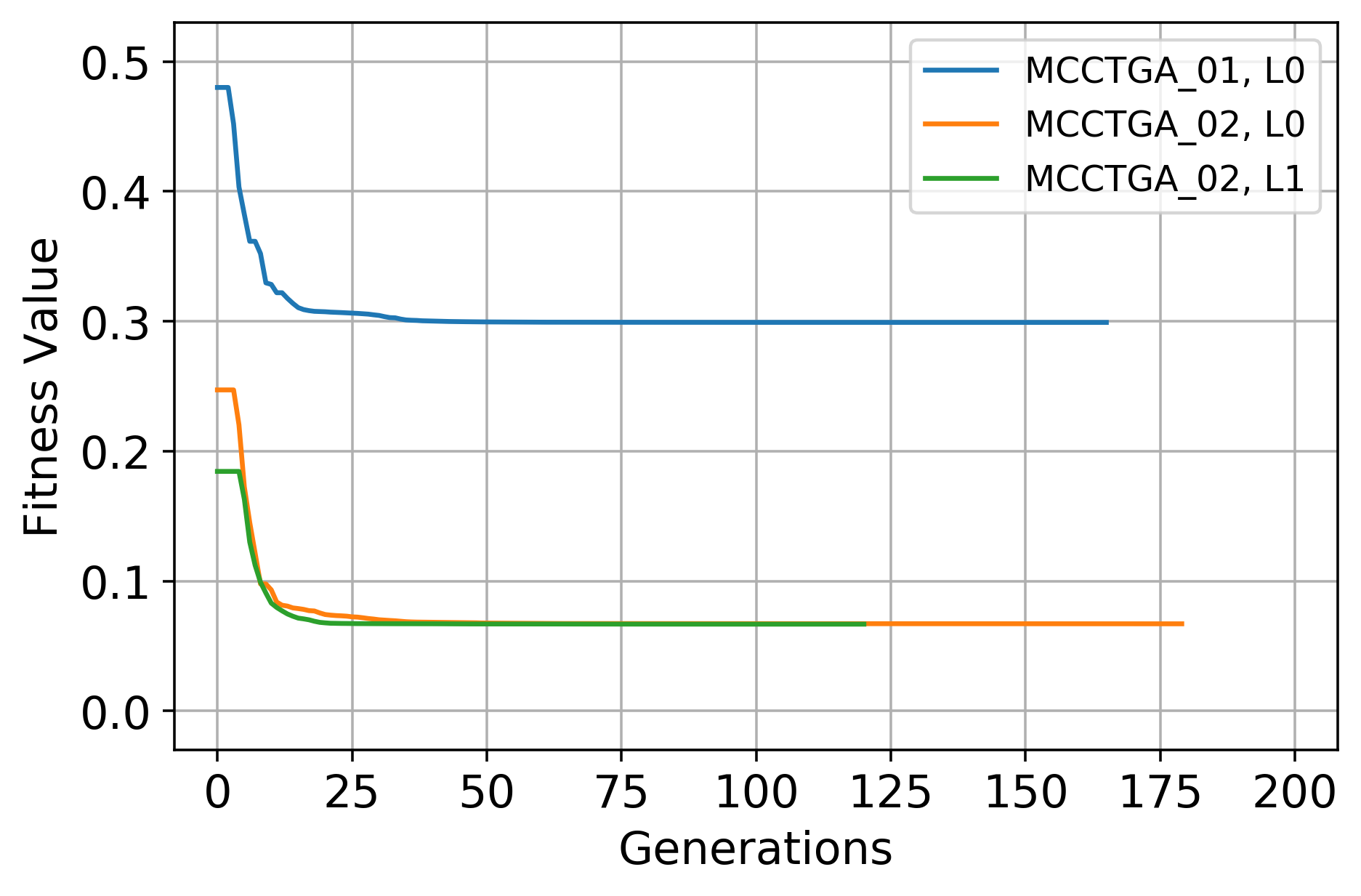}
    \caption{Fitness development of the micro-scale IMP setups. MCCTGA\_02, L1, cut short due to good performance.}
    \label{fig:Micro_Fitness_Combined}
\end{figure}

\begin{figure}[h]
    \centering
    \subfloat[\centering Targets: MCC at 60~K/min, TGA at 10~K/min.
    \label{fig:TGA_10K_best_IMP}]{{\includegraphics[width=0.45\columnwidth]{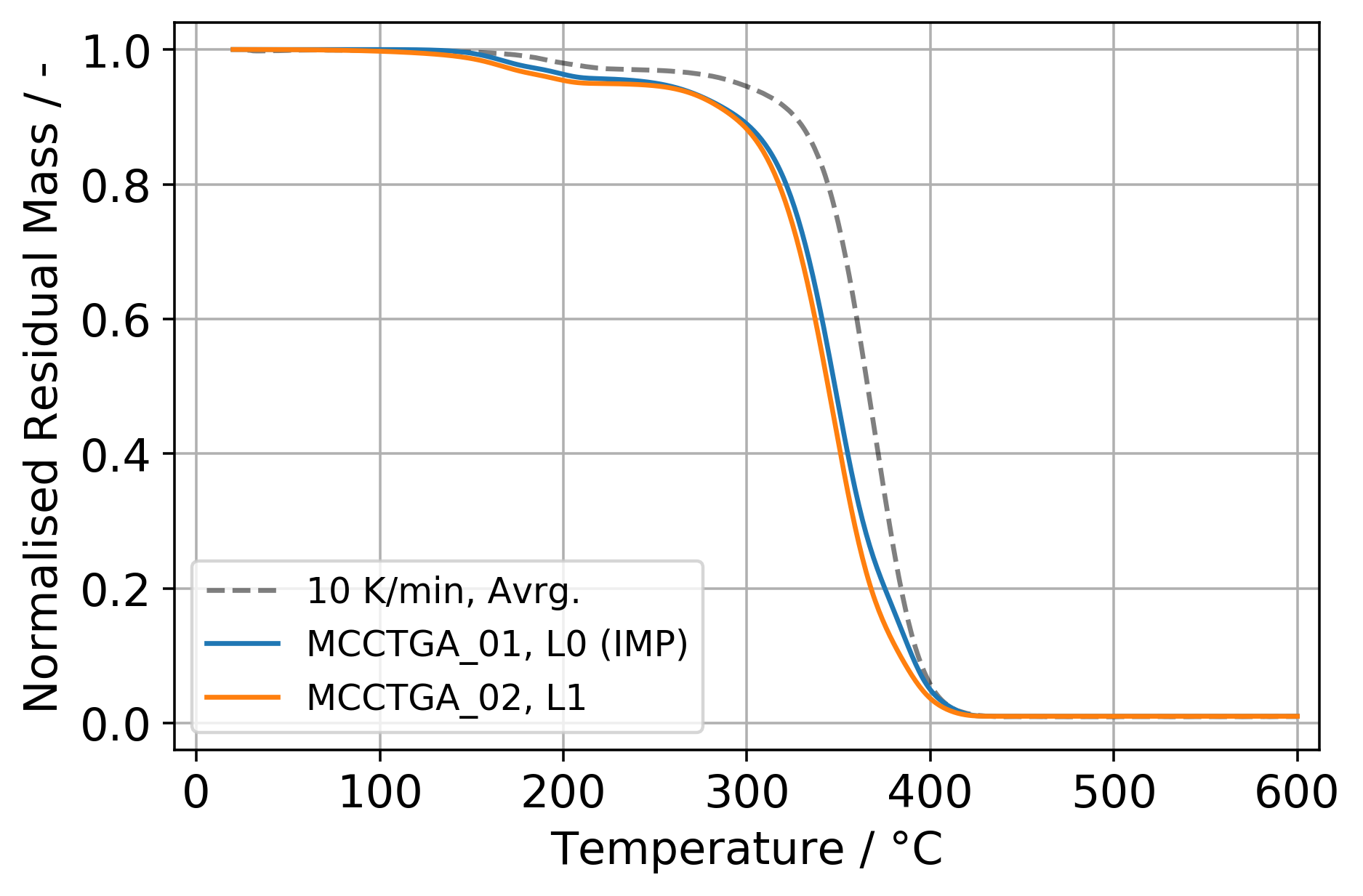} }}%
    \qquad
    \subfloat[\centering Targets: MCC at 60~K/min, TGA at 50~K/min.
    \label{fig:TGA_50K_best_IMP}]{{\includegraphics[width=0.45\columnwidth]{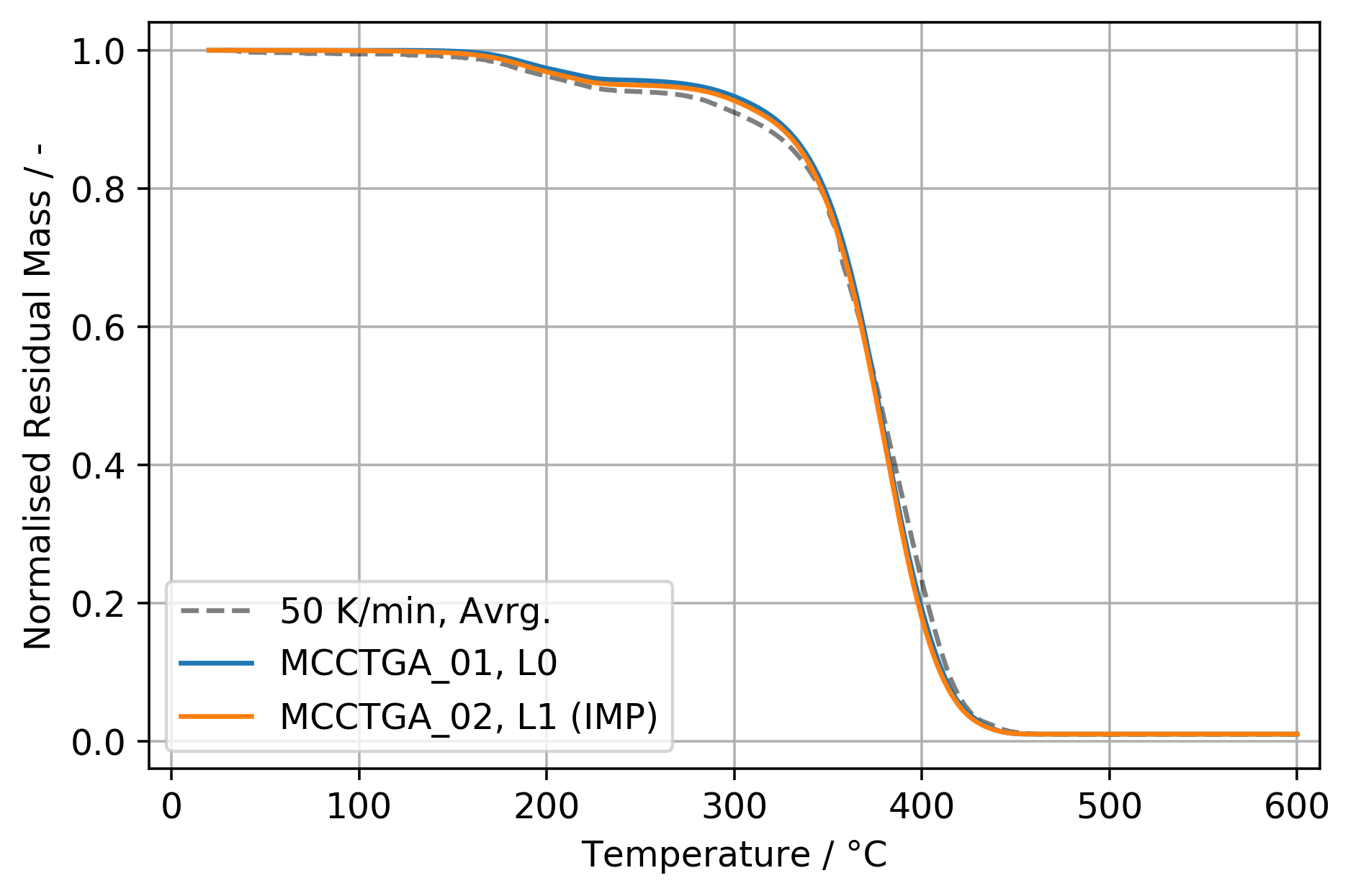} }}%
    \caption{Normalised residual mass from TGA, for heating rates of 10~K/min and 50~K/min. Comparison of response for the best parameter sets, (IMP) indicates the respective target.
    \label{fig:IMP_TGA_Combined}}
\end{figure}

Two IMP runs with different simulation setups have been conducted (table~\ref{tab:MicroScaleIMP}). Both converge to their best fitness values within about 40~generations, see figure~\ref{fig:Micro_Fitness_Combined}.
The simulation responses of the best parameter sets for both IMPs are compared against their TGA target data for both heating rates, see figure~\ref{fig:IMP_TGA_Combined}. The "(IMP)" marks which IMP used the respective target.
The mass loss for a heating rate of 10~K/min happens at lower temperatures in the simulation for both IMPs, with MCCTGA\_01 being slightly closer to its target (figure~\ref{fig:TGA_10K_best_IMP}). For a heating rate of 50~K/min, both yield a very similar response (figure~\ref{fig:TGA_50K_best_IMP}).   

\begin{figure}[h]%
    \centering
    \includegraphics[width=0.45\columnwidth]{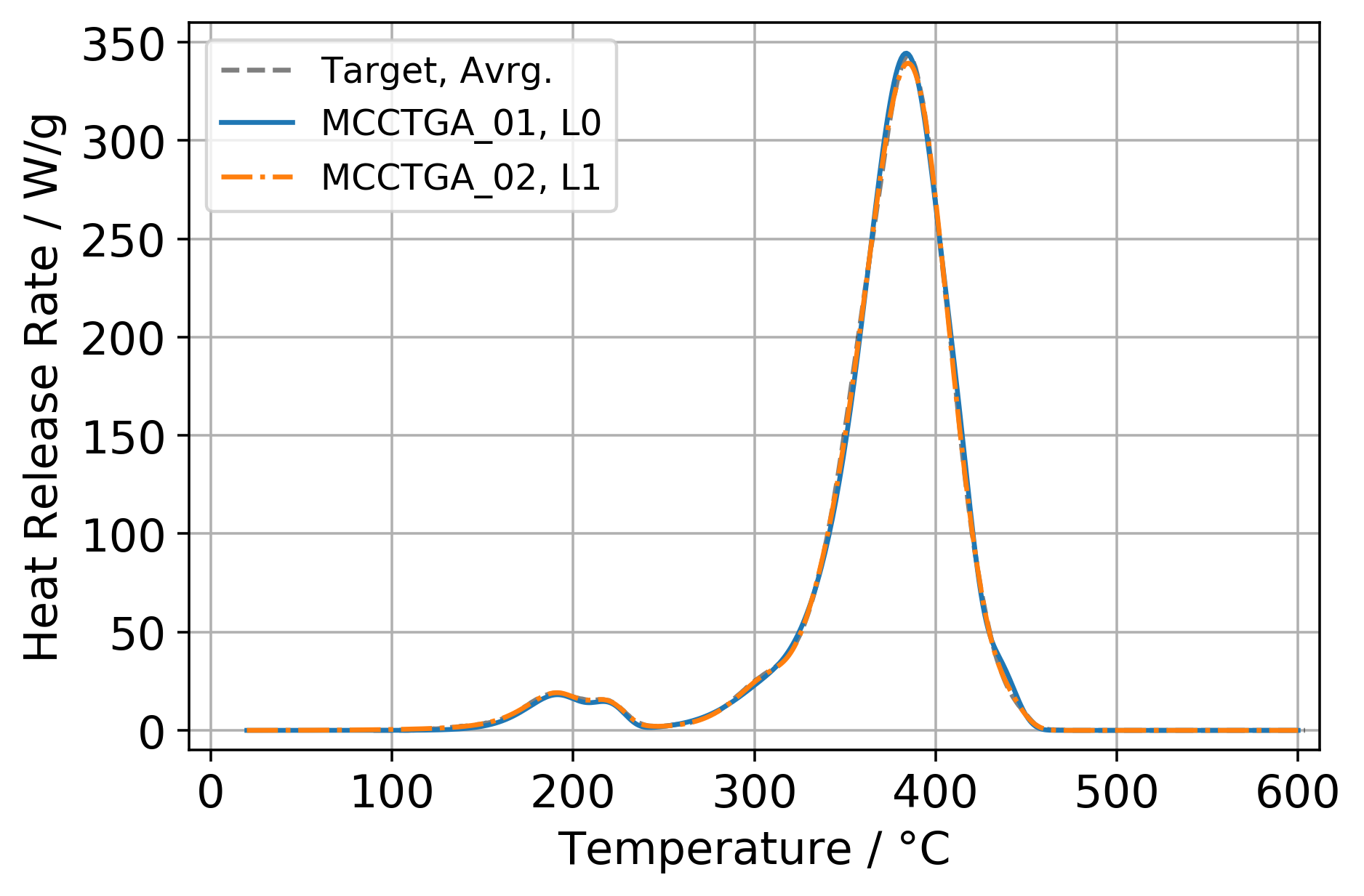}
    \caption{Heat release rate from MCC for a heating rate of 60~K/min. Comparison between experiment and best parameter set from the IMP.}
    \label{fig:IMP_MCC_Combined}
\end{figure}

\begin{figure}[h]
    \centering
    \subfloat[\centering Determined reactions.
    \label{fig:MCCTGA02_L1_IndivReac_locations}]{{\includegraphics[width=0.45\columnwidth]{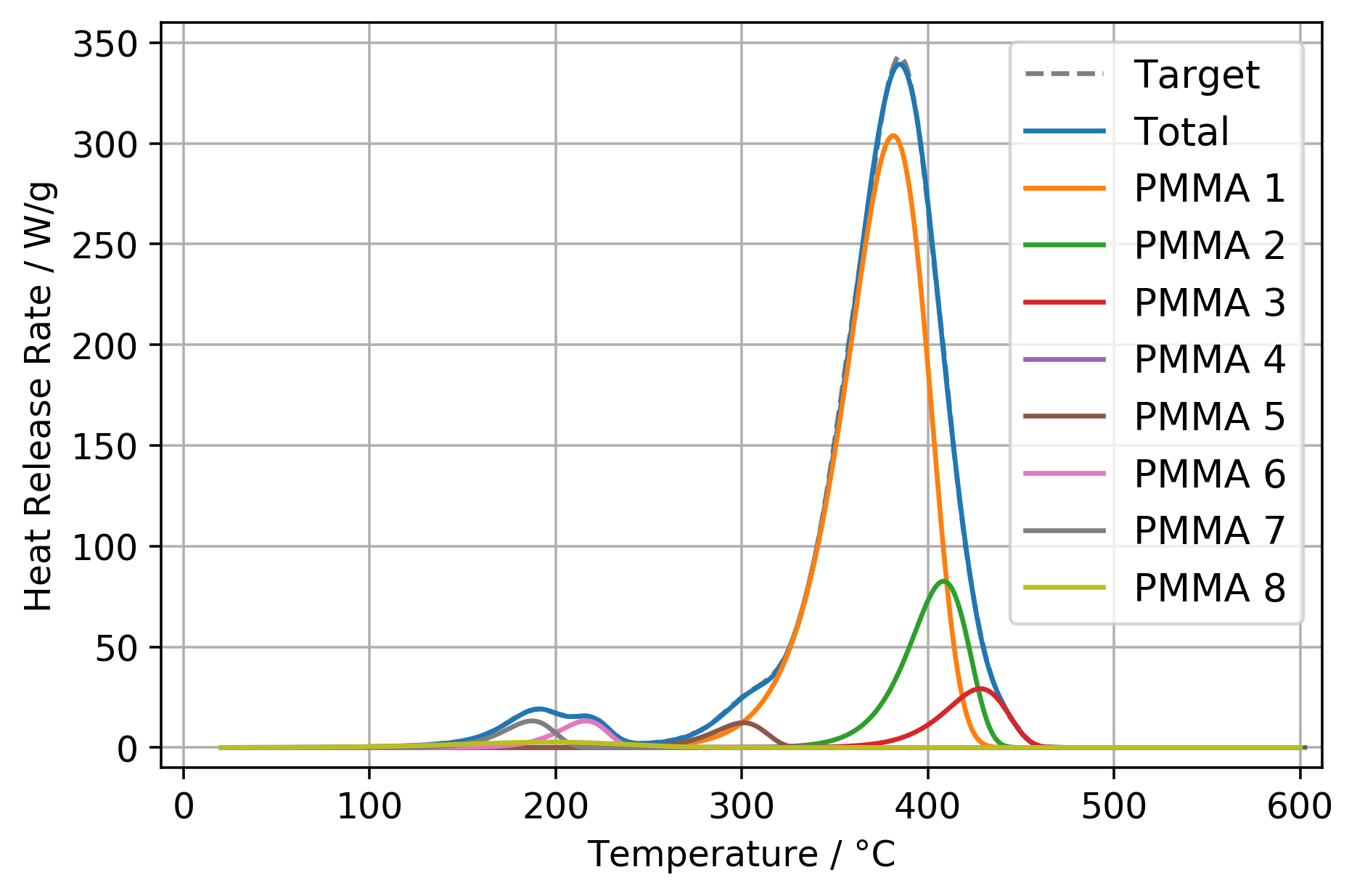} }}%
    \qquad
    \subfloat[\centering Reaction contributions.
    \label{fig:MCCTGA02_L1_IndivReac_contribution}]{{\includegraphics[width=0.45\columnwidth]{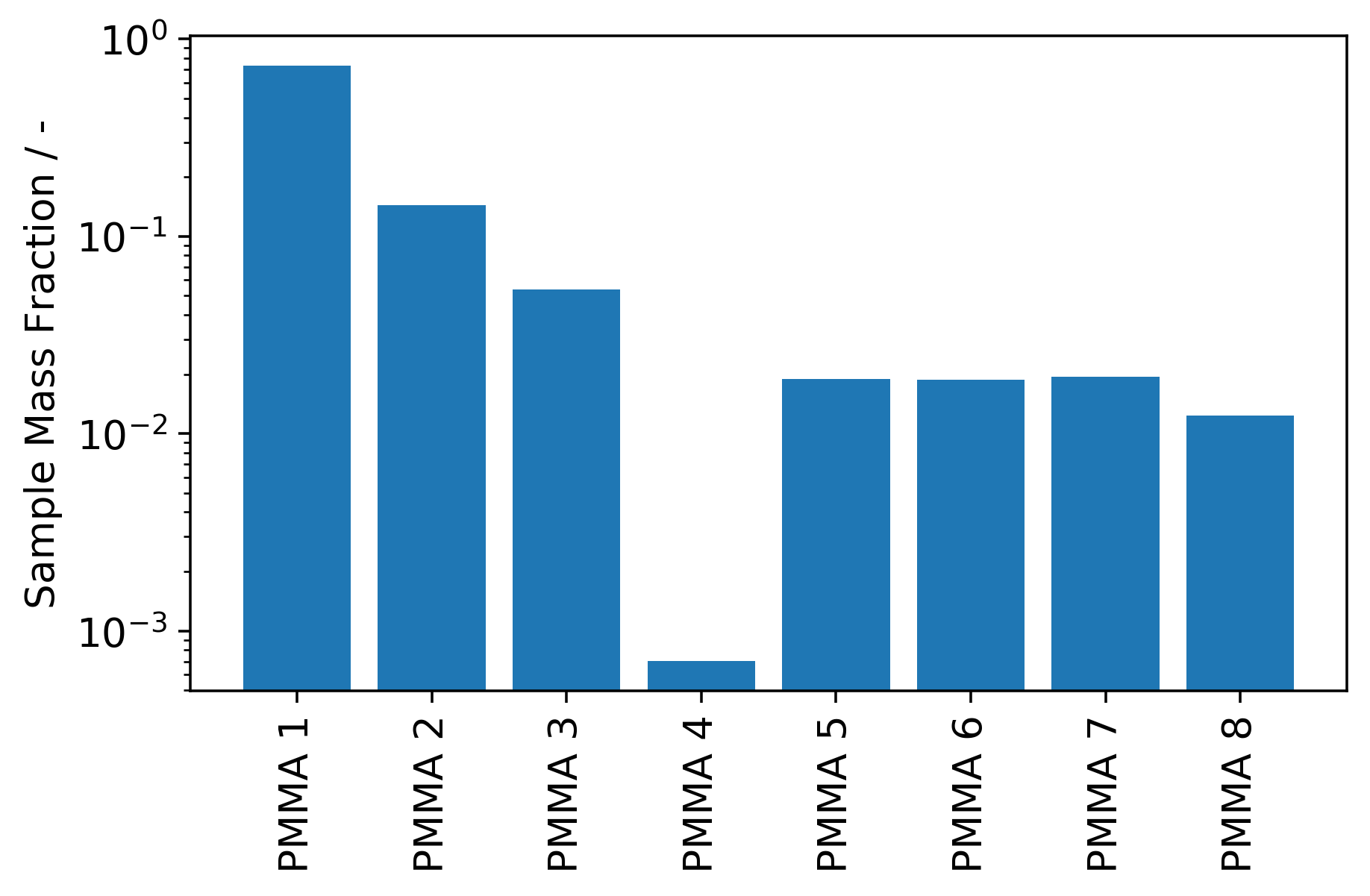} }}%
    \caption{Individual reaction steps (MCCTGA\_02,~L1); residue production excluded.
    \label{fig:MCCTGA02_L1_IndivReac}}
\end{figure}

Figure~\ref{fig:MCCTGA02_L1_IndivReac_locations} shows the eight predefined decomposition reactions, after being fine tuned through the IMP. The tertiary reactions "PMMA 4" and "PMMA 8" provide significantly lower contributions compared to the other reactions (figure~\ref{fig:MCCTGA02_L1_IndivReac_contribution}).

Due to its better fitness value, focus is shifted to MCCTGA\_02 and no further sampling limit adjustment is conducted for MCCTGA\_01. MCCTGA\_02 is used in the following cone calorimeter simulations.

\subsection{Discussion}
\label{subsec:discussion_microscale}

Both IMP setups are able to reproduce the MCC data well (figure~\ref{fig:IMP_MCC_Combined}). The TGA data for a heating rate of 50~K/min is better reproduced than for 10~K/min (figure~\ref{fig:IMP_TGA_Combined}). The deviation between simulation and experiment for 50~K/min is attributed to the non-linear heating rate in the experiment, see figure~\ref{fig:Apdx_MicroScale_HeatingRates_Sandia50K}. Larger differences between experiment and simulation are observable for a heating rate of 10~K/min. Here, MCCTGA\_01 gets slightly closer to its target. Otherwise, the results are similar, yet happen at lower sample temperatures compared to the experiment. With the lower heating rate in the TGA (MCCTGA\_01) the algorithm is not able to find a parameter set suitable for both conditions, i.e. 10~K/min in the TGA and 60~K/min in the MCC. It comes as some surprise that the IMP favours the MCC and does not position the fit somewhere in between both targets. Some bias may have been built into the setup, by manually positioning the first guess reactions based on the MCC data, or by the chosen cost function.

Possibly, some aspects of the apparatus are not captured well enough in the highly simplified micro-scale model. 
We know, from private communication with Karen De Lannoye, that the design of the TGA apparatus has an observable impact on the results~\cite{KarenCorinna_PMMA}. 
Ding et al. reported an observable shift of reaction rate peaks to lower temperatures in MCC experiments when compared to TGA data at the same heating rate~\cite{Stoliarov2019PyrolysisModelDevelopment}. They adjusted the heat release data from the MCC by shifting it to higher temperatures, about 4~K to~20~K. At the point of writing, in the MaCFP repository no data series for both experiments at the same heating rate are reported, which makes a similar assessment here difficult. The shift seems also to depend on the material. More experiment data is necessary for this comparison.

The divergence could also be related to the released gas mixture, which in this contribution assumes an average heat of combustion over the course of the experiment. It is argued~\cite{FIOLA2021103083PMMA_Pyrolysis_Model}, that the first peak at about 187~°C (figure~\ref{fig:MCC_exp_NIST}) could be attributed to residual solvent within the polymer. Furthermore, radically polymerised PMMA is somewhat unstable and starts decomposition at about 220~°C, due to unsaturated end groups~\cite{Zeng_chemreac_burning_PMMA}. Even though the primary decomposition products of PMMA are MMA and carbon dioxide, the MMA is not directly involved in the gas phase combustion~\cite{Zeng_chemreac_burning_PMMA}. Some major compounds involved in the combustion of PMMA are methane, methanol, formaldehyde and acetylene, with ethylene being involved during the acetylene combustion. As an example, for pure compounds the heats of combustion are tabulated in the literature~\cite{Quintire:PrinciplesOfFireBehaviour} with 50.0~MJ/kg for methane, 50.4~MJ/kg for ethylene, 19.8~MJ/kg for methanol and 24.2~MJ/kg for PMMA. 
If these compounds are produced at different, temperature-dependent rates, they could lead to a variable effective heat of combustion over the course of the experiment.
In the work presented here, a mixture of methane, ethylene and carbon dioxide is released during the PMMA pyrolysis. The individual fractions are adjusted during the IMP. However, that leads to a constant average heat of combustion of the PMMA pyrolysis products throughout the simulation. A model allowing for a variable HOC could be an additional degree of freedom in the IMP to match the MCC and TGA data.
Fixing this mixture over all reactions could be too rigid. In future work individual mixtures will be investigated.
The difference for the TGA test at 10~K/min, see figure~\ref{fig:IMP_TGA_Combined}, could be a manifestation of this rigidness, as low and high heating rates need to be reproduced simultaneously. A possible solution would be to use multiple gas mixtures. In an effort to reduce complexity downstream, i.e. definition of multiple chemical reactions and solving more transport equations, the mixture could be generated on-the-fly by releasing a single species per reaction. It might also be sufficient to mix only methane and carbon dioxide. Each decomposition reaction in figure~\ref{fig:MCCTGA02_L1_IndivReac} could be doubled. One would release methane and the other carbon dioxide, the mixture would then be controlled from the mass fraction in the surface definition. This strategy is to be investigated in future work.

Here, no IMP target is provided to explicitly match the heats of reaction for the individual decomposition reactions. This could be accomplished by using experiment data from differential scanning calorimetry (DSC). Alonso et al.~\cite{Alonso2019_LLDPEKineticPropertiesEstimation} used TGA and DSC data as targets in their IMP setup, changing the contribution of each to the overall fitness assessment. With this, the target of higher importance is reproduced better to the detriment of the other. The employed decomposition scheme uses two consecutive decomposition reactions forming an intermediate material and a residue. A parallel decomposition scheme, as is proposed here, could be able to capture more gradual changes in the heats of reaction across the temperature range of the experiment. This could improve the overall performance of the parameter sets generated here and should be investigated in future work.

The proposed approach using gas mixtures allows to model more sophisticated technical materials. Specifically, the behaviour of fire retardant materials could be reproduced. Non-combustible gas could be released early on, which cannot be captured with a single surrogate fuel. This makes it necessary to take MCC and TGA data into account simultaneously. This could even be expanded further, by adjusting which reaction contributes most to the production of the residue. Also considering different residues, for example for intumescent materials.

Arguably, the goal to use as little decomposition reactions as possible is not achieved. Looking at figure~\ref{fig:MCCTGA02_L1_IndivReac}, PMMA~4 ($0.07\%$) could be removed, possibly also PMMA~8 ($1.23\%$) -- even tough it is not too far off of PMMA~6 ($1.88\%$). 
It should be noted that there is an error in the definition of the pyrolysis reaction input for PMMA~4. Its heating rate is set to 80~K/min instead of the desired 60~K/min. Since the contribution of PMMA~4 is negligible, it is regarded inconsequential here. This is confirmed with a corrected IMP, see data repository (MCCTGA\_2b)~\cite{zenodo:ArticleDataset}, which virtually yields the same result.


\section{Bench-Scale Setup  ---  Thermophysical Material Parameters}
\label{sec:benchscale}

\subsection{Methods}
\label{subsec:methods_benchscale}

The second IMP step is used to determine the thermophysical parameters of the PMMA material model. They are determined in the bench-scale from cone calorimeter experiments. 
The cone calorimeter experiment data, provided by Aalto University, is taken from MaCFP~\cite{macfp_matl_git}. According to the reported information, the experiments have been conducted at a radiative heat flux of 65~kW/m², without a retainer frame.
Square samples of cast black PMMA, with an edge length of 10~cm and a thickness of 0.6~cm, were used. These samples were placed onto the sample holder, atop a layer of insulating ceramic wool, with a thickness of about 2.0~cm.

\begin{figure}[h]%
    \centering
    \subfloat[\centering Heat release.\label{fig:ConeExpAalto_err}]{{\includegraphics[width=0.45\columnwidth]{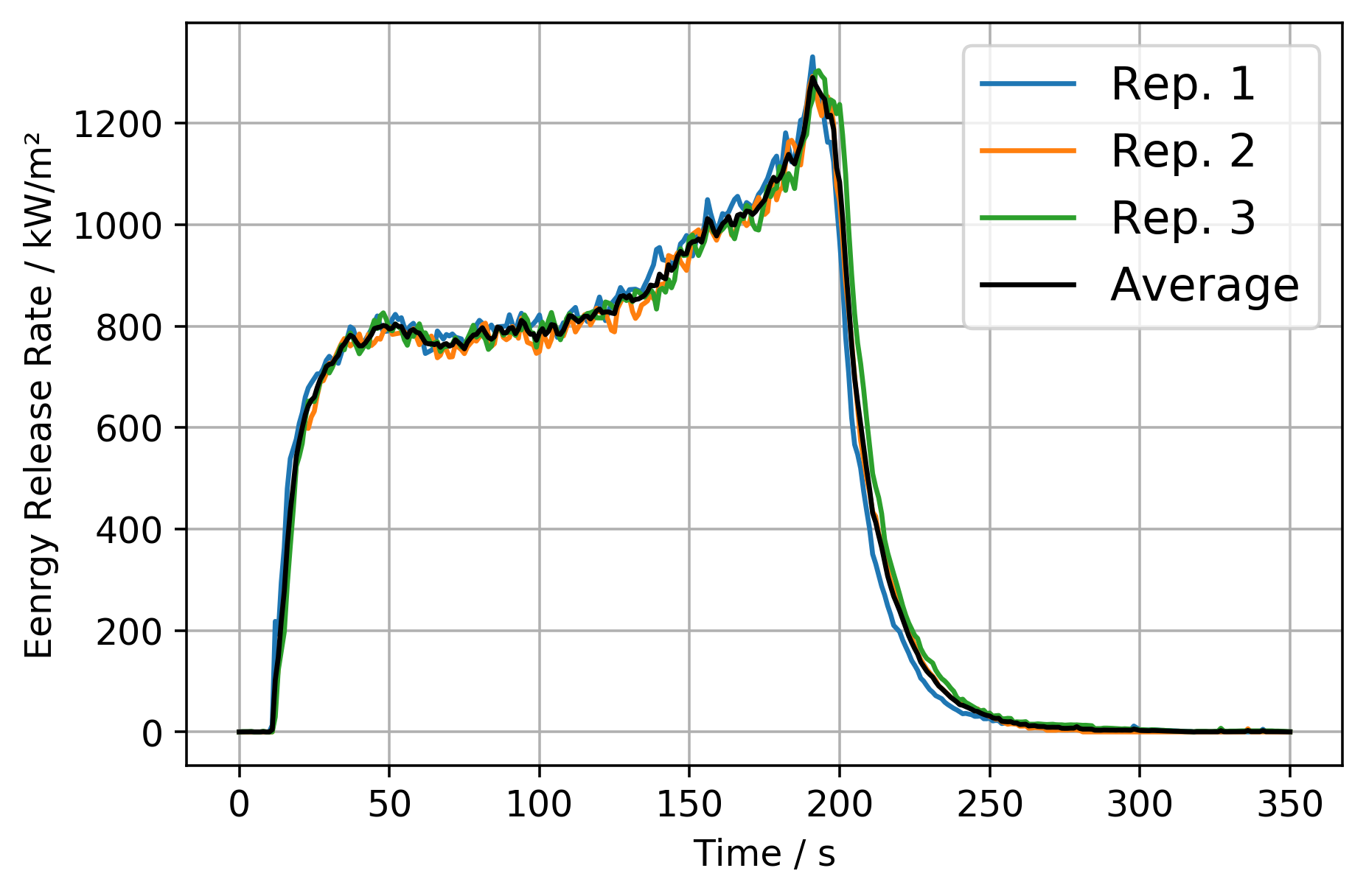} }}%
    \qquad
    \subfloat[\centering Back face temperature.\label{fig:ConeExpAalto_temp}]{{\includegraphics[width=0.45\columnwidth]{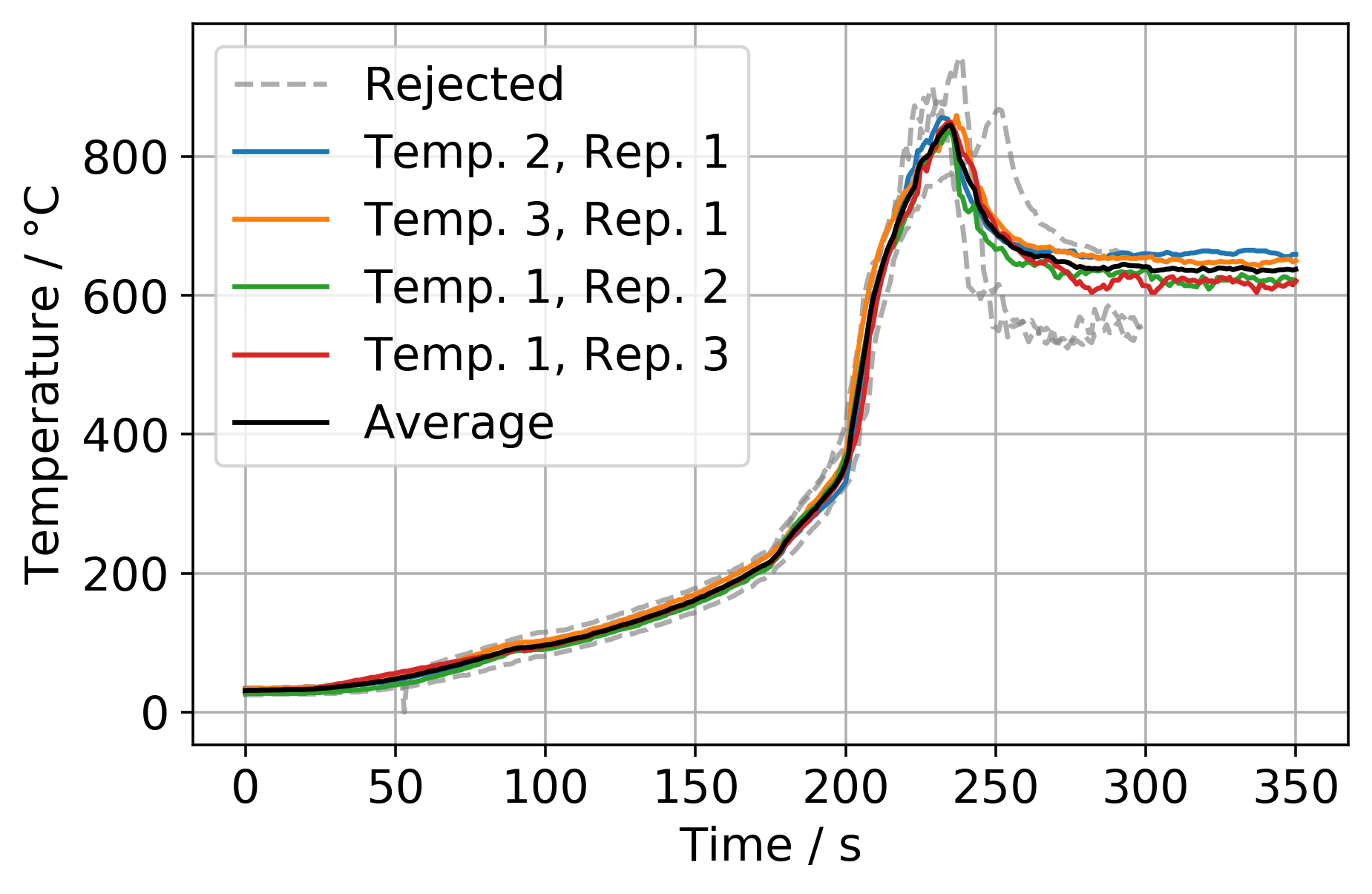} }}%
    
    \caption{Cone calorimeter experiment results, for 65~kW/m² radiative flux condition (MaCFP~\cite{macfp_matl_git}, Aalto). "Temp\_1" values are recorded at centre, "Temp\_2/3" are recorded 1.5~cm to the sides. Averages are used as IMP target.}%
    \label{fig:ConeExpAalto}%
\end{figure}

As target data for this inverse modelling step the heat release rate and the back face temperature of the sample are chosen. 
They are processed similarly as described in section~\ref{subsec:methods_microscale}. The experiments show good repeatability, as demonstrated by the heat release rate in figure~\ref{fig:ConeExpAalto_err}.
Thermocouples have been placed between the insulation layer and PMMA sample to record the back face temperature of the sample. Temperatures have been recorded at the centre of the back face (Temp\_1) and at two opposite locations with a radial distance of 1.5~cm around the face centre (Temp\_2/3).
Regardless of their location, most thermocouples show very similar temperature development, see figure~\ref{fig:ConeExpAalto_temp}. Few diverge and are rejected here, due to assumed sample deformation, see discussion in section~\ref{subsec:discussion_benchscale}.

The simplified cone calorimeter simulation setup, as outlined below, is designed to enable gas phase combustion, while keeping the IMP in a manageable time frame (table~\ref{tab:ParametersIMP}). 
It allows the formation of a flame, which is based on the combustible mass released from the sample.
This in contrast to an approach without the gas phase, where the flame heat flux would need to be prescribed and would effectively define a static imaginary flame with no connection to the gas phase model. 
Simulation setups neglecting the gas phase combustion altogether impact the parameter estimation. In principle, surrogate fuels could be changed between setups, because FDS would maintain the heat release by using the heat of combustion value, set in the material definition (\texttt{MATL}), see section~\ref{subsec:methods_gas_phase_combustion}. However, the heat feedback from the flame is important to take into account. For example, different surrogate fuels can have different radiative fractions (see table~16.1 in~\cite{fdsUserGuide676}). The influence of the flame, formed from the chosen surrogate, will impact the determined material parameter set.
In the presented work, care is taken to use the same surrogate fuel throughout all scales, to not introduce errors when changing between setups.

During the IMP, the cone calorimeter apparatus is simulated in a simplified way. The simulation mode in FDS is set to \texttt{LES} (Large Eddy Simulation) to facilitate the transfer to the real-scale setup, by maintaining model consistency across scales. The computational domain is a single mesh and computed by a single computing core.  
The sample is assumed to be a square with an edge length of 10~cm.
The domain extents 30~cm in the x- and y-directions, with the sample centred. From the top of the sample, the domain extents 60~cm in the positive and two cells in the negative z-direction. 
Throughout this document, the fluid mesh resolutions are referred to by the number of fluid cells dividing a sample top face edge. For example, consider that each edge is divided by~3, thus~$3\times 3$ cells cover the sample top surface. This is referred to as "C3". It results in an edge length for the individual fluid cells of 3.33~cm. Consequently, for C5 the cell edge length is 2.0~cm and $5\times 5$~cells are covering the sample surface. The primary simulation setup for the IMP uses the "C3" resolution. Figure~\ref{fig:app_MeshLayout} in the appendix provides a graphical overview over the domain layouts.

The sample holder assembly is reproduced as a boundary condition (\texttt{SURF}) with two layers. The top layer is associated with the PMMA material, the bottom layer with the insulation material. In an effort to separate the sample from the insulation and holder, the values of the insulation material are adjustable during the IMP. This leads to an artificial material that can compensate some influence of the boundary conditions in the experiments. It is called "backing" throughout this text.

The separation between sample and backing is to be achieved by using the back face temperature as IMP target. Regardless of their recording location in the experiment, the temperatures are very close to each other (figure~\ref{fig:ConeExpAalto_temp}). 
With "C3" being the primary resolution, all temperature locations from the experiment fall into the face area of the single centre fluid cell in the simulation. 
Thus, in the simulation it is not possible to distinguish between these locations.

\begin{figure}[h]
    \centering
    \includegraphics[width=0.45\columnwidth]{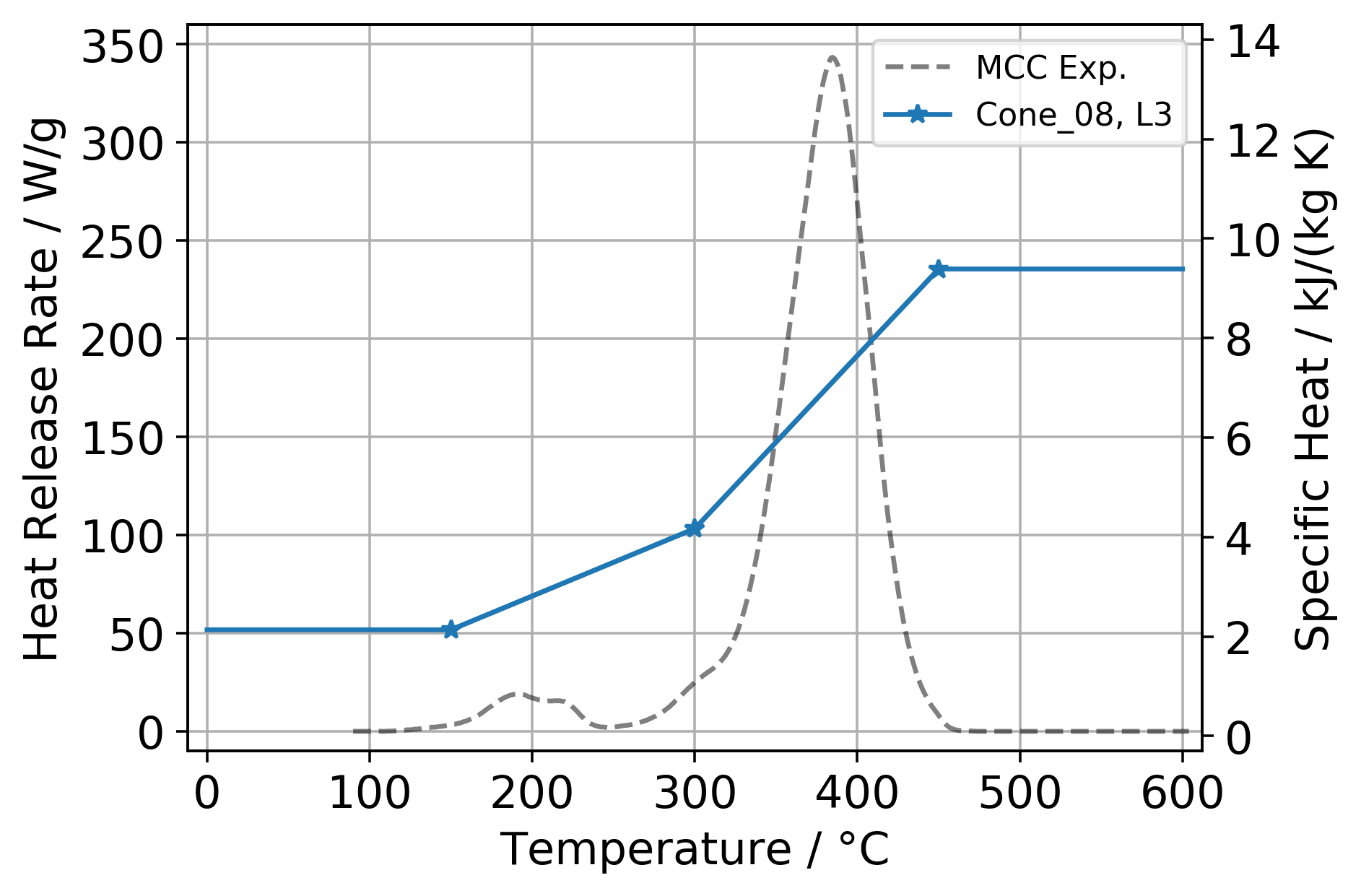}
        \caption{Example of a temperature dependence of the specific heat, realised as a ramp in FDS (Cone\_06 and 07). The chosen temperature values (150~°C, 300~°C and 450~°C) are based on the significant temperature interval in the MCC experiment. The same temperature references are used for the heat conductivity.}
        \label{fig:RAMP_Temperatures}
\end{figure}

To study the impact of different IMP setups, a couple variations of IMP targets and simulation setups are used.
They vary with respect to the temperature-dependent specific heat and thermal conductivity definitions, IMP targets, as well as fluid cell sizes. 
These IMP variations are built on a base case, labelled "Cone\_01"  (table~\ref{tab:SimpleConeIMP}).
For it, the PMMA density ($\rho$) is computed to about 1201.72~kg/m$^3$, based on the reported sample mass and dimensions. The remaining thermophysical and optical parameters are: emissivity, absorption coefficient, refractive index, specific heat capacity ($c$) and thermal conductivity ($k$). They are solely determined during the IMP. A summary of the adjusted parameters for the base case is provided in table~\ref{tab:SimpleConeIMP_Paras}. Their initial sampling ranges are guessed and changed with successive limit adjustments, see section~\ref{subsec:IMP}. Parameters for the pyrolysis and combustion are taken from the micro-scale IMP (MCCTGA\_02), see section~\ref{subsec:results_microscale}.
Thermal conductivity and specific heat for the sample material are represented as temperature dependent values (\texttt{RAMP}). 
The parameter values are adjusted during the IMP, while the temperature points are fixed. In Cone\_01 and Cone\_03 to~05 the three temperature points are arbitrarily chosen to be 150~°C, 480~°C and 800~°C. 
In Cone\_06 and Cone\_07 the temperature values are determined based on the significant temperature interval of the MCC measurement, see stars in figure~\ref{fig:RAMP_Temperatures}.
The chosen values are 150~°C, 300~°C and 450~°C to represent this interval.
They are also used for Cone\_08, but here the conductivity for PMMA and the backing material are in addition using the low temperature data reported by DBI/Lund from the MaCFP materials database~\cite{macfp_matl_git}.

In general, the backing material and the residue are treated as unknown. The density of the backing material is set to 65~kg/m³, taken of the insulation material from the Aalto contribution~\cite{macfp_matl_git}. 
The density of the residue is chosen arbitrarily to be 2500~kg/m³. 
For backing and residue, their individual emissivity, thermal conductivity and specific heat are adjusted during the IMP. The respective sampling ranges are guessed. This is intended to provide some degrees of freedom, in an attempt to separate the sample behaviour from the boundary conditions of the experiment. 
The back face temperature is used as IMP target to achieve this separation. For one IMP setup (Cone\_04) only the energy release is used as target to serve as comparison.
With respect to fluid cell sizes, the default is C3. However, Cone\_05 uses the C5 and Cone\_07 the C2 setup. An overview of all investigated setups is outlined in table~\ref{tab:SimpleConeIMP}.

\begin{table*}[h]
    \caption{Overview over the different simplified cone calorimeter IMP setups. PMMA density is 1201.72~kg/m$^3$ for all cases. 
    \label{tab:SimpleConeIMP}}
    \centering
    \begin{tabular}{l l l l}
        \toprule
        IMP Setup  & Details & Resolution & Target   \\
        \midrule
        Cone\_01  & Base case & C3 (3.3~cm) & HRR, Temperature \\
        Cone\_02  & Temperatures for the specific heat and conductivity  & C3 (3.3~cm) & HRR, Temperature \\
                  & \texttt{RAMP}s based on FDS parallel panel validation case &   &   \\ 
        Cone\_03  & PMMA slab thickness set to 6.1 mm & C3 (3.3~cm) & HRR, Temperature \\
        Cone\_04  & Base case & C3 (3.3~cm) & HRR \\
        Cone\_05  & Base case & C5 (2.0~cm) & HRR, Temperature \\
        Cone\_06  & Temperatures of conductivity and spec. heat \texttt{RAMP}s   & C2 (3.3~cm) & HRR, Temperature \\
                  & based on MCC plot (figure~\ref{fig:RAMP_Temperatures}) &   &  \\
        Cone\_07  & Like Cone\_06  & C2 (5.0~cm) & HRR, Temperature \\
        Cone\_08  & Like Cone\_06, lower temperature data added to \texttt{RAMP}s & C3 (3.3~cm) & HRR, Temperature \\ 
                  & for conductivity of PMMA and insulation, from DBI~\cite{macfp_matl_git} \\
        \bottomrule
    \end{tabular}
\end{table*}

\begin{table*}[h]
    \caption{Overview over the parameters that are adjusted in the bench-scale IMP step, base case. 
    \label{tab:SimpleConeIMP_Paras}}
    \centering
    \begin{tabular}{l l}
        \toprule
        Material  & Parameter    \\
        \midrule
        PMMA    & Emissivity \\
                & Absorption coefficient \\ 
                & Refractive index \\ 
                & Conductivity at 150~°C \\
                & Conductivity at 480~°C \\
                & Conductivity at 800~°C \\
                & Specific heat at 150~°C \\
                & Specific heat at 480~°C \\
                & Specific heat at 800~°C \\
        Residue & Emissivity \\
                & Conductivity \\
                & Specific heat \\
        Backing & Emissivity \\
                & Conductivity \\
                & Specific heat \\
        \bottomrule
    \end{tabular}
\end{table*}

The PMMA sample and the insulation layer below it are defined as a layered boundary condition (\texttt{SURF}). The cell size for the solid phase is determined by FDS automatically, based on the density, thermal conductivity and specific heat capacity (thermal diffusivity: $k / (\rho c)$). By default the cell size changes across a layer thickness, providing the highest resolution towards the surfaces. For the PMMA layer the cell size is forced to be uniform (\texttt{STRETCH\_FACTOR=1}). The resolution is further increased by a factor of~10 (\texttt{CELL\_SIZE\_FACTOR=0.1}). Since the thermal conductivity and specific heat capacity are adjusted during the inverse modelling, the solid cell size changes for any given PMMA material parameter set during the estimation process. The insulation layer uses the FDS default settings.

\begin{figure}[h]
    \centering
    \includegraphics[width=0.45\columnwidth]{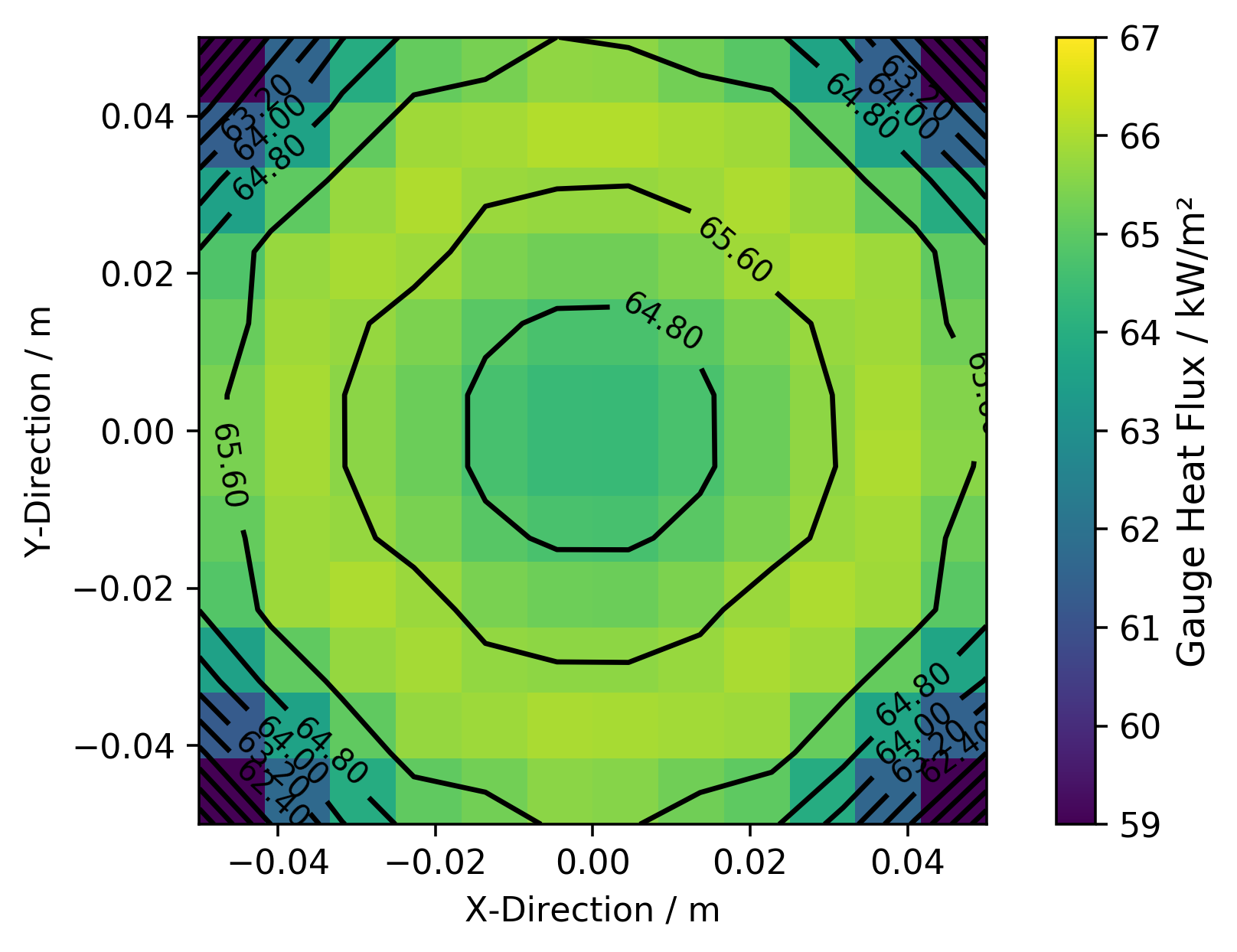}
    \caption{Gauge heat flux on the top surface of the sample, from a simulation with geometrical heater model (\texttt{GEOM}) for a target of 65~kW/m². Fluid cell resolution C12, i.e. 0.83~cm.}
    \label{fig:ConeGaugeFlux12CellsContour}
\end{figure}
\begin{figure}[h]
    \centering
    \subfloat[\centering  Resolution: 2 by 2 cells (C2, i.e. 5.0~cm).\label{fig:ConeFluxMapped_C2}]{{\includegraphics[width=0.45\columnwidth]{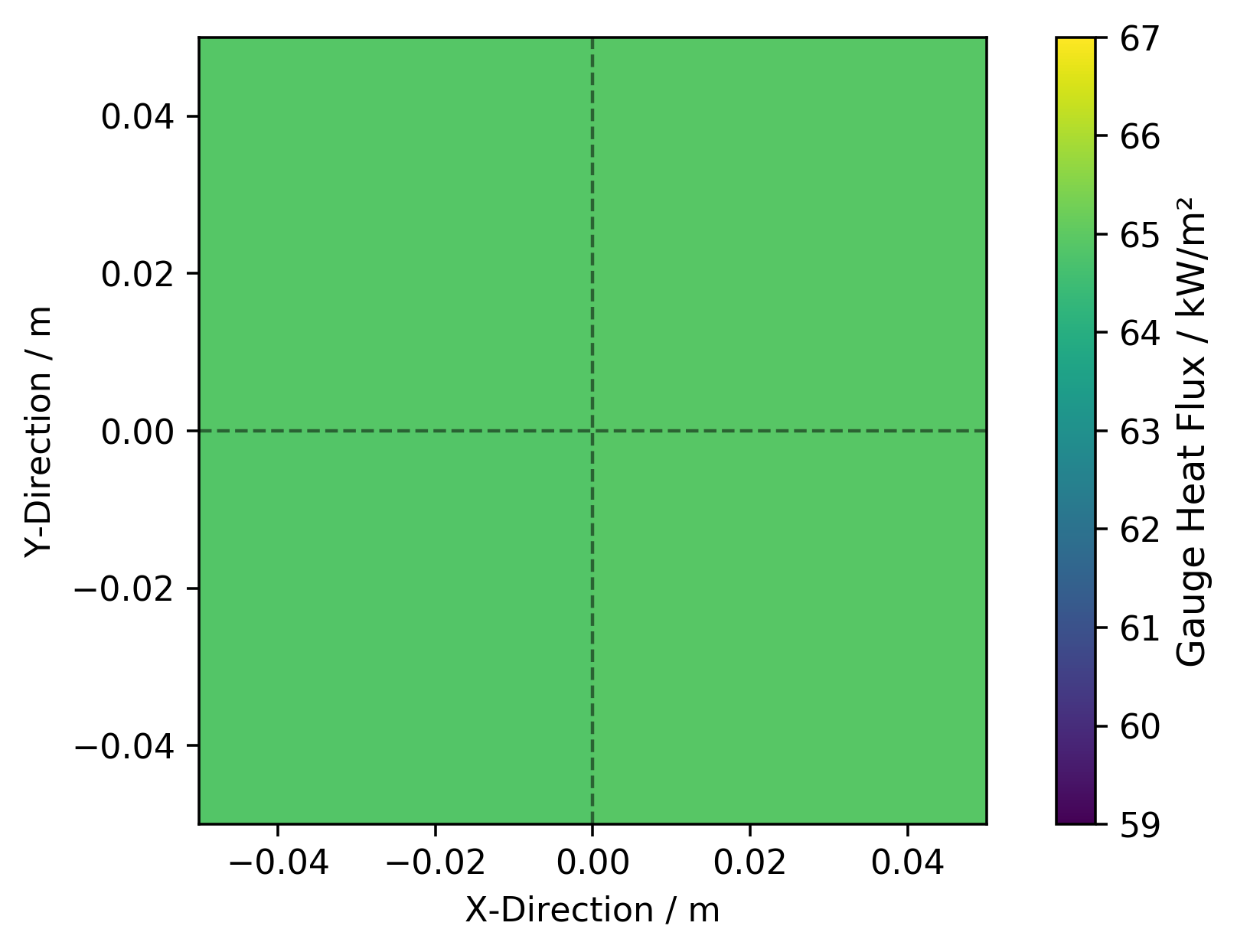} }}%
    \qquad
    \subfloat[\centering  Resolution: 3 by 3 cells (C3, i.e. 3.3~cm).\label{fig:ConeFluxMapped_C3}]{{\includegraphics[width=0.45\columnwidth]{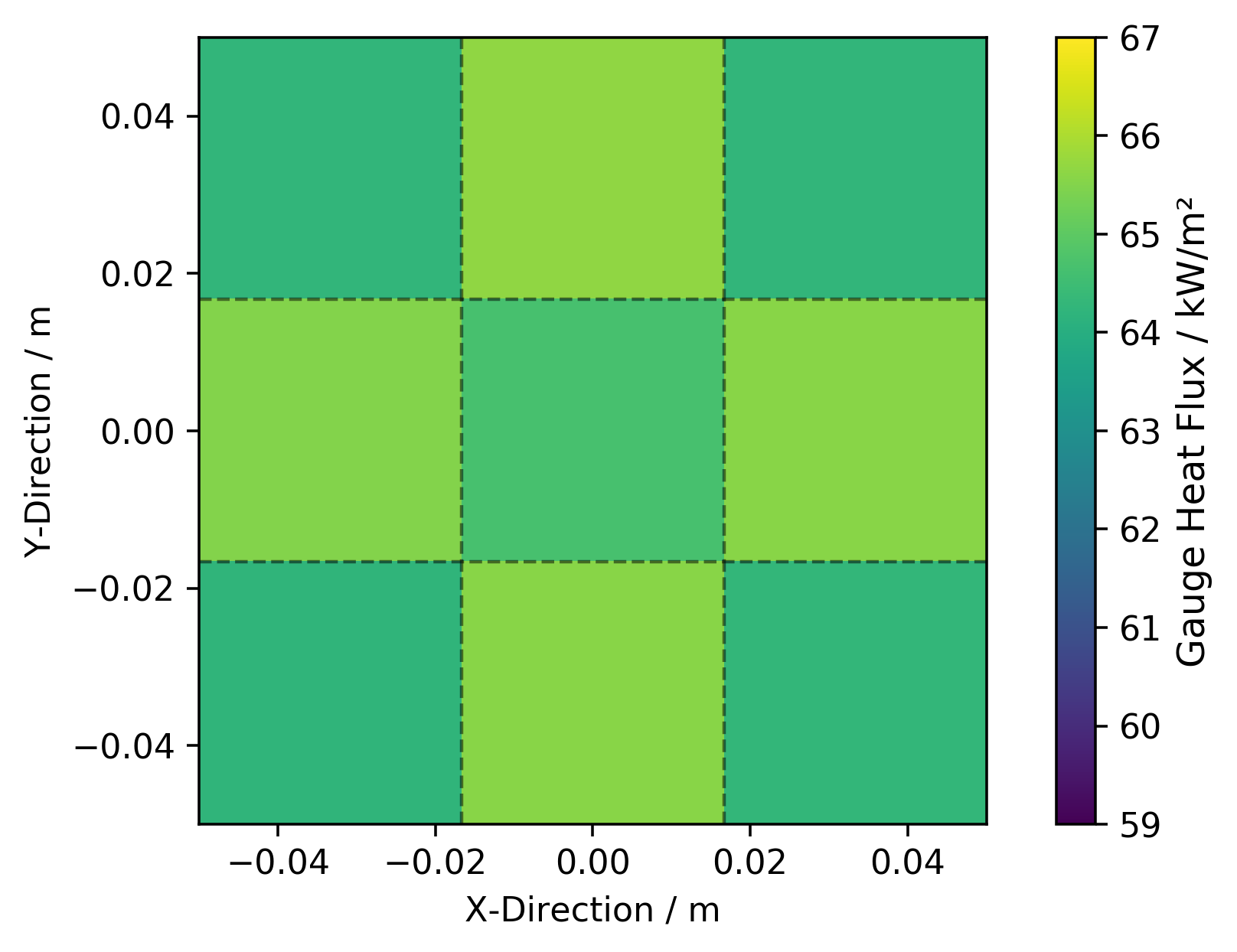} }}%
    \qquad
    \subfloat[\centering Resolution: 5 by 5 cells (C5, i.e. 2.0~cm).\label{fig:ConeFluxMapped_C5}]{{\includegraphics[width=0.45\columnwidth]{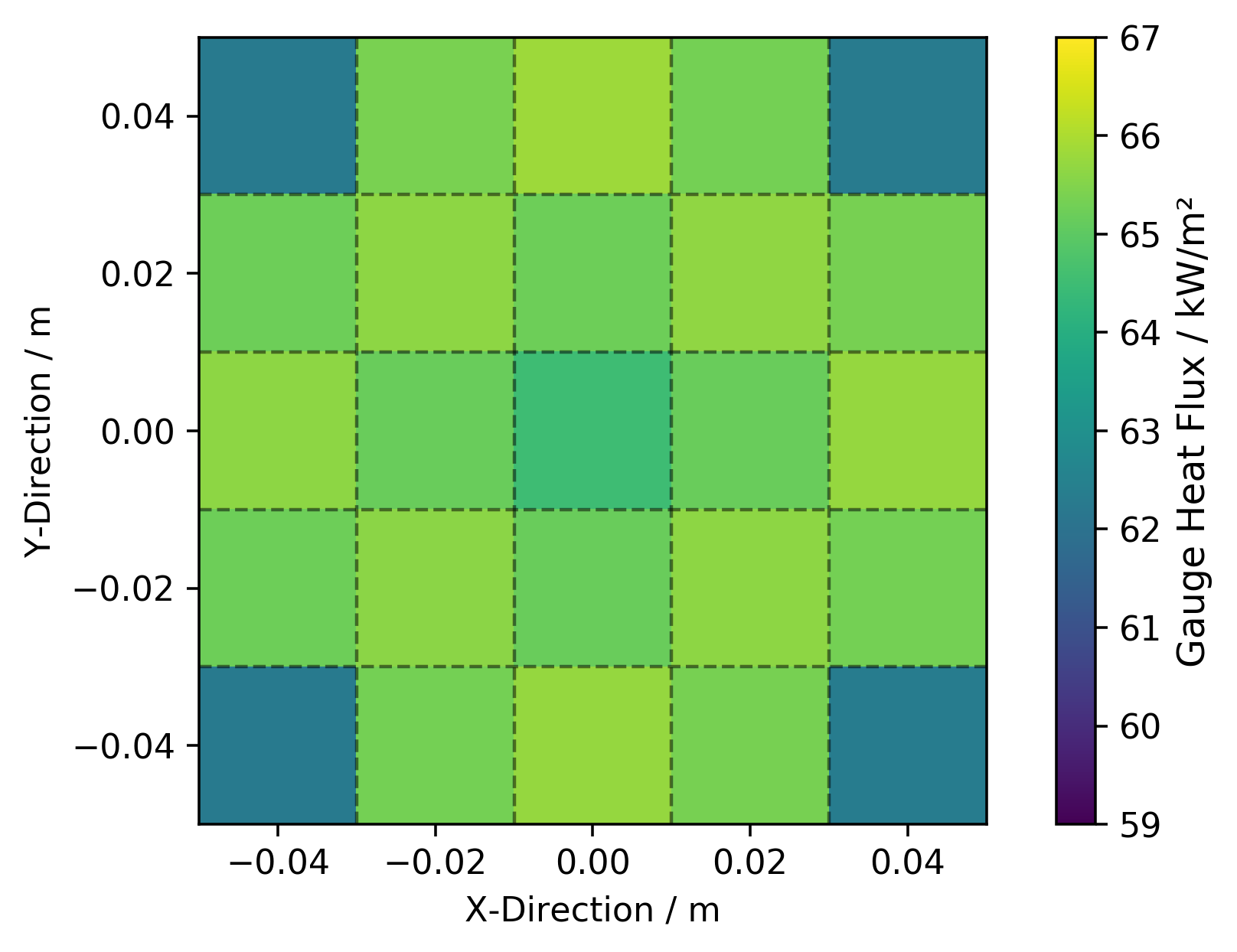} }}%
    
    \caption{Coarse gauge heat fluxes mappings for the IMP, constructed from a high resolution simulation (C12, figure~\ref{fig:ConeGaugeFlux12CellsContour}).}%
    \label{fig:ConeGaugeFluxesMapped}%
\end{figure}

The radiative heat flux of the heater is imprinted to the sample surface in the low resolution setups, such that a heater model can be neglected. This radiative heat flux is determined by employing a high resolution simulation (C12) containing a geometrical model of the heater, see appendix section~\ref{Appx:ConeGEOM}. The model is designed based on information in the literature~\cite{BabrauskasConeCalorimeter, iso5660_1}.
The resulting heat flux distribution on the sample surface is recorded (\texttt{GAUGE HEAT FLUX}), see figure~\ref{fig:ConeGaugeFlux12CellsContour}. It is observable, that the radiative flux is not uniform across the sample surface, as was also reported earlier~\cite{BOULET201253, HostikkaAxelssonRadiativeFeedbackFlamesConeCalorimeter,braennstroem_cone_radiation_Interflam}. Based on the fluid cell size of the respective the IMP (C2, C3 and C5), low resolution maps are computed, see figure~\ref{fig:ConeGaugeFluxesMapped}. These maps are implemented, using multiple surface definitions with different heat flux values (\texttt{EXTERNAL\_FLUX}) for the individual sample surface cells.


\subsection{Results}
\label{subsec:results_benchscale}

\begin{figure*}[h]%
    \centering
    \subfloat[\centering Fitness development.
    \label{fig:IMPConeBestParaC3_fit}]{{\includegraphics[width=0.45\columnwidth]{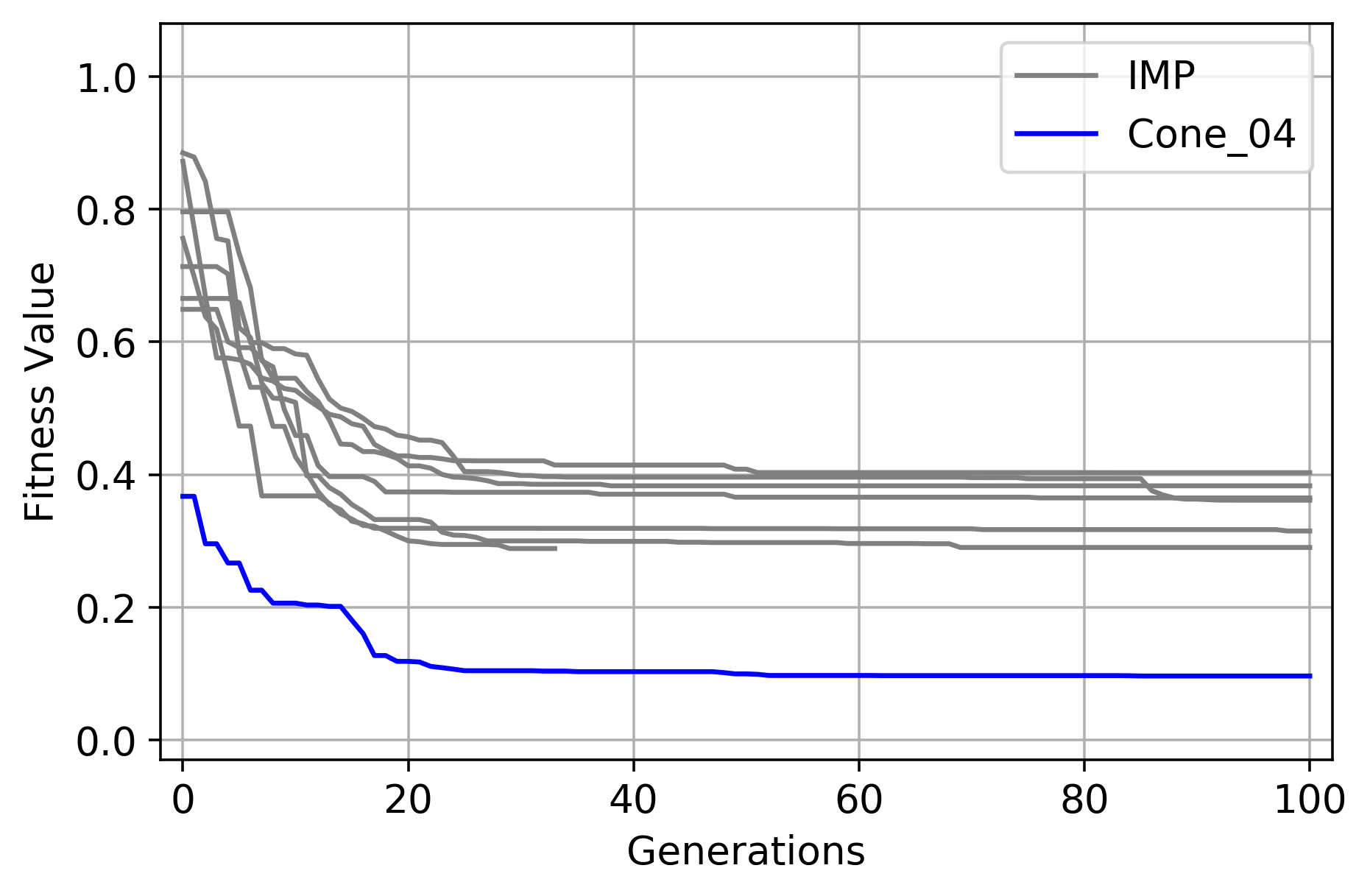} }}%
    \qquad
    \subfloat[\centering Energy release.
    \label{fig:IMPConeBestParaC3_err}]{{\includegraphics[width=0.45\columnwidth]{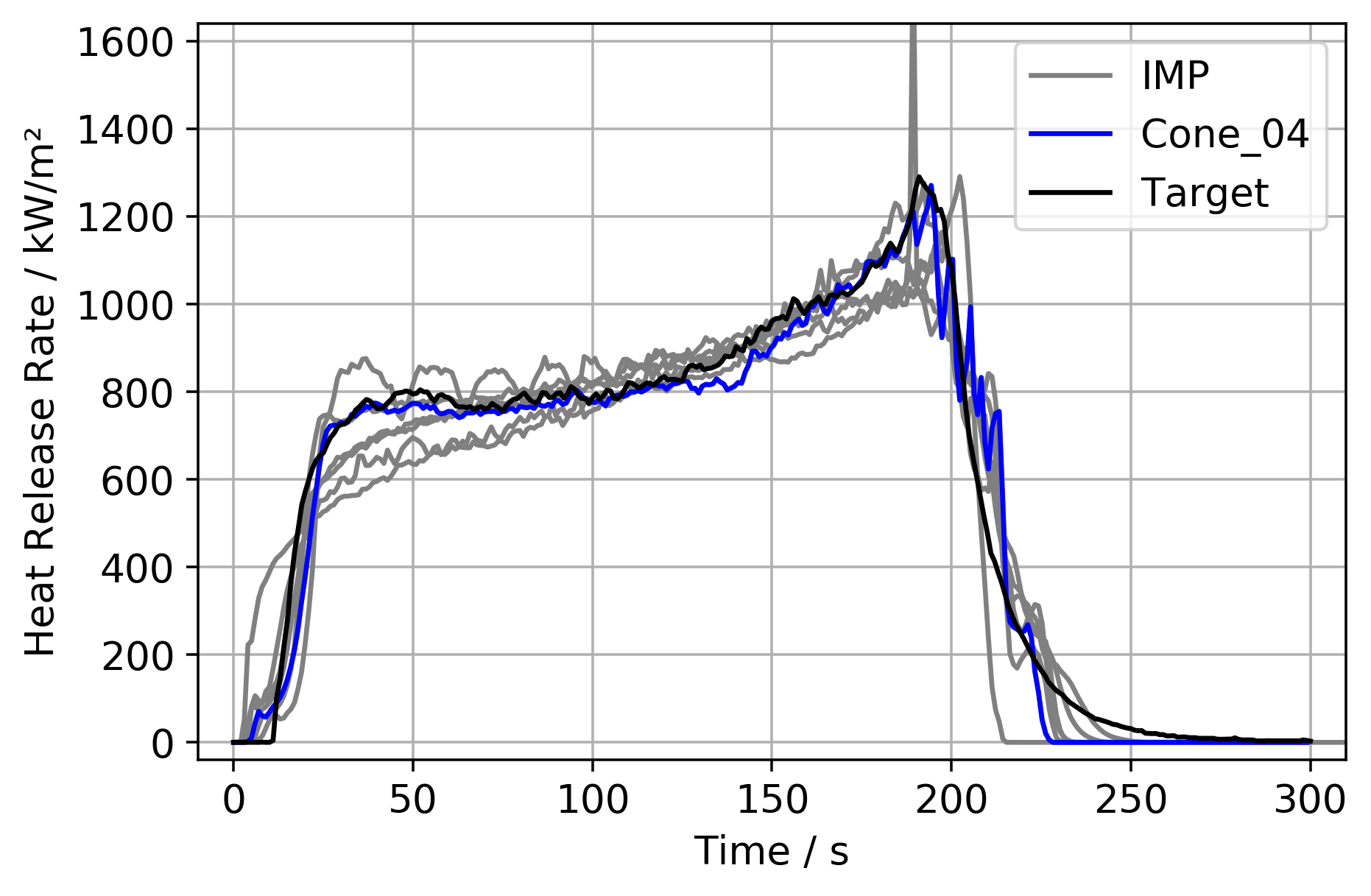} }}%
    \qquad
    \subfloat[\centering Back face temperature.
    \label{fig:IMPConeBestParaC3_bft}]{{\includegraphics[width=0.45\columnwidth]{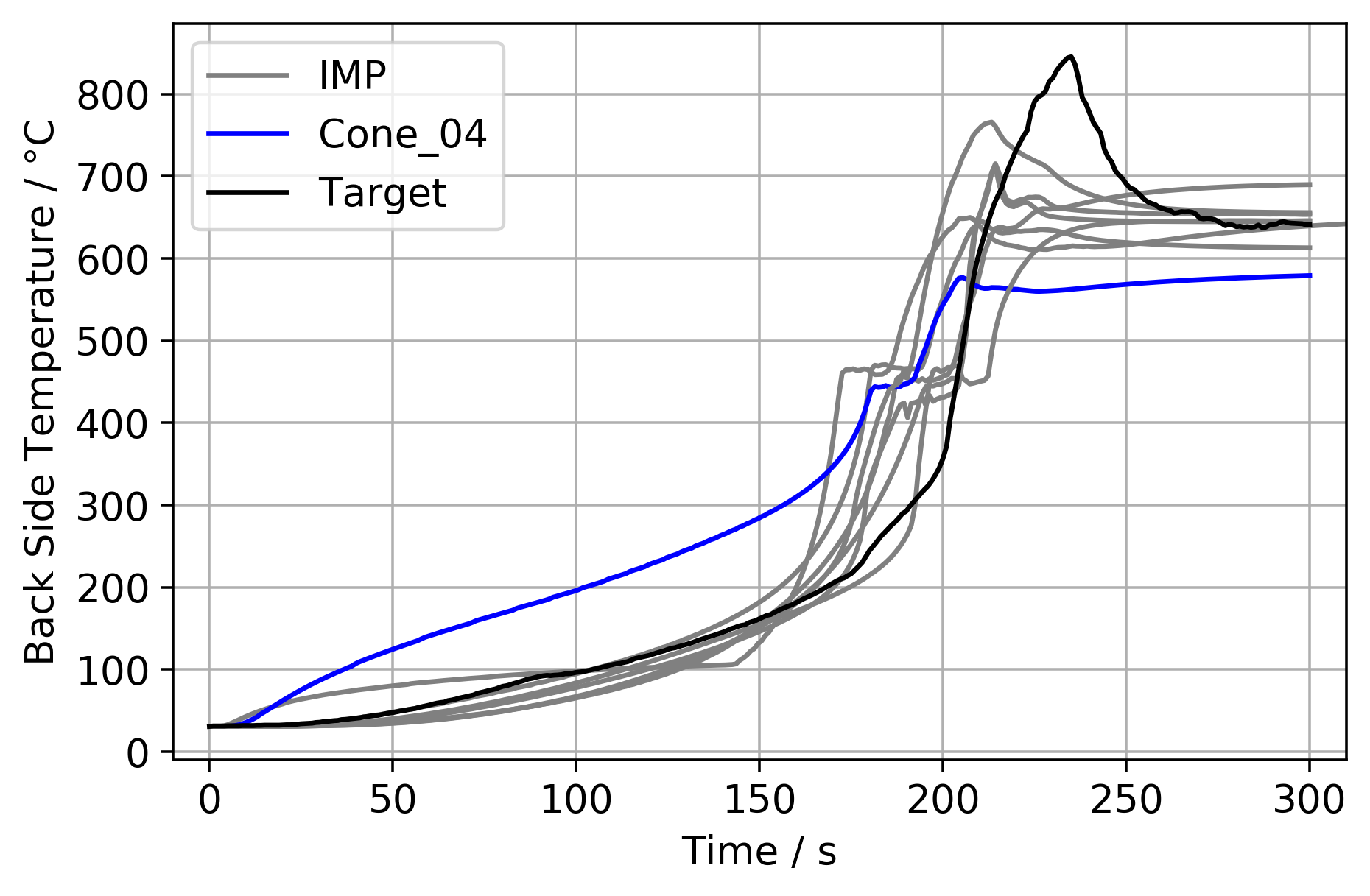} }}%
    \qquad
    \subfloat[\centering Sample mass.
    \label{fig:IMPConeBestParaC3_mass}]{{\includegraphics[width=0.45\columnwidth]{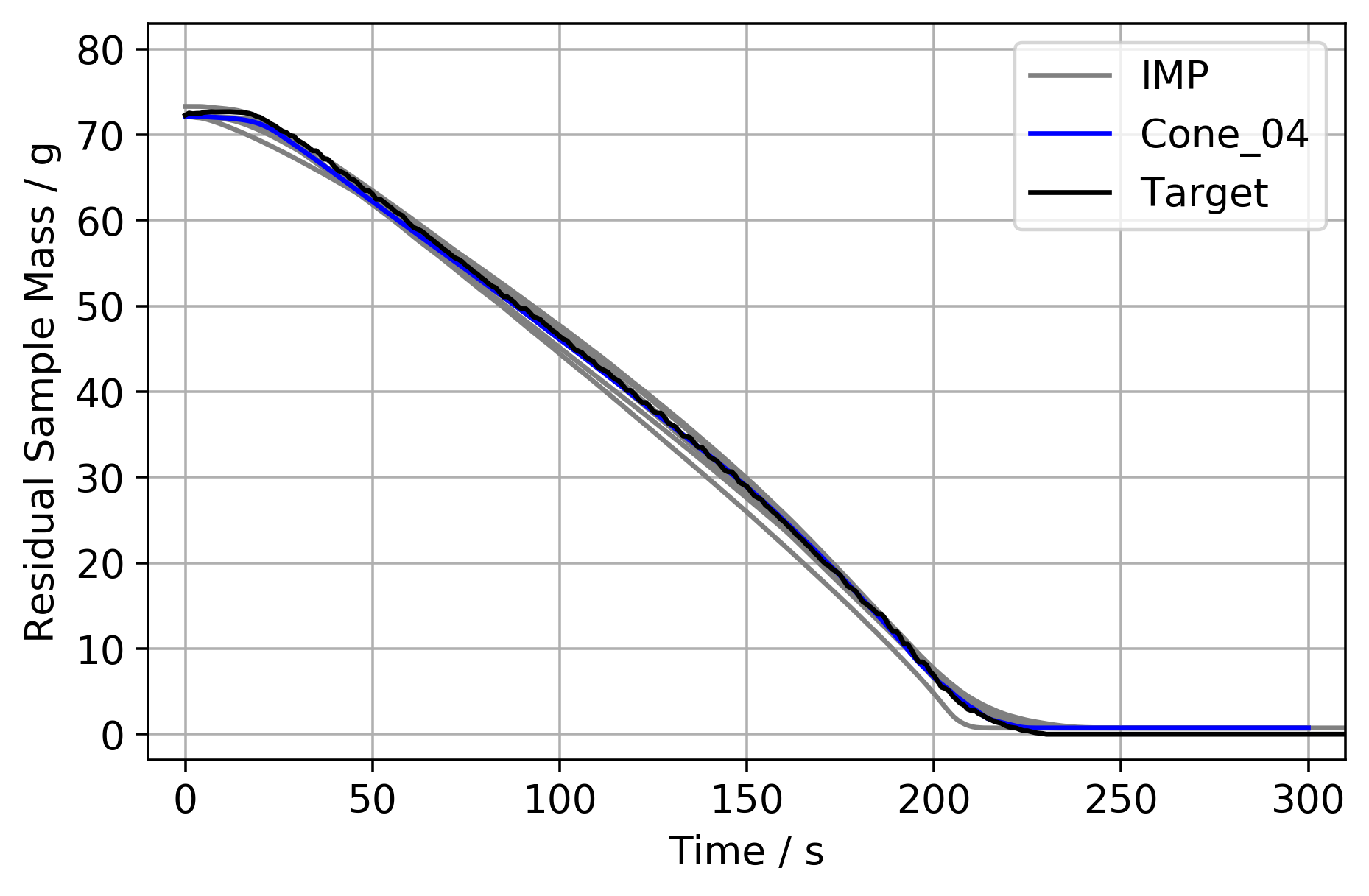} }}%
    
    \caption{Condensed results of the IMPs. Cone\_04 uses only energy release as target. Parameter set of the FDS parallel panel validation case for reference (Vali. PP). Full data provided in appendix~\ref{appendix_simplecone}.}%
    \label{fig:IMPConeBestParaC3}%
\end{figure*}

Here, an overview of the simple cone calorimeter IMP results of the best parameter sets is presented. The full data is provided in appendix~\ref{appendix_simplecone} for completeness.  
The best fitness values of the IMP runs are summarised in table~\ref{tab:SimpleConeIMP_BestPara}. 
For each IMP run, 100 generations have been completed, see figure~\ref{fig:IMPConeBestParaC3_fit} and figure~\ref{fig:ConeSimBestParaFitness_Aalto}. 
Cone\_04-L3 has the lowest fitness value, but its target is only the energy release, thus it is not directly comparably to the remaining IMPs.

\begin{table}[h]
    \caption{Best parameter sets of the different simple cone calorimeter IMP setups. Note: Cone\_04 is lower, since it only uses energy release as target, while all others also solve for back face temperature.
    \label{tab:SimpleConeIMP_BestPara}}
    \centering
    \begin{tabular}{l l l l}
        \toprule
        IMP Setup  & Limits & Repetition & Fitness Value    \\
        \midrule
        Cone\_01  & L2 & 43718 & 0.361 \\
        Cone\_02  & L2 & 38608 & 0.402 \\
        Cone\_03  & L3 & 17857 & 0.383 \\
        Cone\_04  & L3 & 39571 & \underline{0.096} \\
        Cone\_05  & L3 & 13568 & \underline{0.288} \\
        Cone\_06  & L3 & 45893 & 0.314 \\
        Cone\_07 & L2 & 35532 & 0.364 \\
        Cone\_08  & L3 & 32262 & 0.290 \\
        \bottomrule
    \end{tabular}
\end{table}

Figure~\ref{fig:IMPConeBestParaC3} shows the responses of the best parameter sets of all the different IMP setups. For all cases, the optimiser is able to find a parameter set that reproduces the experiment data relatively well. This is emphasised by drawing all their responses without distinction, including different fluid cell resolutions (C2, C3 and C5). Cone\_04 is highlighted, because it uses only the heat release as target.

With respect to the heat release, difficulties exist in reproducing the first bump at around 20~s to 50~s, the final peak at about 190~s and the following decay phase. In some cases, pronounced steps are visible towards the end of the simulations. See for example 
figure~\ref{fig:ConeSimBestParaHRR_Aalto} for Cone\_06. These steps are associated with the burn-out of the individual cells. Furthermore, in all cases a small peak is visible in the beginning of the cone calorimeter simulations, at about 15~s. Cone\_04 and Cone\_08 capture the heat release profile best (figure~\ref{fig:ConeSimBestParaHRR_Aalto}).

Reproducing the temperature recorded at the back face of the sample proves to be challenging, see figure~\ref{fig:IMPConeBestParaC3_bft} and figure~\ref{fig:ConeSimBestParaBackTemp_Aalto}. For cases where it is a target, the temperature development during the first about 160~s can be reproduced. Between about 160~s to about 270~s departures are visible, with a pronounced step around 200~s. A peak can be observed towards the end, yet less pronounced as in the experiment. Cone\_04 does not have the temperature development target and is not able to reproduce a similar behaviour on its own.

The residual sample masses during the simulation are close to the experiment data, see figure~\ref{fig:IMPConeBestParaC3_mass} and figure~\ref{fig:ConeSimBestParaSampleMass_Aalto}.

\subsection{Discussion}
\label{subsec:discussion_benchscale}

In terms of the fitness value, Cone\_08-L3 and Cone\_05-L3 performed best across all IMP setups, see table~\ref{tab:SimpleConeIMP_BestPara} and  figure~\ref{fig:ConeSimBestParaFitness_Aalto}. This is excluding Cone\_04-L3, because it neglects the back side temperature. Adjusting the sampling limits leads mostly to better parameter sets. Occasionally, the IMPs do not find better sets within the given amount of generations. For example, L1 of Cone\_03 shows worse fitness values throughout, compared to L0, see~\ref{fig:ConeSimBestParaFitness_Aalto}. The likely reason is that with each adjustment the process starts anew, combined with the randomness for choosing the individual values. With more generations, better parameter sets may be found. The impact of smaller fluid cells during the IMP is not clear. It might lead to better parameters for the Cone\_05 series, yet its enormous runtime makes it not feasible to wait its completion during this work. As of now, it shows fitness values that are just marginally better than Cone\_08-L3 (table~\ref{tab:SimpleConeIMP_BestPara}).

\begin{figure}[h]
    \centering
    \includegraphics[width=0.45\columnwidth]{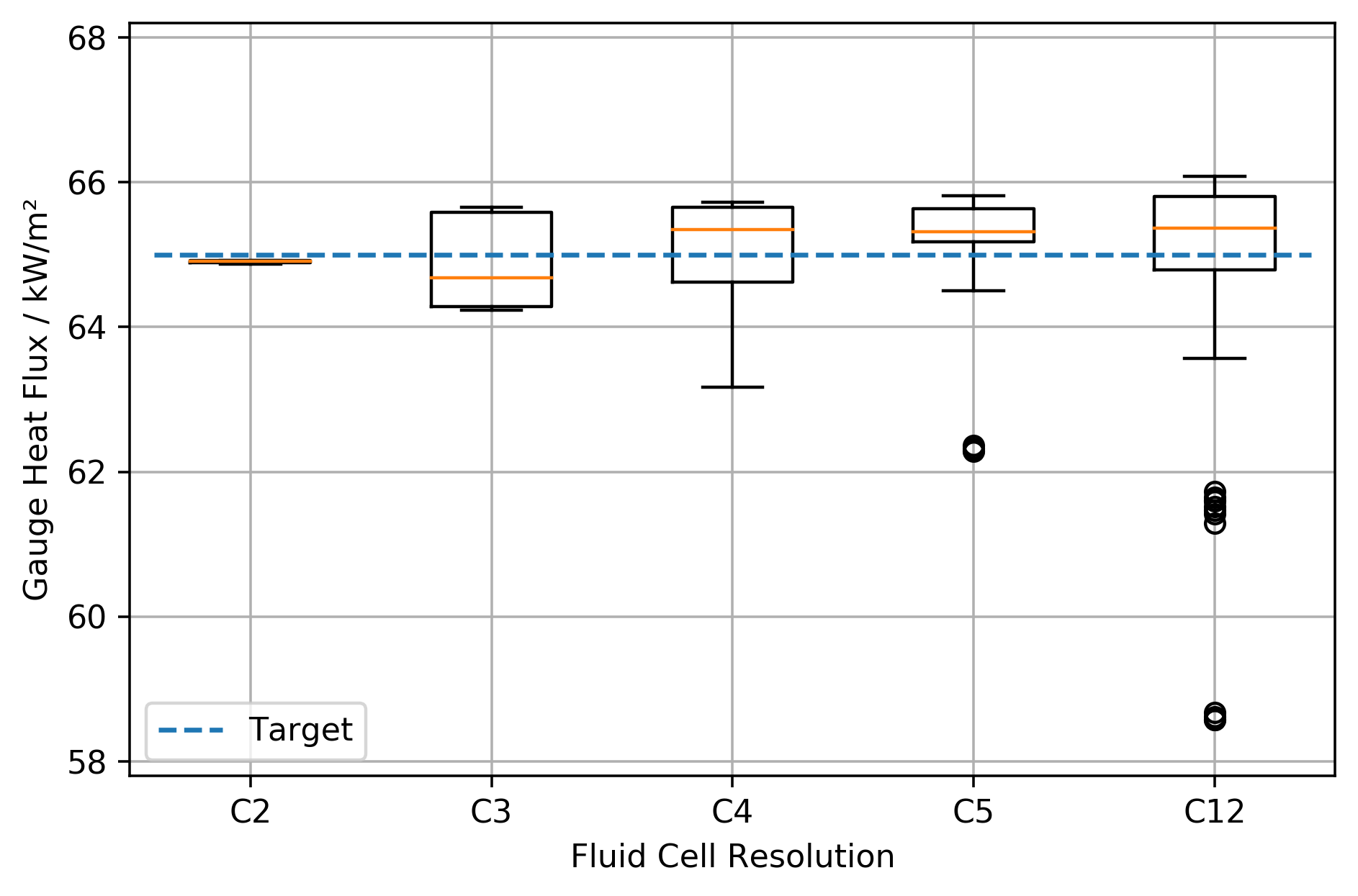}
    \caption{Gauge heat flux distribution for different resolutions (C2 to C12) across all surface cells.}
    \label{fig:ConeGaugeFluxBoxplots}
\end{figure}

Compared with previous work, e.g.~\cite{AlexandraViitanen_CableTrays, Hehnen_fire3030033}, here, higher fluid cell resolutions are used to cover the cone calorimeter sample. With C3 and above, the cells become distinguishable between corners, edges and centre cells --- compare the flux maps for two, three and five cells (figure~\ref{fig:ConeGaugeFluxesMapped}). For the C2 configuration, each cell has essentially the same value -- the average over the whole surface. This is summarised in figure~\ref{fig:ConeGaugeFluxBoxplots}. Higher resolutions capture the ring-shaped pattern of higher heat flux in the centre and the substantially lower heat flux in the corners better, see figure~\ref{fig:ConeGaugeFlux12CellsContour}.
The steps in the simulations during the decay phase might be a result of it, see figure~\ref{fig:ConeSimBestParaHRR_Aalto} between about 200~s to 250~s for Cone\_06\_L0. Two processes control the decay: cells burning out and the local burning behaviour, depending on the material parameters. Combining both can smooth out the decay phase. This combining is difficult to achieve without an optimiser, because the parameter set needs to behave as if three-dimensional heat transfer is possible even though it is not. The C2 cases are primarily controlled by the parameters and show a steep drop at the end (Cone\_07), because uneven sample consumption cannot be covered well. In the C3 setups the formation of pronounced steps is visible in some L0 cases, but higher limit adjustments can show a smoother decay. 
During the IMP a fluid cell resolution of C3 seems to be beneficial. It allows to capture the uneven heat flux to the sample surface, distinguish between cell locations and also provides a better resolution of the flow and radiation fields. Higher resolutions are computationally very expensive, specifically within the constrains provided by the SCE.

This behaviour is also reflected in the back side temperature. During the decay period a spike is visible (figure~\ref{fig:ConeExpAalto_temp}), until the temperature settles at a constant value, from around 280~s onward. This constant value primarily shows the influence from the heating element, without the sample. During the experiment, the sample material is consumed and at some point the thermocouples below are exposed, starting from the centre. 
Thus, they are able to receive the heat radiation from the flame of the surrounding sample directly, in addition to the heater radiation. Due to thermal tension, they may also bend towards the heater. All these aspects together lead to the formation of the spike.

\begin{figure}[h]%
    \centering
    \subfloat[\centering Overview over the sample configuration, at about 2:25 min after experiment start.\label{fig:PMMAsampleDeformationOverview}]{{\includegraphics[width=\textwidth]{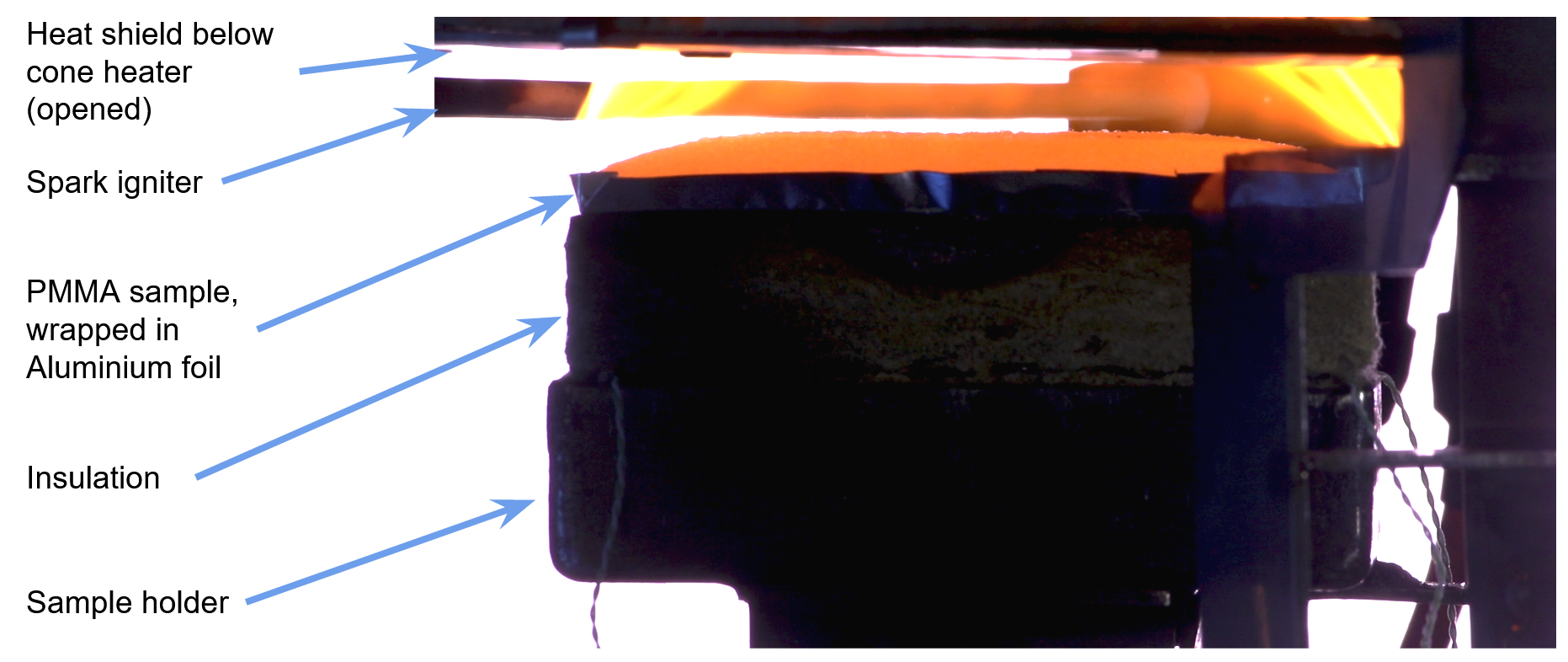} }}%
    \qquad
    \subfloat[\centering Begin of sample deformation at about 1:51 min after experiment start (left). Peak sample deformation at about 2:41 min after experiment start (right). Images cropped to highlight the deformation.\label{fig:PMMAsampleDeformationCropped}]{{\includegraphics[width=\textwidth]{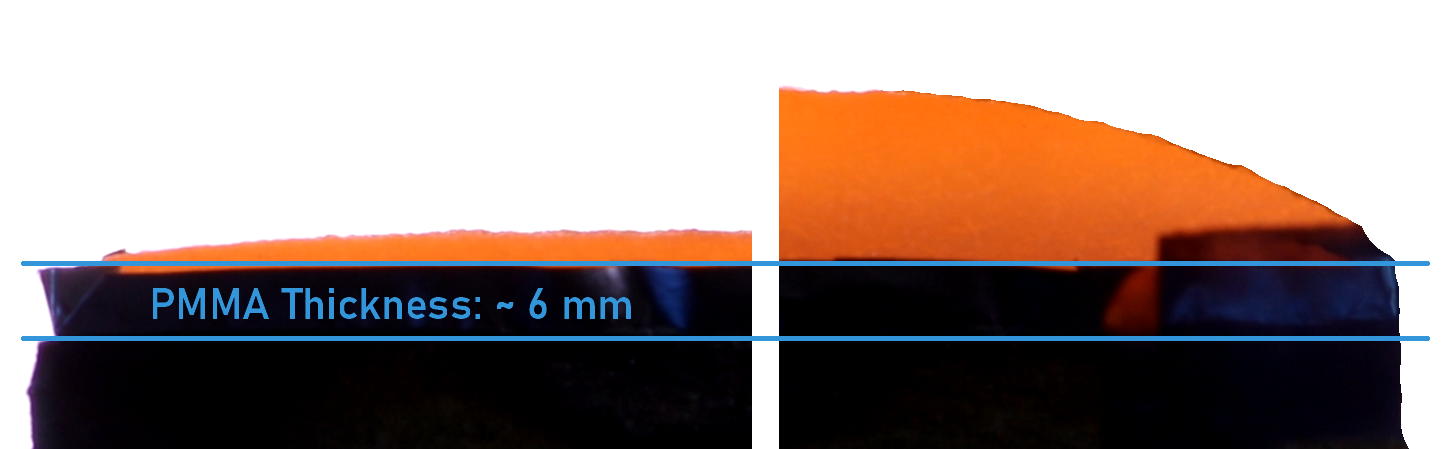} }}%
    
    \caption{Photographs of a cone calorimeter test (50~kW/m$^2$) of a PMMA sample with 6~mm thickness (side view). Orange colour is caused by the reflection of the flame. Images provided by Karen De Lannoye via private communication, see also~\cite{Karen_ConeDeformationASTM2023}.}%
    \label{fig:PMMAsampleDeformation}%
\end{figure}

The optimiser has difficulties to capture both, the heat release rate and the material temperature. With lower fluid cell resolution (C2, Aalto\_6b in figure~\ref{fig:ConeSimBestParaBackTemp_Aalto}) the temperature peak between 200~s and 250~s cannot be reproduced. This is associated to the near-uniform consumption of the sample material, see above. In the other setups, the peak can be captured for the initial sampling limits, but mostly disappears with further adjustments. Certainly, neglecting three dimensional heat conduction inside the sample, is influencing the outcome as well. On the other hand, the sample shape can change significantly during the experiment, see figure~\ref{fig:PMMAsampleDeformation}. Relatively early on, it creates a foam layer and starts to bend towards the heater. In similar experiments performed by Karen De Lannoye, the maximum height of the bump was observed to extend approximately two sample thicknesses above the original surface of a sample with a thickness of 6~mm (figure~\ref{fig:PMMAsampleDeformationCropped}, right), but the behaviour can change depending on the experiment conditions~\cite{Karen_ConeDeformationASTM2023}. 
This relatively symmetrical bump can change its shape significantly during its decay. While the sample material is consumed, its surface retracts further away from the heater, compared to the beginning of the experiment. Thus, in the experiment, the received radiative flux from the heater should change. These deformations are not replicated in the simulations presented here -- the component of the radiative flux of the heater stays constant, by construction. Only the heat flux component from the flame can change. The sample deformation is likely misinterpreted as a change in mass and energy release during the inverse modelling.


Given the deformation, it is fundamentally unclear where the back face temperature is actually recorded.  
Furthermore, changes in the rigidity of the PMMA sample (e.g. melting) such that the thermocouple tip can move into the sample, thermal expansion of the sample holder assembly and associated movement between individual components, mechanical tension on the thermocouples and others are contributing to the uncertainty. Assuming they would be tightly attached to the backing material, an air gap could form between them and a bent sample. Such a gap would be interpreted as lower thermal conductivity during the IMP. As an example, compare Cone\_04 with the other IMP results in figure~\ref{fig:IMPConeBestParaC3_bft}. 
It only uses the energy release as target, not the back side temperature. The temperatures are higher throughout the simulation.
Curiously, in the parallel panel simulation setup this behaviour appears to be beneficial, as discussed later in section~\ref{subsec:discussion_realscale}. This seems to hint at incorrect temperature readings during the cone calorimeter experiment. If the sample separates from the thermocouple due to deformation, its recorded temperature should be lower than the actual back face temperature. Consequently, higher back face temperatures are visible in Cone\_04 (figure~\ref{fig:IMPConeBestParaC3_bft}), since in the simulation it is recorded at the back face of the sample by construction. 
From the above, the uncertainty of the recorded temperature increases during the run time of the experiment. Reliable temperatures might only be able to be recorded for low sample temperatures during the beginning. Specifically considering the deformation, melting and consumption of the sample. Therefore, the intended separation between sample behaviour and the boundary conditions could not be achieved. 

In the future, it would be interesting to assess the surface deformation of the sample during cone calorimeter tests. Maybe one could leverage methodologies used in assessing the performance of intumescent coating, e.g.~\cite{ElliottIntumescentCoatings}. This could then be used to adjust the prescribed radiative heat flux to the sample surface over time. 

The conductivity and specific heat change with sample temperature in the simulation (\texttt{RAMP}). This can account for the sample deformation to some degree, when an air gap forms between sample and thermocouple. The temperatures are arbitrarily chosen and used for both parameters, except for Cone\_02. At high temperatures Cone\_01,~03~to~05 get relatively high values assigned by the optimiser, see figures~\ref{fig:ConeSimBestParaCond_Aalto} and~\ref{fig:ConeSimBestParaSpecHeat_Aalto}. Since the material is consumed way before the 800~°C could be reached, only a very small fraction of the ramp piece between the last two points can meaningfully contribute. Thus, ramp values for 800~°C are poorly chosen and should be disregarded. The cases where the temperature points are chosen based on the MCC experiment data, see figure~\ref{fig:RAMP_Temperatures}, lead to more reasonable results for the conductivity, which is different for the specific heat. The final point is at a temperature close to the maximum the sample material can reach. 
Still, it seems to be a useful approach for unknown materials to align both temperature dependent values to the micro-scale data. Maybe the highest temperature value of the ramp could be chosen to be about 20~K lower than the highest meaningful value in the experiment data. This is an attempt to prevent confusion of the optimiser with temperatures that are impossible to reach, because the material is consumed.
Another strategy could be to run an IMP solely to determine the ramps for some best parameter set. Thus, more parameters could be spent on the ramp alone without getting too large generation sizes.

With data provided from DBI/Lund~\cite{macfp_matl_git}, extending the conductivity ramps in Cone\_08-L3 seems not to change the temperature development significantly, see figure~\ref{fig:ConeSimBestParaCond_Aalto}. For most of the IMP setups, values for the lower temperatures are determined that are already in the vicinity of the experiment data. 

Overall, the sample mass loss during the cone calorimeter simulation is relatively close to the experiment data, see figure~\ref{fig:ConeSimBestParaSampleMass_Aalto}. This behaviour is an emergent phenomenon of the steps taken with the gas mixture and gives confidence into the proposed method, since it is not an explicit target of the IMP. 
However, the flame height and gas temperature change with the release of the different surrogate fuel species, see appendix~\ref{App:FlameHeight}. This is likely due to dilution of the surrogate fuel with carbon dioxide. Which of the presented centre line gas temperatures in figure~\ref{fig:FlameHeight_Temperatures} is more realistic in context of the PMMA cone calorimeter experiment studied here is unclear for now. Still, it is worth noting that the difference exists, because higher flames might have an impact on the fire spread in a simulation.

\begin{figure*}[h]%
    \centering
    \subfloat[\centering Cone\_04, L0.\label{fig:SimpleCone_CellSize_Comparison_Aalto4L0}]{{\includegraphics[width=0.45\columnwidth]{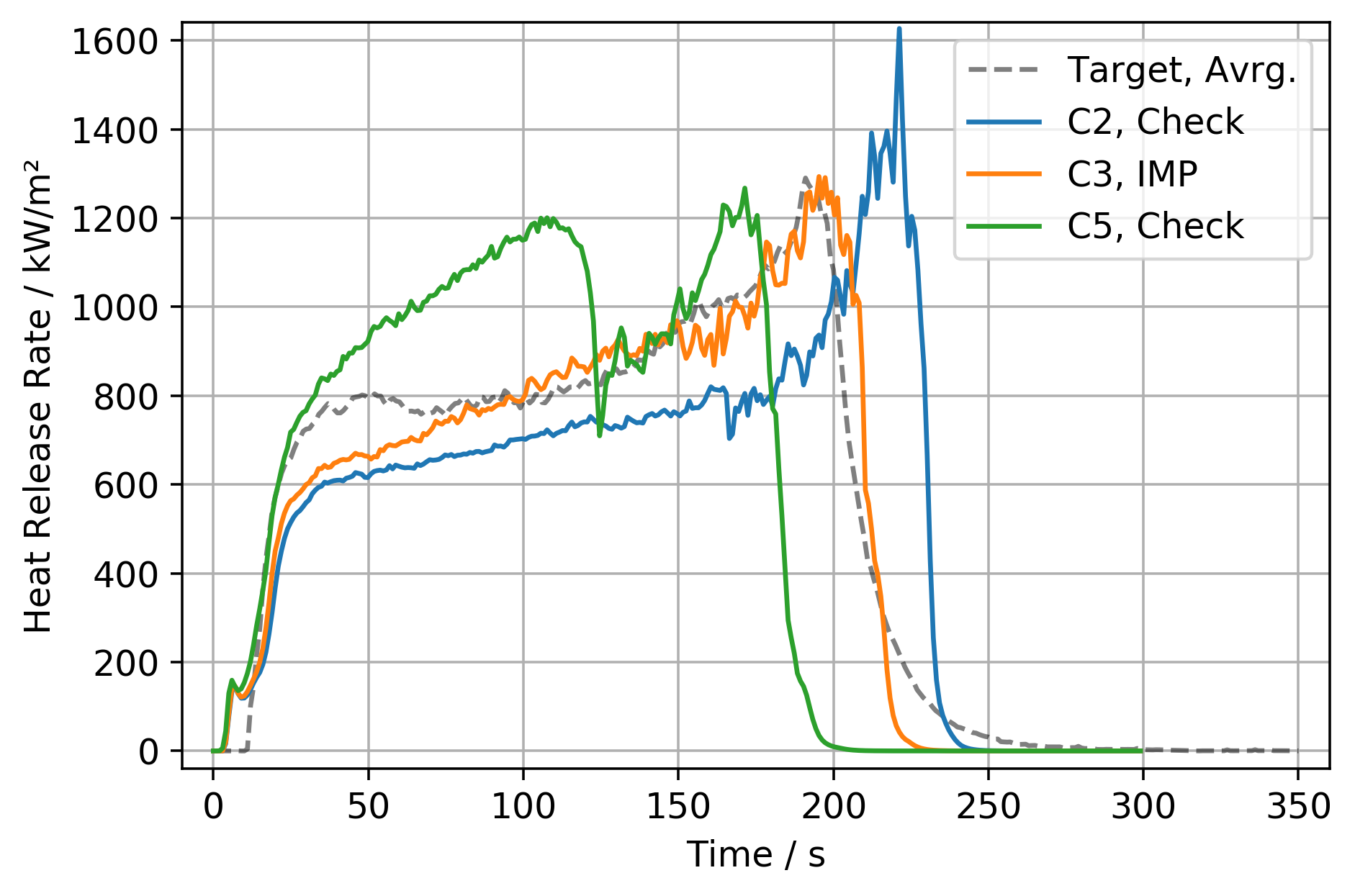} }}%
    \qquad
    \subfloat[\centering Cone\_04, L3.\label{fig:SimpleCone_CellSize_Comparison_Aalto4L3}]{{\includegraphics[width=0.45\columnwidth]{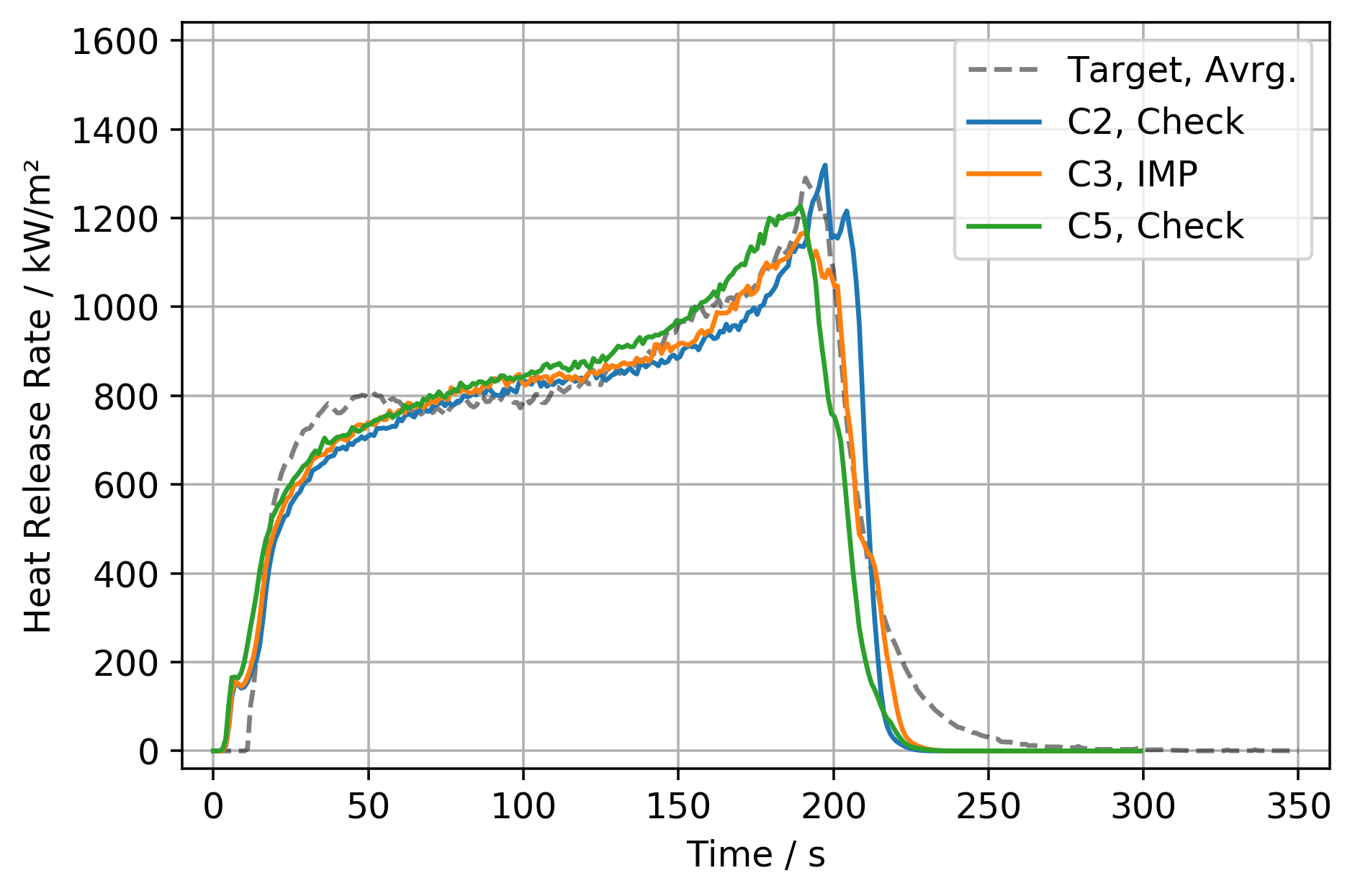} }}%
    \qquad
    \subfloat[\centering Cone\_05, L0.\label{fig:SimpleCone_CellSize_Comparison_Aalto5L0}]{{\includegraphics[width=0.45\columnwidth]{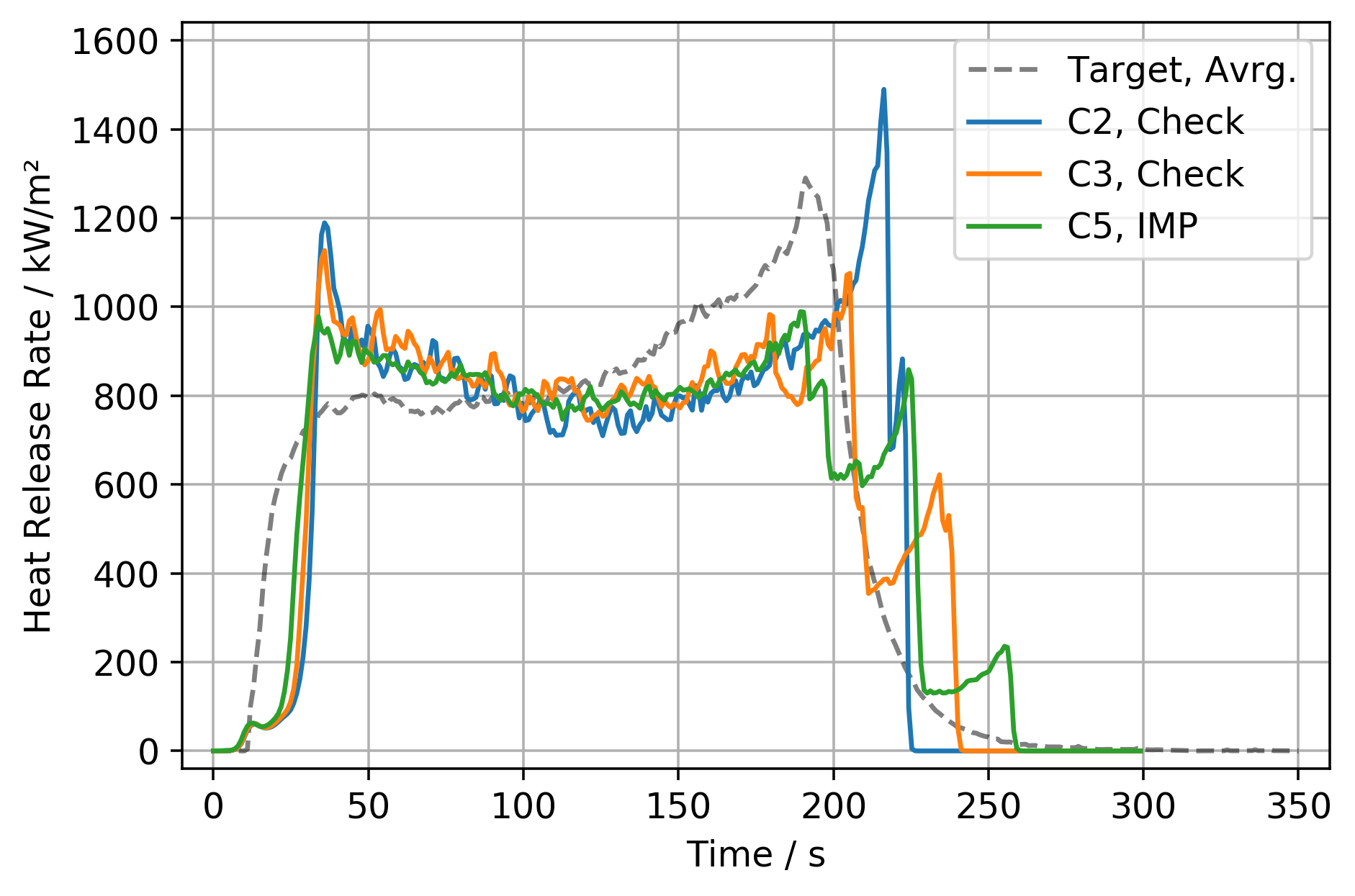} }}%
    \qquad
    \subfloat[\centering Cone\_05, L3.\label{fig:SimpleCone_CellSize_Comparison_Aalto5L3}]{{\includegraphics[width=0.45\columnwidth]{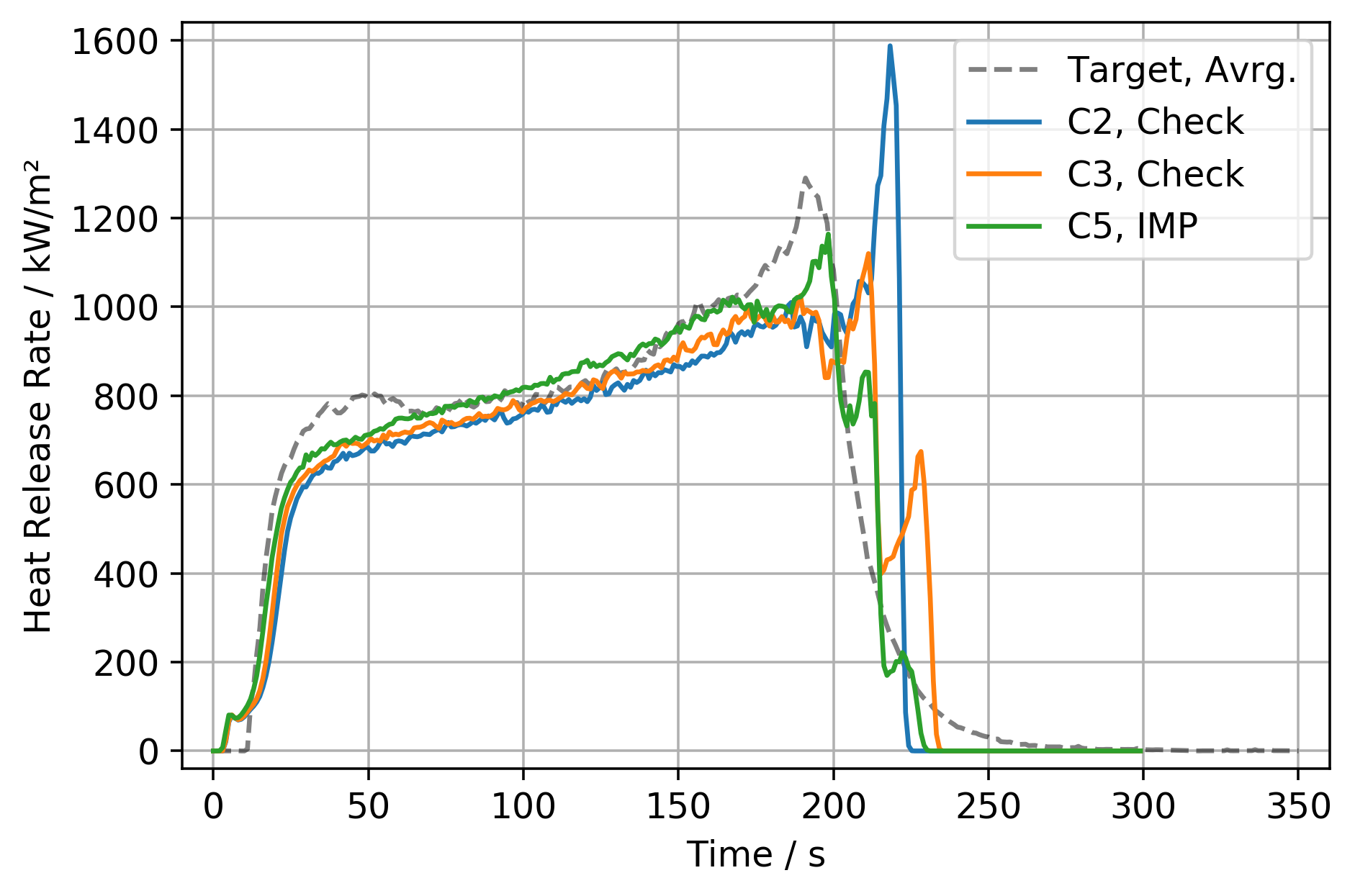} }}%
    
    \caption{Comparison of the energy release of a best parameter set (IMP) across different fluid cell resolutions (Check) for the simple cone calorimeter setup.}%
    \label{fig:SimpleCone_CellSize_Comparison}%
\end{figure*}

For the IMP, the fluid cell resolution of 3.3~cm (C3) seems to be beneficial. It is able to capture the uneven sample consumption, yet it runs relatively fast. However, the question arises, how will the parameters perform at higher resolutions. For this, figure~\ref{fig:SimpleCone_CellSize_Comparison} demonstrates an exemplary comparison: a parameter set determined at a given resolution (labelled "IMP") is used at an other resolution (labelled "Check"). Some of the investigated cases show similar behaviour across all limit adjustments, see figure~\ref{fig:SimpleCone_CellSize_Comparison_Aalto5L0} and figure~\ref{fig:SimpleCone_CellSize_Comparison_Aalto5L3}. Others converge towards the higher resolution over the course of multiple limit adjustments, see figure~\ref{fig:SimpleCone_CellSize_Comparison_Aalto4L0} and figure~\ref{fig:SimpleCone_CellSize_Comparison_Aalto4L3}. All limits are provided in appendix~\ref{appendix_SimpleConeComparison}. Given the reduction in computational demand, this is promising.

The overall runtime of this IMP step is a notable limitation for practical usage. Given the development of the fitness values for each of the IMP runs, one could consider to run less generations. Possibly~60 to~80 generations could be sufficient. Furthermore, using less parameters during the IMP has a strong impact on the amount of simulations necessary, see equation~\ref{formula:GenerationSize}. Maybe the parameters of the backing could be determined in an individual step. Using data from differential scanning calorimetry, the specific heat capacity could be determined in the less costly first IMP step. Optimising this second IMP step needs further investigation.


\section{Real-Scale Setup  ---  Validation}

\subsection{Methods}
\label{subsec:realscale}

\begin{figure}[h]%
    \centering
    \includegraphics[width=0.45\columnwidth]{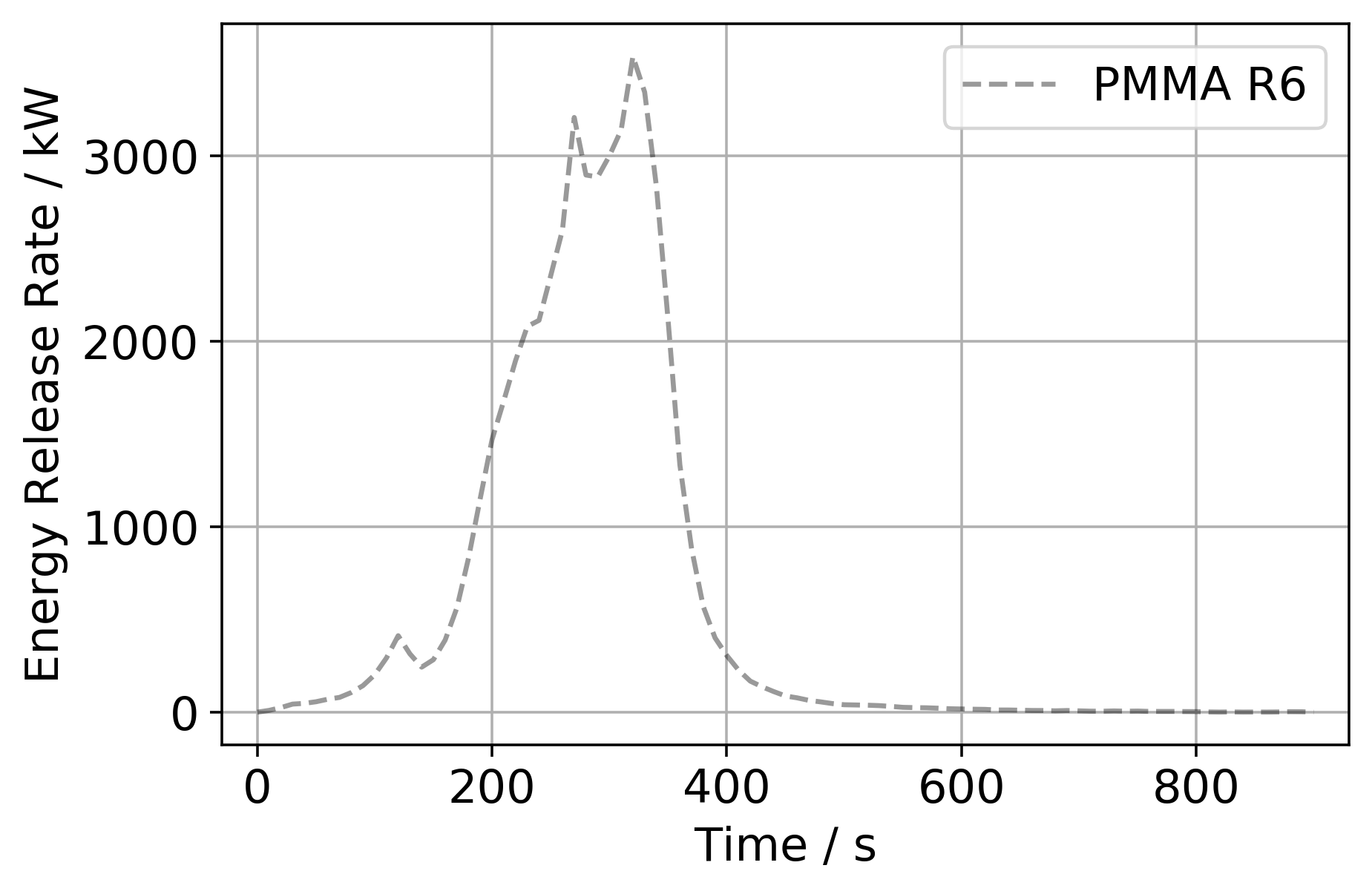}
    \caption{Fire development over PMMA panels in the parallel panel experiment.}
    \label{fig:PP_Exp_Data}
\end{figure}

As a validation step, the material parameter sets (table~\ref{tab:SimpleConeIMP_BestPara}) are used in a real-scale simulation setup of a parallel panel test. The results are compared to the heat release rate measured in the experiment, see figure~\ref{fig:PP_Exp_Data}. The data set used here is "Test\_7\_PMMA\_R6", labelled "PMMA~R6"~\cite{Leventon2022ParallelPanel}.

The parallel panel test consists of two 0.61~m wide panels facing each other with a separation of 0.3~m. In between both, at the bottom, is a gas burner located with a width of 0.3~m and a length of 0.61~m, see validation guide~\cite{fdsValiGuide676} cases "FM Parallel Panel Experiments" and "NIST/NRC Parallel Panel Experiments", as well as the MaCFP data base~\cite{Leventon2022ParallelPanel}. The combustible sample, attached to the inside of the panels, extents 2.44~m above the burner surface. The burner is fed with propane gas. It reaches a quasi-steady energy release of about 60~kW, between 60~s to~80~s after its ignition. After the sample is confirmed burning (sustained flaming across the panel walls), the burner is shut off. This happens about 120~s
after the start of the experiment. Shutting the burner off slows down the fire development for about half a minute, see figure~\ref{fig:PP_Exp_Data} between about 120~s to 150~s. The sample material is the same cast black PMMA used throughout the MaCFP test campaign.

The simulations are conducted for the three different fluid cell resolutions introduced in section~\ref{subsec:methods_benchscale}. The computational domain spans a volume of 1.2~m~$\times$~0.8~m~$\times$~4.8~m  and is divided into multiple sub-domains (\texttt{MESH}), with the following divisions along the x, y and z axis: (3, 1, 12). 
A graphical overview over the domain layout is provided in appendix~\ref{Appx:MeshLayout}, figure~\ref{fig:app_MeshLayoutPP}. Thus, the individual mesh dimensions are multiples of 10~cm and can be nicely divided following the scheme outlined in section~\ref{subsec:methods_benchscale}. This number of meshes allows the simulation to be run on a single computing node, utilising its 64 cores. The simulation mode is set to LES. 

The surface definitions are taken from the "NIST/NRC Parallel Panel Experiments" validation case, associated with \texttt{FDS6.7.6-810-ge59f90f-HEAD}. The sample material definitions are built from the parameter sets created within this work. Propane is used as fuel species for the gas burner in the simulation. This differs to the FDS validation setup "NIST/NRC Parallel Panel Experiments", where the combustion reaction for MMA is used for the burner and the sample. Note, the respective case was updated recently for FDS~6.8 and up.
In contrast to the experiments, during the simulation the gas burner is kept at a continuous heat release of 60~kW throughout, releasing a mass flux of about 0.00732~kg/(m²~s). Shutting the burner off earlier leads to fire extinction relatively fast, see discussion in section~\ref{subsec:discussion_realscale}.

\subsection{Results}
\label{subsec:results_realscale}

\begin{figure*}[h!]%
    \centering
    \subfloat[\centering Fluid cell resolution: C5.\label{fig:PP_BestHRR_20mm_RF30}]{{\includegraphics[width=0.75\columnwidth]{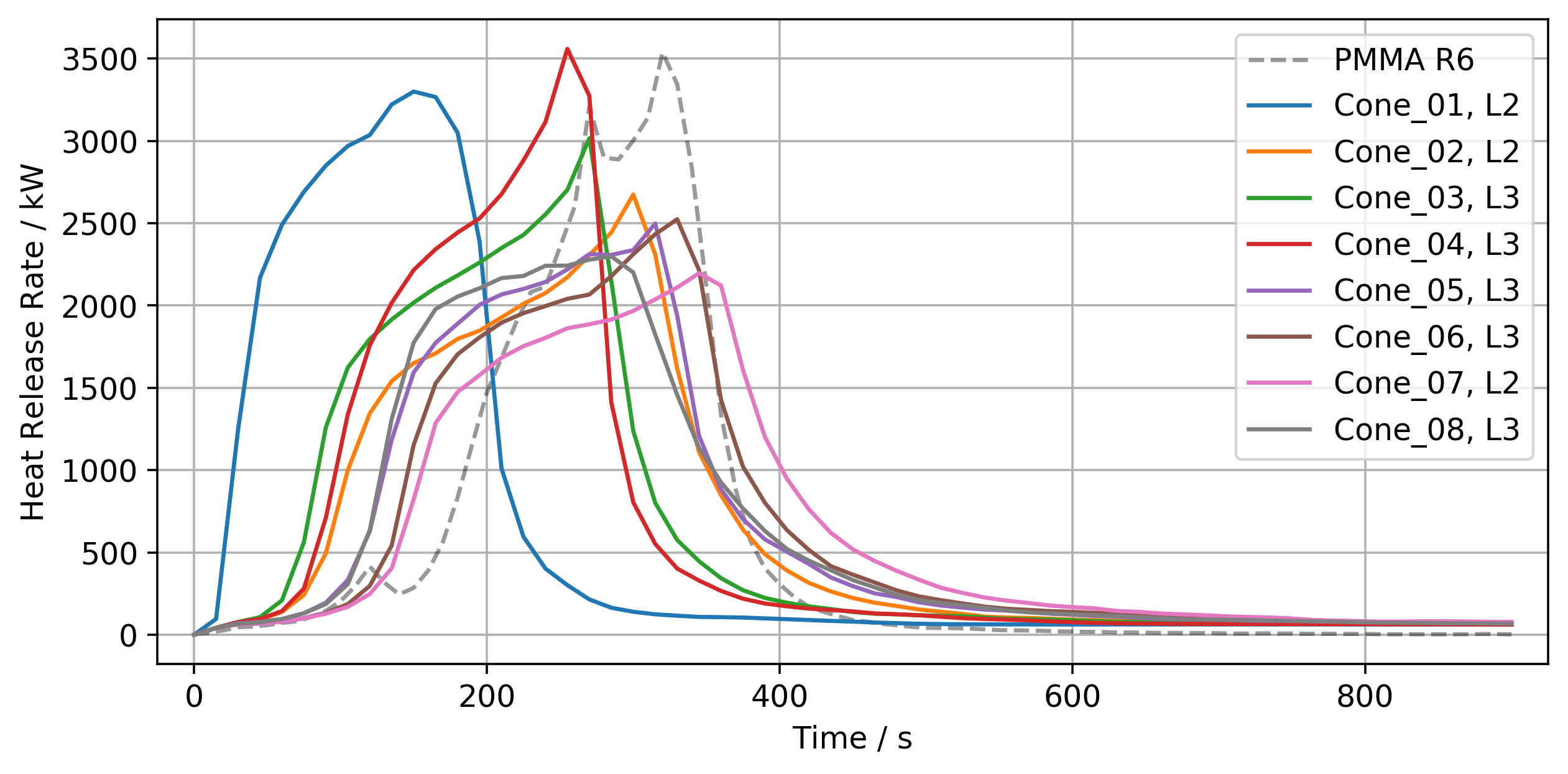} }}

    \subfloat[\centering Fluid cell resolution: C3.\label{fig:PP_BestHRR_50mm_RF30_C3}]{{\includegraphics[width=0.75\columnwidth]{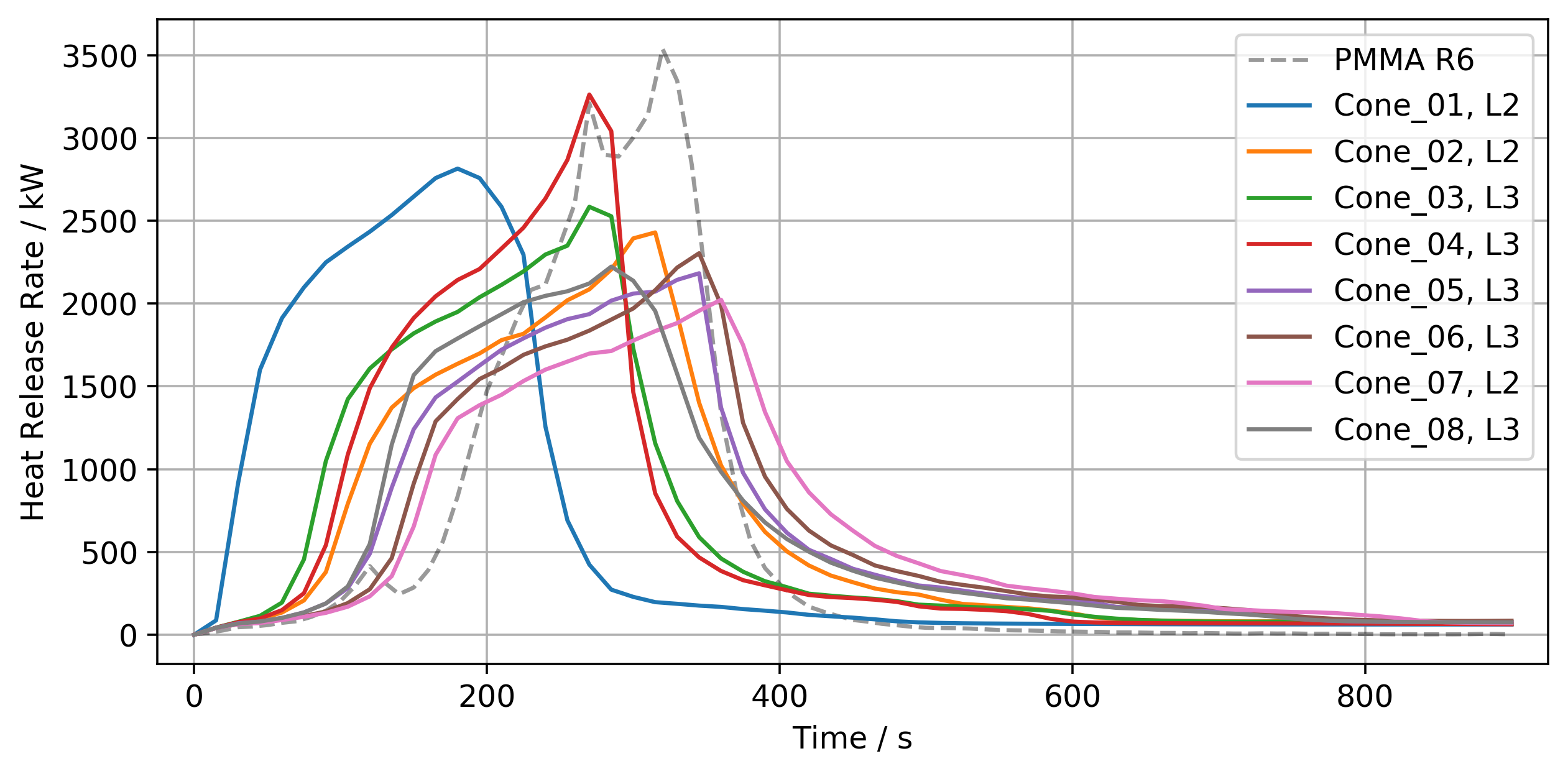} }}

    \subfloat[\centering Fluid cell resolution: C2.\label{fig:PP_BestHRR_50mm_RF30_C2}]{{\includegraphics[width=0.75\columnwidth]{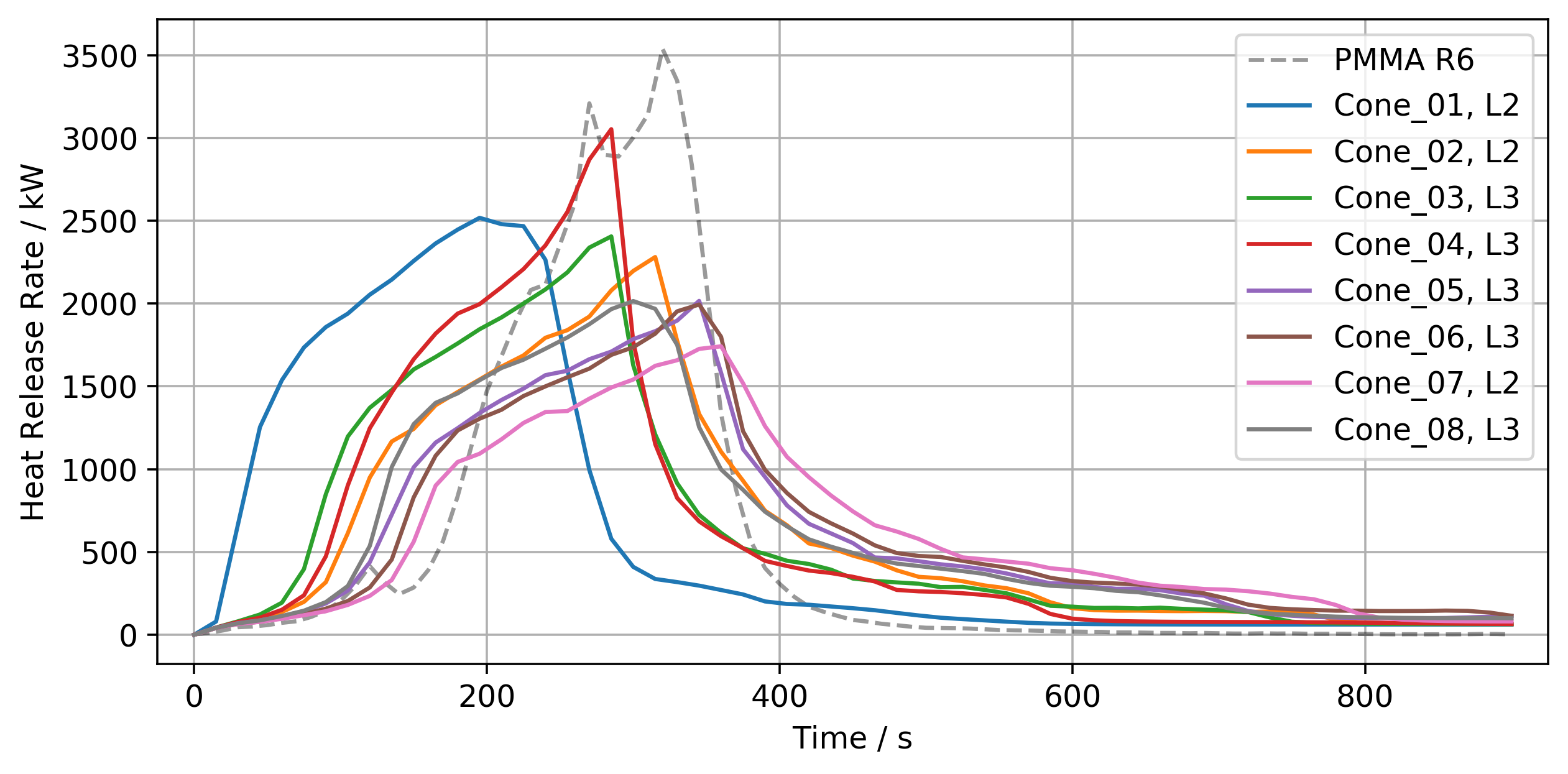} }}
    
    \caption{Comparison between experiment and simulation of the parallel panel setup for different fluid cell sizes.}
    \label{fig:PPSimBestParaHRR_MESH_B_20mm_Summary}%
\end{figure*}

The heat release rates in the parallel panel simulation setups are presented in figure~\ref{fig:PPSimBestParaHRR_MESH_B_20mm_Summary}.

\begin{figure}[h]%
    \centering
    \subfloat[\centering Time to reach 1~MW.\label{fig:PP_BaseAssessment_1MW}]{{\includegraphics[width=0.45\columnwidth]{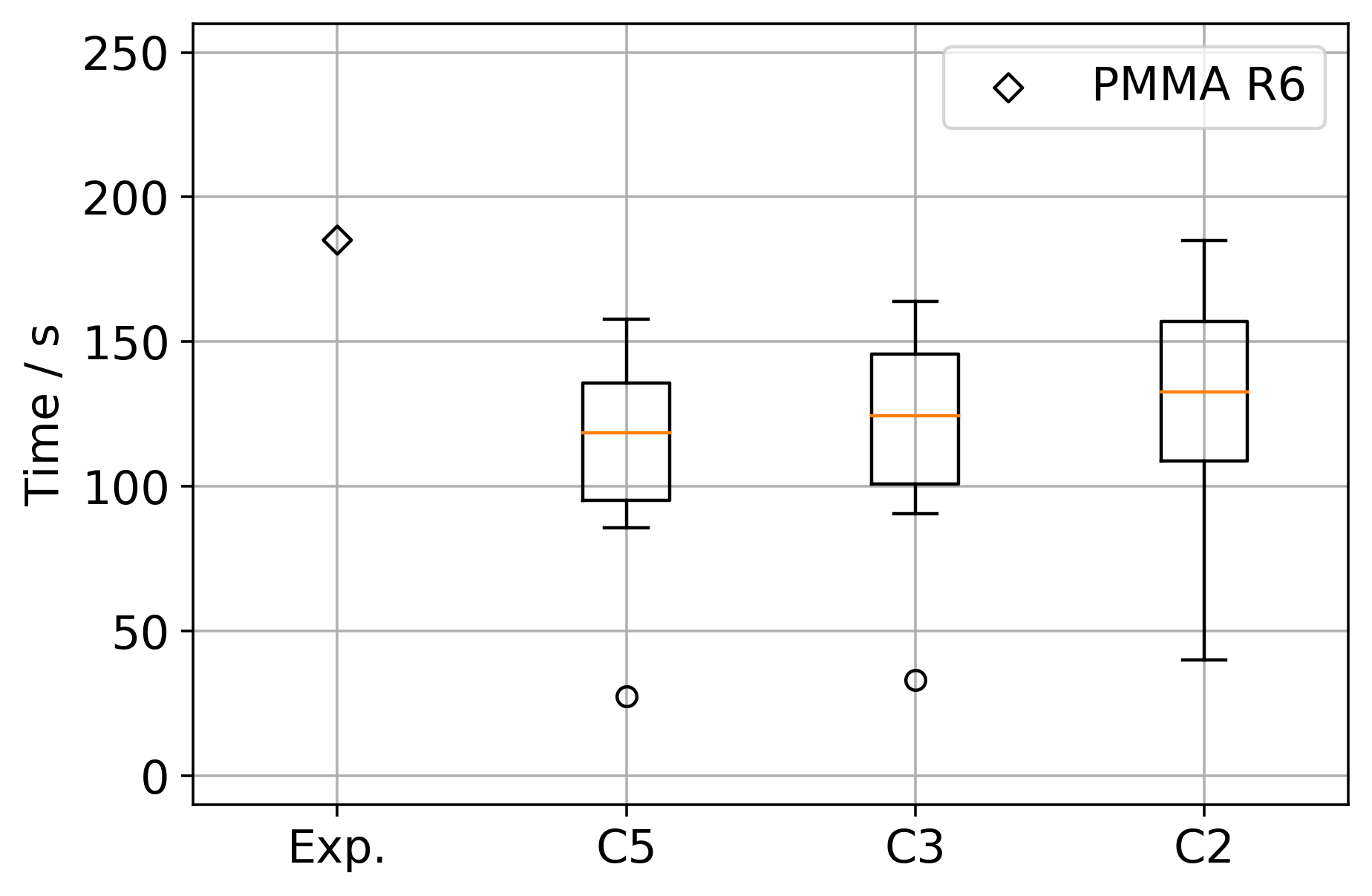} }}%
    \qquad
    \subfloat[\centering Peak heat release rate.\label{fig:PP_BaseAssessment_Peak}]{{\includegraphics[width=0.45\columnwidth]{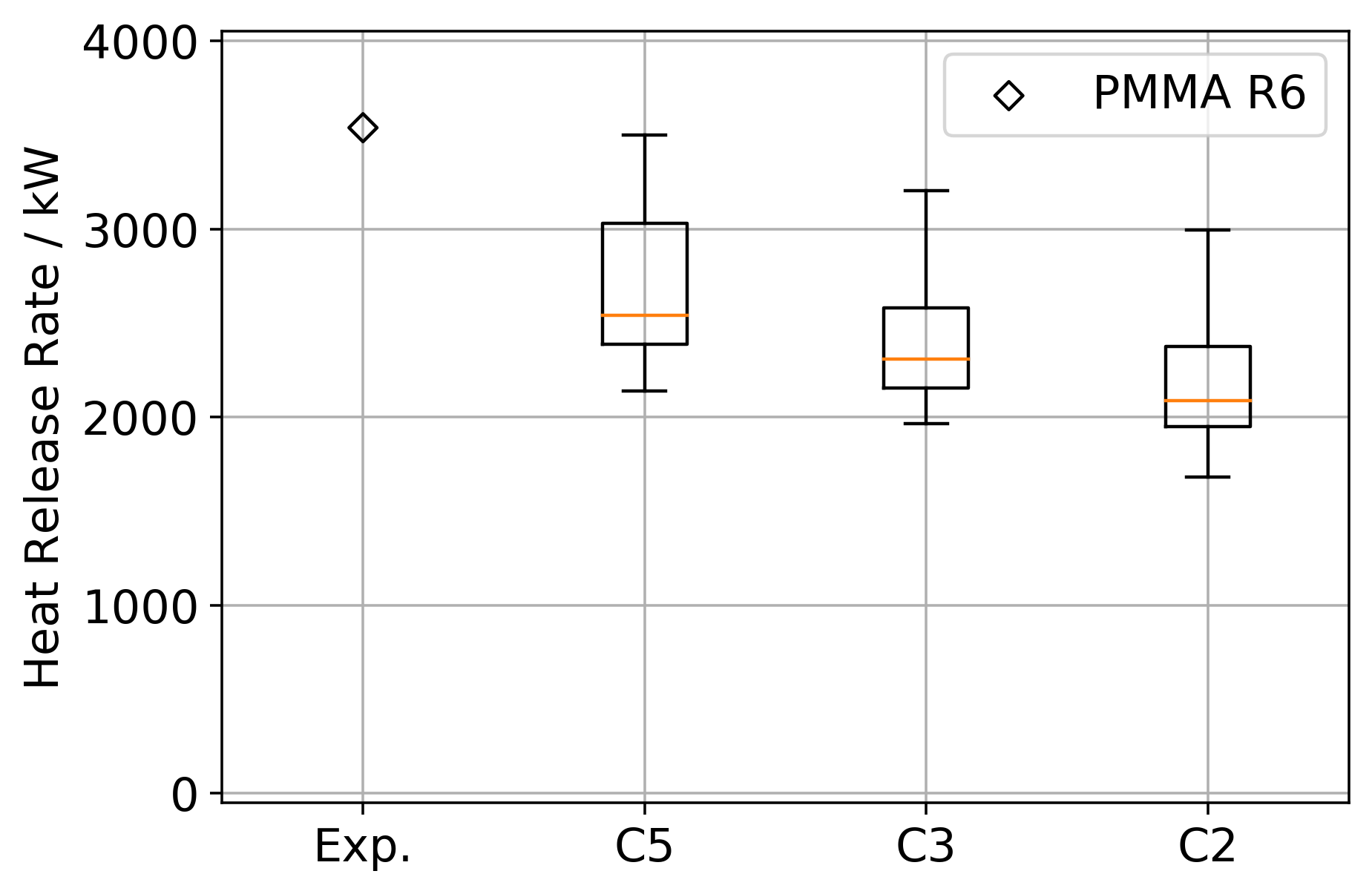} }}%
    
    \caption{Peak energy release and time to reach 1~MW in parallel panel simulation setup, using the best parameter sets from IMP. Comparison between experiment (Exp) and simulation data with different radiative fractions (RF) for the gas burner reaction.
    }%
    \label{fig:PP_BaseAssessment}%
\end{figure}

\begin{figure}[h]%
    \centering
    \includegraphics[width=0.45\columnwidth]{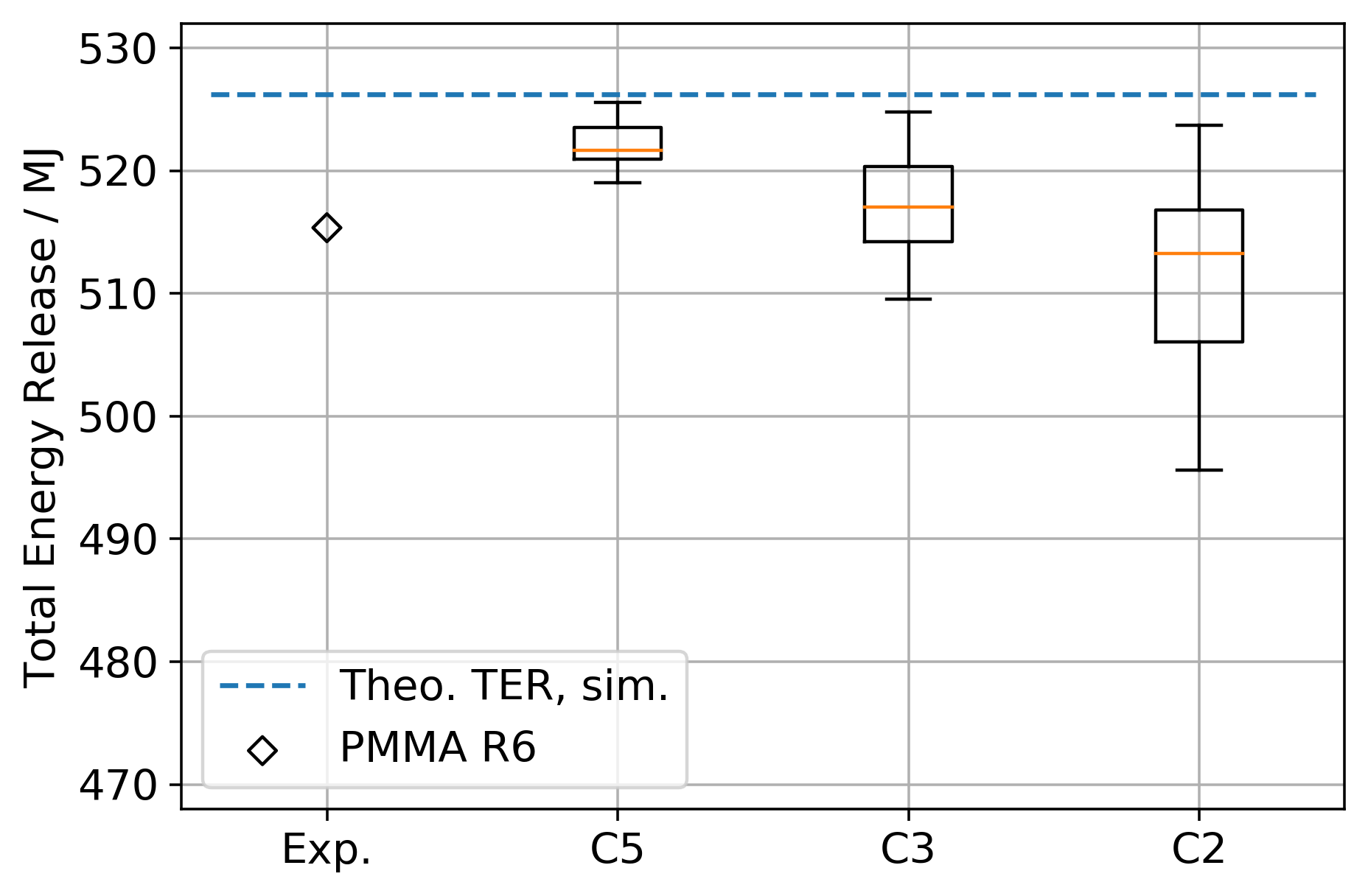}
    
    \caption{Total energy release of the best parameter set in parallel panel test for all IMP setups. Comparison between experiment (Exp) and simulation data with different fluid cell sizes. Simulation response with gas burner reaction of propane. Dashed line indicates theoretical total energy release of the sample in the simulation.}
    \label{fig:PP_TER}
\end{figure}

For the smallest fluid cells (C5, figure~\ref{fig:PP_BestHRR_20mm_RF30}), the peaks are overall narrower and taller compared to the largest cells  (C2, figure~\ref{fig:PP_BestHRR_50mm_RF30_C2}). This is emphasised by comparing the peak energy release, see figure~\ref{fig:PP_BaseAssessment_Peak}. In the simulation the fire develops overall faster compared to the experiment emphasised by the time a HRR of 1~MW is reached, see figure~\ref{fig:PP_BaseAssessment_1MW}. Larger cells slow the development down slightly. 
In general, faster fire development leads to higher energy release, see figure~\ref{fig:PP_TER}. The total energy release (TER) of all parallel panel simulations with different radiative fractions of the propane combustion reaction is provided in figure~\ref{fig:PP_TER_Complete} in the appendix.

\subsection{Discussion}
\label{subsec:discussion_realscale}

With the real-scale setup, the performance of the best parameter sets of the different IMPs (table~\ref{tab:SimpleConeIMP_BestPara}) is assessed. 

The fire development in the simulation is faster than the experiment for all parameter sets and fluid cell sizes, see figure~\ref{fig:PPSimBestParaHRR_MESH_B_20mm_Summary}. This could be related to the faster ramp-up of the burner in the simulation. It takes 10~s to reach the desired heat release, compared to the 60~s to~80~s in the experiment. Also, the burner is kept burning throughout the simulation, because the fire would extinguish otherwise (see below). Shutting the burner off in the experiment leads to a visible delay in fire development, see figure~\ref{fig:PP_Exp_Data} between about 100~s to 200~s. With larger fluid cells the overall fire development is more drawn out. This could be related to a poorer resolution in the radiation field, since for example the radiation angles are not adjusted, using the FDS default of 104 solid angles. This is also reflected in the TER, indicating that less sample material is consumed than in the experiment, see figure~\ref{fig:PP_TER}. 
Overall, smaller fluid cell sizes, specifically C5, lead to HRR peak shapes with closer resemblance of the experiment data. With the chosen model settings here, the fluid cell resolution should be C5 or higher. Throughout all parallel panel simulations, Cone\_07 performs the worst. It uses the largest fluid cell size during the IMP. However, its performance may also be influenced by its IMP target, i.e. using the HRR and the back face temperature. This should be investigated in more detail in future work.

The peak heat release is about 30\% to 40\% lower in the simulation, depending on fluid cell size and parameter set, see figure~\ref{fig:PP_BaseAssessment_Peak}. A notable exception is the parameter set of Cone\_04. Its performance stands out, by most closely resembling the shape of experiment "PMMA R6" and reaching a similar peak HRR. This behaviour seems to be associated to neglecting the back face temperature as IMP target in the simplified cone calorimeter setup. In future work, it is worth to look in more detail at Cone\_04. Removing the back face temperature constraint seems to be beneficial for the real-scale and may be improved with better chosen temperature values for the ramps.

\begin{figure}[h]
    \centering
    \includegraphics[width=0.45\columnwidth]{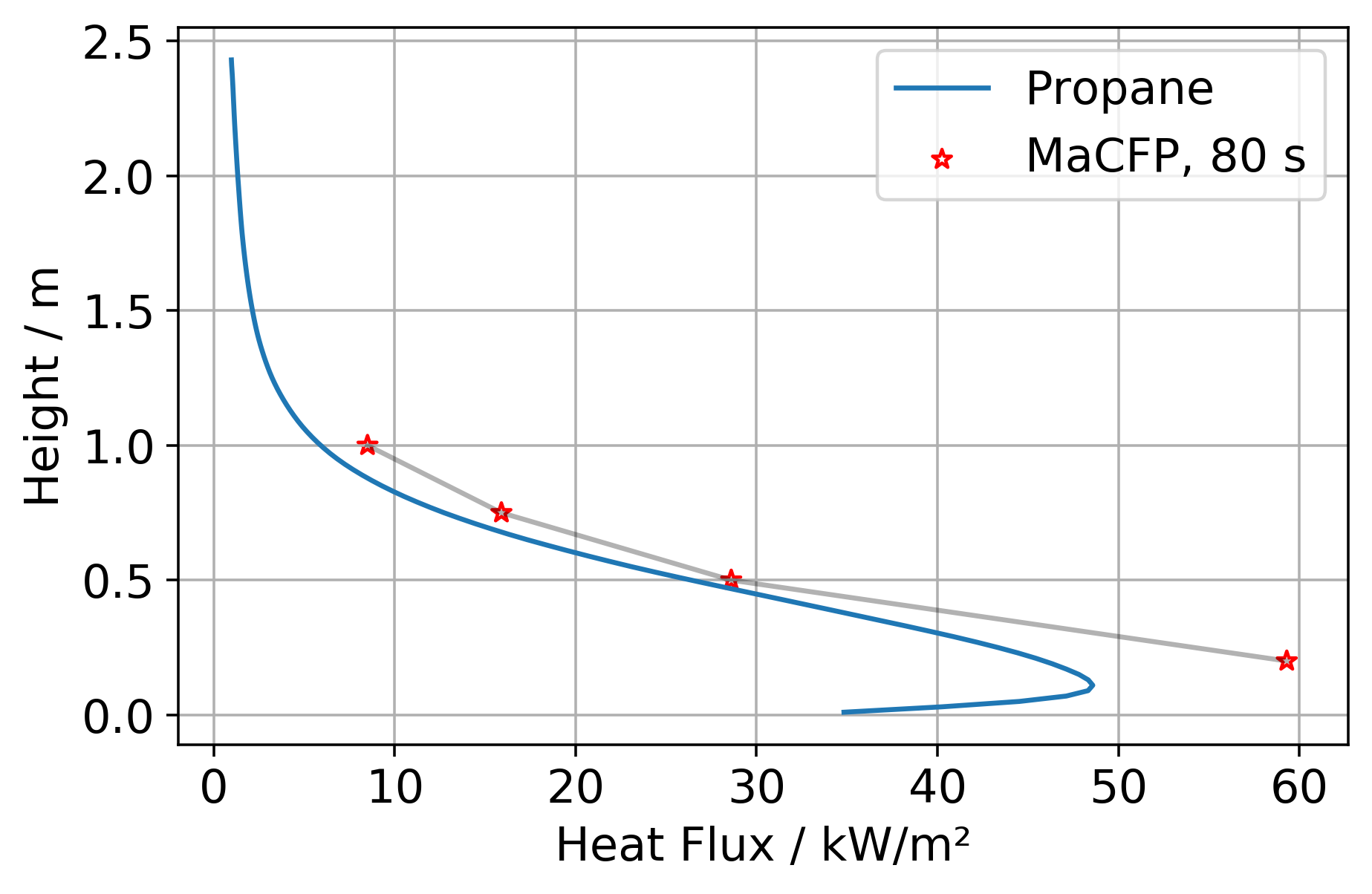}
    \caption{Centre line heat flux, from propane gas burner to empty panel.}
    \label{fig:PP_Sim_EmptyPanel}
\end{figure}

\begin{figure*}[h]%
    \centering
    \subfloat[\centering Experiment~\cite{Leventon2022ParallelPanel}.\label{fig:PP_HeatFluxMapExp}]{{\includegraphics[width=0.45\columnwidth]{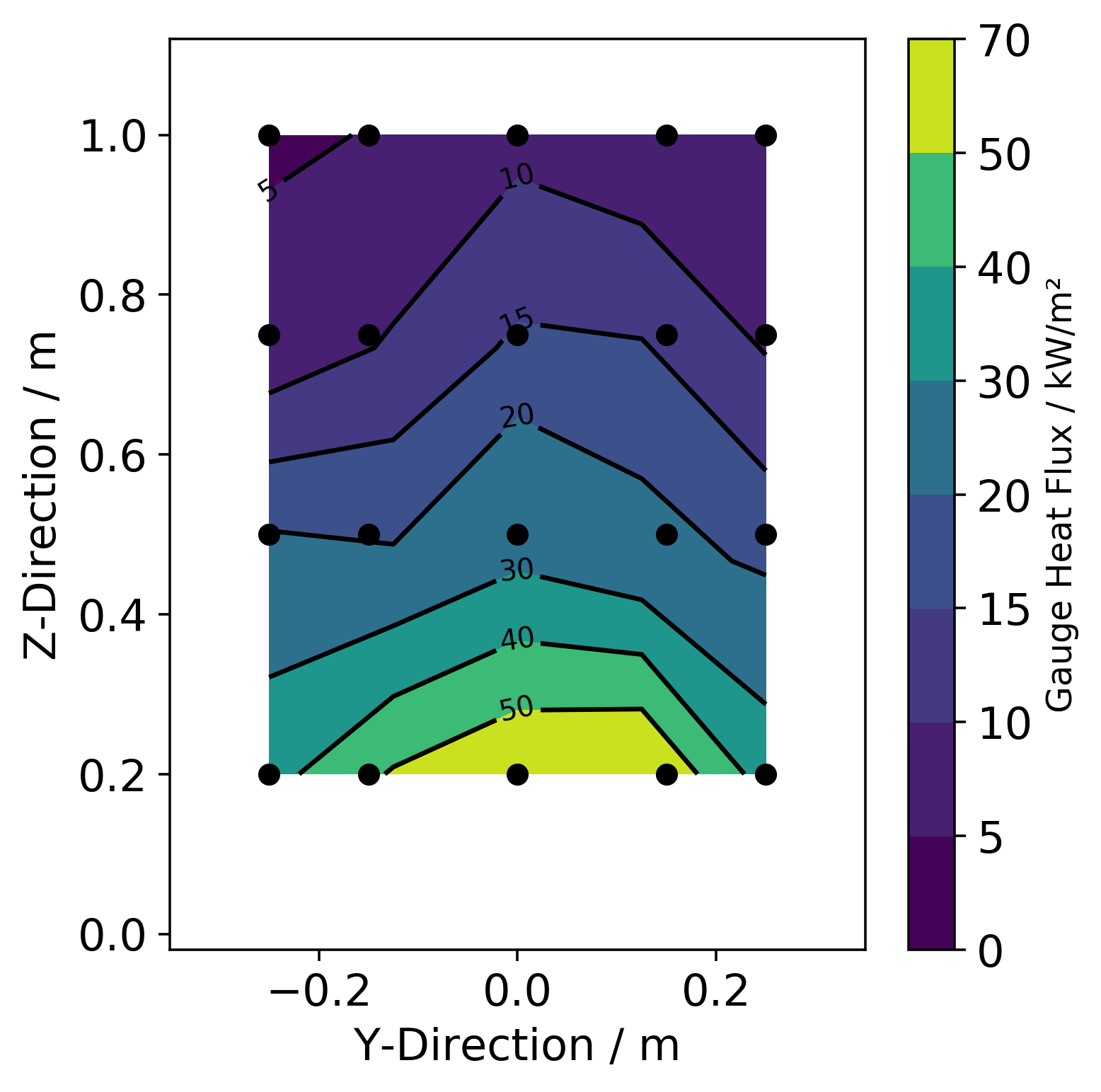} }}%
    \qquad
    \subfloat[\centering Simulation (C5).\label{fig:PP_HeatFluxMapSimC5}]{{\includegraphics[width=0.45\columnwidth]{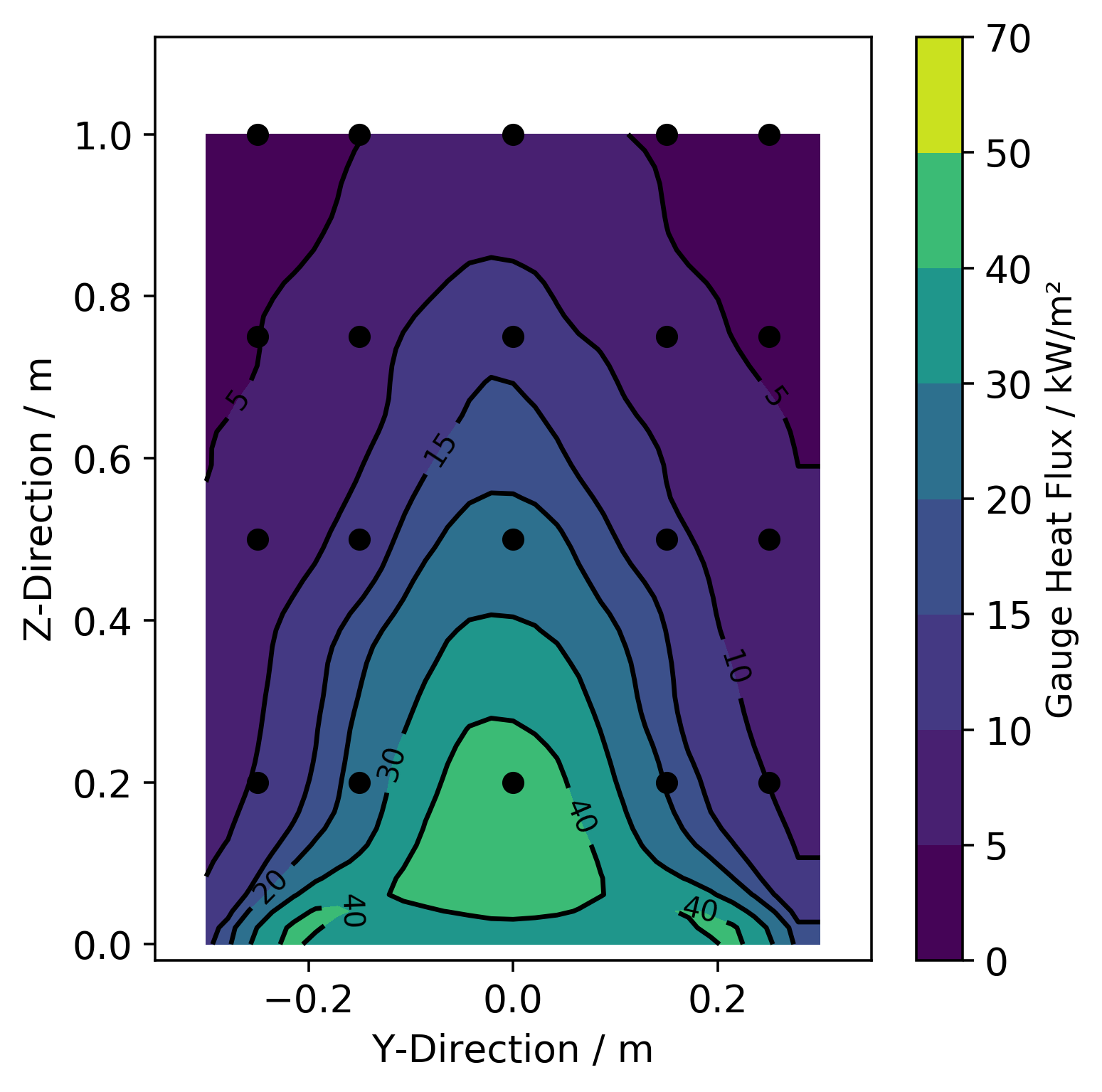} }}%
    \qquad
    \subfloat[\centering Simulation (C3).\label{fig:PP_HeatFluxMapSimC3}]{{\includegraphics[width=0.45\columnwidth]{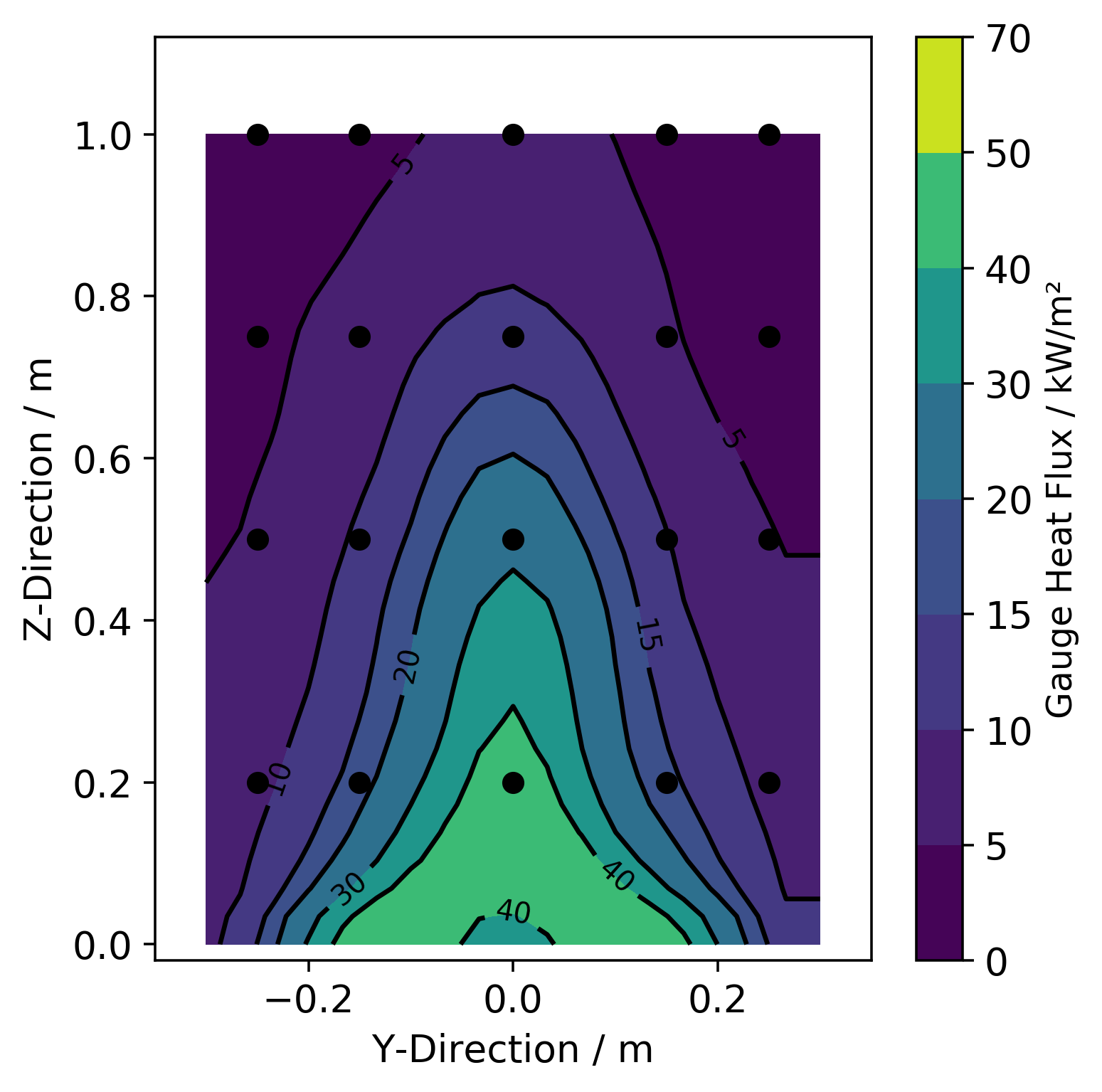} }}%
    \qquad
    \subfloat[\centering Simulation (C2).\label{fig:PP_HeatFluxMapSimC2}]{{\includegraphics[width=0.45\columnwidth]{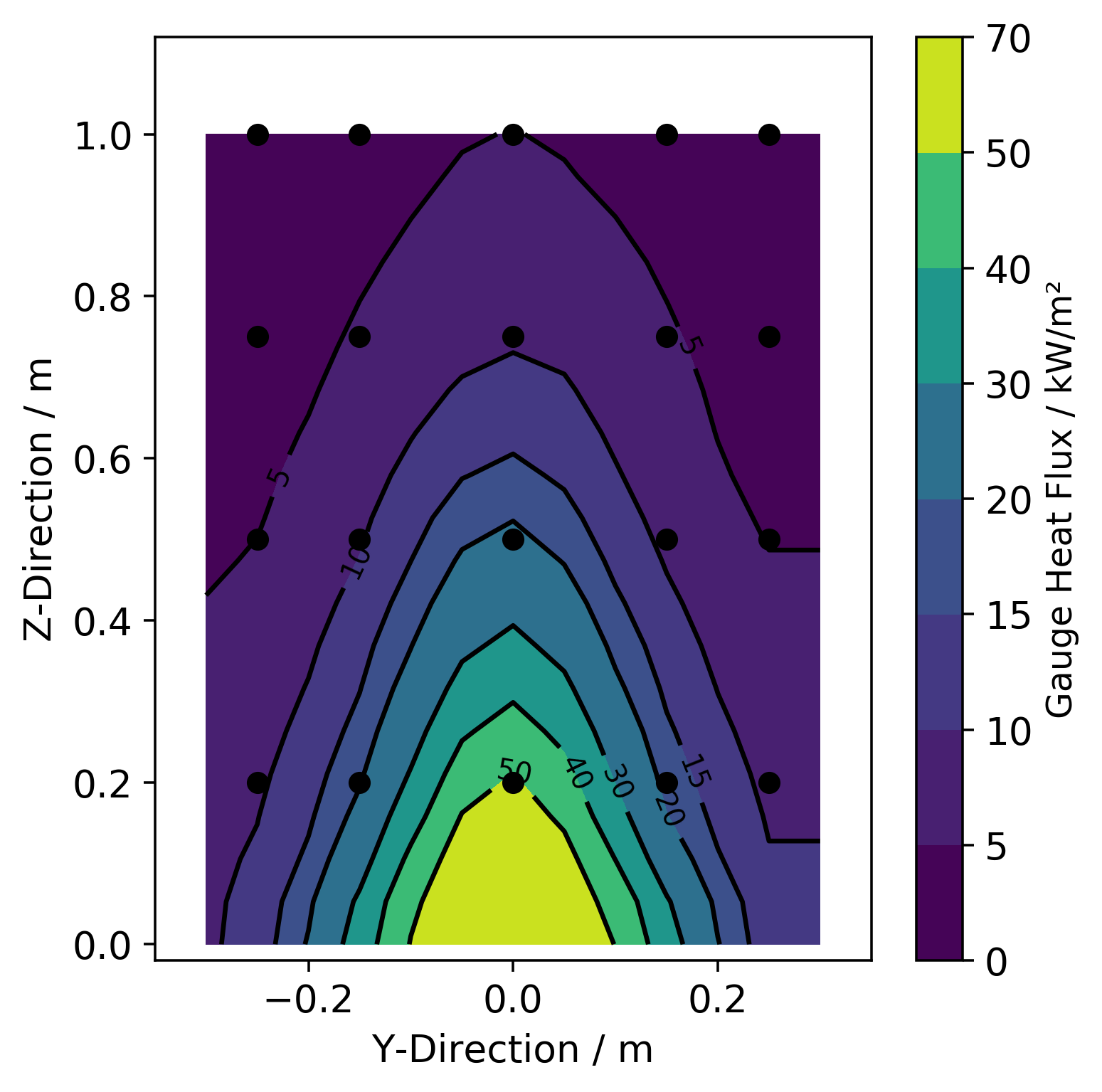} }}%
    
    \caption{Flame heat flux to an empty panel, for different fluid cell resolutions. The dots indicate locations of heat flux gauges during the experiment. Simulation data extracted from boundary (\texttt{GAUGE HEAT FLUX}). All show 20~s average during steady-state flaming.}%
    \label{fig:PP_HeatFluxMap}%
\end{figure*}

During the work presented here, only the flame heat flux along the vertical centre line of the empty panels is taken into account as starting condition (figure~\ref{fig:PP_Sim_EmptyPanel}). More detailed data was recently made available via MaCFP~\cite{Leventon2022ParallelPanel}. Unfortunately, this was too late to be taken into account for the conducted simulations here. Even though the simulations could not be changed anymore, the flame heat flux data across the lower part of the empty panels is compared against simulation responses for the different fluid cell resolutions, see figure~\ref{fig:PP_HeatFluxMap}. All four plots show flux data averaged over 20~s during the steady-state. The dots show the device locations during the experiment. From the simulation, the heat flux is extracted from the solid boundary directly (\texttt{GAUGE HEAT FLUX}). In the experiment, the flux is spread out nearly horizontally along the panels (figure~\ref{fig:PP_HeatFluxMapExp}). While in the simulation, it is more focused towards the centre line, which coincides with the location of the simulated flame (figure~\ref{fig:PP_HeatFluxMapSimC5}). With larger fluid cells, the heat flux is more concentrated at the lower centre line (figure~\ref{fig:PP_HeatFluxMapSimC2}). This indicates that the initial conditions are not reproduced well. It is not sufficient to simply match the heat flux to the vertical centre line of the panels. To properly assess the performance of the parameter sets, the burner itself needs to be accurately modelled first.
For future work, it is necessary to develop a more comprehensive representation of the gas burner setup. Further investigations should incorporate the impact of simulation parameters like soot production, cell sizes, parameters of the radiation model and material parameters of the burner top face and empty panels.

\begin{figure*}[h]%
    \centering
    \subfloat[\centering 500 kW HRR~\cite{Leventon2022ParallelPanel}.\label{fig:PP_BurnerCutOff_Exp} ]{{\includegraphics[width=1.77cm]{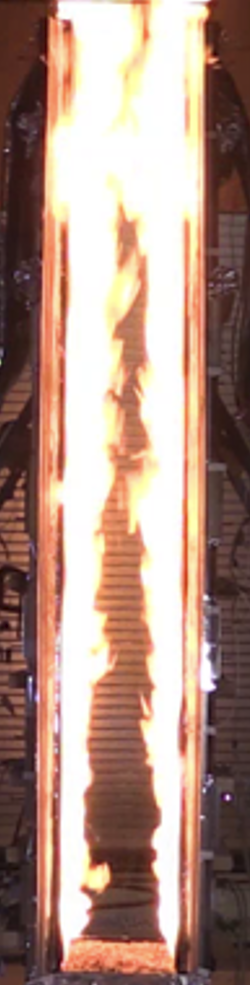} }}
    \hfill
    \subfloat[\centering 120 s, side.\label{fig:PP_BurnerCutOff_StartSide}]{{\includegraphics[width=1.93cm]{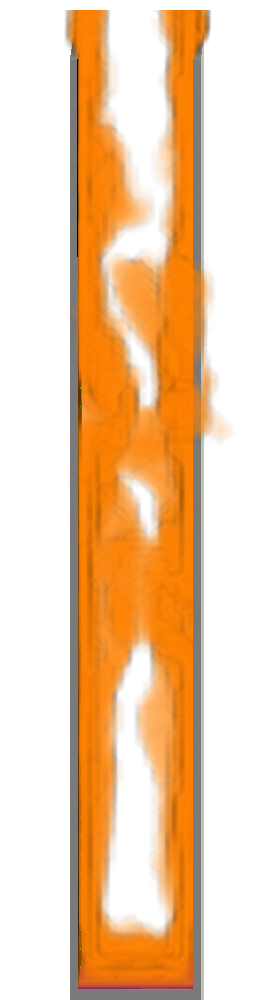} }}%
    \hfill
    \subfloat[\centering 120 s.]{{\includegraphics[width=3.07cm]{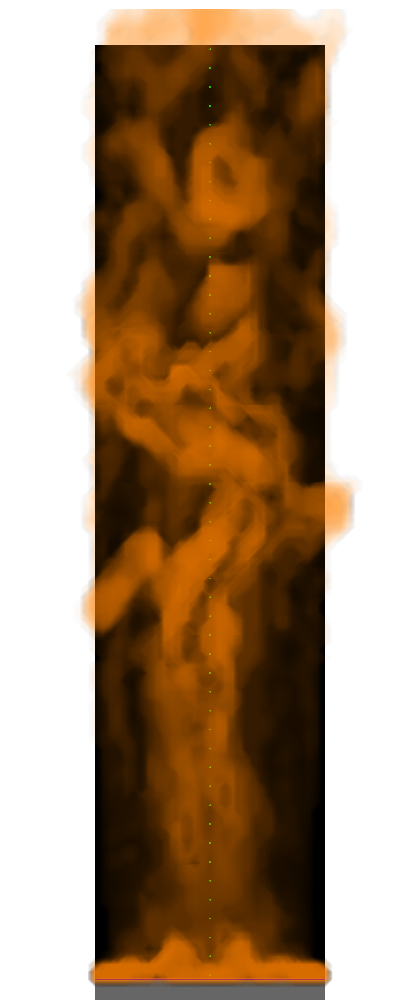} }}%
    \hfill
    \subfloat[\centering 140 s.]{{\includegraphics[width=3.07cm]{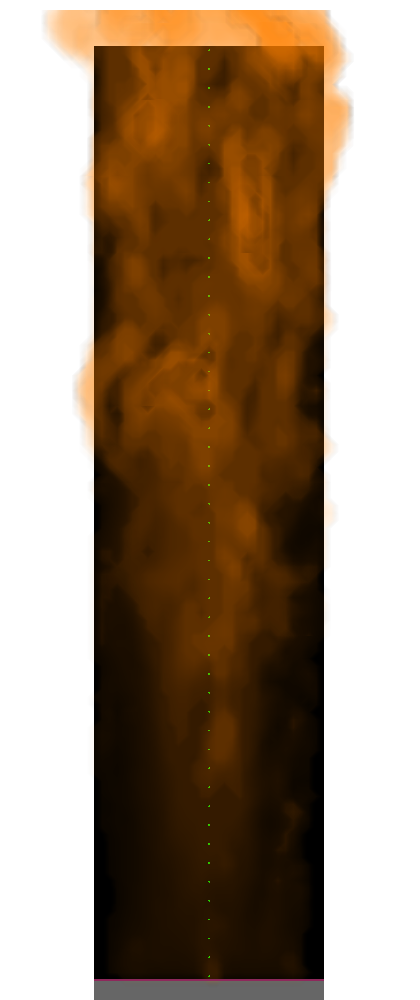} }}%
    \hfill
    \subfloat[\centering 160 s.]{{\includegraphics[width=3.07cm]{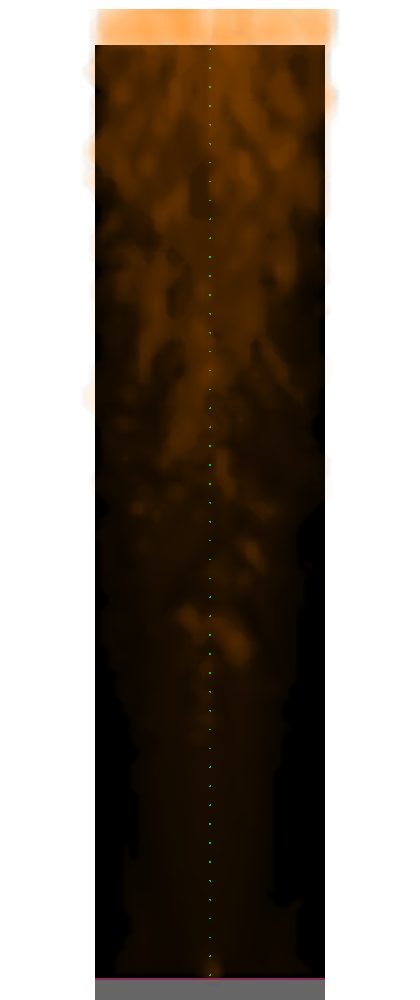} }}%
    \hfill
    \subfloat[\centering 180 s.]{{\includegraphics[width=3.07cm]{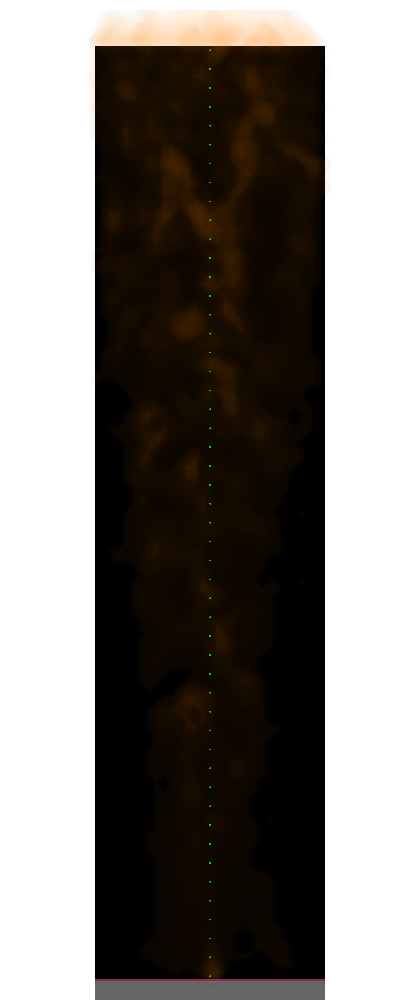} }}%
    
    \caption{Flame extinction after burner cut-off, at 120~s and ramp down over 6~s. Pilot flame of one cell in the centre at the bottom of each panel. (Cone\_03, L3, propane RF=0.15), Photograph from experiment (figure~\ref{fig:PP_BurnerCutOff_Exp}) at HRR of 500~kW, close to value of burner cutoff in the simulation (figures~\ref{fig:PP_BurnerCutOff_StartSide}), cropped out from~\cite{Leventon2022ParallelPanel}.}%
    \label{fig:PP_BurnerCutOff}%
\end{figure*}

Simulations with different burner cut-off times have been conducted to assess the capability to predict self-sustained fire development.
Burner cut-off times of 120~s and 220~s are chosen.
In all cases the fire is not able to recover, see figure~\ref{fig:PPSimBestParaHRR_BurnerOff}, even though the peak energy release in some cases is in excess of 1~MW. 
As an example, figure~\ref{fig:PP_BurnerCutOff} shows an image of the experiment "PMMA R6"~\cite{Leventon2022ParallelPanel} and an image series captured in Smokeview, covering 60~s after the burner is shut off. The flame region is flat against the panels, about one to two cells. This might interfere with the radiative heat transfer to cells below the lower edge of the flame, as well as to the sides. A closer look at the parameters of the FDS radiation model might be necessary, for example the path length or number of radiation angles. Smaller fluid cells might be beneficial as well, due to a better resolution of the resulting temperature distribution in the combustion region. It should be noted further, that the radiative fraction for combustion reaction of the PMMA pyroylsis products is treated here as unknown and the FDS default is used, i.e.\ 35~\%. 

Overall, there is a clear need to investigate the conditions necessary for self-sustained fire spread in the real-scale simulation and the parameter transfer from micro- to bench-scale and further from bench- to real-scale. In the given setups here, the model seems to struggle to provide meaningful energy transfer to the cells around the combustion reaction zone to keep the flame. Additionally, the impact of environmental effects on the experiment needs to be quantified or reduced, e.g.\ by conducting multiple repetitions of the same experimental setups.

\section{General Discussion}
\label{subsec:discussion_general}

All IMP results are able to reproduce the chosen micro-scale setups and cone calorimeter data well, see figure~\ref{fig:IMP_TGA_Combined},~\ref{fig:IMP_MCC_Combined} and figure~\ref{fig:IMPConeBestParaC3}. Yet, in the parallel panel simulation, differences become apparent, see figure~\ref{fig:PPSimBestParaHRR_MESH_B_20mm_Summary}. Just from the cone calorimeter simulation results alone, it is not obvious how the parameter sets perform in the real-scale. 
Preliminary investigations indicate that individual parameters of the material model may be differently sensitive to the simulation and experiment setup~\cite{Tassia_Sensitivity_SFPE2023}.
In the cone calorimeter setup fire spread is negligible, specifically for higher radiative heat fluxes. This might mask the behaviour of some parameters. For example, the emissivity has certainly a high impact on the received energy and therefore how the sample material heats up. However, in the cone calorimeter the sample receives a constant and high flux, which might be high enough to heat up the sample fast, regardless of which value for the emissivity is chosen.

In~\cite{CHAUDHARI2021109433PMMA_RoomCorner_Simulation} the authors discuss similar observations. They state that for their real-scale simulations, the prediction of the fire development is of lower accuracy, despite the good reproduction of the micro- and bench-scale seups~\cite{FIOLA2021103083PMMA_Pyrolysis_Model}. They state further that small changes in the specific heat capacity of the PMMA material have a large impact to the HRR in their model, specifically in the uncoupled setup. They mitigate this by adjusting the specific heat capacity, as well as the imprinted fractions of convective and radiative heat transfer for the individual surface elements.

The energy transfer to the sample has a significant impact on the fire development. Specifically for the cone calorimeter setup the thermal radiation is important. More detailed investigation of the impact of the radiation model parameters in this setup is necessary. This assessment should also take the performance in the real-scale simulation into account, to ensure that the model for both setups is the same. More care should be taken when setting up the gas burner simulation model. Using individual gas phase combustion reactions for burner and sample should allow to simulate the initial sample ignition more precisely, without compromising the overall sample behaviour.

In this work, the energy release is assumed to solely take place as a gas phase combustion reaction. This may be a sufficient model for PMMA. For other materials that show significant surface reactions, like wood, this assumption might not hold.

Furthermore, experimental campaigns should incorporate medium-scale setups that focus on self-sustained fire spread. This allows to test the model specifically on this aspect, which a cone calorimeter cannot provide due to its design and severe condition. It would also be useful to provide more information on the burners themselves. Specifically, their surface temperature development and emissivities over the course of the experiment.

\section{Conclusions}

In general, it seems attainable to simulate fire propagation in FDS, based on material parameters. 
Although the material parameter sets lead to similarly good representation in the micro- and bench-scale, they lead to significantly different results in the real-scale. 
It is highlighted, that the experiments at this scale introduce further modelling parameters, e.g. the characteristics of the ignition source. Thus, the system is not only dependent on the performance of the material parameters alone.

The reasons for the observed deviations in the real-scale simulations may be:

\begin{itemize}
    \item Compensation or trade-off effects during the inverse modelling due to the simplified simulation setups at the micro- and bench-scale.

    \item Not all necessary physical processes being captured well enough in the real-scale setup. This includes experimental as well as modelling aspects.

    \item Parameter transfer from small-scale to real-scale. The simulation setups used in the IMP may be differently sensitive to individual parameters compared to the real-scale setups. If the real-scale is sensitive to parameters which are insensitive in the small-scale, then the fundamental assumption of transferability is compromised.

\end{itemize}

Employing a gas mixture allows to capture the MCC and TGA experiments and prevents FDS from scaling the mass introduced into the fluid domain. Thus, it is a step towards more physical parameter sets. The higher fluid cell resolution in the simple cone setup, compared to similar studies, can account for uneven radiative heat flux and sample consumption. 

A clear limitation is the enormous computational cost and runtime for determining the material parameter sets. We are currently investigating different strategies to significantly reduce the runtime, by using different sampling methods and artificial-intelligence based approaches.

Overall, it seems clear that many parameters on all levels of this endeavour are important, and their influence needs more cohesive investigation. It is not sufficient to focus on the bench- and micro-scale experiments alone. Despite good performance during these simulations, it is not obvious how the parameters perform in the real-scale.
It is necessary to look into the whole chain of setups, to understand how well the parameters eventually translate over to the real-scale.
Furthermore, the impact of other model parameters, like the radiative fraction, or the radiation model in general, needs further investigation. Finally, the landscape of experimental data is fractured specifically for real-scale setups with the same sample material as in the smaller scales. 
Within these constrains, the parameter set should yield a response close to the observations in the experiments at all scales. This ultimately means the parameter set needs to compensate for the simulation model and experimental shortcomings, which can hardly be accomplished by a "physical" parameter set, thus it needs to be an "effective" representation.

\section*{Data Availability}
The experiment data is available from the MaCFP git repositories~\cite{macfp_matl_git, Leventon2022ParallelPanel}. 
The input for the inverse modelling, the results of the IMPs and the scripts used for data processing 
are provided in a (open access) Zenodo data repository~\cite{zenodo:ArticleDataset}. A video series on how the data processing and inverse modelling is setup and used is provided on YouTube~\cite{firesimandcoding_playlist}. 

\section*{Acknowledgements}
The authors thank Karen De Lannoye, for discussions on conducting micro- and small-scale experiments and the provided images. 
The authors thank Isaac Leventon, for discussions on conducting the parallel panel experiments. 
We gratefully acknowledge the computing time granted through JARA (project jjsc27) on the supercomputer JURECA \cite{JURECA} at Forschungszentrum Jülich and through the project on the CoBra-system, funded by the German Federal Ministry of Education and Research with the grant number 13N15497. This research was partially funded by the German Federal Ministry of Education and Research with the grant number 13N15497.

\section*{CRediT Authorship Contribution Statement}

\textbf{Tristan Hehnen:} conceptualisation, data curation, formal analysis, investigation, methodology, software, validation, visualisation, writing -- original draft preparation, writing -- review and editing

\textbf{Lukas Arnold:} conceptualisation, funding acquisition, methodology, project administration, resources, software,  supervision, validation, writing -- review and editing

\appendix

\section{Cone Calorimeter Simulation Setup}
\label{Appx:ConeGEOM}

\begin{figure}[h]
    \centering
    \includegraphics[width=0.55\columnwidth]{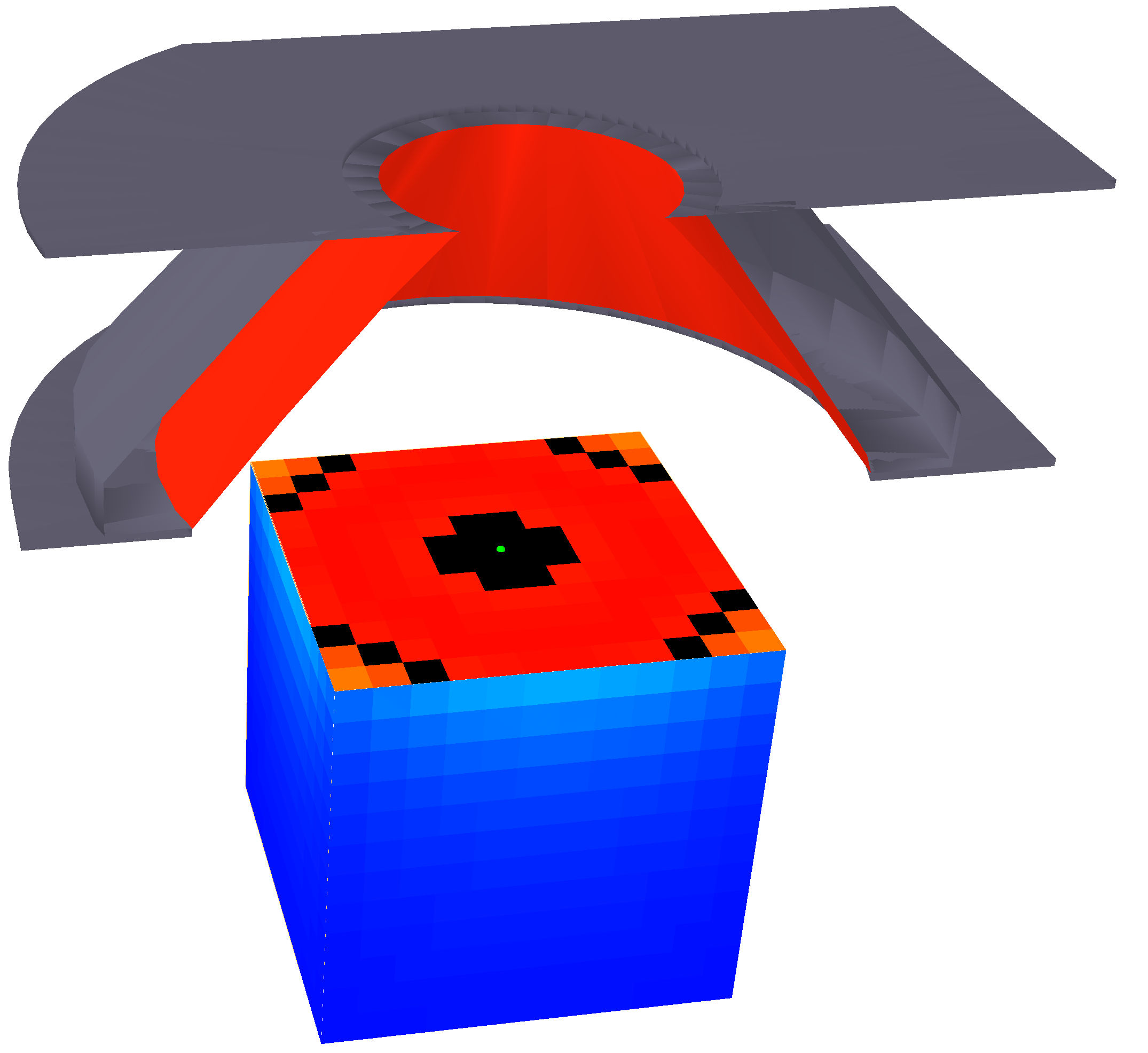}
    \caption{Geometrical model of the cone calorimeter heater, using the \texttt{GEOM} namelist. Heater surface idealised as a simple conical shape. Black areas on the sample surface receive a radiative heat flux of about 65~kW/m².
    \label{Apdx_SimpleConeMedium_GaugeFluxGEOM}}
\end{figure}

Based on Babrauskas'~\cite{BabrauskasConeCalorimeter} original report on the development of the cone calorimeter, a simplified geometrical representation of the heating element is created. The simplification is primarily focused on the heating coil, which is represented as a smooth conical surface and not as a wound wire, see figure~\ref{Apdx_SimpleConeMedium_GaugeFluxGEOM}. The fluid cell resolution was chosen, such that the sample surface (10 cm by 10 cm) was covered with 12 by 12 cells. The geometry itself was built in Blender, using the BlenderFDS addon by Emanuele Gissi.

\begin{figure}[h]
    \centering
    \includegraphics[width=0.55\columnwidth]{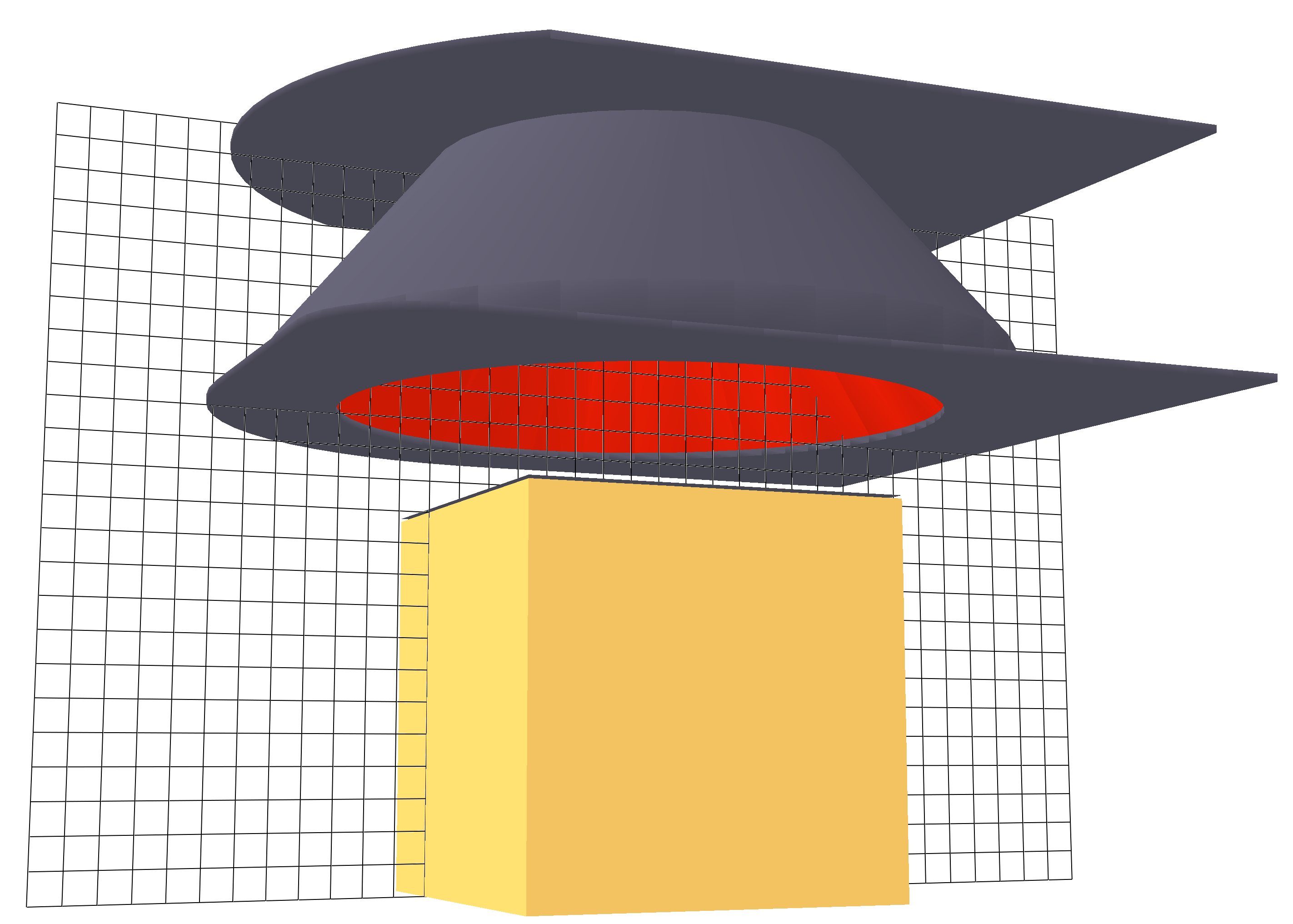}
    \caption{Geometrical model of the cone calorimeter heater, using the \texttt{GEOM} namelist. Sample surface resolved as C12.
    \label{Apdx_SimpleConeMedium_GEOM}}
\end{figure}

The heater calibration procedure is mimicked in the simulation, to determine the parameters of the boundary condition (\texttt{SURF}) of the heater. A device (\texttt{DEVC}) with the \texttt{GAUGE HEAT FLUX GAS} quantity is located in the centre of where the sample surface is supposed to be during the test. A Python script is used to automatically find an emissivity value for the heater boundary condition that leads to the 65 kW/m² at the device. Refer to the \texttt{FindTMP\_FRONT.py} in the \texttt{ConeRadiationAssessment} directory of the data set~\cite{zenodo:ArticleDataset}. The heater temperature is set, based on the temperature reported by Babrauskas, but linearly interpolated between the two enveloping values to get to the desired radiative flux. The simulation includes the gas phase, thus interactions between the radiation and the air are taken into account. The radiative flux is assessed over 20~s, after reaching a quasi-steady state. It is averaged over this time span.

Afterwards, a simulation is conducted in which an obstruction (\texttt{OBST}) is introduced to represent the sample and its holder. There is a distance of 25~mm between the sample surface and the bottom of the cone heater assembly. From the top boundary of the obstruction the radiative heat flux is recorded (\texttt{GAUGE HEAT FLUX}). Per cell, it is averaged over 20~s after reaching a quasi-steady state, same as in the previous step.

\section{Computational Domain Layouts}
\label{Appx:MeshLayout}




\begin{figure}[h]
    \centering
    \subfloat[\centering  Resolution: 5~cm fluid cells (C2).\label{fig:app_MeshLayoutC2}]{{\includegraphics[width=0.45\columnwidth]{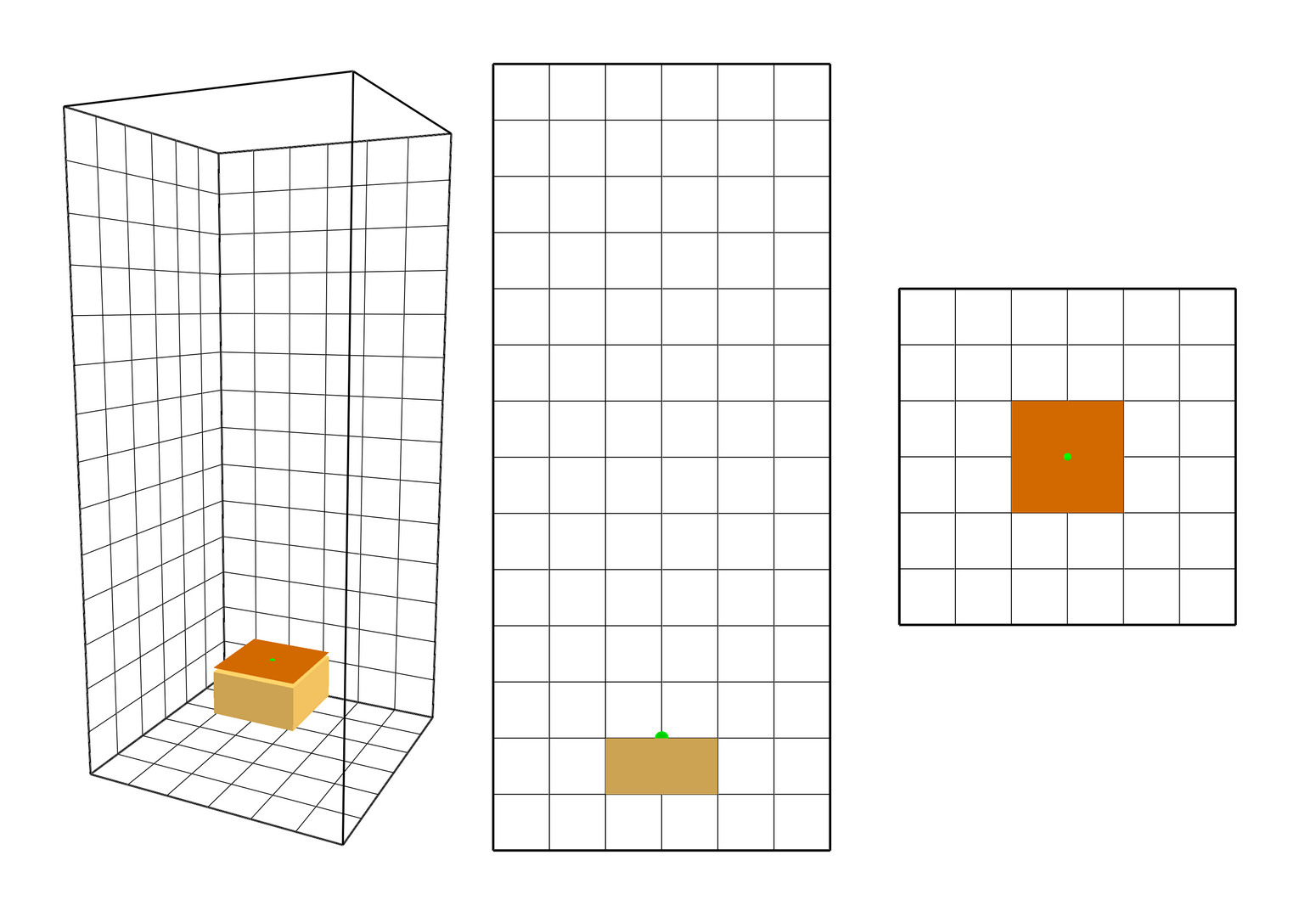} }}%
    \qquad
    \subfloat[\centering  Resolution: 3.3~cm fluid cells (C3).\label{fig:app_MeshLayoutC3}]{{\includegraphics[width=0.45\columnwidth]{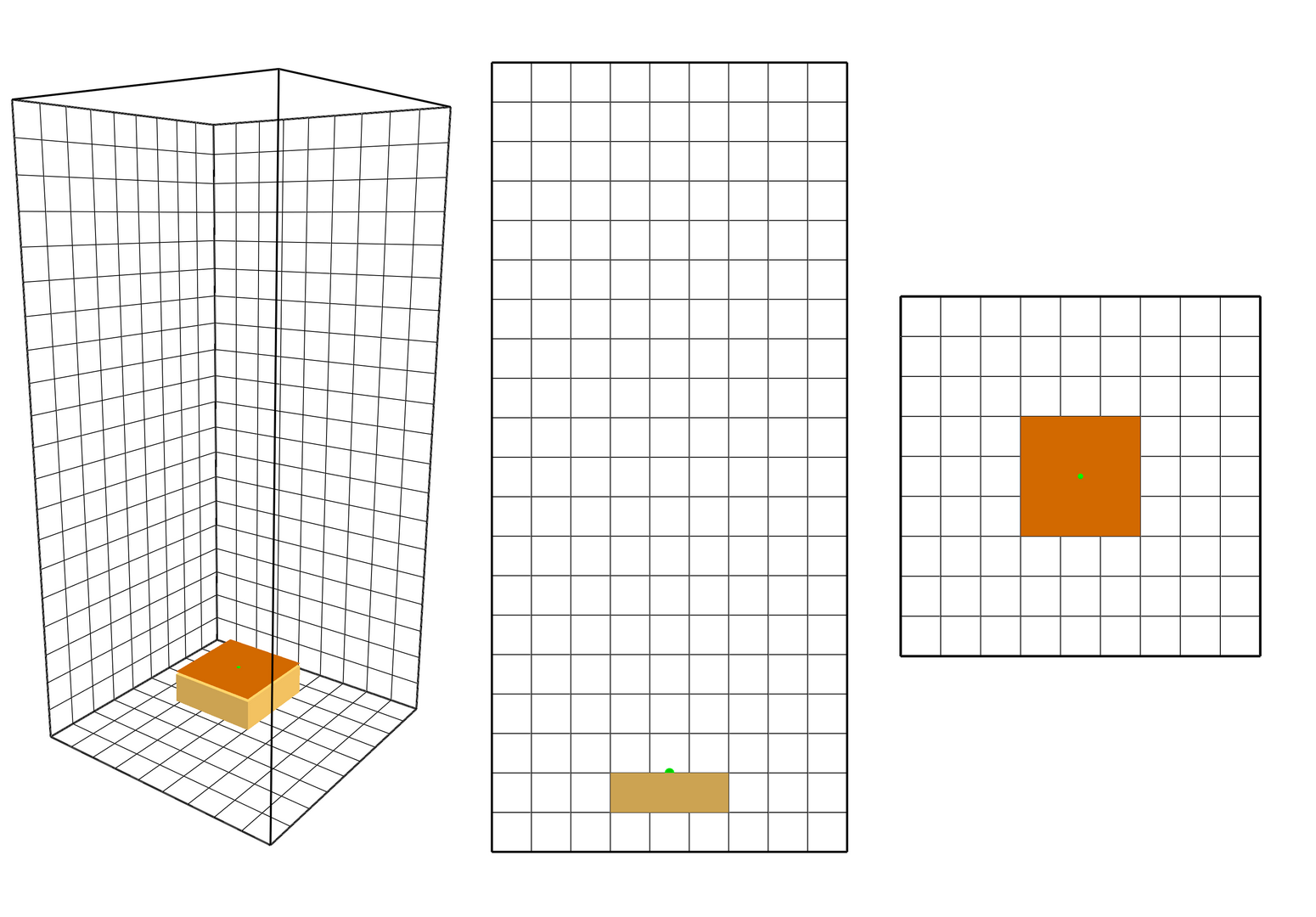} }}%
    \qquad
    \subfloat[\centering Resolution: 2~cm fluid cells (C5).\label{fig:app_MeshLayoutC5}]{{\includegraphics[width=0.45\columnwidth]{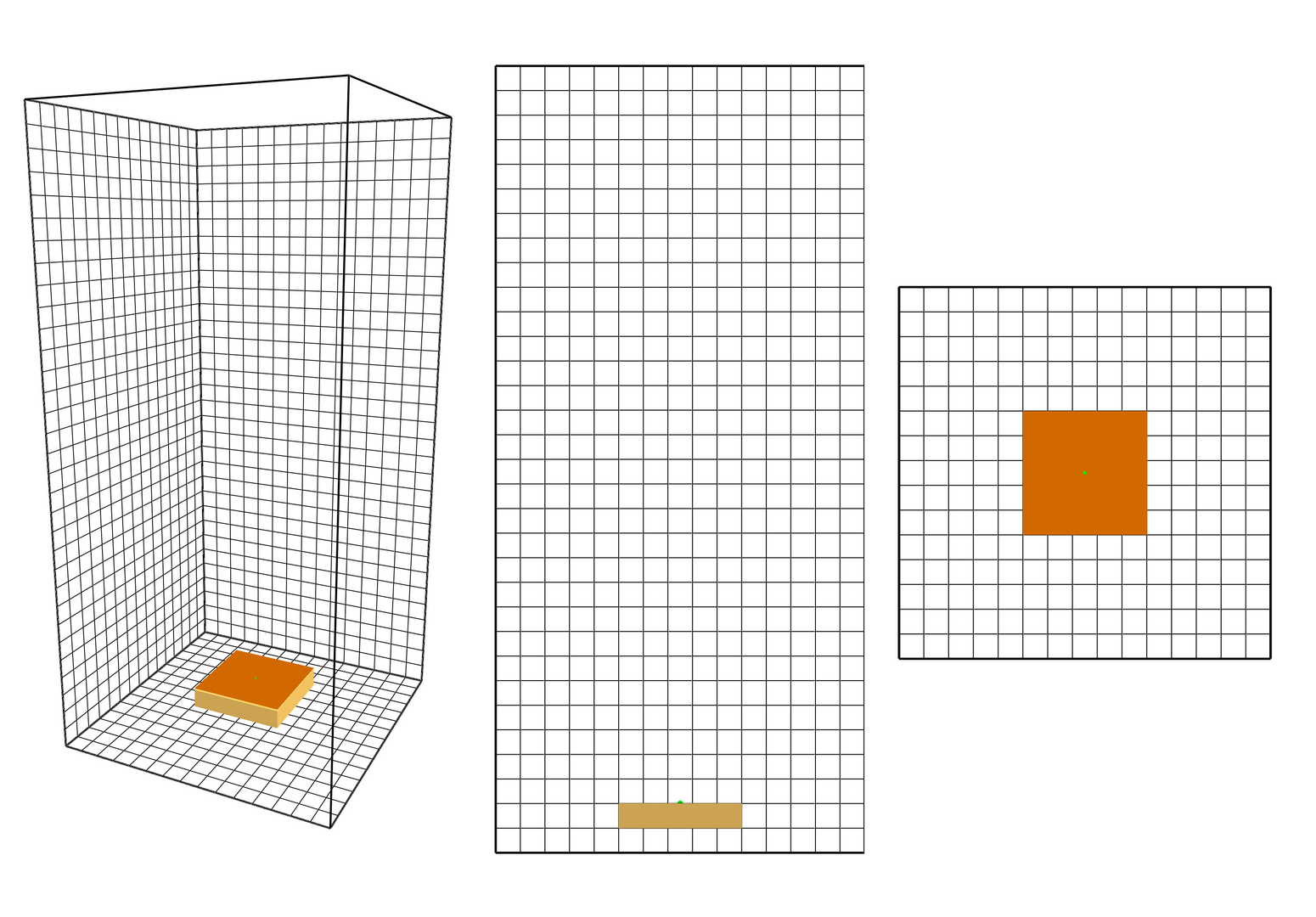} }}%
    
    \caption{Computational domain layout for the simplified cone calorimeter simulation setups. Each setup uses a single domain.}%
    \label{fig:app_MeshLayout}%
\end{figure}

\begin{figure}[h]
    \centering
    \includegraphics[width=0.7\columnwidth]{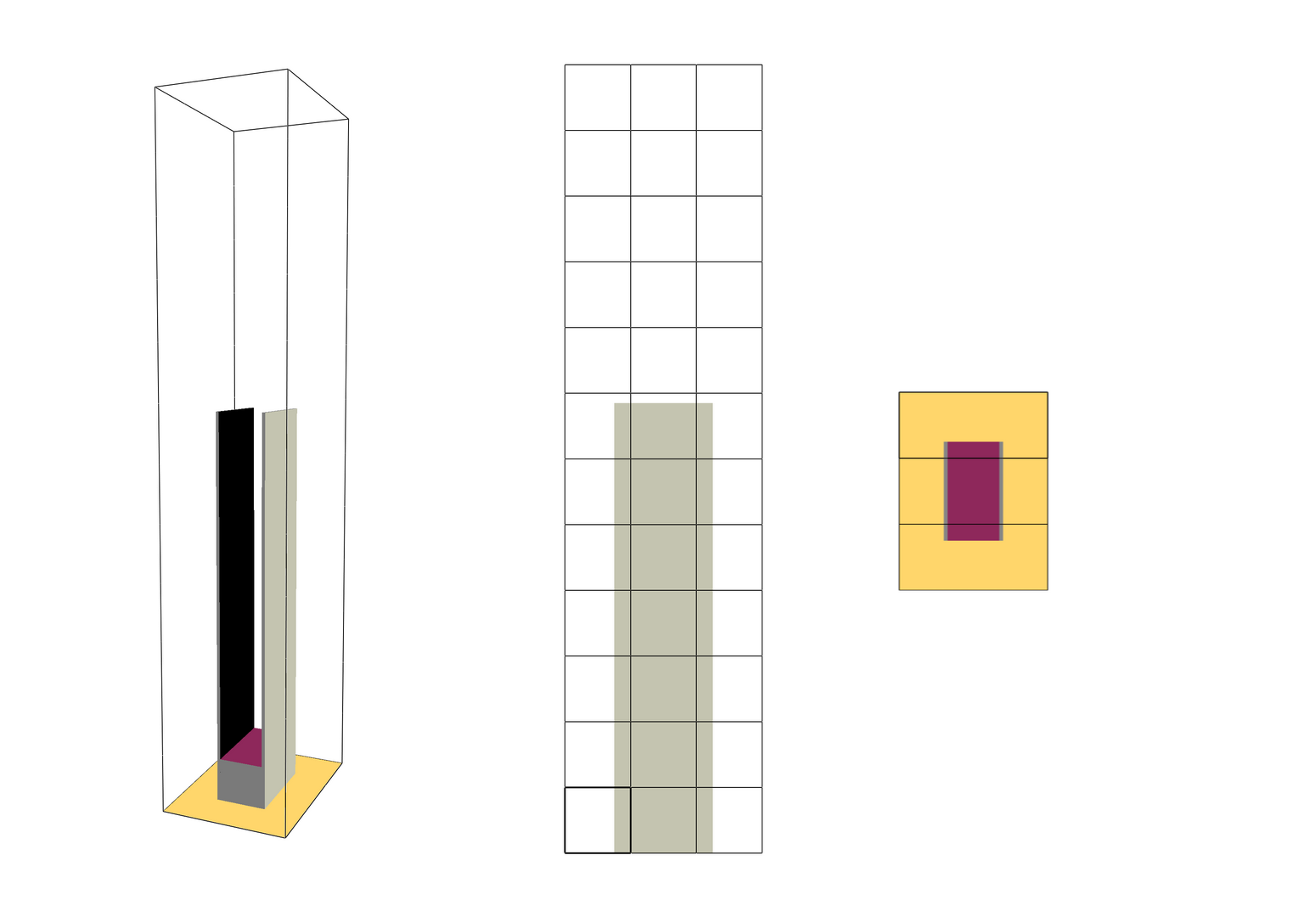}
    \caption{Computational domain layout for the parallel panel simulation setups. Only sub-domains are shown for clarity.
    \label{fig:app_MeshLayoutPP}}
\end{figure}

\section{Complex Chemistry}
\label{App:ComplexChem}

During the simple chemistry FDS calculates the stoichiometry itself and provides the results in the \texttt{CHID.out} file. This, however, only works if FDS is used regularly, meaning not with \texttt{TGA\_ANALYSIS}. A Python script was designed that automatically finds the recent best parameter set (lowest RMSE, see section~\ref{subsec:IMP}) from the micro-scale IMP and creates a new FDS input file to generate a respective \texttt{CHID.out} file. After manual execution of this simulation the script can extract the needed stoichiometry information from the \texttt{CHID.out} file, build the appropriate input lines and write them to a new FDS input file. These steps are handled in the \texttt{GetChemicalReaction.ipynb} notebook, which can be found in the data repository~\cite{zenodo:ArticleDataset}. To ensure consistency, the repetition information of the best parameter set is written to the respective FDS input files as well. For the parallel panel simulations this information can be copied over manually.

\section{Mass Losses and Flame Heights}
\label{App:FlameHeight}

\begin{figure}[h]
    \centering
    \subfloat[\centering Solid phase simulation.
    \label{fig:FlameHeight_MassLoss_Solid}]{{\includegraphics[width=0.45\columnwidth]{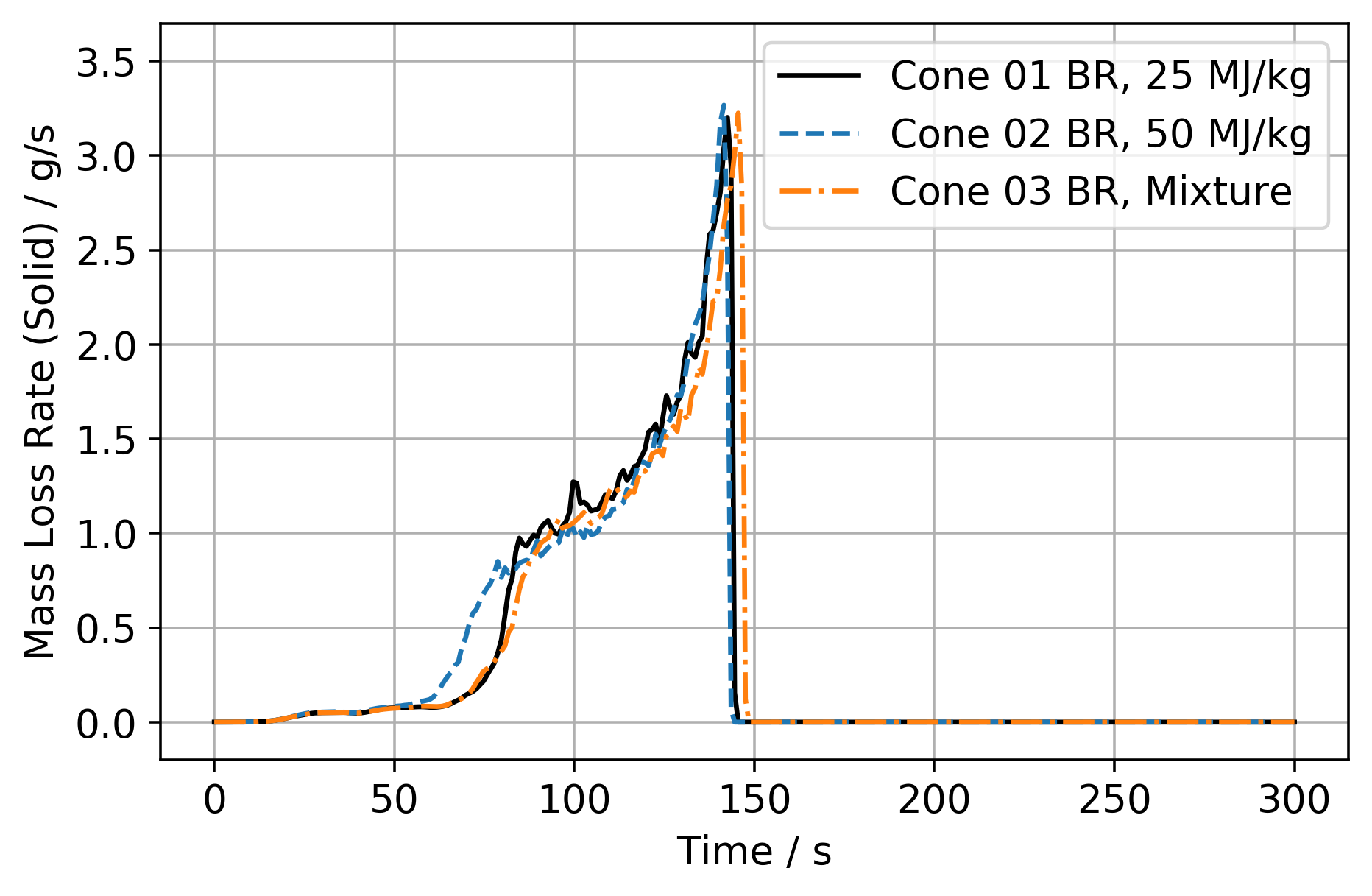} }}%
    \qquad
    \subfloat[\centering Gas phase simulation.
    \label{fig:FlameHeight_MassLoss_Gas}]{{\includegraphics[width=0.45\columnwidth]{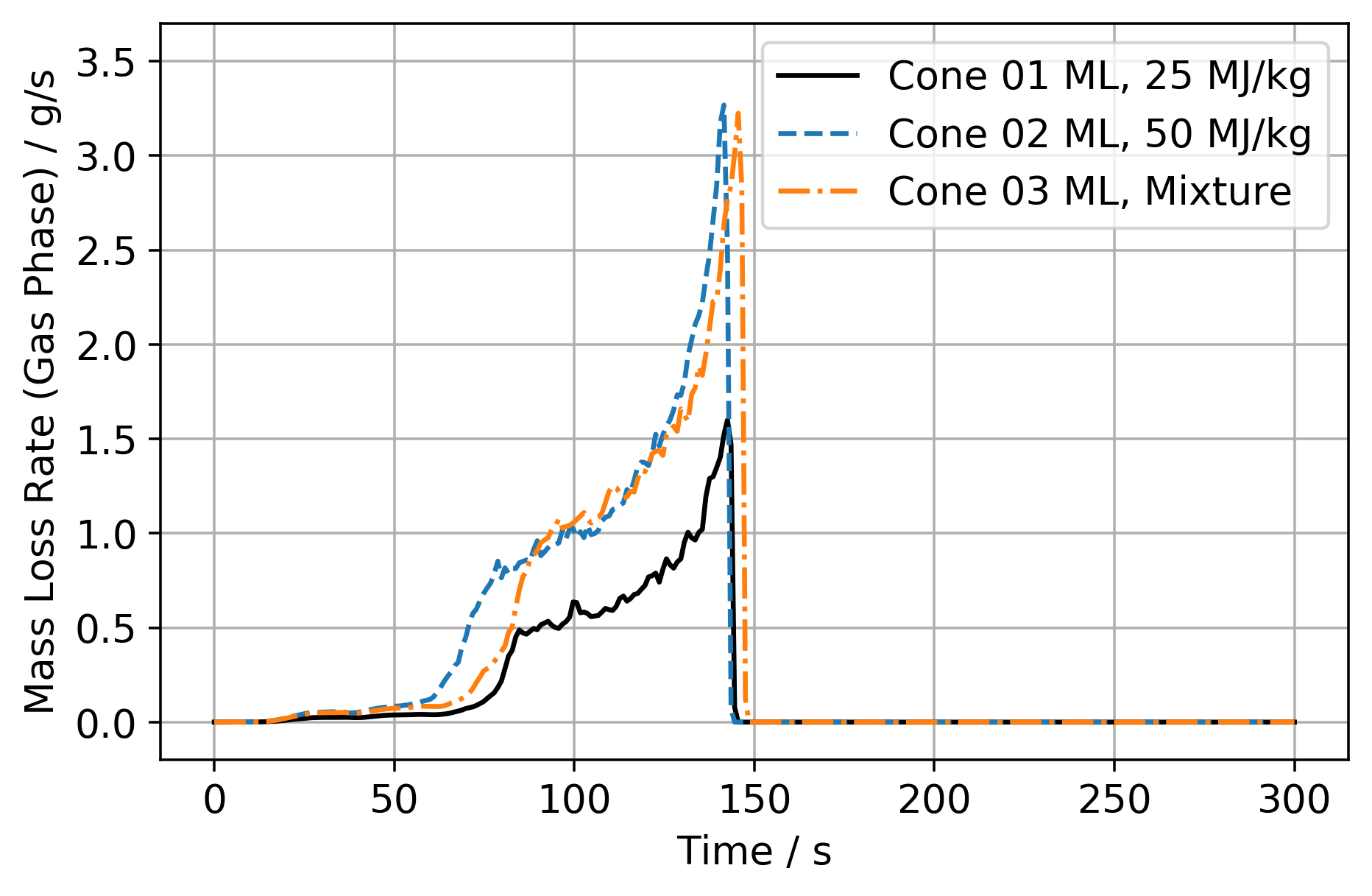} }}%
    \caption{Mass loss rates for different surrogate fuel species (pyrolysis) at different locations in the same simulation. Visualising the energy release based scaling. Different heats of combustion are used as indicated, "Mixture" has a HOC of about 25.5~MJ/kg.
    \label{fig:FlameHeight_MassLoss_Combined}}
\end{figure}

\begin{figure}[h]
    \centering
    \includegraphics[width=0.45\columnwidth]{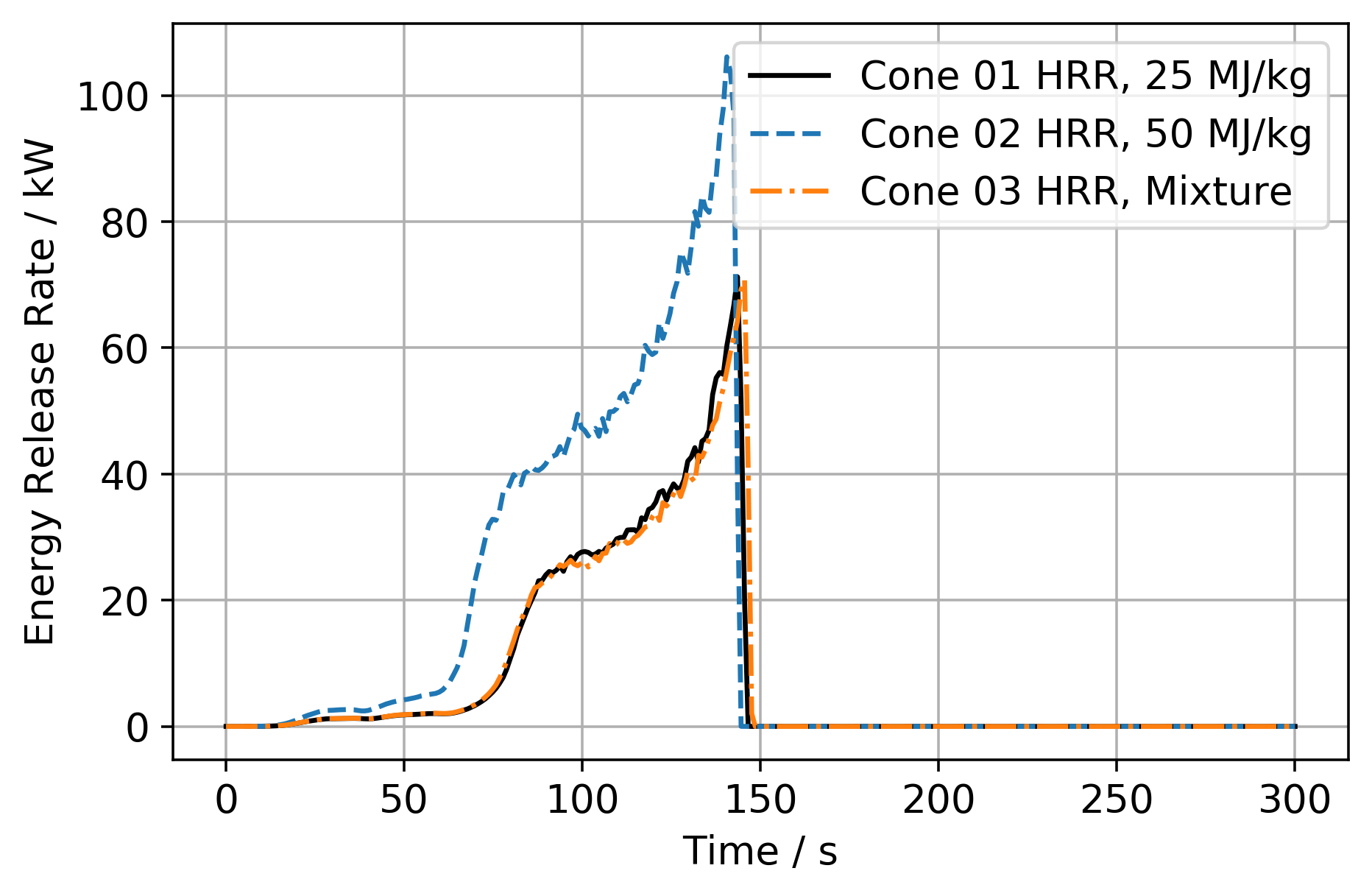}
        \caption{Energy release for different surrogate fuel species. Different heats of combustion are used as indicated, "Mixture" has a HOC of about 25.5~MJ/kg.}
        \label{fig:FlameHeight_EnergyRelease}
\end{figure}

An example parameter set is used to demonstrate the different mass loss rates and energy release. The cases "Cone 01" to "Cone 03" in figures~\ref{fig:FlameHeight_MassLoss_Solid},~\ref{fig:FlameHeight_MassLoss_Gas}~and~\ref{fig:FlameHeight_EnergyRelease} all use the same pyrolysis scheme. "Cone 01" and "Cone 02" use only methane as surrogate fuel. For "Cone 01" the heat of combustion in the material definition is set to 25~MJ/kg, which is about half of the value of pure methane. No heat of combustion value is provided for "Cone 02", thus it is the predefined value of 50~MJ/kg of methane. "Cone 03" uses a surrogate fuel gas mixture that consists of 26 volume percent of carbon dioxide and 74 volume percent of methane. This leads to roughly the same average heat of combustion than the 25~MJ/kg of "Cone 01". Also, the radiative fraction of the gas mixture was set to 0.20 to match the value of pure methane. Again, no HOC value is provided in the material definition, thus the released mass in the solid is transported directly to the gas domain. This highlights the distinction between the solid and gas phase side of the FDS simulation. FDS uses the heat of combustion parameter provided in the material definition to scale the mass of fuel that is introduced into the gas domain. Figure~\ref{fig:FlameHeight_MassLoss_Solid} shows the mass loss in the solid. All three cases experience a very similar development. In figure~\ref{fig:FlameHeight_MassLoss_Gas} it can be observed that the mass introduced into the gas domain is about half for "Cone 01" compared to the others, due to the scaling of the HOC. Consequently, figure~\ref{fig:FlameHeight_EnergyRelease} shows about double the energy release for case "Cone 02" than the others.

\begin{figure}[h]%
    \centering
    \subfloat[\centering Mass loss rate.\label{fig:FlameHeight_MassLoss}]{{\includegraphics[width=0.45\columnwidth]{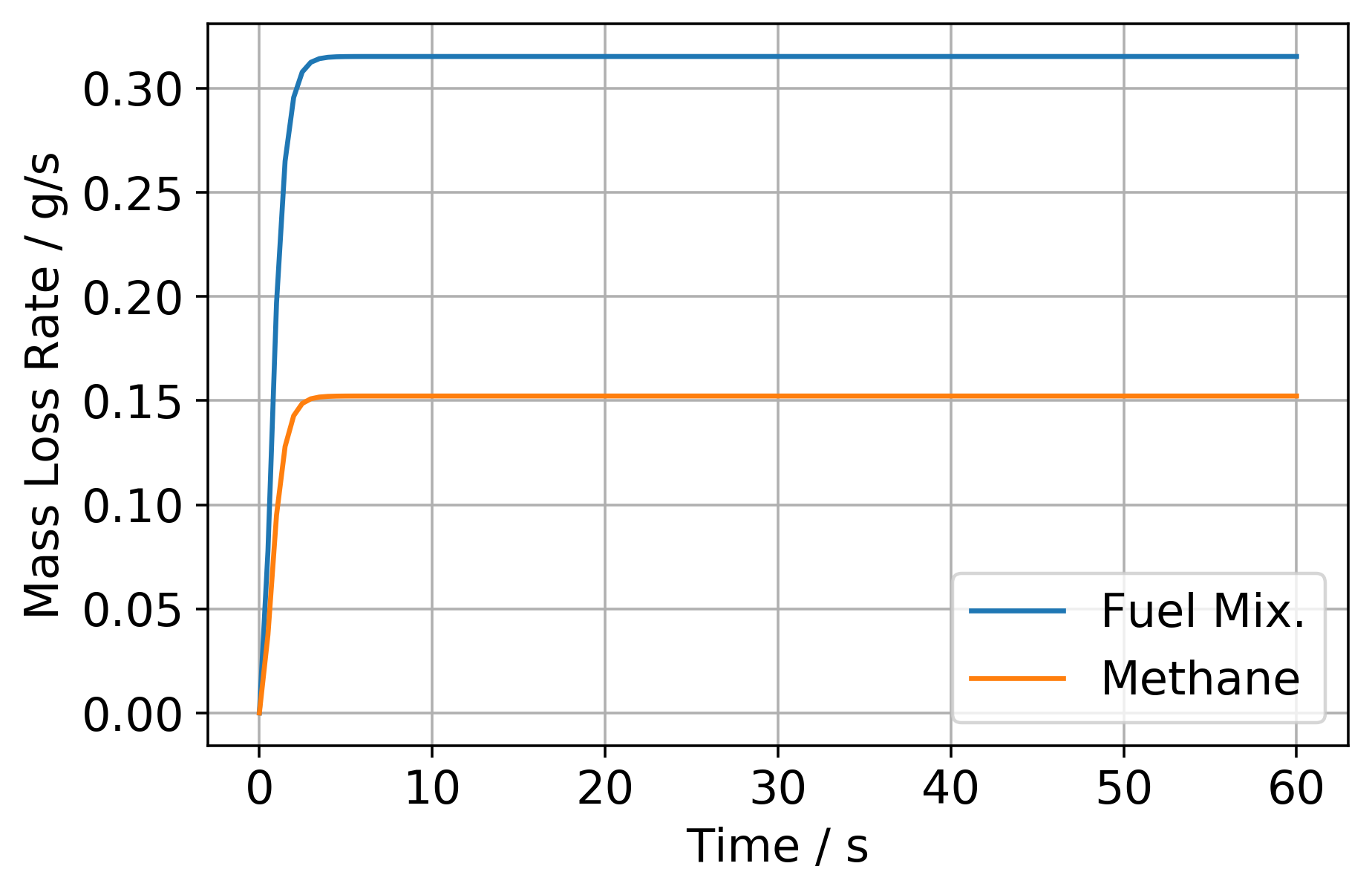} }}%
    \hfill
    \subfloat[\centering Centre line temperature.\label{fig:FlameHeight_Temperatures}]{{\includegraphics[width=0.45\columnwidth]{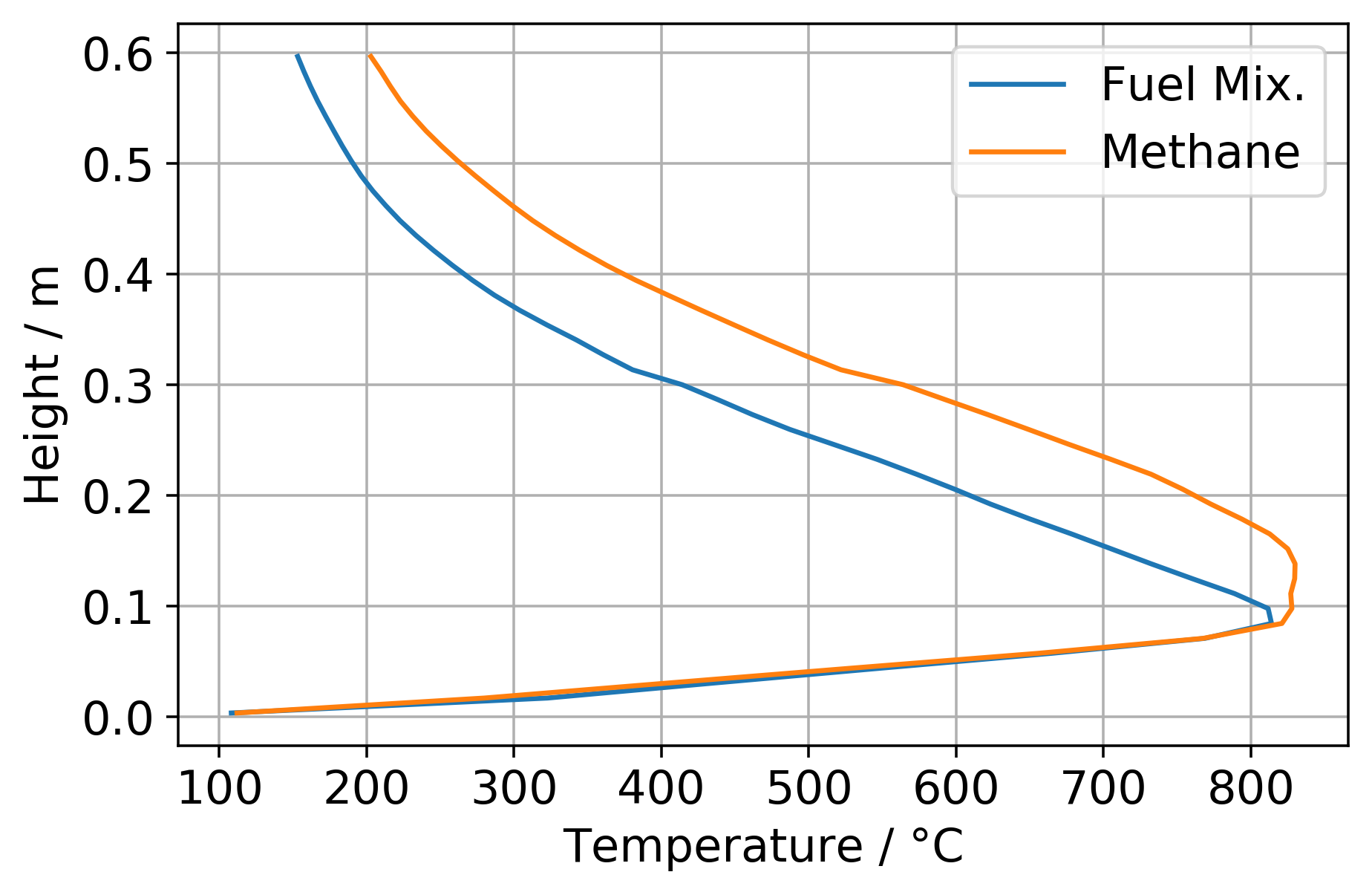} }}%
    
    \caption{Simplified cone calorimeter simulation (C15) for different surrogate fuel species to compare flame heights. Mass flux adjusted to get the same energy release. Fuel mixture: methane, ethene and carbon dioxide.}%
    \label{fig:FlameHeights}%
\end{figure}

The different surrogate fuel strategies from above, "Cone 01" and "Cone 03", lead also to different flames. Two simulations are conducted with a constant mass release (\texttt{HRRPUA}), shown in figure~\ref{fig:FlameHeight_MassLoss} to mimic both setups. Gas temperatures are recorded on the vertical centre line of the flame and averaged over the second half of the simulation (30~s). This leads to differences in the flame structure as shown in figure~\ref{fig:FlameHeight_Temperatures}. No claim is made here as to which one is more "realistic", just the difference pointed out.

\clearpage
\section{Example Limit Adjustments}
\label{App:LimitAdjustmentExample}

\begin{figure}[h]
    \centering
    \includegraphics[width=0.45\columnwidth]{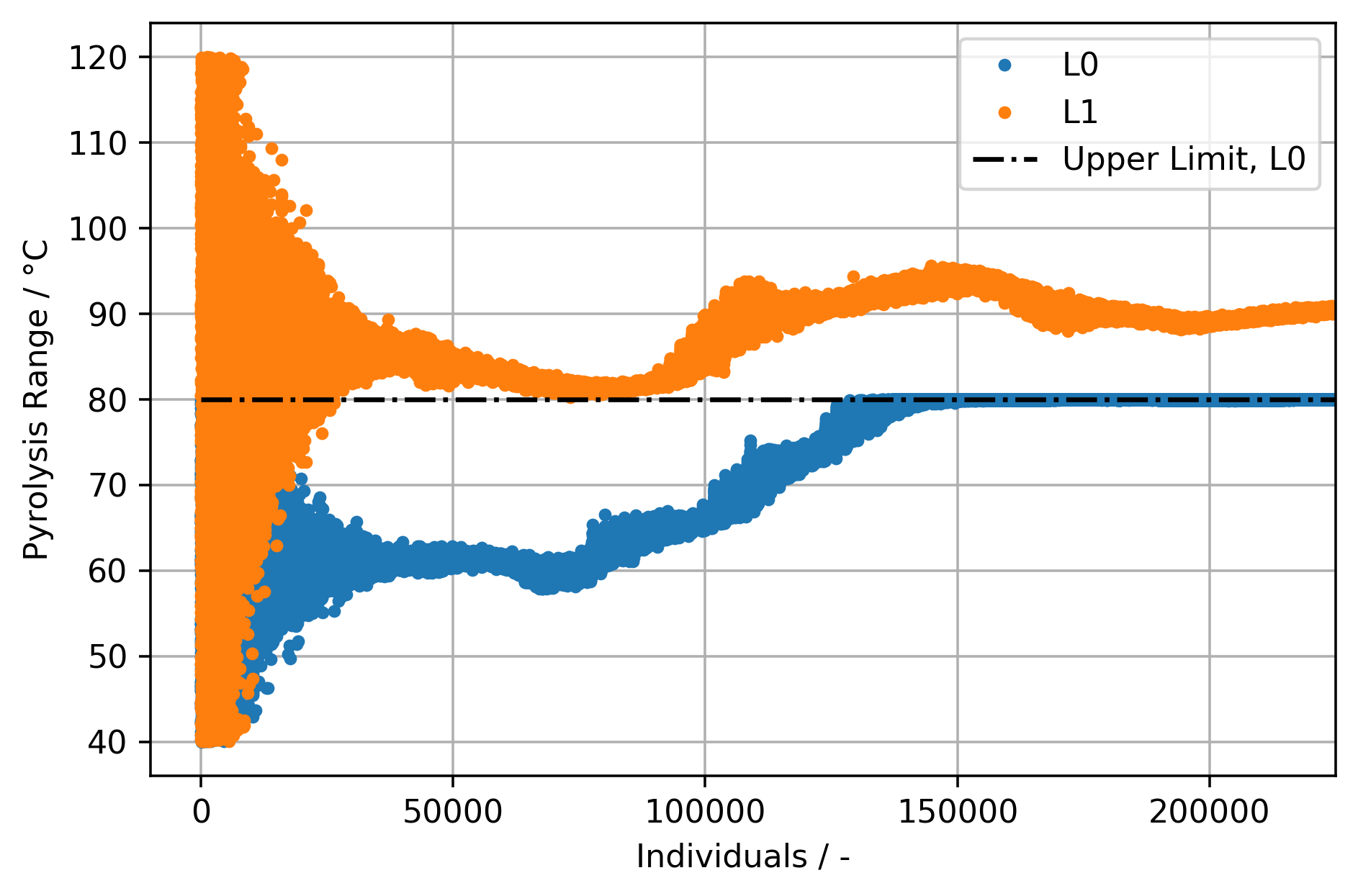}
    \caption{Example for sampling limit adjustment.}
    \label{fig:LimitAdjustmentExample}
\end{figure}

During the IMP individual parameters can get stuck at their limits. Figure~\ref{fig:LimitAdjustmentExample} shows an example of this. During the initial limit definition (L0) the pyrolysis range parameter got stuck at its upper limit. Another IMP run was set up, with an expanded range (L1). Note: only the upper limit was adjusted and the lower limit was kept at its original value. Thus, the sampling space only grows larger over multiple adjustments. In the beginning, both developments are different, because not only this parameters limits are adjusted for this new run, but also for others that were stuck.

\clearpage
\onecolumn

\section{Micro-Scale Tests}
\label{App:MicroScaleTest}

Data sets from the micro-scale tests. NIST reported that the equipment for the TGA was a Netzsch F1 Jupiter and a FAA microscale combustion calorimeter for the MCC, Sandia used a Netzsch F3 Jupiter~\cite{macfp_matl_git}.

The TGA experiments by NIST were conducted in nitrogen atmosphere. For the MCC experiments the pyrolysis chamber was fed with nitrogen and oxygen was mixed into the effluents before entering the combustion chamber. The TGA experiments by Sandia were conducted in argon atmosphere.

\begin{figure}[h]%
    \centering
    \subfloat[\centering MCC 60~K/min, NIST.]{{\includegraphics[width=0.45\columnwidth]{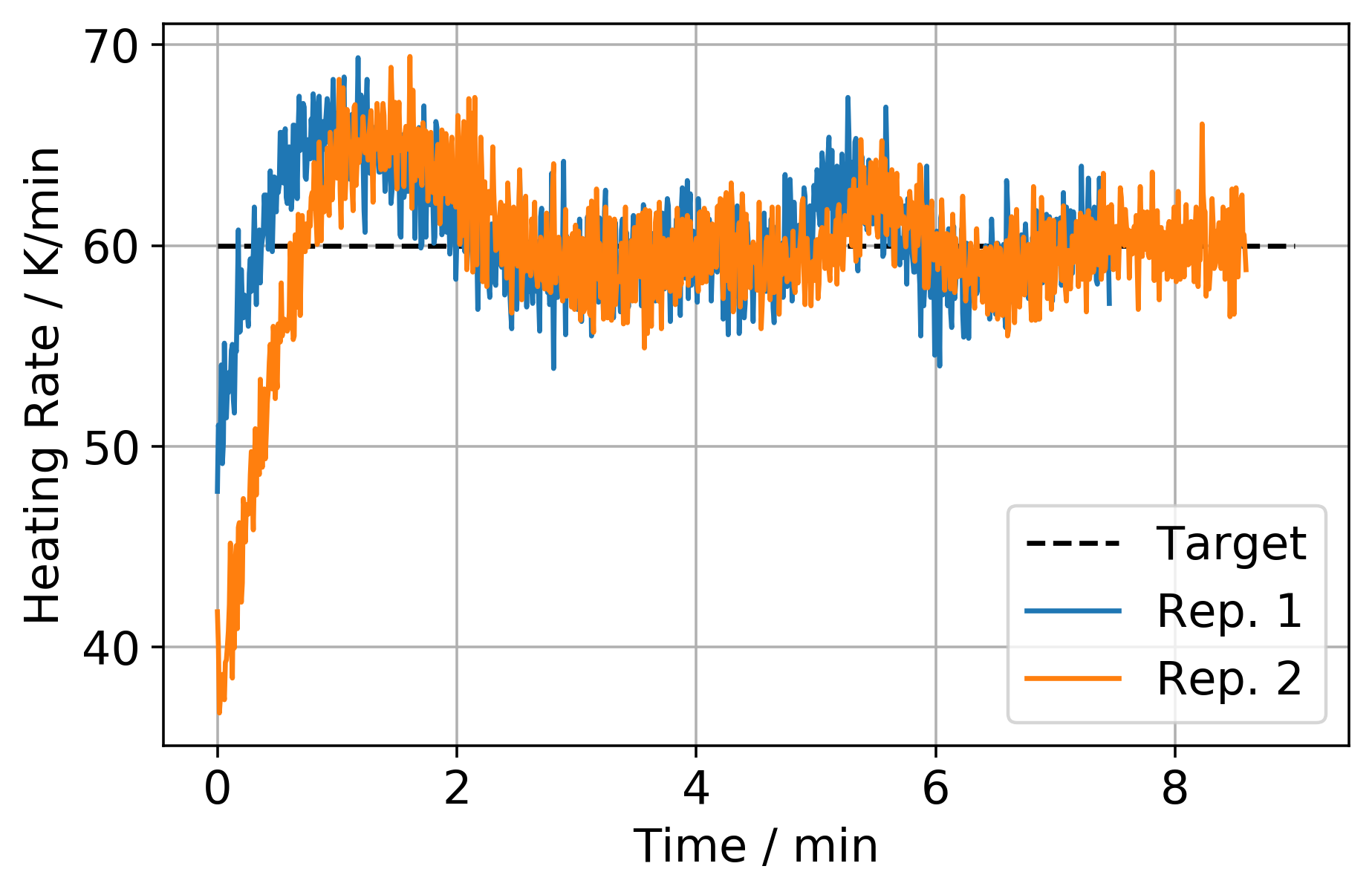} }}%
    \qquad
    \subfloat[\centering TGA 10~K/min, NIST.
    \label{fig:Apdx_MicroScale_HeatingRates_NIST10K}]{{\includegraphics[width=0.45\columnwidth]{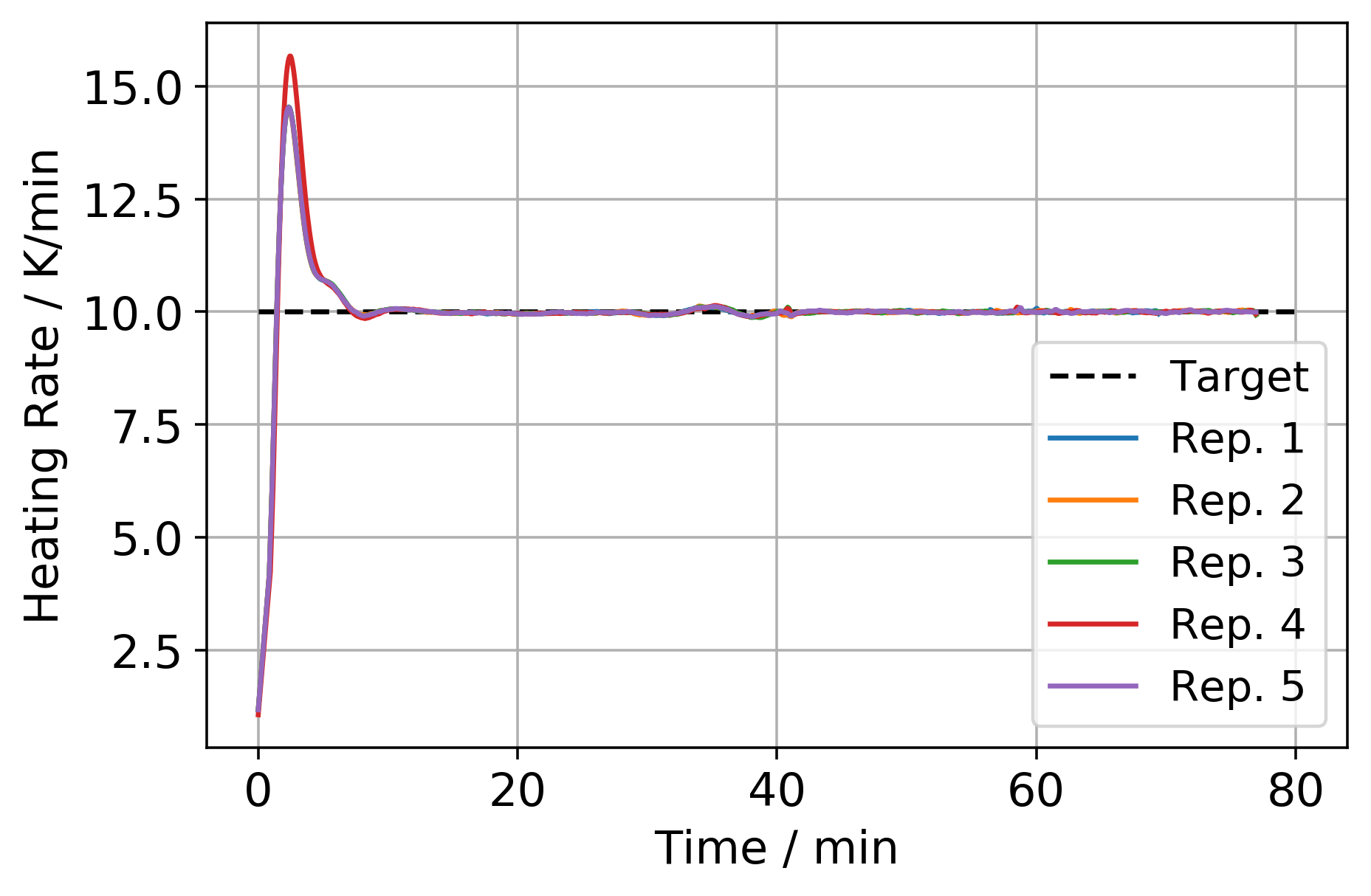} }}%

    \subfloat[\centering TGA 10~K/min, Sandia.]{{\includegraphics[width=0.45\columnwidth]{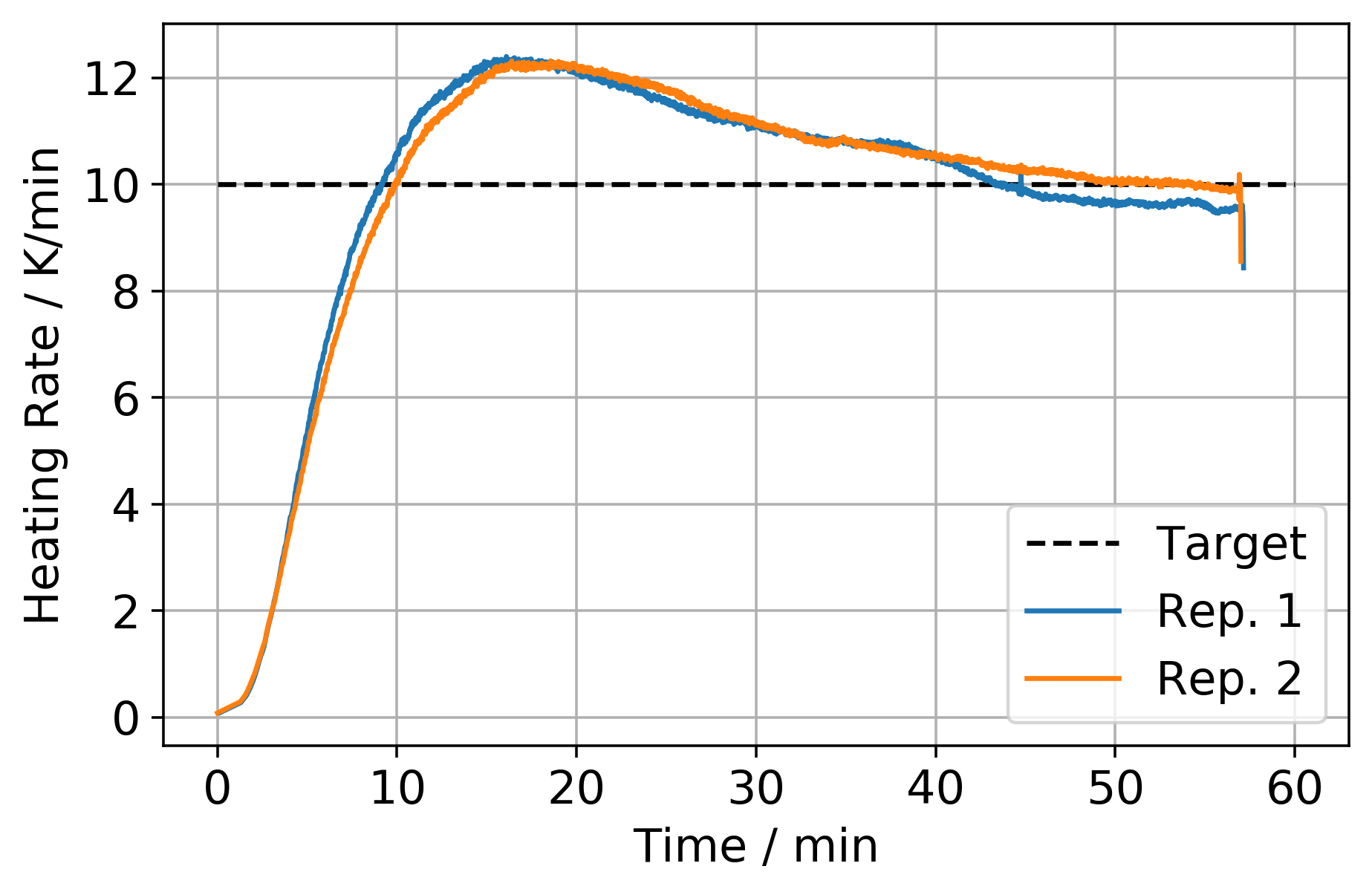} }}%
    \qquad
    \subfloat[\centering TGA 50~K/min, Sandia.    \label{fig:Apdx_MicroScale_HeatingRates_Sandia50K}]{{\includegraphics[width=0.45\columnwidth]{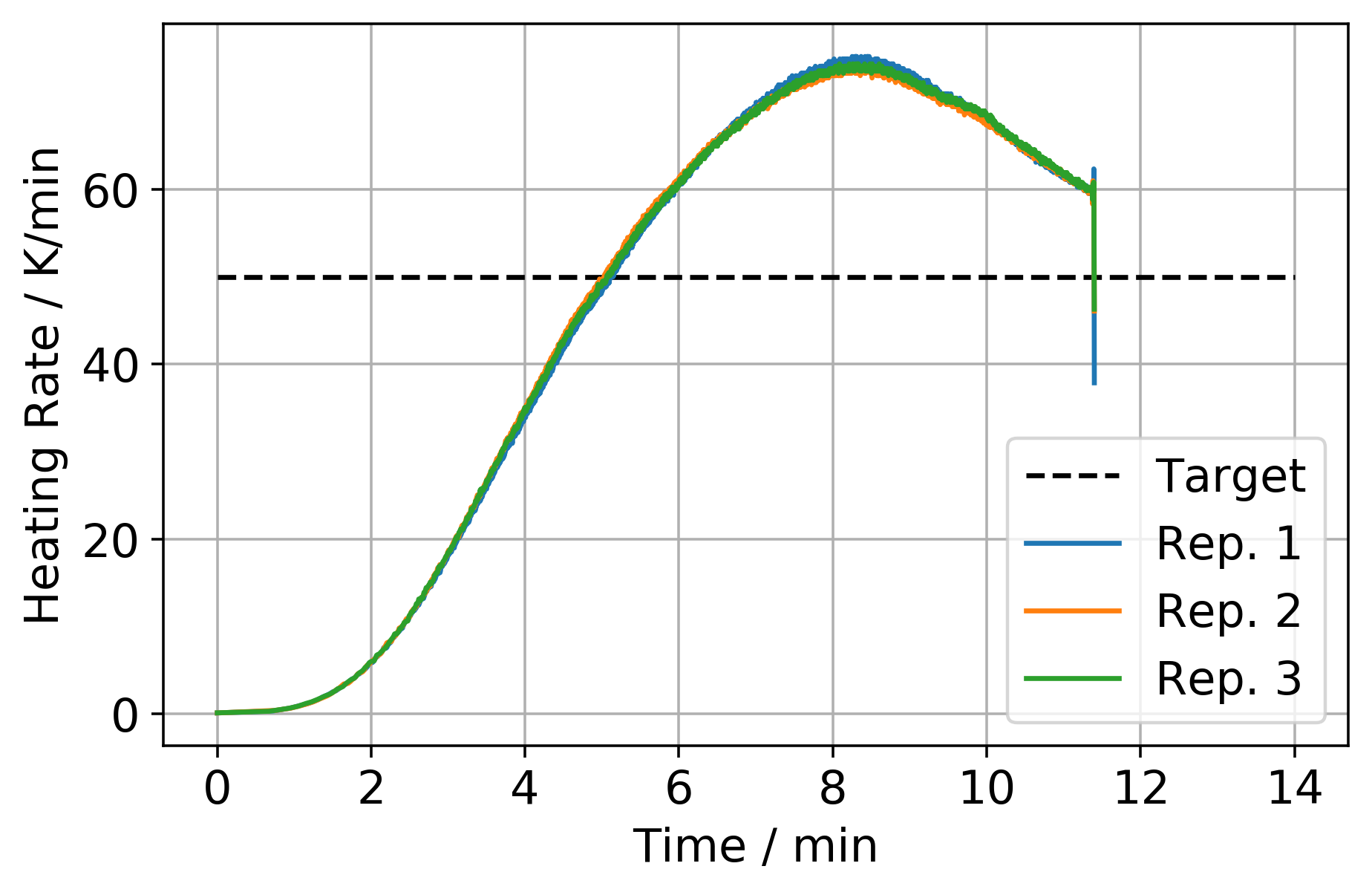} }}%
    
    \caption{Different heating rates of micro-scale experiments from MaCFP data base~\cite{macfp_matl_git}}%
    \label{fig:Apdx_MicroScale_HeatingRates}%
\end{figure}

\clearpage
\section{Simple Cone Calorimeter Simulation Results}
\label{appendix_simplecone}

\subsection{IMP Fitness Development}

\begin{figure}[h]%
    \centering
    \subfloat[\centering IMP Setup Cone\_01.]{{\includegraphics[width=0.395\columnwidth]{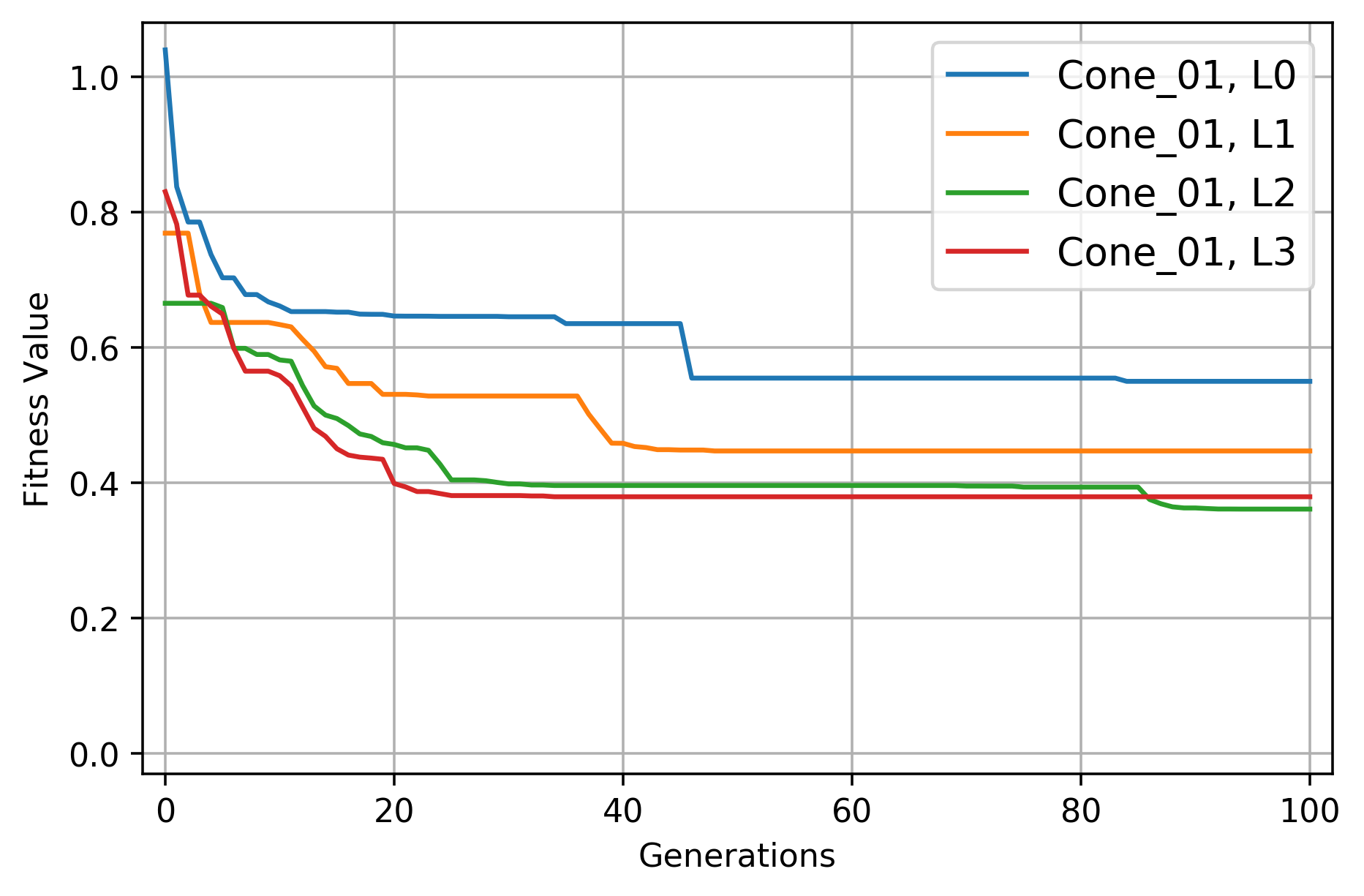} }}%
    \hfill
    \subfloat[\centering IMP Setup Cone\_02.]{{\includegraphics[width=0.395\columnwidth]{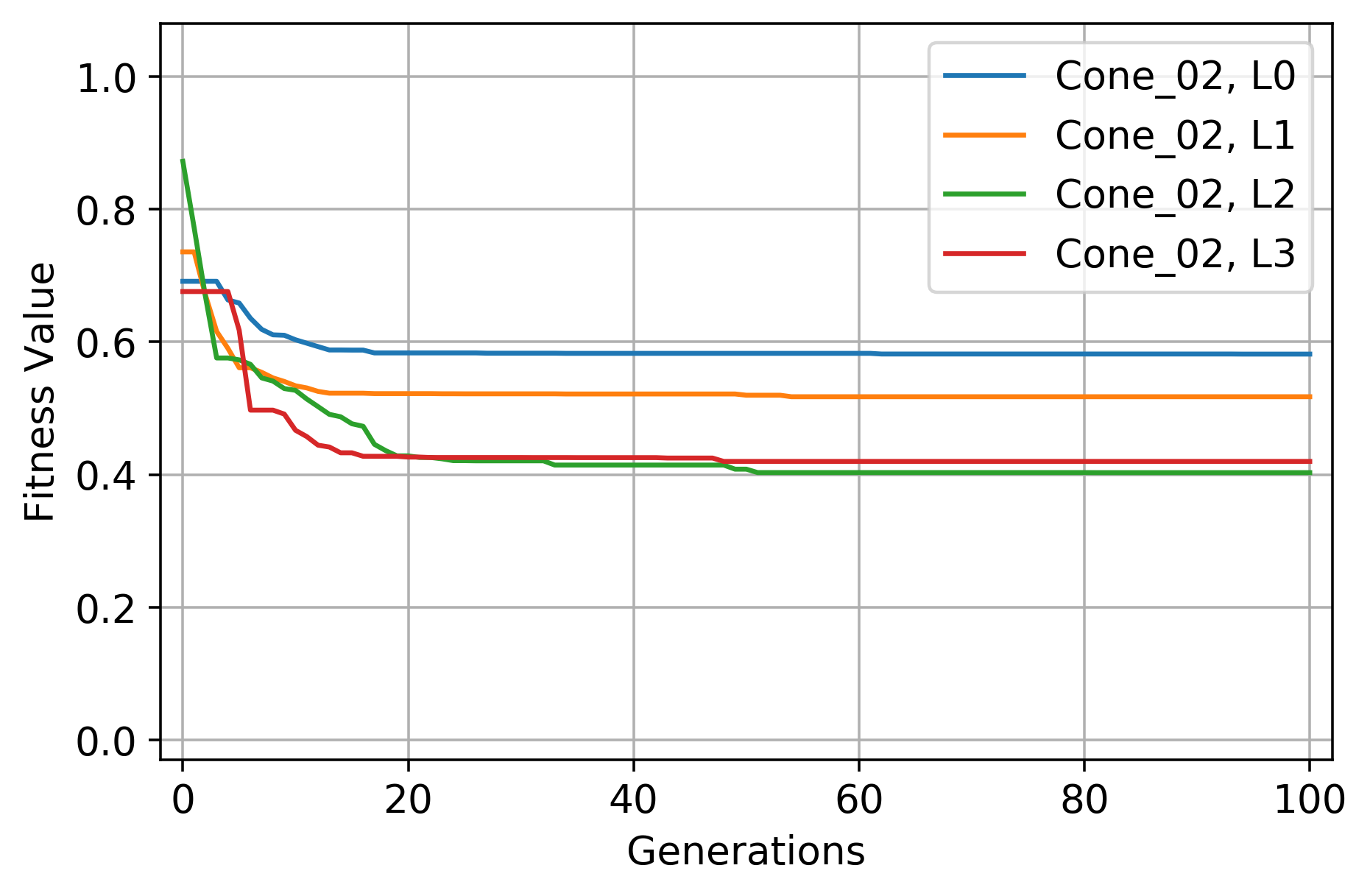} }}%
    \hfill
    \subfloat[\centering IMP Setup Cone\_03.]{{\includegraphics[width=0.395\columnwidth]{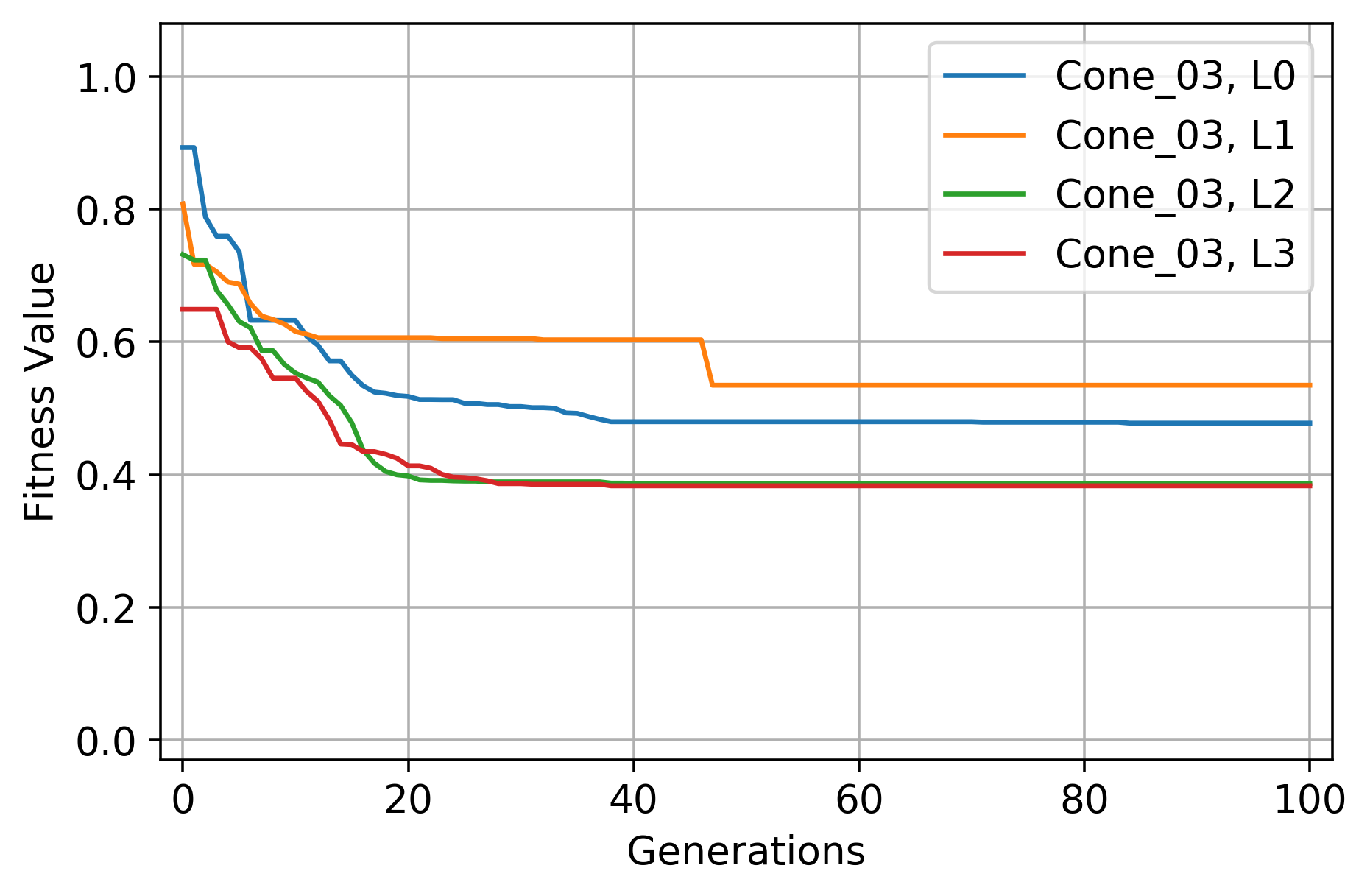} }}%
    \hfill
    \subfloat[\centering IMP Setup Cone\_04.]{{\includegraphics[width=0.395\columnwidth]{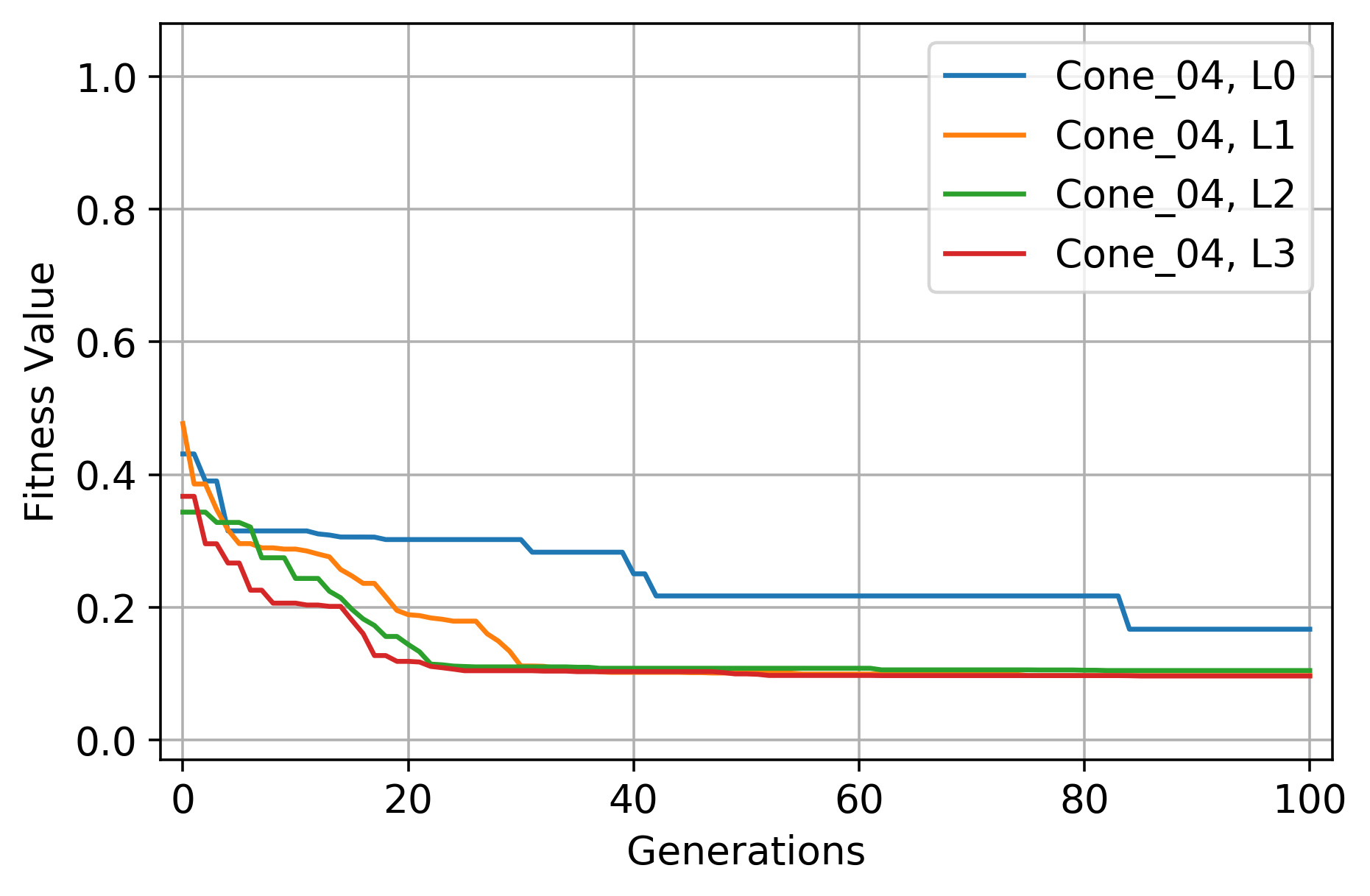} }}%
    \hfill
    \subfloat[\centering IMP Setup Cone\_05.]{{\includegraphics[width=0.395\columnwidth]{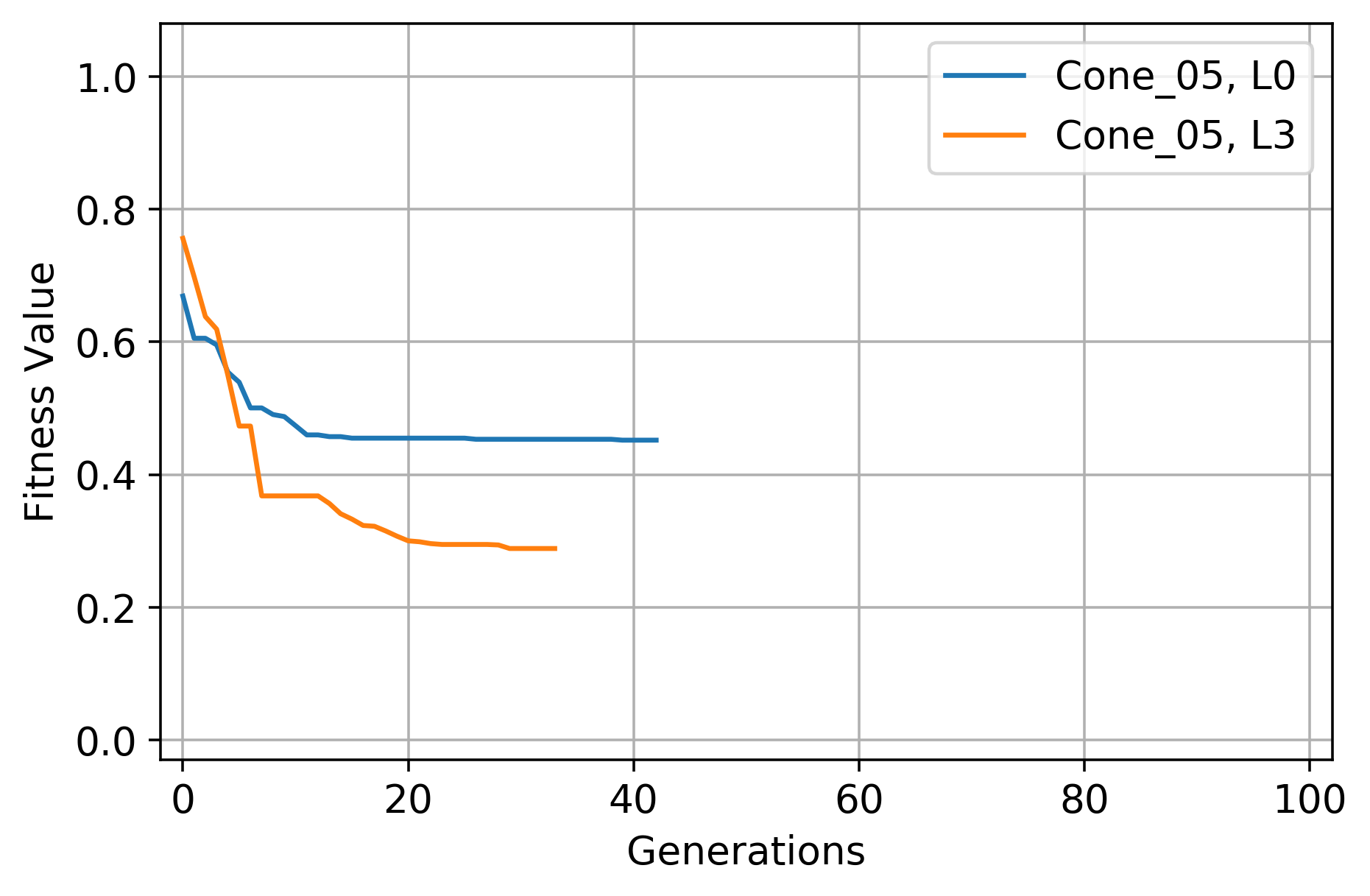} }}%
    \hfill
    \subfloat[\centering IMP Setup Cone\_06.]{{\includegraphics[width=0.395\columnwidth]{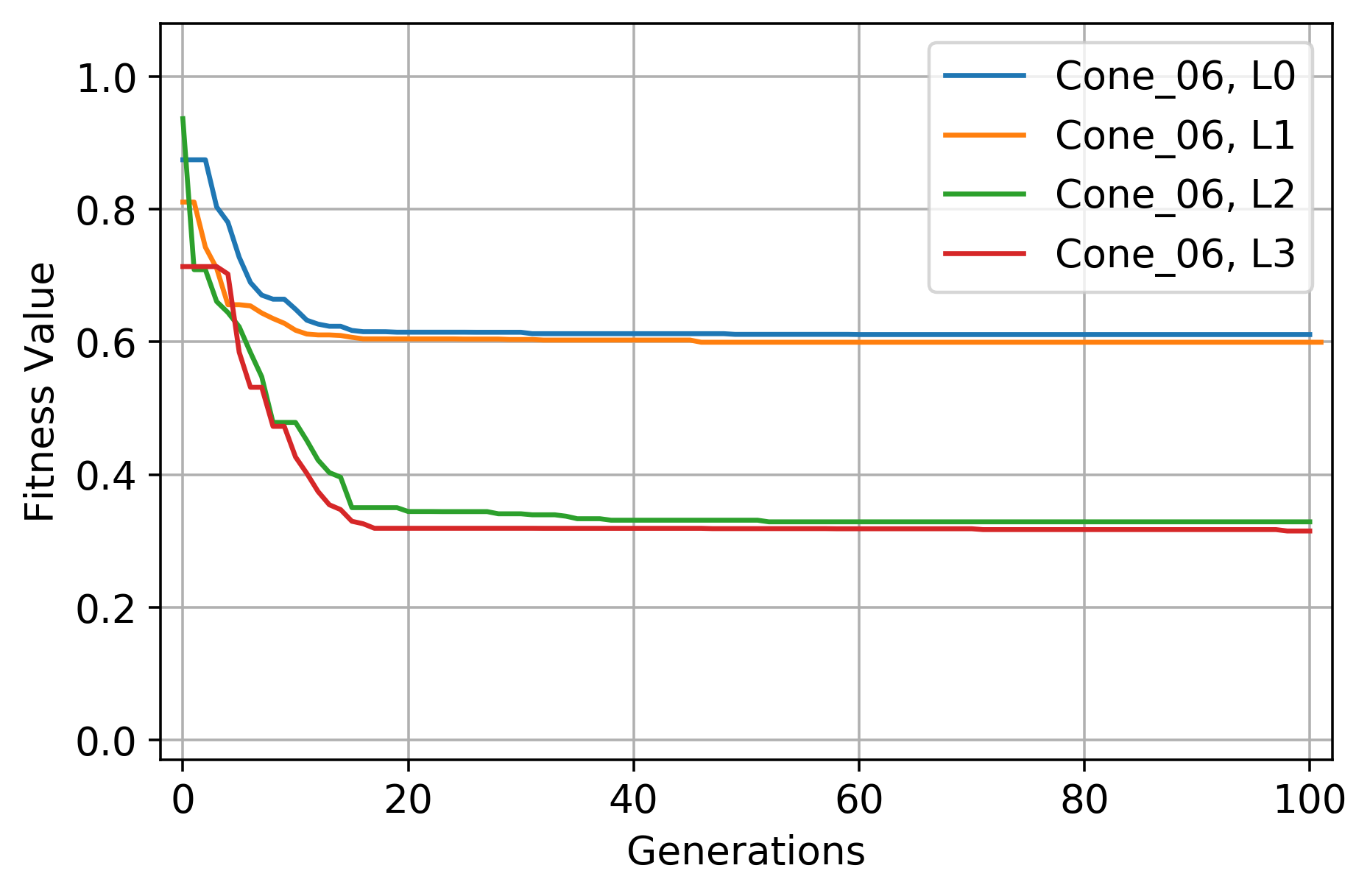} }}%
    \hfill
    \subfloat[\centering IMP Setup Cone\_07.]{{\includegraphics[width=0.395\columnwidth]{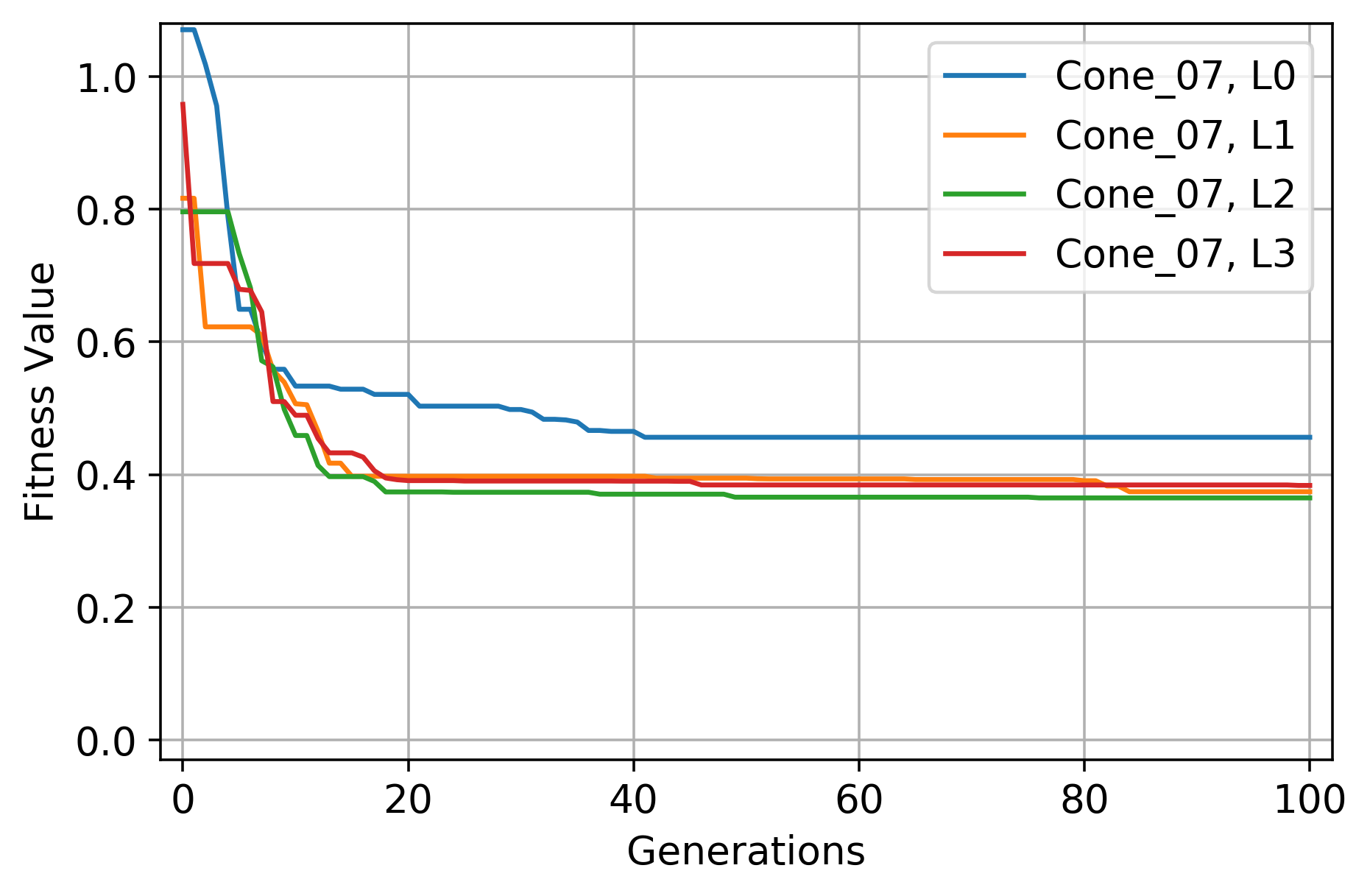} }}%
    \hfill
    \subfloat[\centering IMP Setup Cone\_08.]{{\includegraphics[width=0.395\columnwidth]{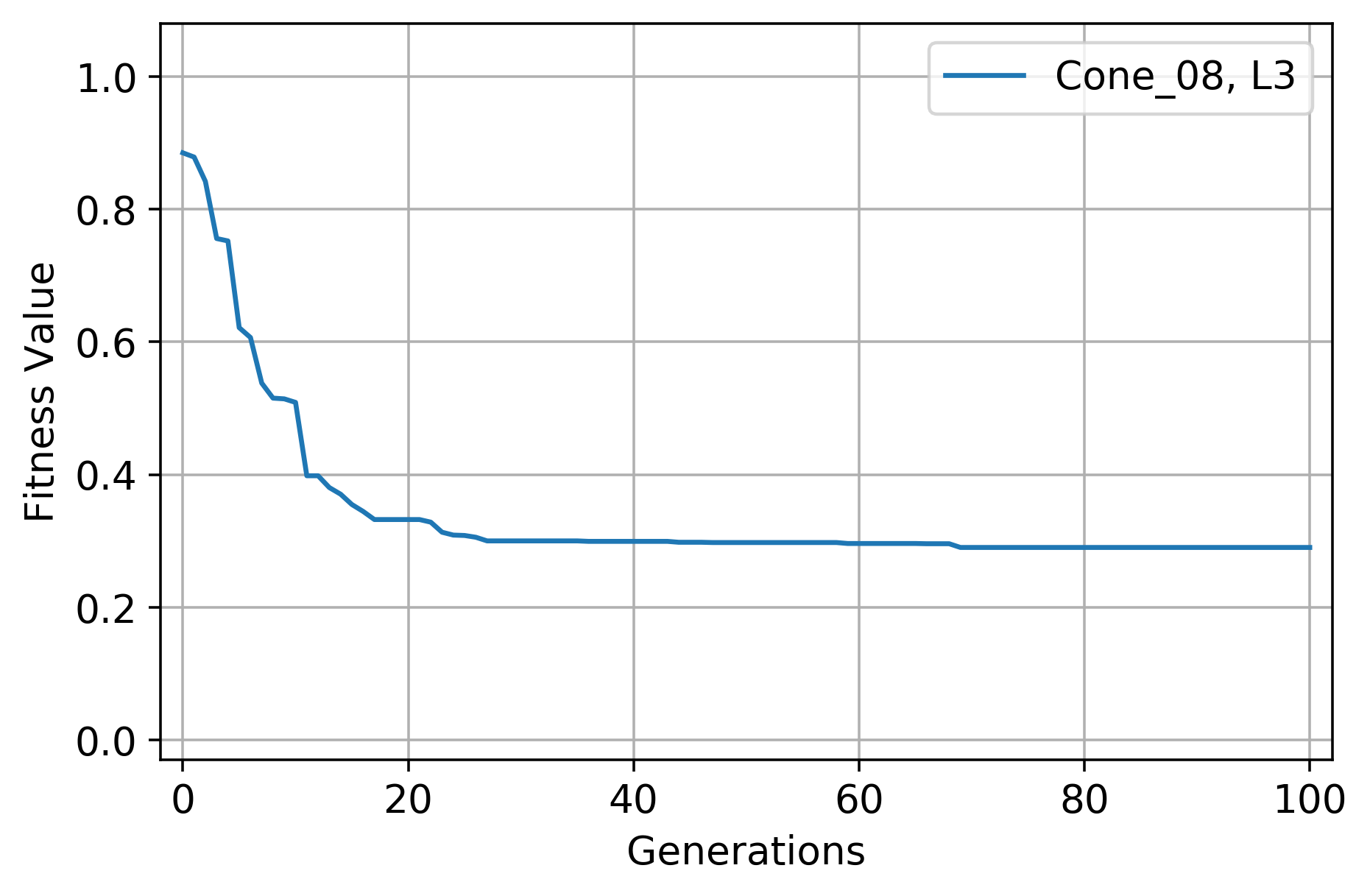} }}%
    
    \caption{IMP fitness development from simplified cone calorimeter simulations at 65 kW/m².}%
    \label{fig:ConeSimBestParaFitness_Aalto}%
\end{figure}

\clearpage
\subsection{Energy Release Rates}

\begin{figure}[h]%
    \centering
    \subfloat[\centering IMP Setup Cone\_01.]{{\includegraphics[width=0.395\columnwidth]{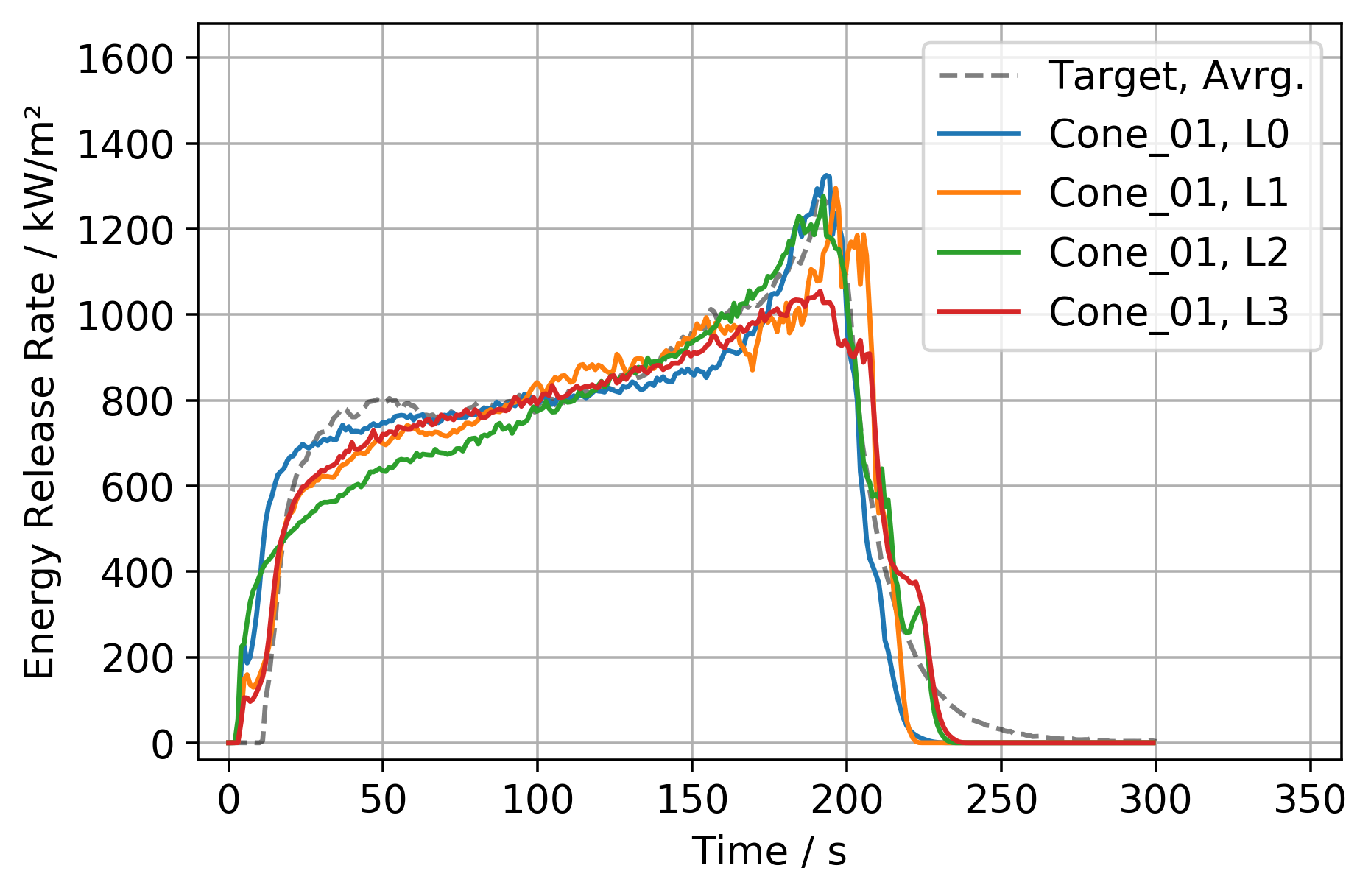} }}%
    \hfill
    \subfloat[\centering IMP Setup Cone\_02.]{{\includegraphics[width=0.395\columnwidth]{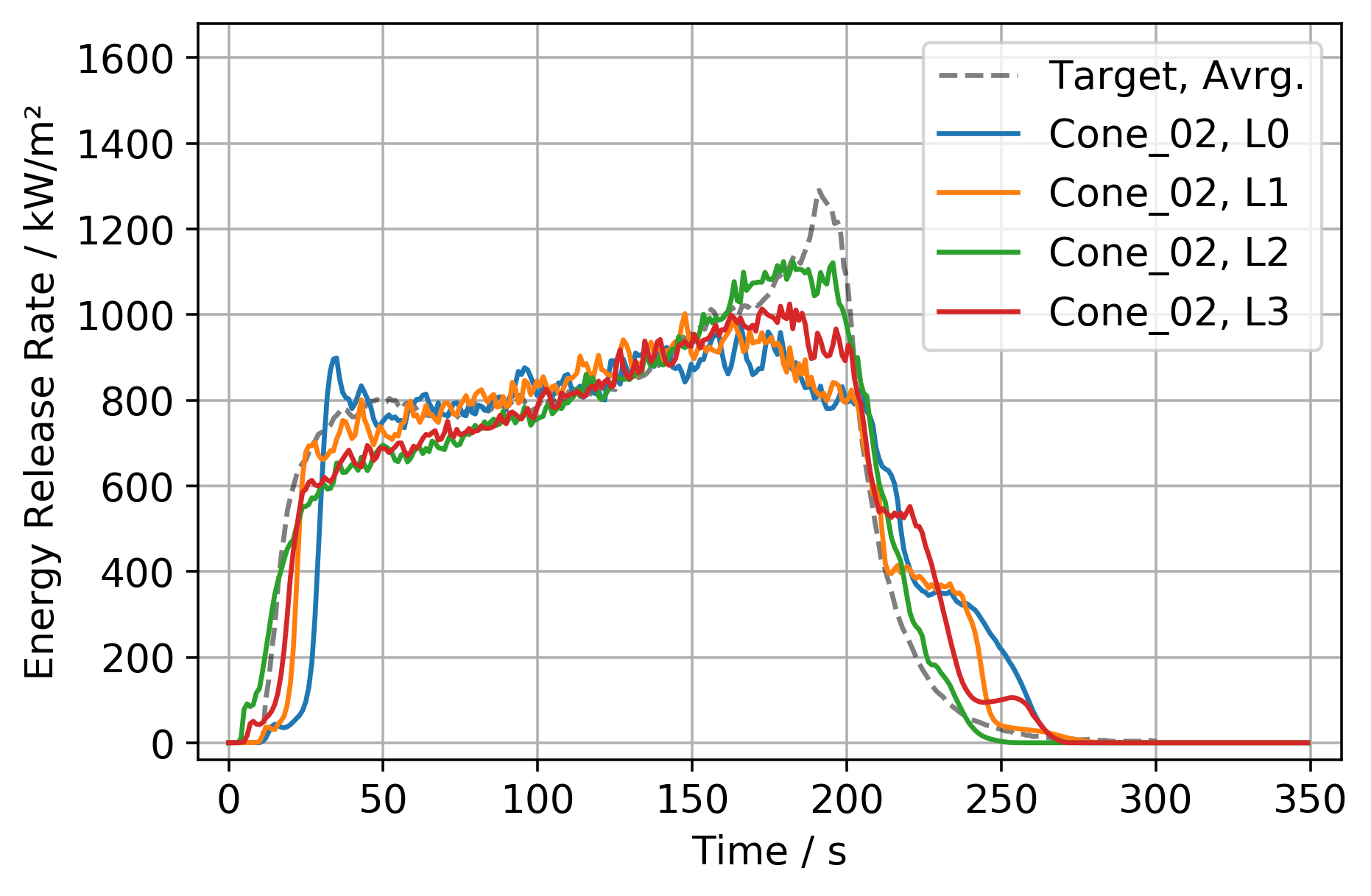} }}%
    \hfill
    \subfloat[\centering IMP Setup Cone\_03.]{{\includegraphics[width=0.395\columnwidth]{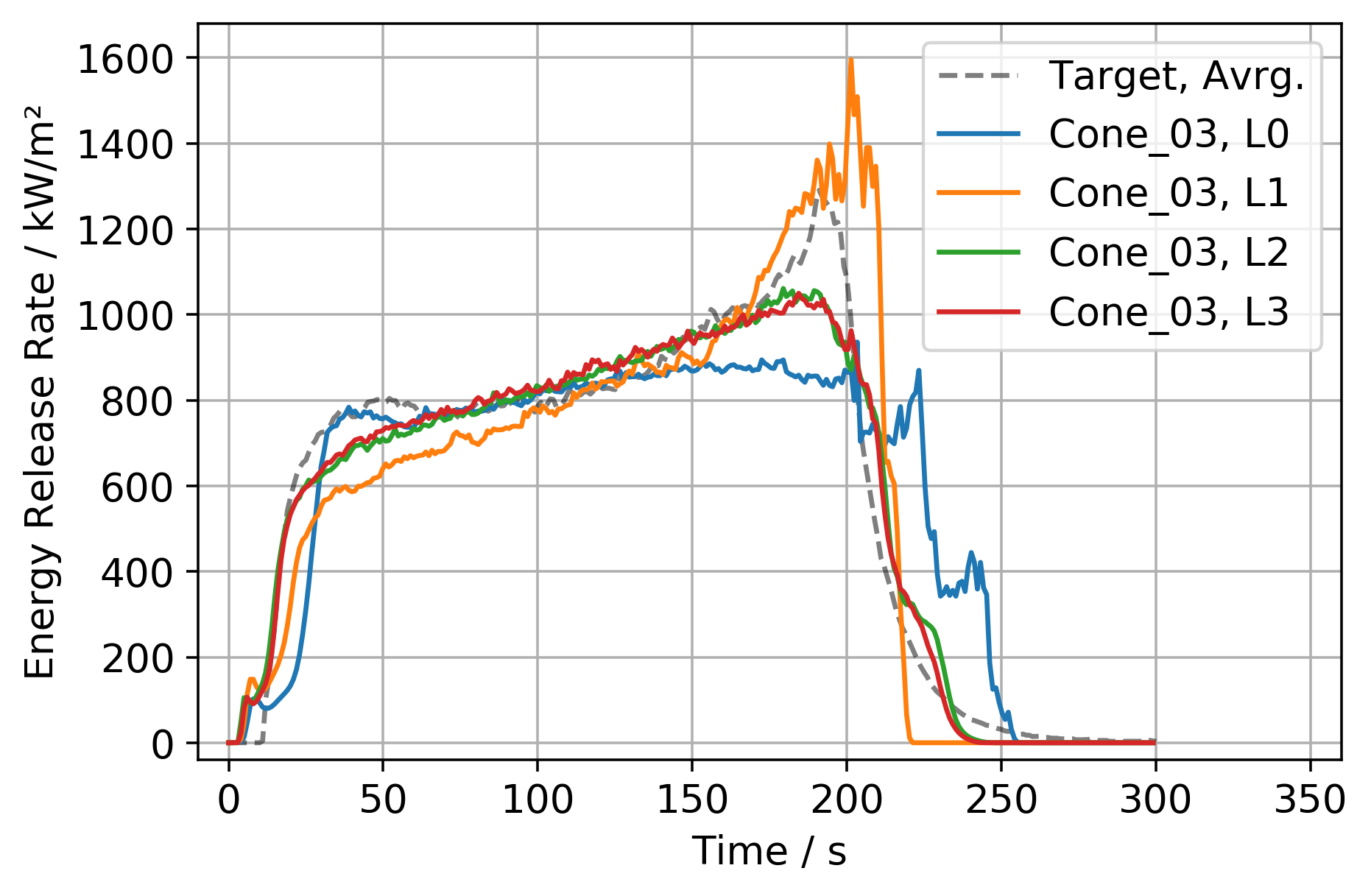} }}%
    \hfill
    \subfloat[\centering IMP Setup Cone\_04.]{{\includegraphics[width=0.395\columnwidth]{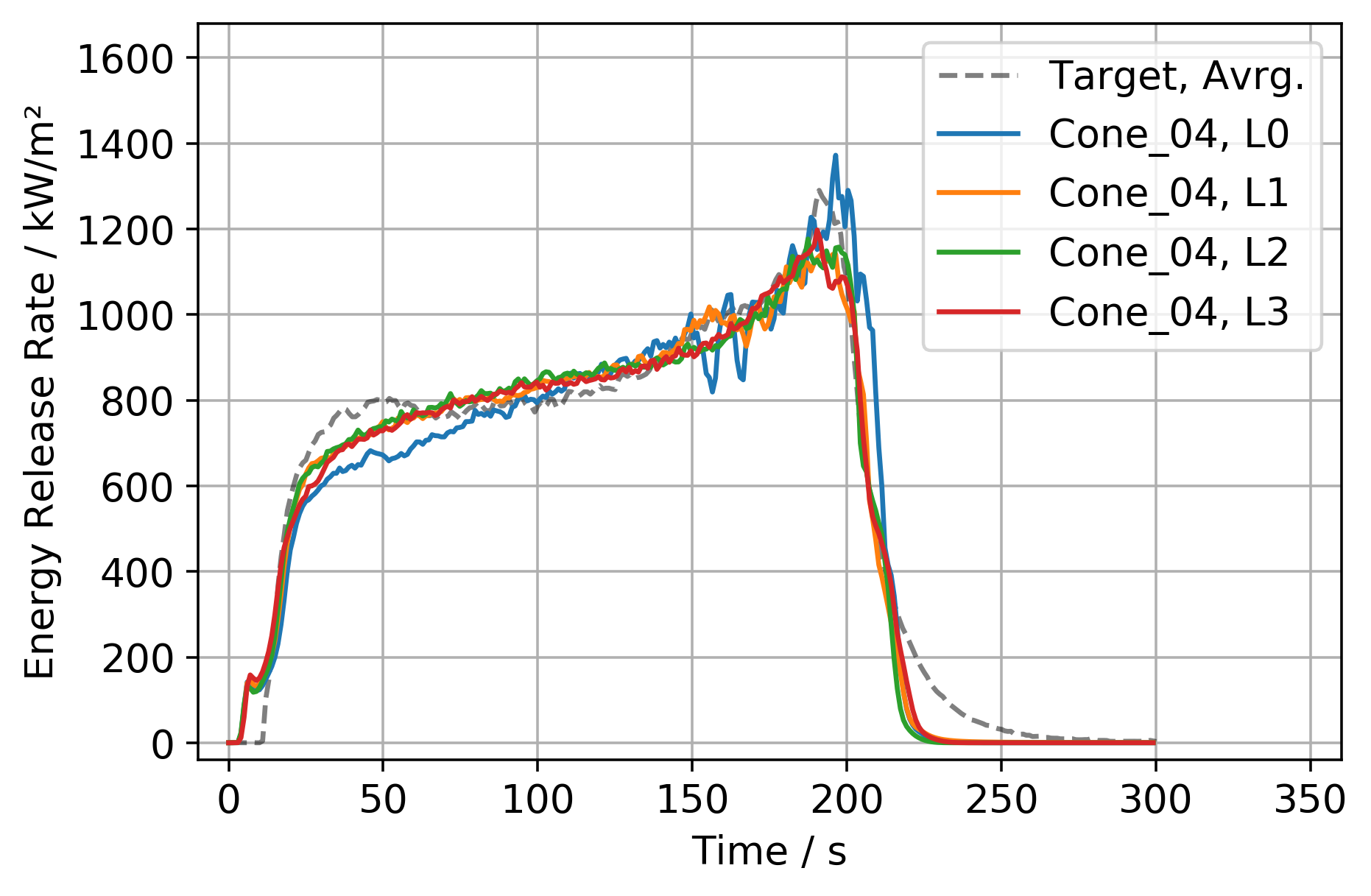} }}%
    \hfill
    \subfloat[\centering IMP Setup Cone\_05.]{{\includegraphics[width=0.395\columnwidth]{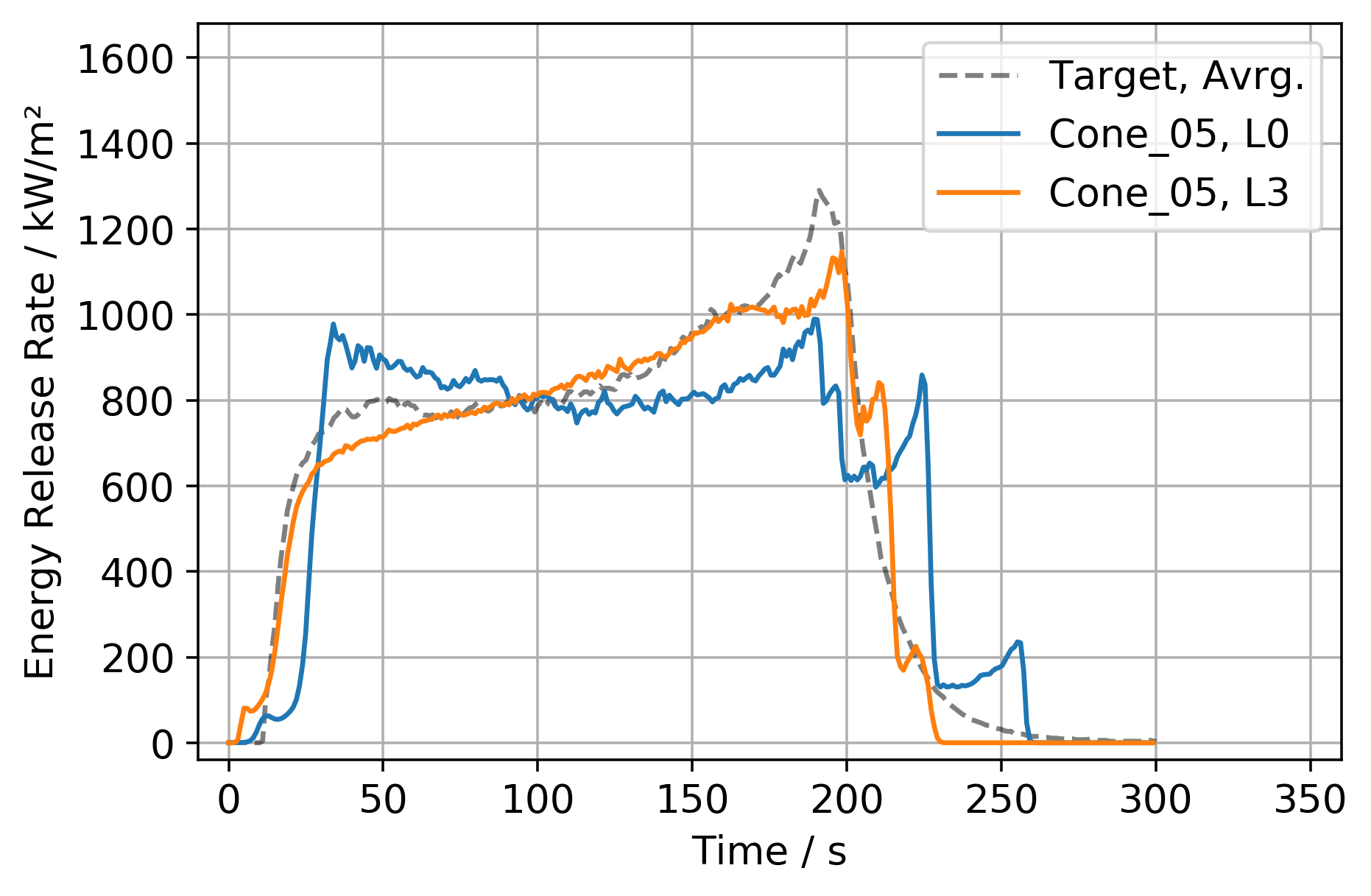} }}%
    \hfill
    \subfloat[\centering IMP Setup Cone\_06.]{{\includegraphics[width=0.395\columnwidth]{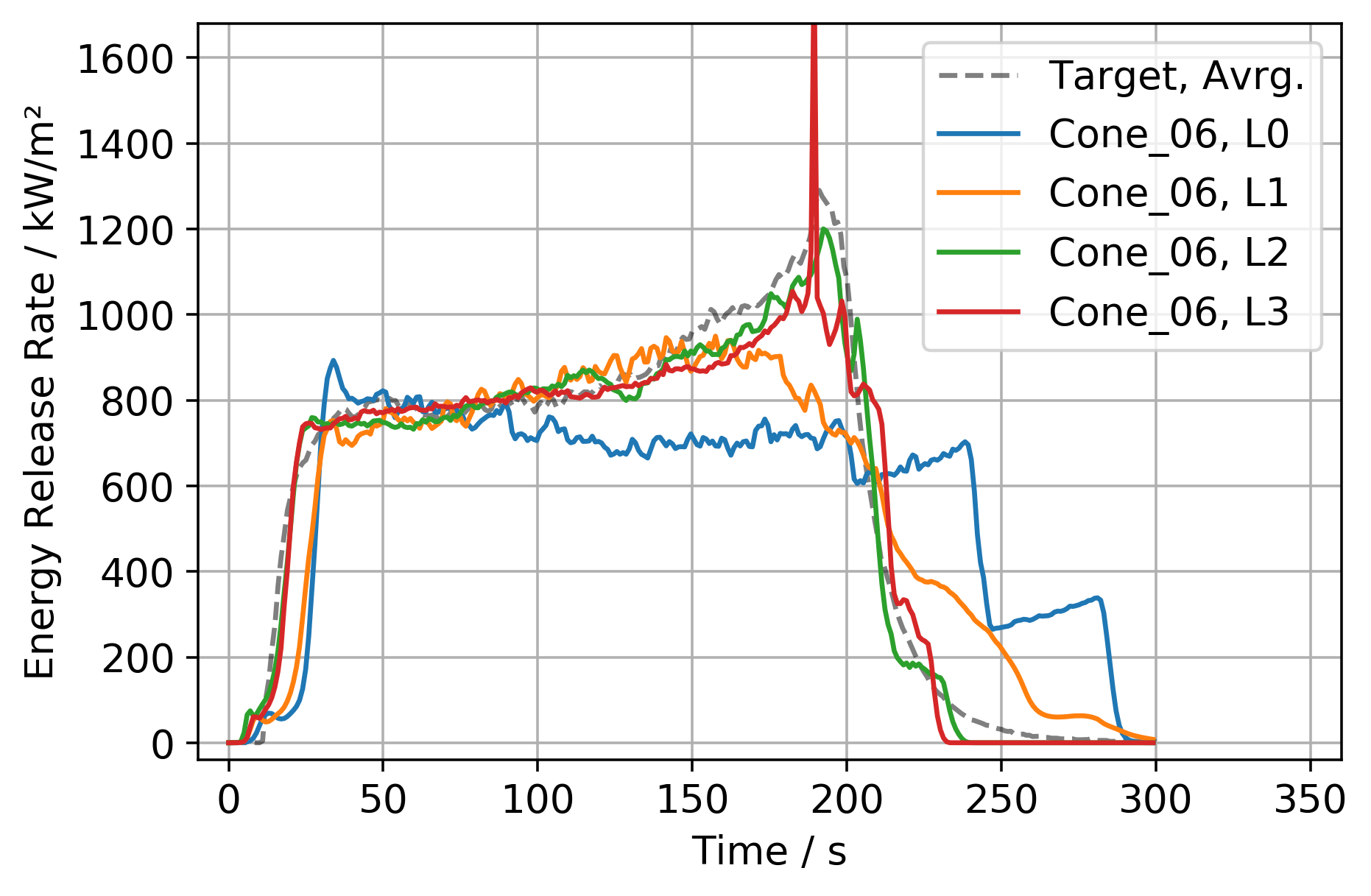} }}%
    \hfill
    \subfloat[\centering IMP Setup Cone\_07.]{{\includegraphics[width=0.395\columnwidth]{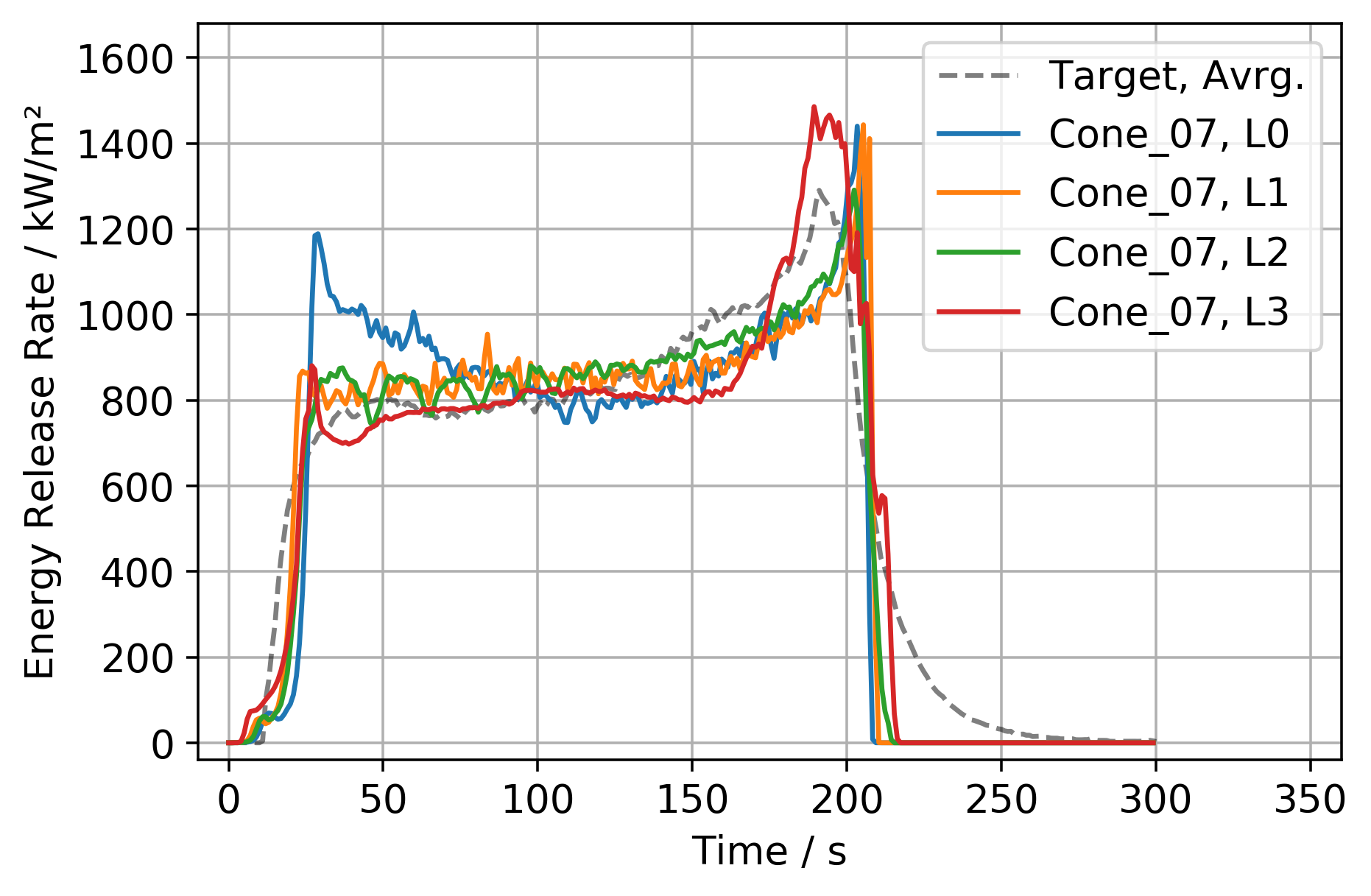} }}%
    \hfill
    \subfloat[\centering IMP Setup Cone\_08.]{{\includegraphics[width=0.395\columnwidth]{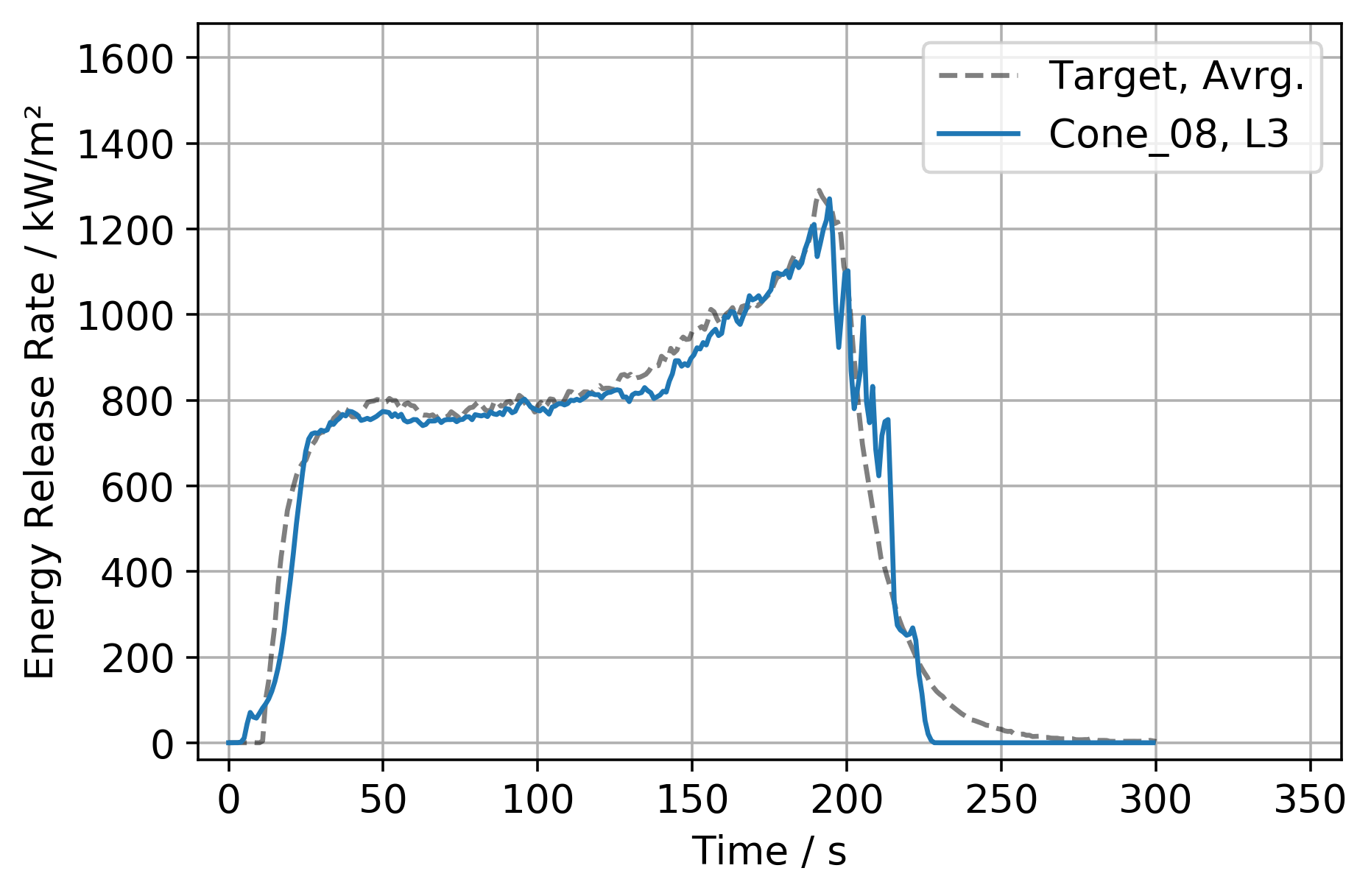} }}%
    
    \caption{Energy release in simplified cone calorimeter simulation at 65 kW/m².}%
    \label{fig:ConeSimBestParaHRR_Aalto}%
\end{figure}

\clearpage
\subsection{Back Side Temperatures}

\begin{figure*}[h]%
    \centering
    \subfloat[\centering IMP Setup Cone\_01.]{{\includegraphics[width=0.395\columnwidth]{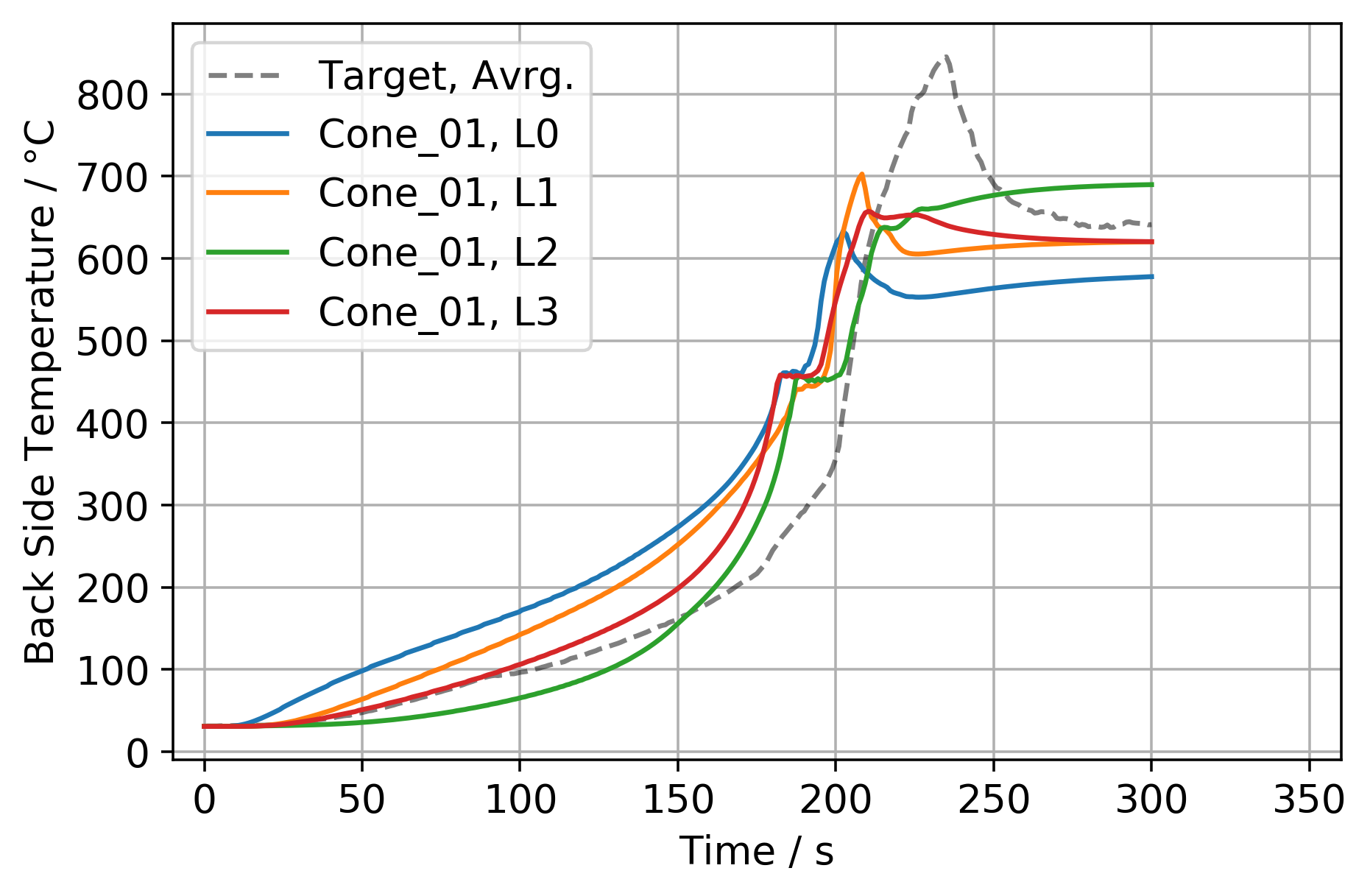} }}%
    \hfill
    \subfloat[\centering IMP Setup Cone\_02.]{{\includegraphics[width=0.395\columnwidth]{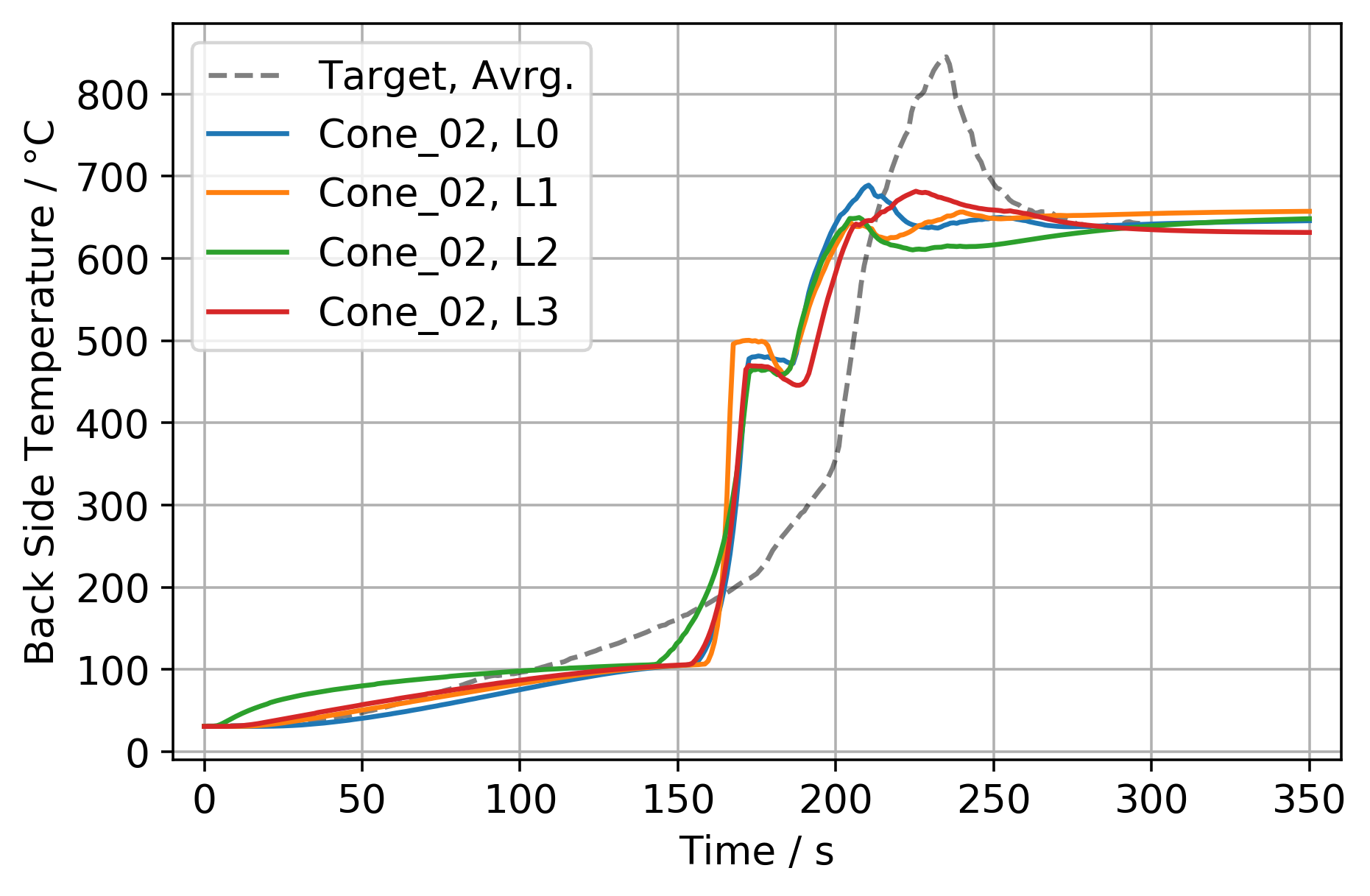} }}%
    \hfill
    \subfloat[\centering IMP Setup Cone\_03.]{{\includegraphics[width=0.395\columnwidth]{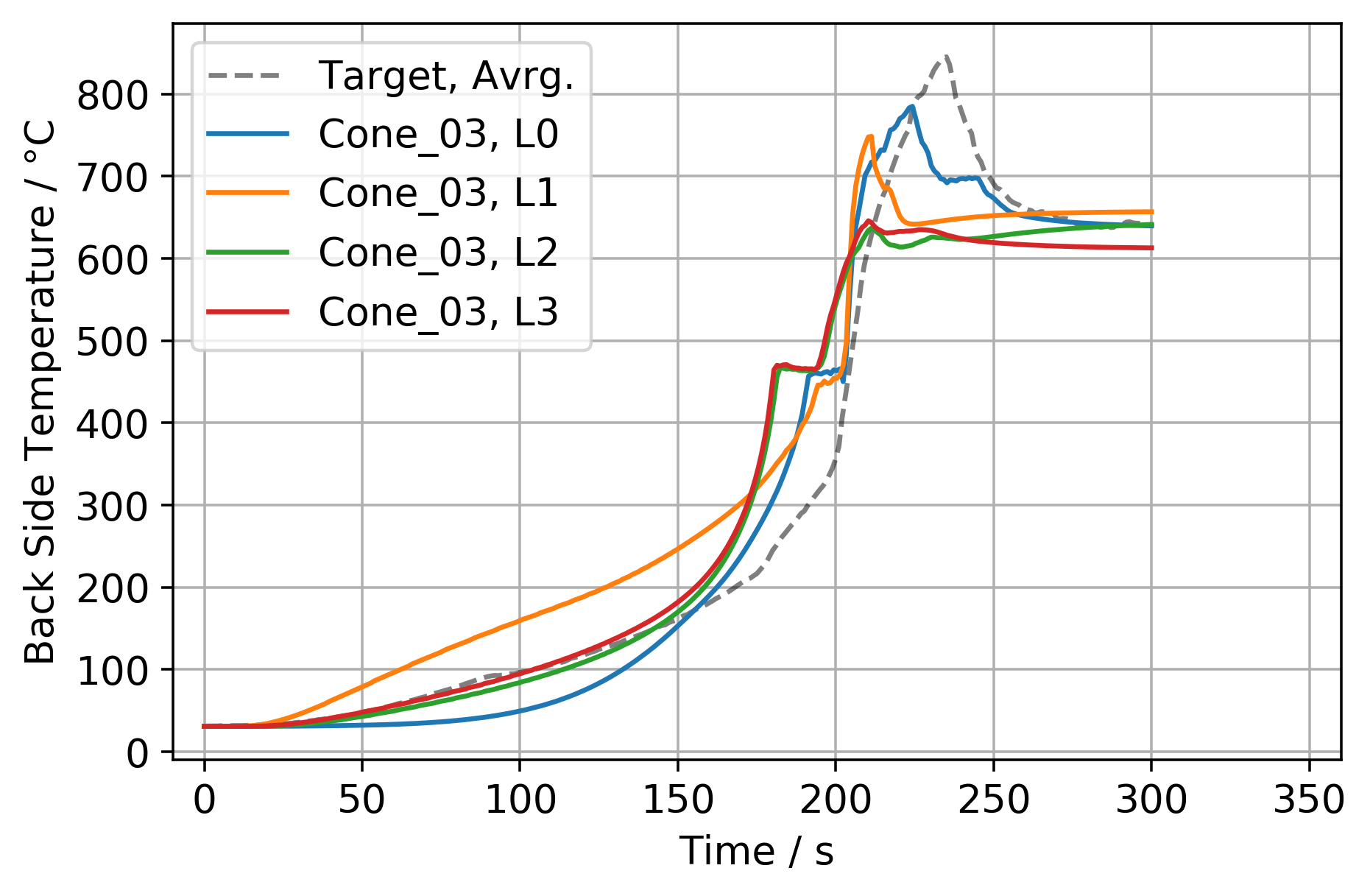} }}%
    \hfill
    \subfloat[\centering IMP Setup Cone\_04.]{{\includegraphics[width=0.395\columnwidth]{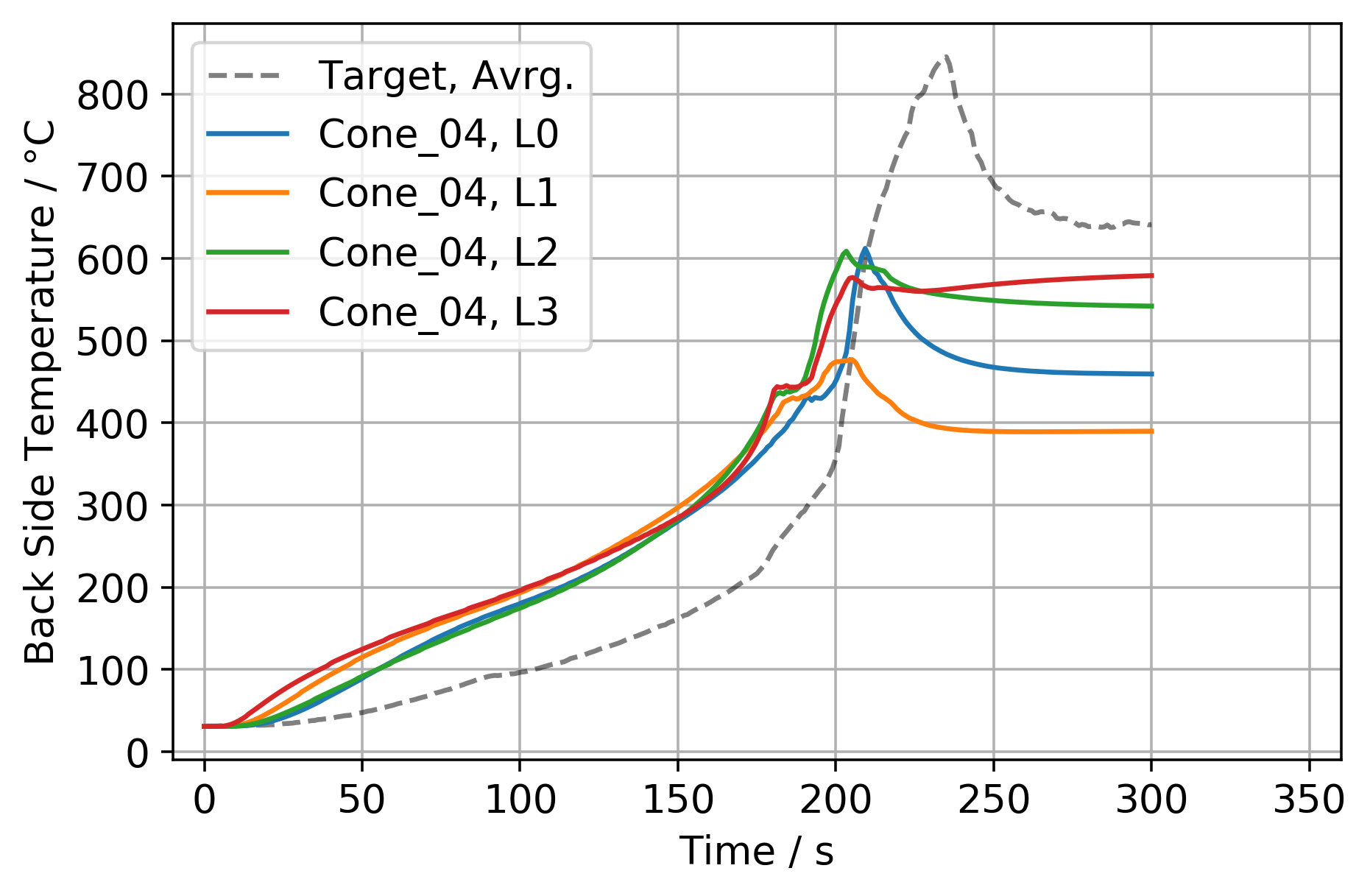} }}%
    \hfill
    \subfloat[\centering IMP Setup Cone\_05.]{{\includegraphics[width=0.395\columnwidth]{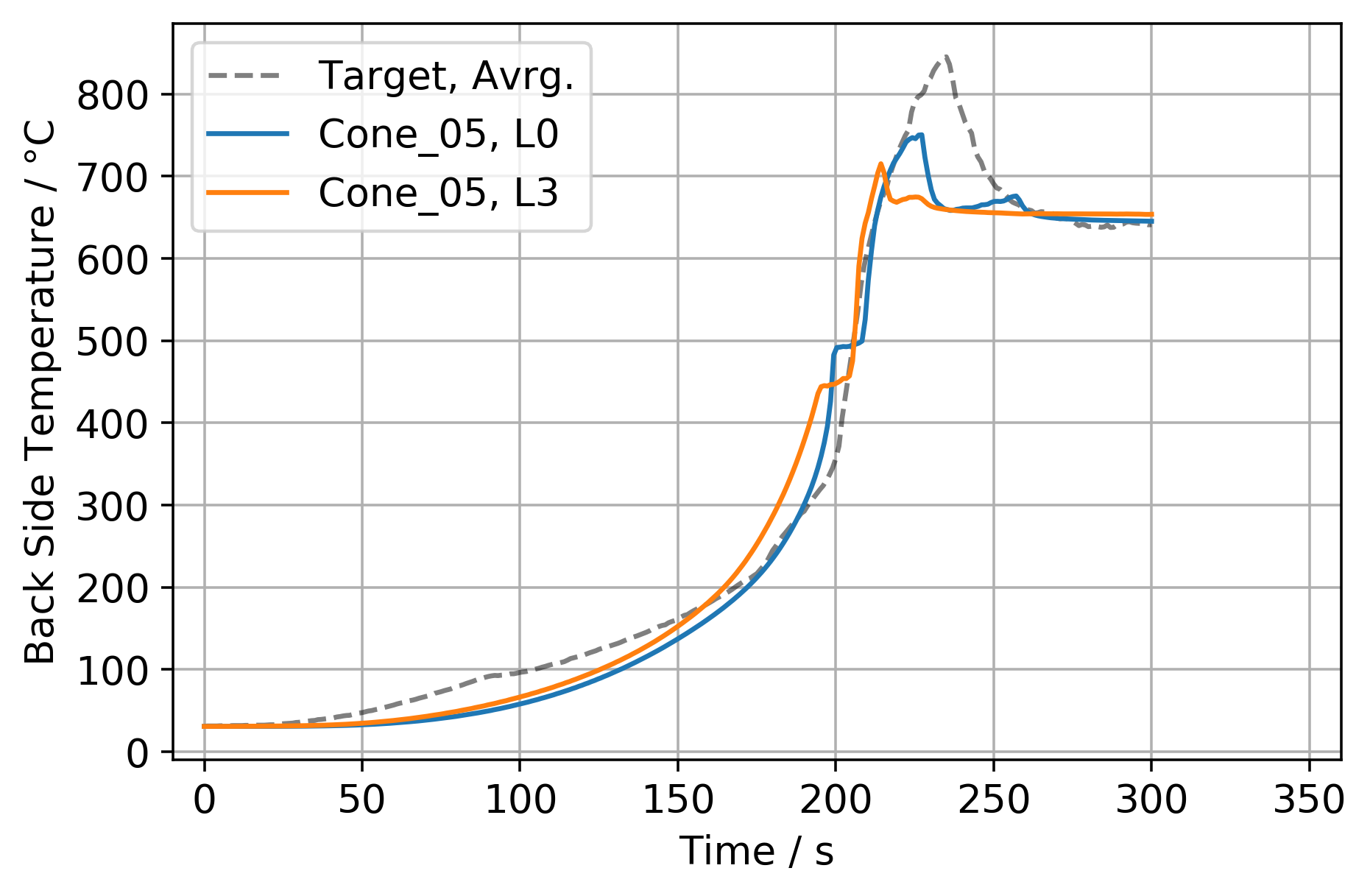} }}%
    \hfill
    \subfloat[\centering IMP Setup Cone\_06.]{{\includegraphics[width=0.395\columnwidth]{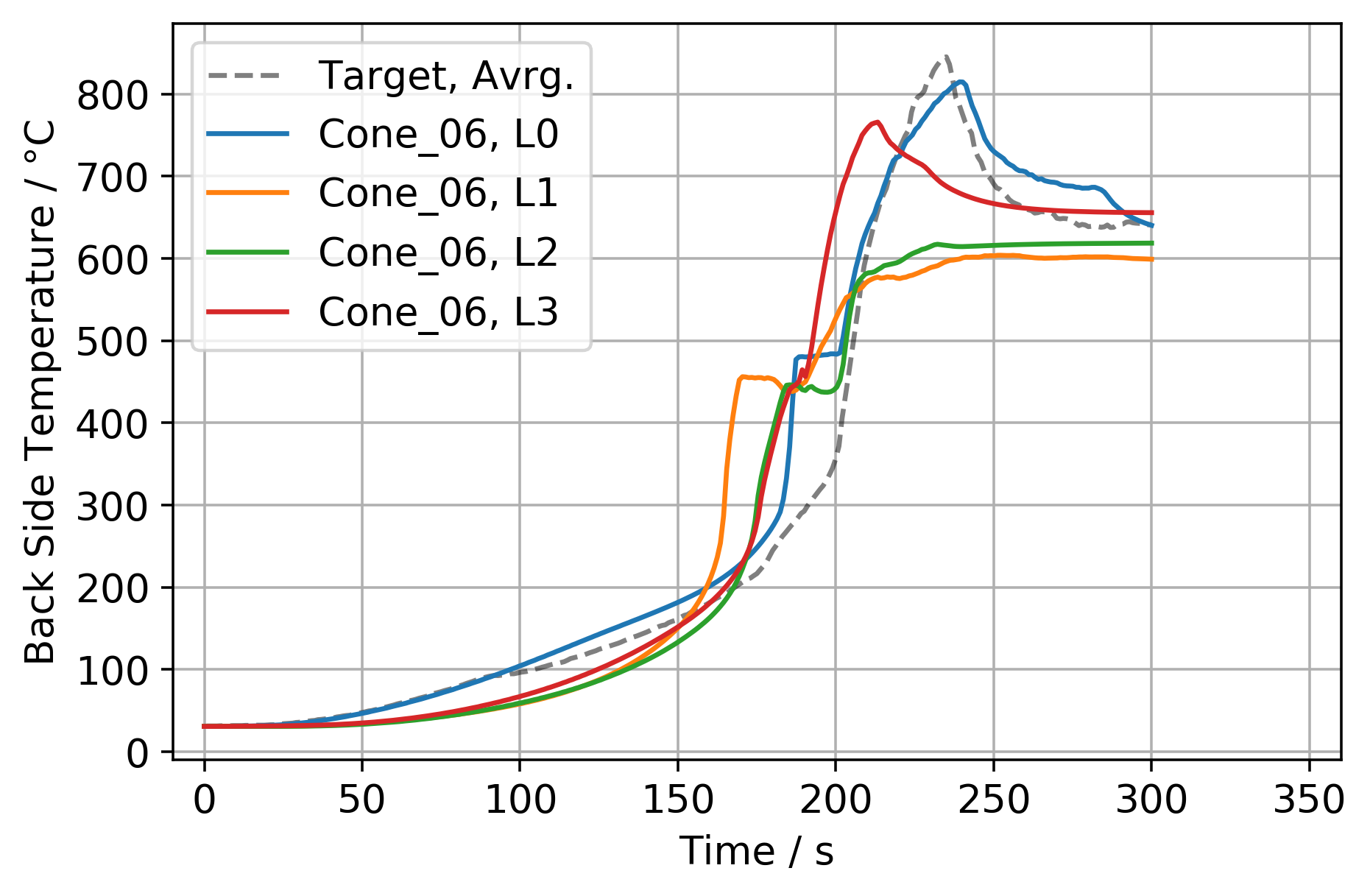} }}%
    \hfill
    \subfloat[\centering IMP Setup Cone\_07.]{{\includegraphics[width=0.395\columnwidth]{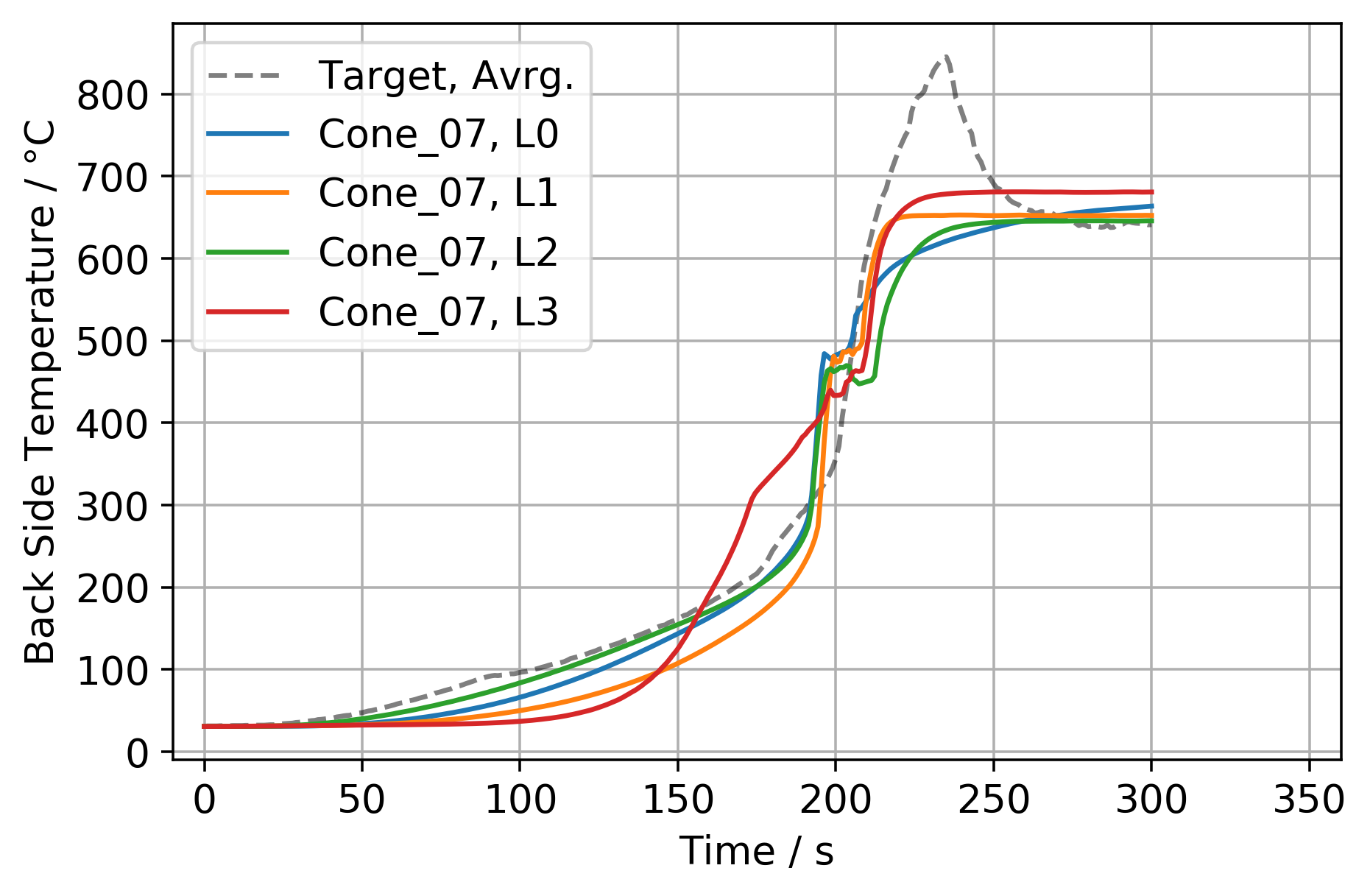} }}%
    \hfill
    \subfloat[\centering IMP Setup Cone\_08.]{{\includegraphics[width=0.395\columnwidth]{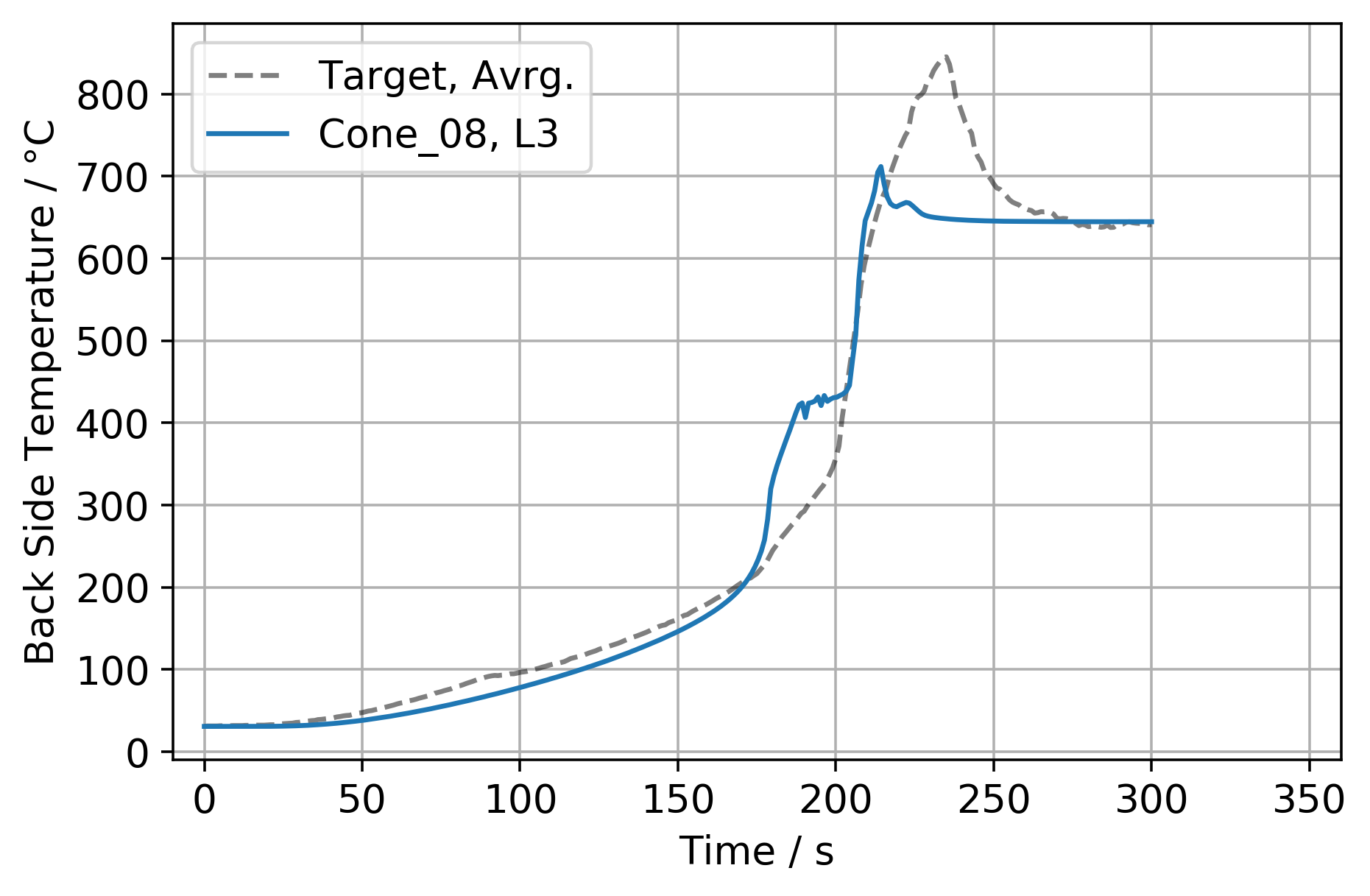} }}%
    
    \caption{Back side temperature in simplified cone calorimeter simulation at 65 kW/m².}%
    \label{fig:ConeSimBestParaBackTemp_Aalto}%
\end{figure*}

\clearpage
\subsection{Thermal Conductivity}
\label{apx_sec:conductivity}

\begin{figure}[h]%
    \centering
    \subfloat[\centering IMP Setup Cone\_01.]{{\includegraphics[width=0.395\columnwidth]{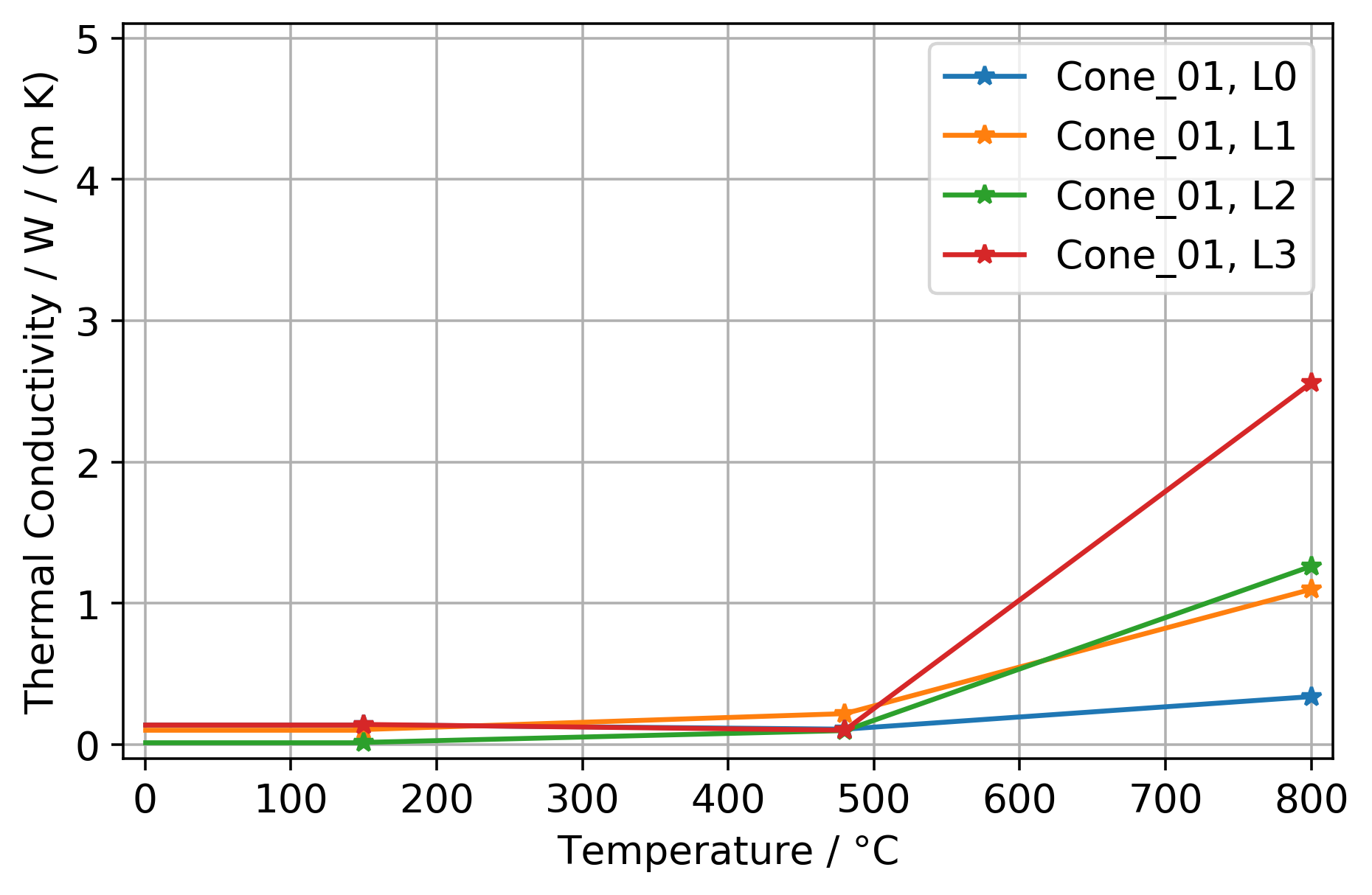} }}%
    \hfill
    \subfloat[\centering IMP Setup Cone\_02.]{{\includegraphics[width=0.395\columnwidth]{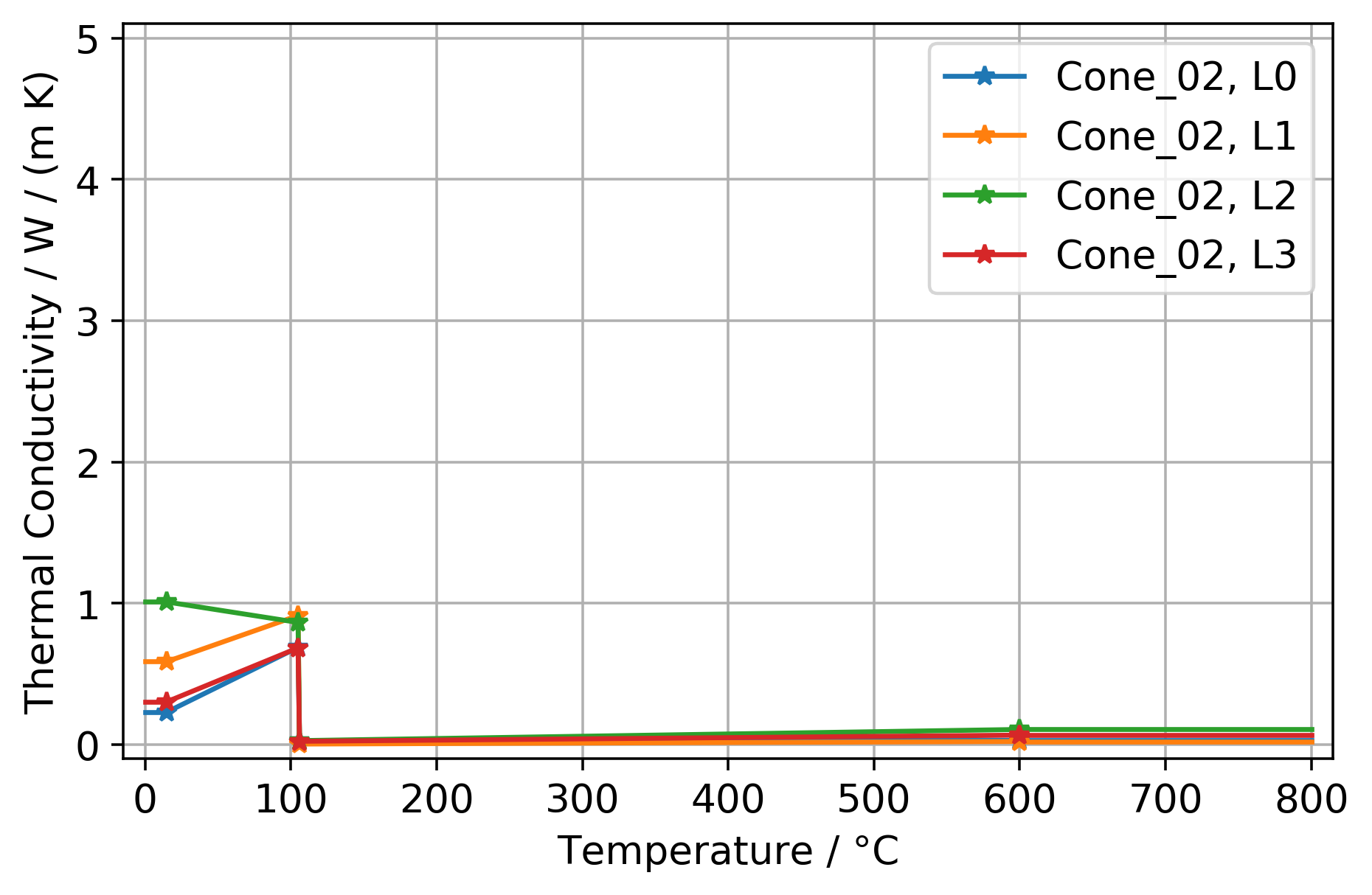} }}%
    \hfill
    \subfloat[\centering IMP Setup Cone\_03.]{{\includegraphics[width=0.395\columnwidth]{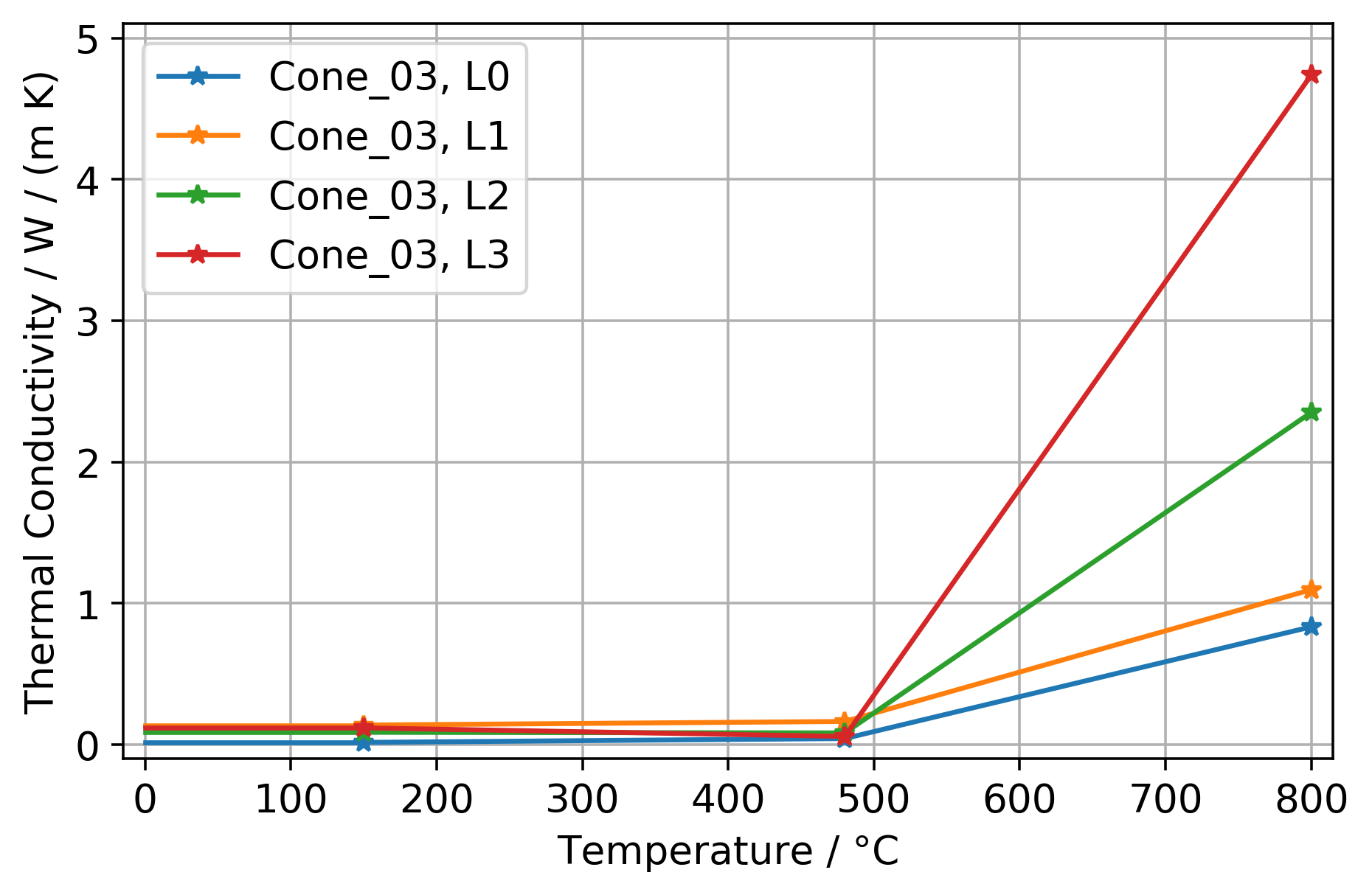} }}%
    \hfill
    \subfloat[\centering IMP Setup Cone\_04.]{{\includegraphics[width=0.395\columnwidth]{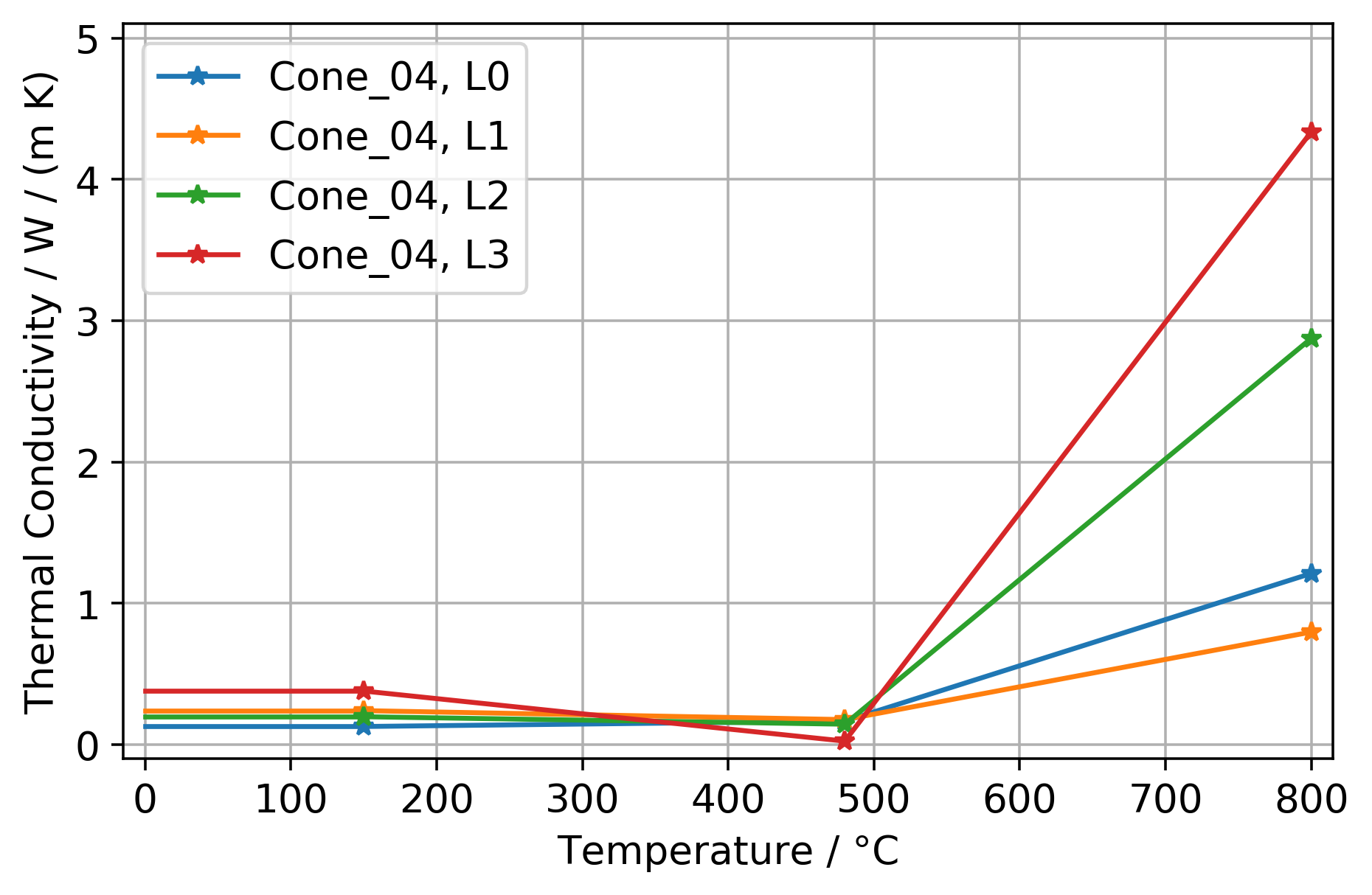} }}%
    \hfill
    \subfloat[\centering IMP Setup Cone\_05.]{{\includegraphics[width=0.395\columnwidth]{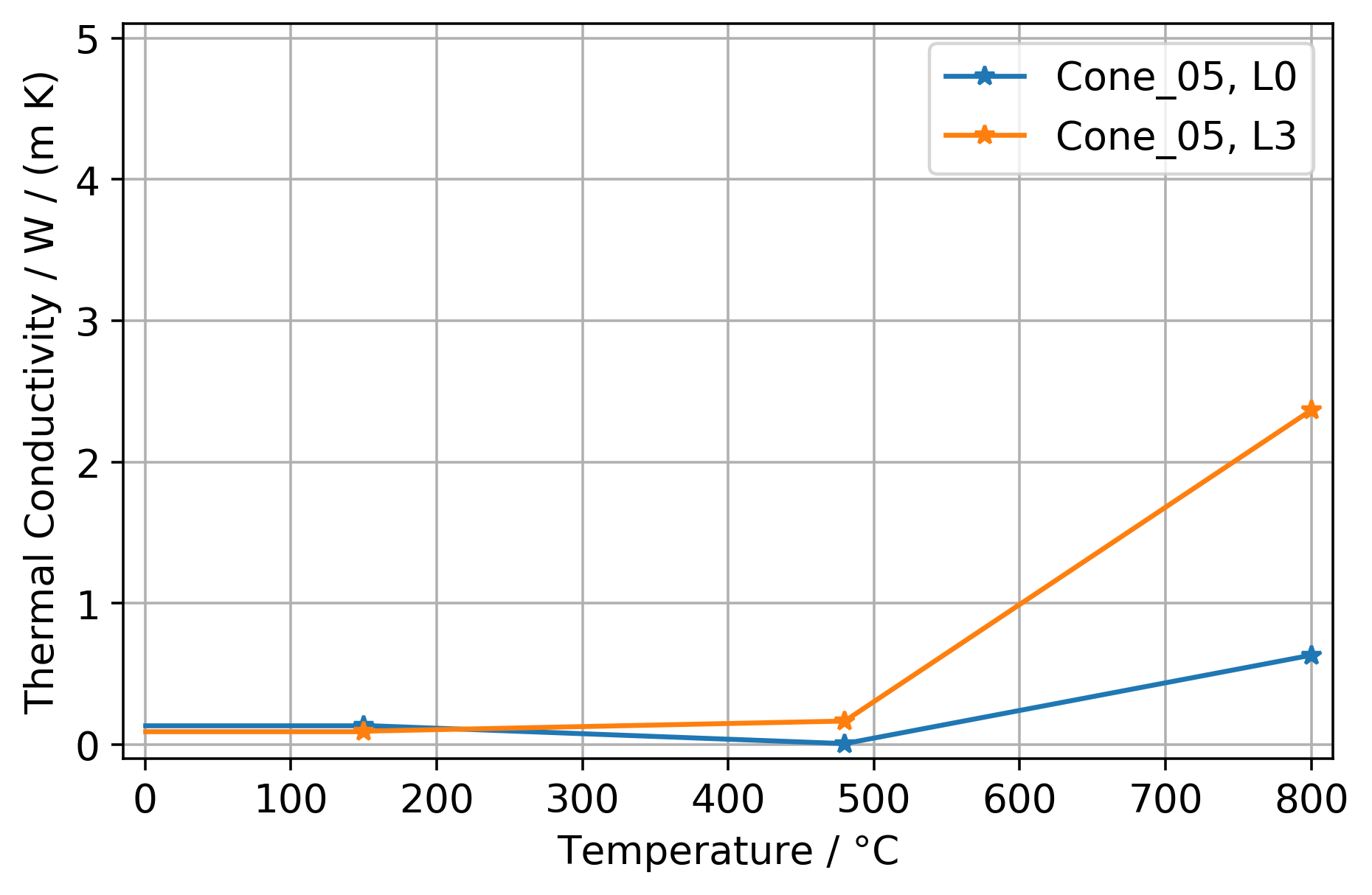} }}%
    \hfill
    \subfloat[\centering IMP Setup Cone\_06.]{{\includegraphics[width=0.395\columnwidth]{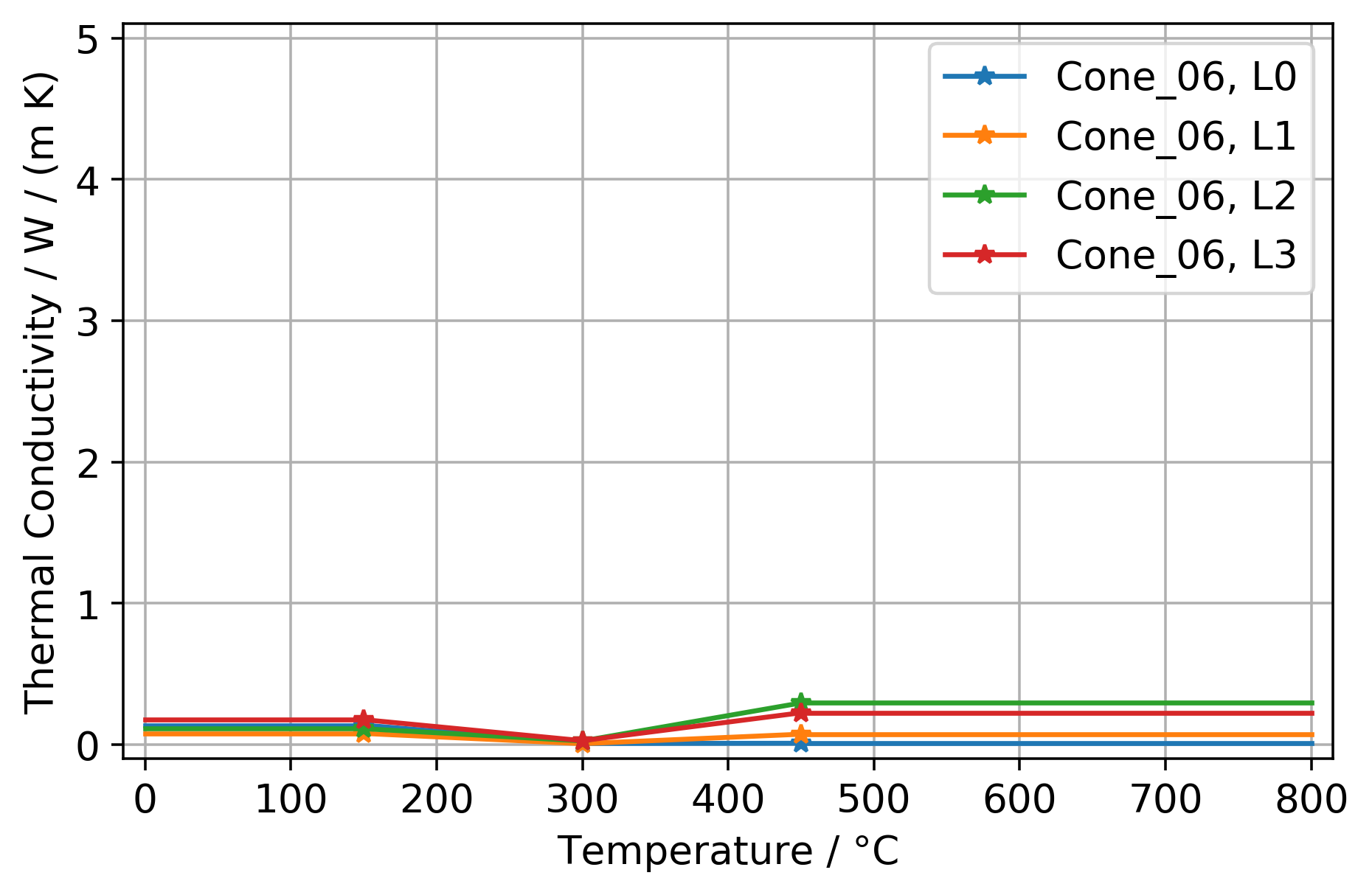} }}%
    \hfill
    \subfloat[\centering IMP Setup Cone\_07.]{{\includegraphics[width=0.395\columnwidth]{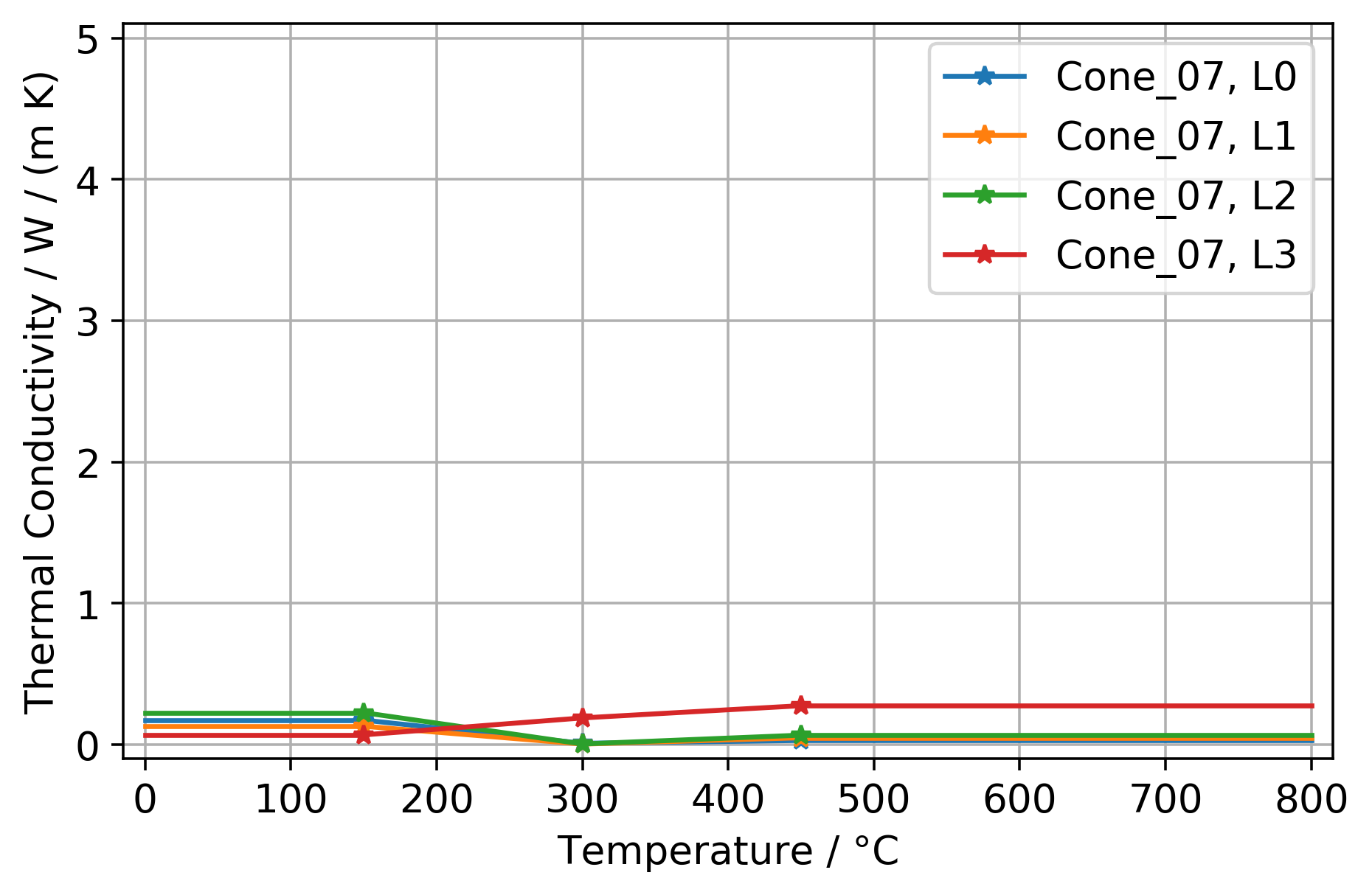} }}%
    \hfill
    \subfloat[\centering IMP Setup Cone\_08.]{{\includegraphics[width=0.395\columnwidth]{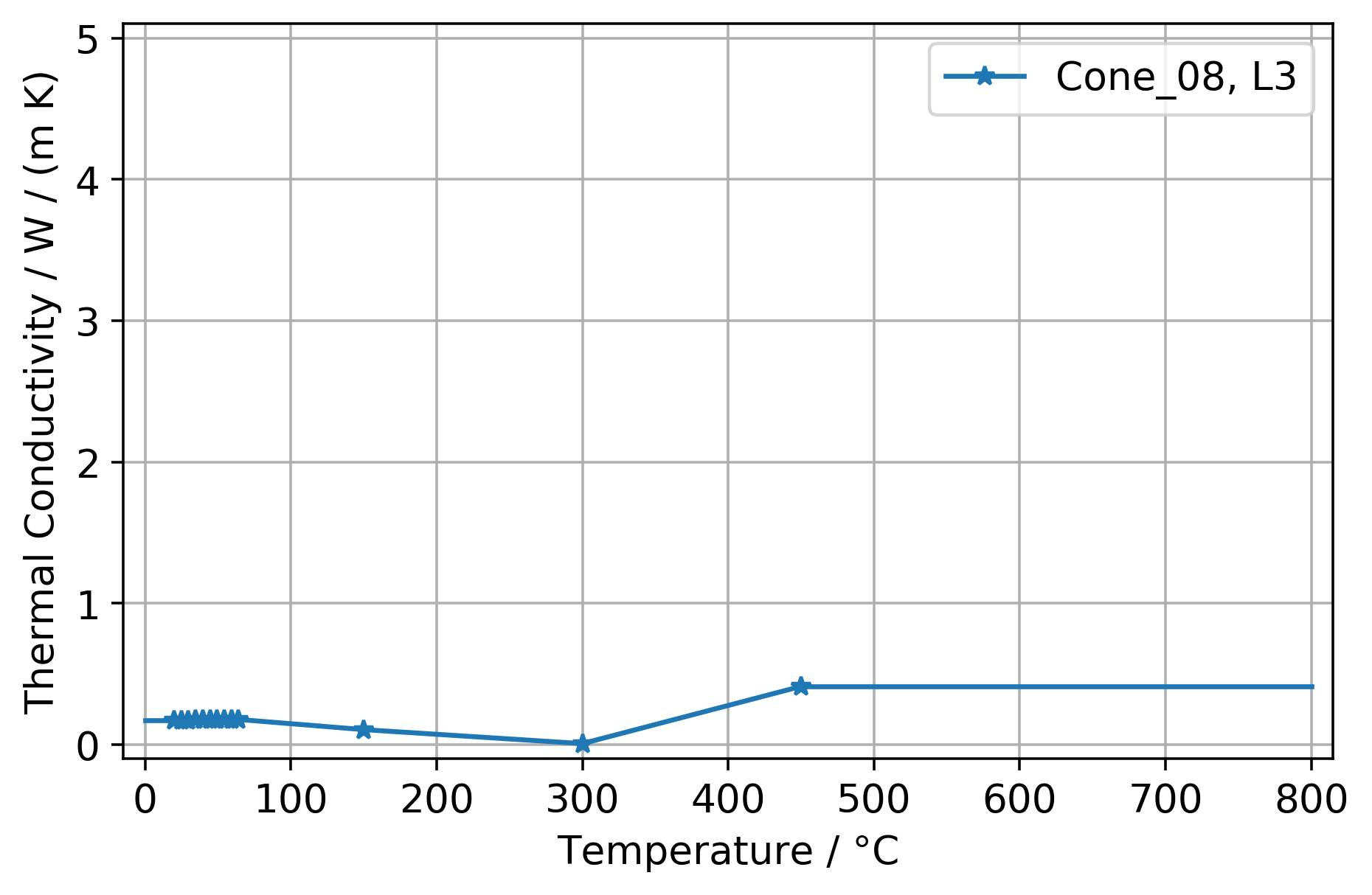} }}%
    
    \caption{Thermal conductivity in simplified cone calorimeter simulation at 65~kW/m².}%
    \label{fig:ConeSimBestParaCond_Aalto}%
\end{figure}

\clearpage
\subsection{Specific Heat}
\label{apx_sec:spec_heat}

\begin{figure}[h]%
    \centering
    \subfloat[\centering IMP Setup Cone\_01.]{{\includegraphics[width=0.395\columnwidth]{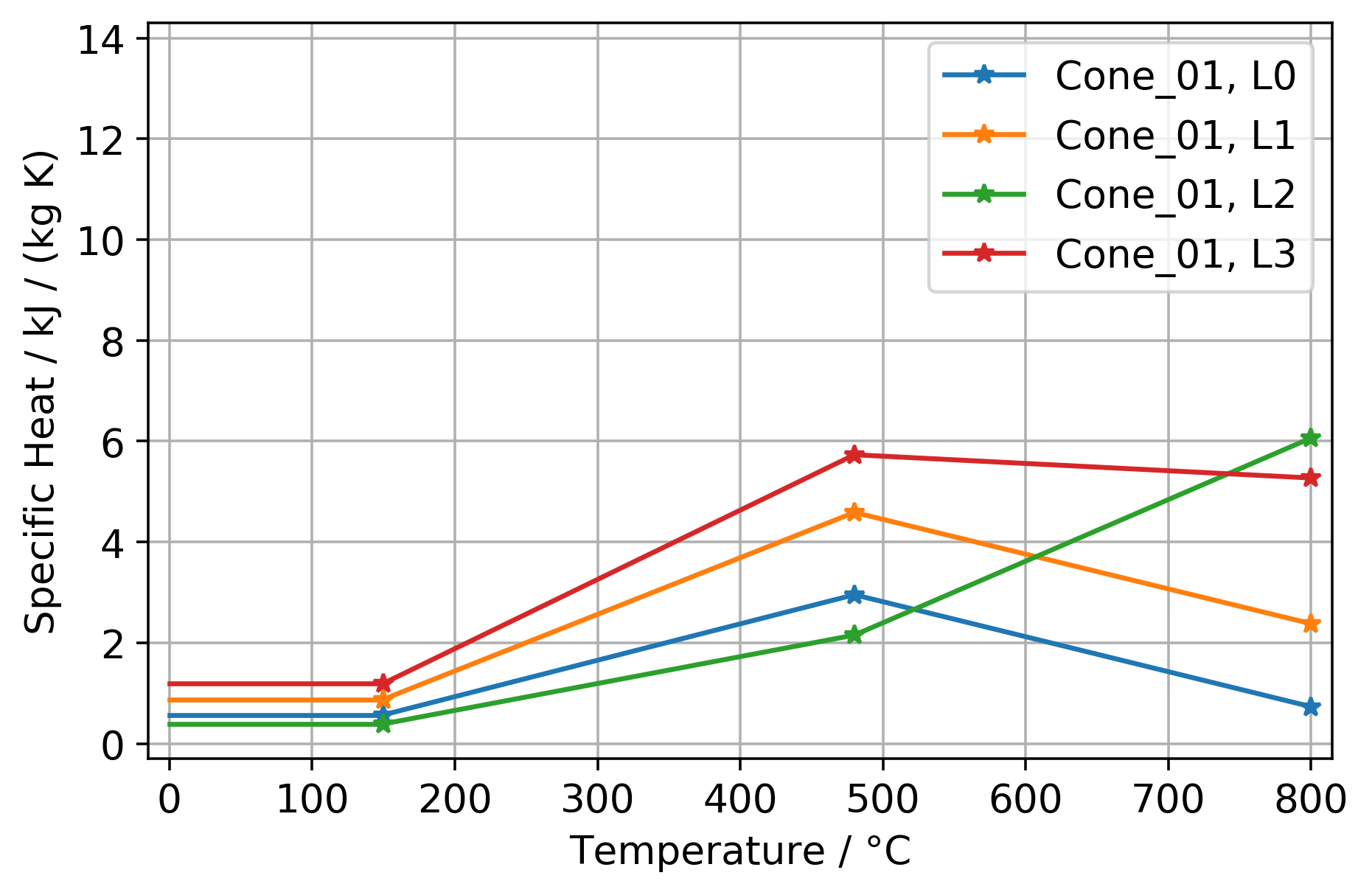} }}%
    \hfill
    \subfloat[\centering IMP Setup Cone\_02.]{{\includegraphics[width=0.395\columnwidth]{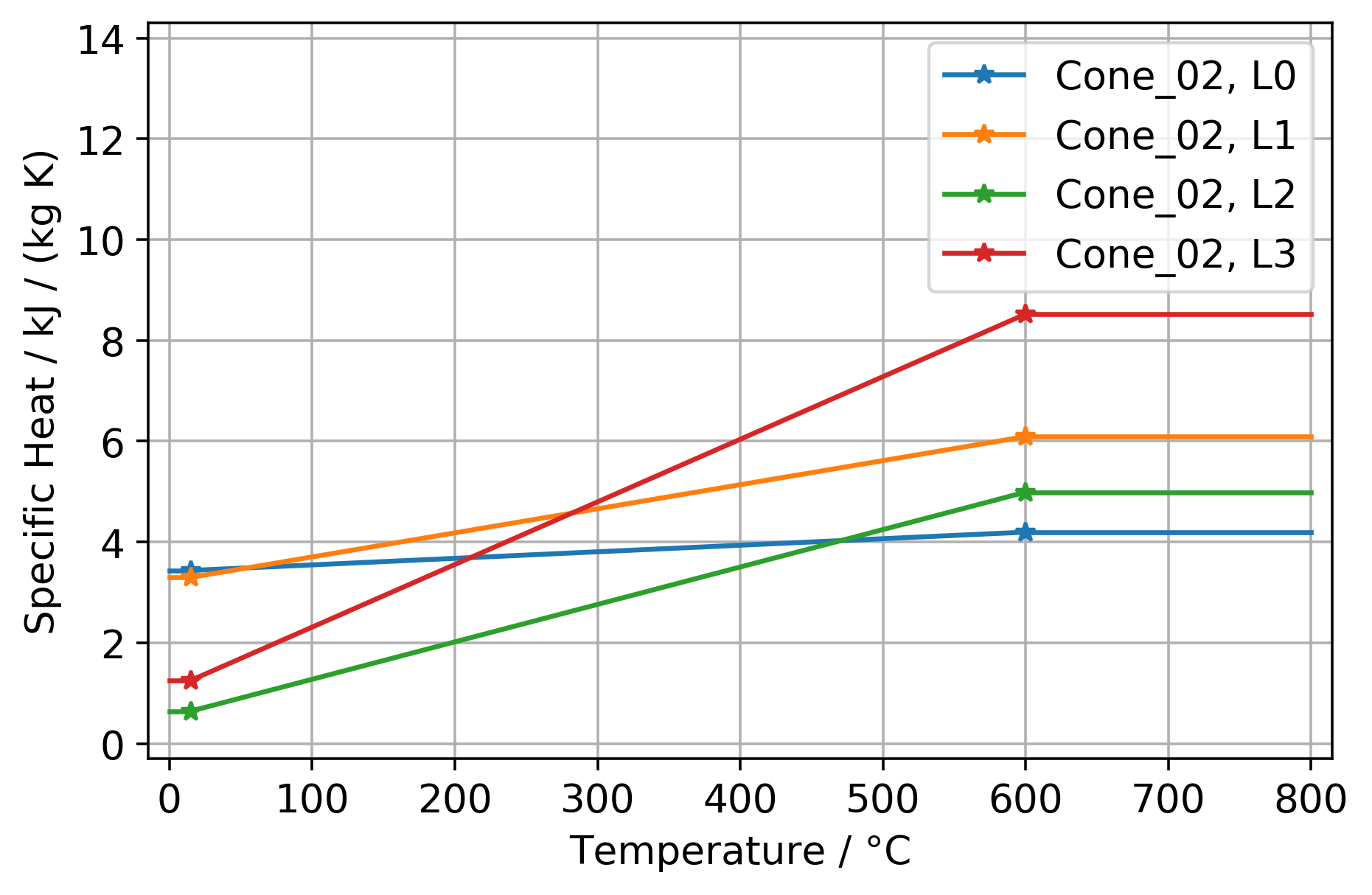} }}%
    \hfill
    \subfloat[\centering IMP Setup Cone\_03.]{{\includegraphics[width=0.395\columnwidth]{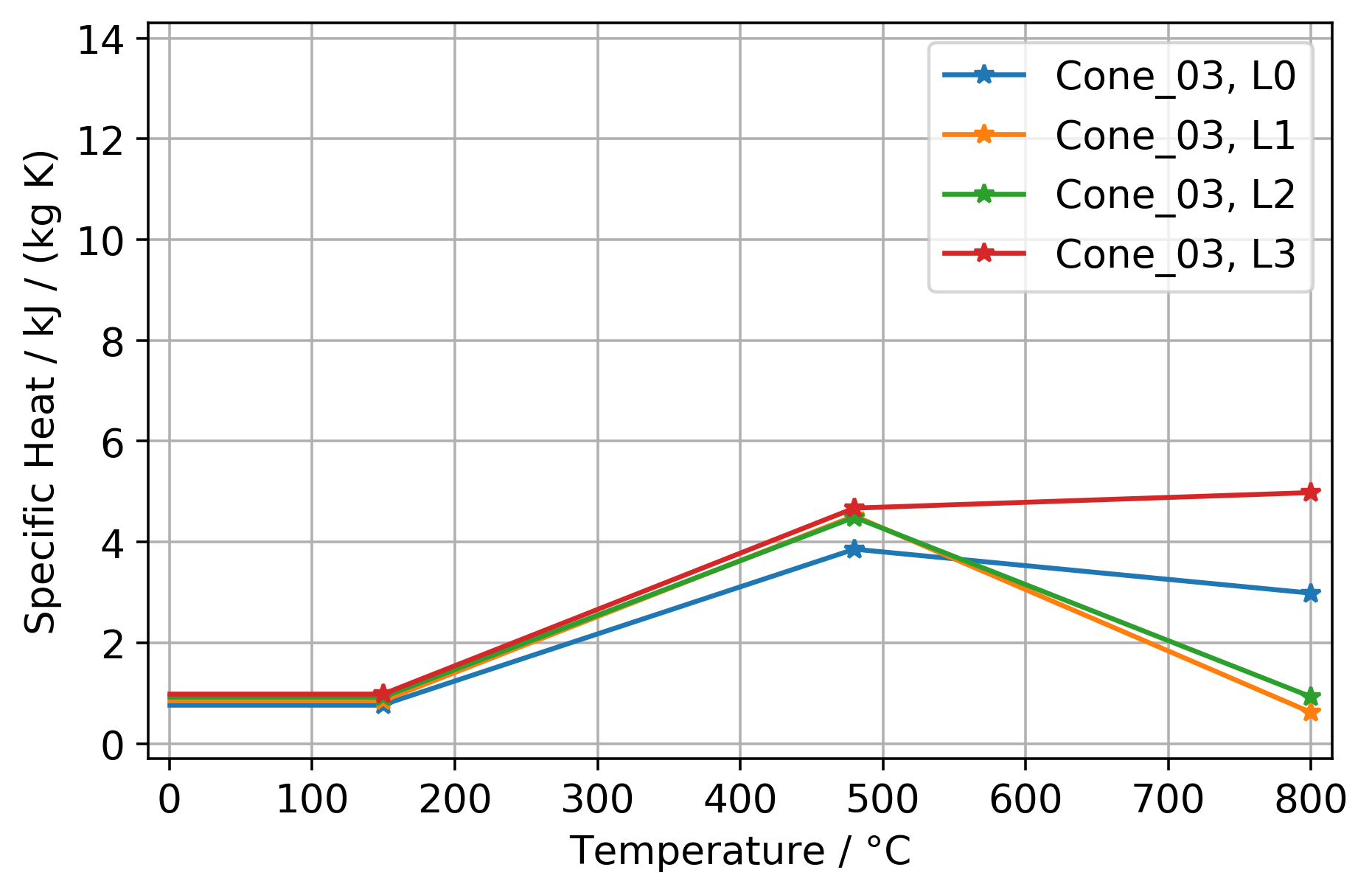} }}%
    \hfill
    \subfloat[\centering IMP Setup Cone\_04.]{{\includegraphics[width=0.395\columnwidth]{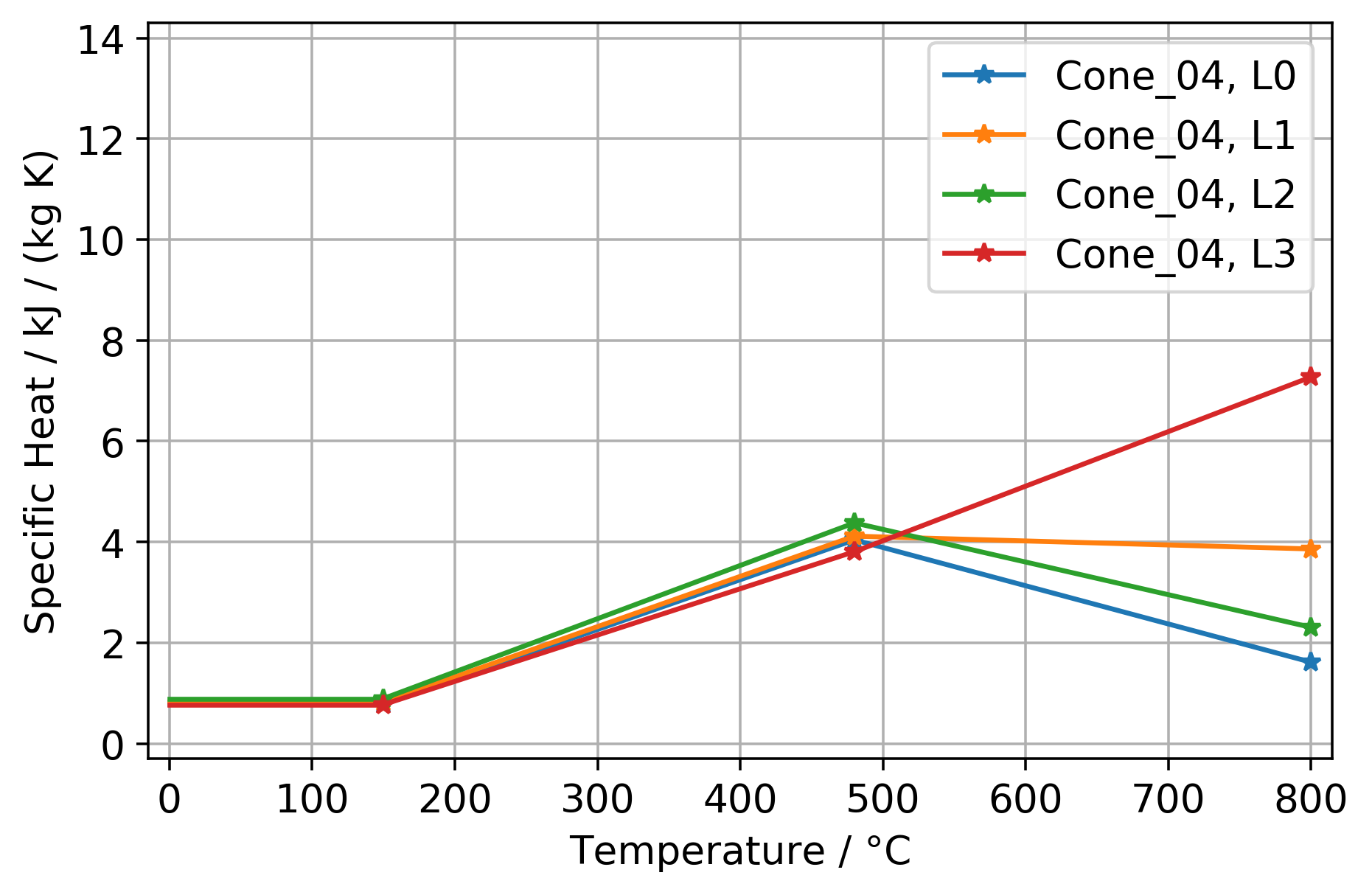} }}%
    \hfill
    \subfloat[\centering IMP Setup Cone\_05.]{{\includegraphics[width=0.395\columnwidth]{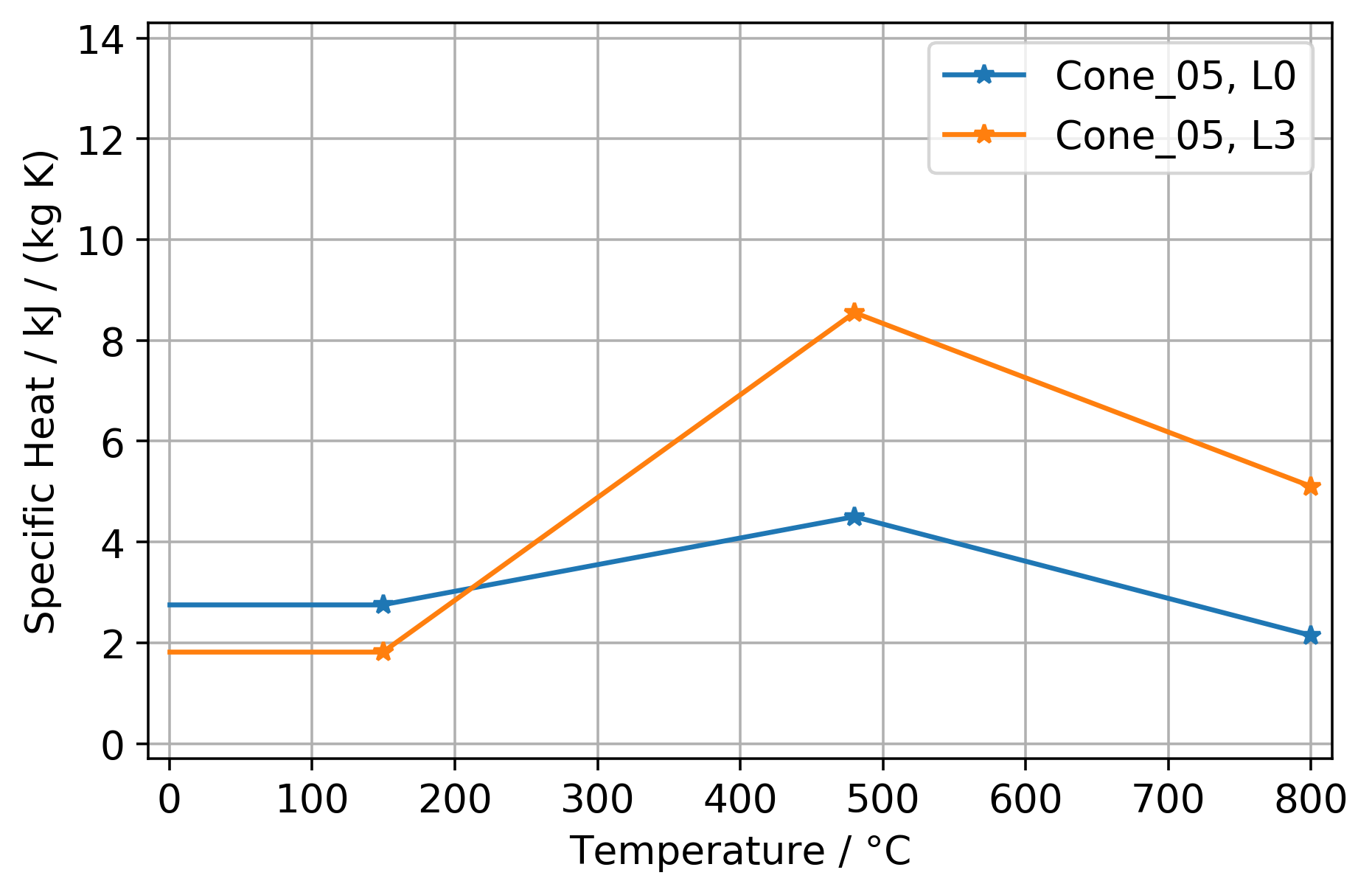} }}%
    \hfill
    \subfloat[\centering IMP Setup Cone\_06.]{{\includegraphics[width=0.395\columnwidth]{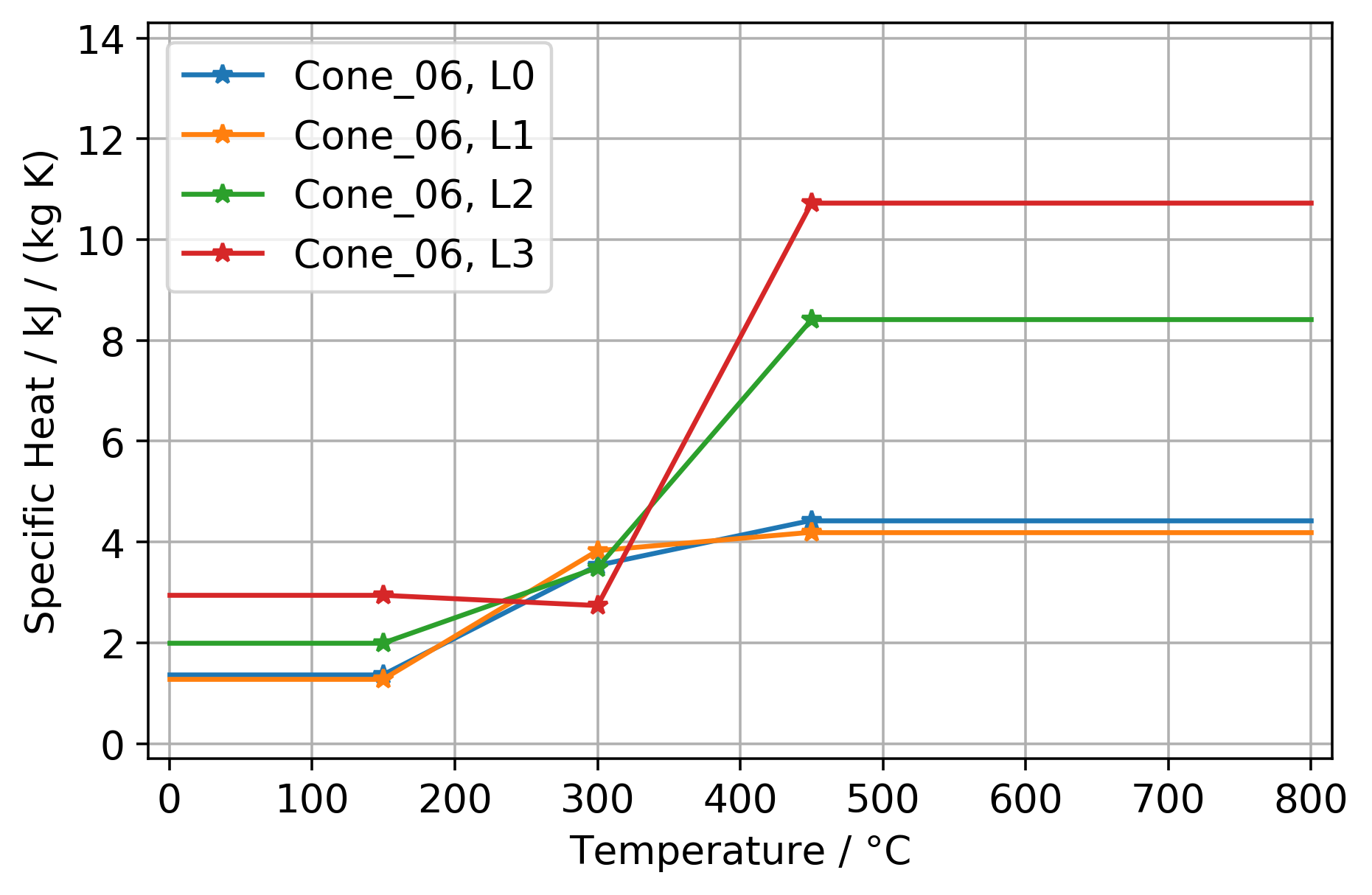} }}%
    \hfill
    \subfloat[\centering IMP Setup Cone\_07.]{{\includegraphics[width=0.395\columnwidth]{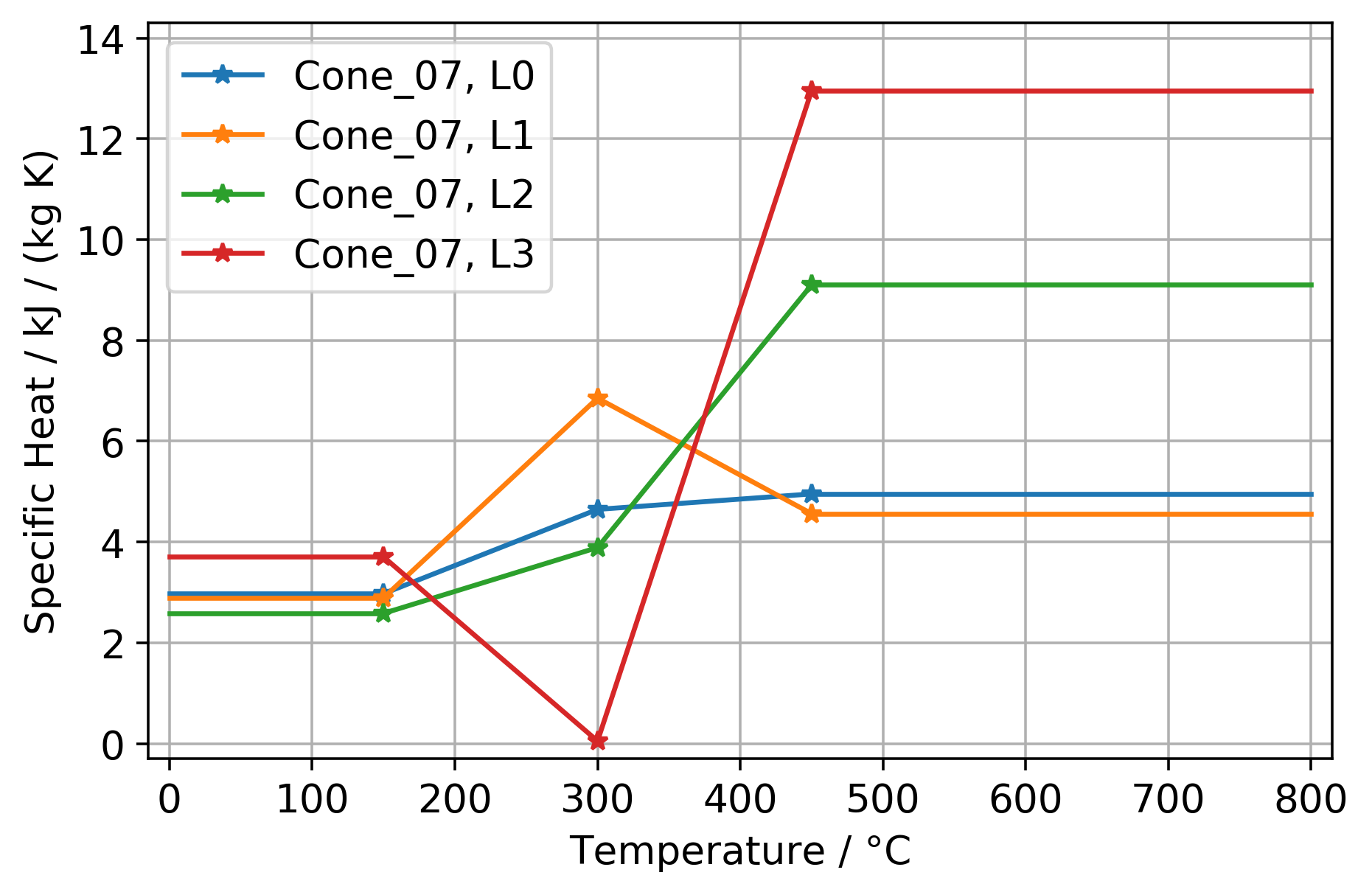} }}%
    \hfill
    \subfloat[\centering IMP Setup Cone\_08.]{{\includegraphics[width=0.395\columnwidth]{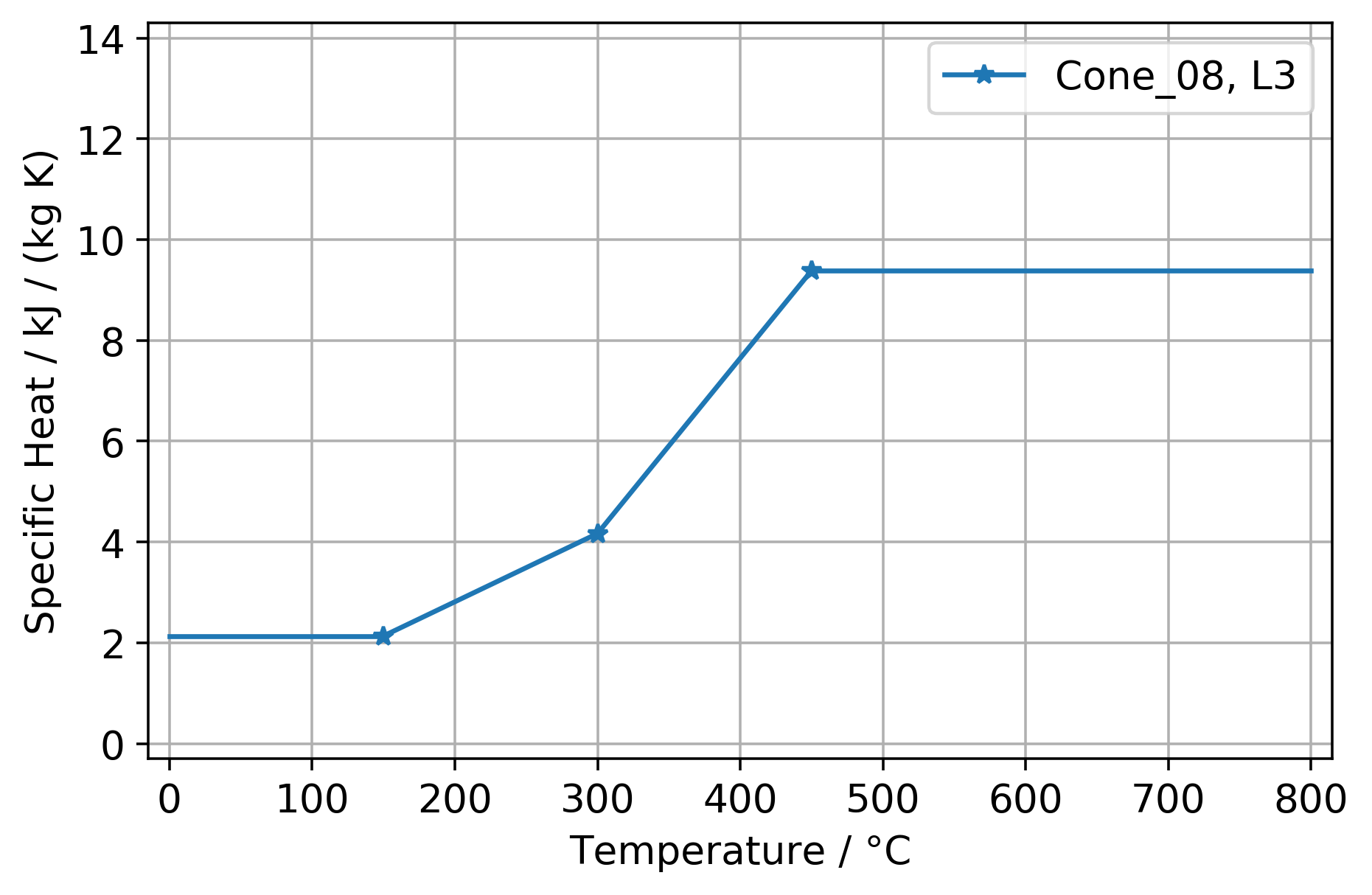} }}%
    
    \caption{Specific heat in simplified cone calorimeter simulation at 65 kW/m².}%
    \label{fig:ConeSimBestParaSpecHeat_Aalto}%
\end{figure}

\clearpage
\subsection{Residual Sample Mass}

\begin{figure*}[h]%
    \centering
    \subfloat[\centering IMP Setup Cone\_01.]{{\includegraphics[width=0.395\columnwidth]{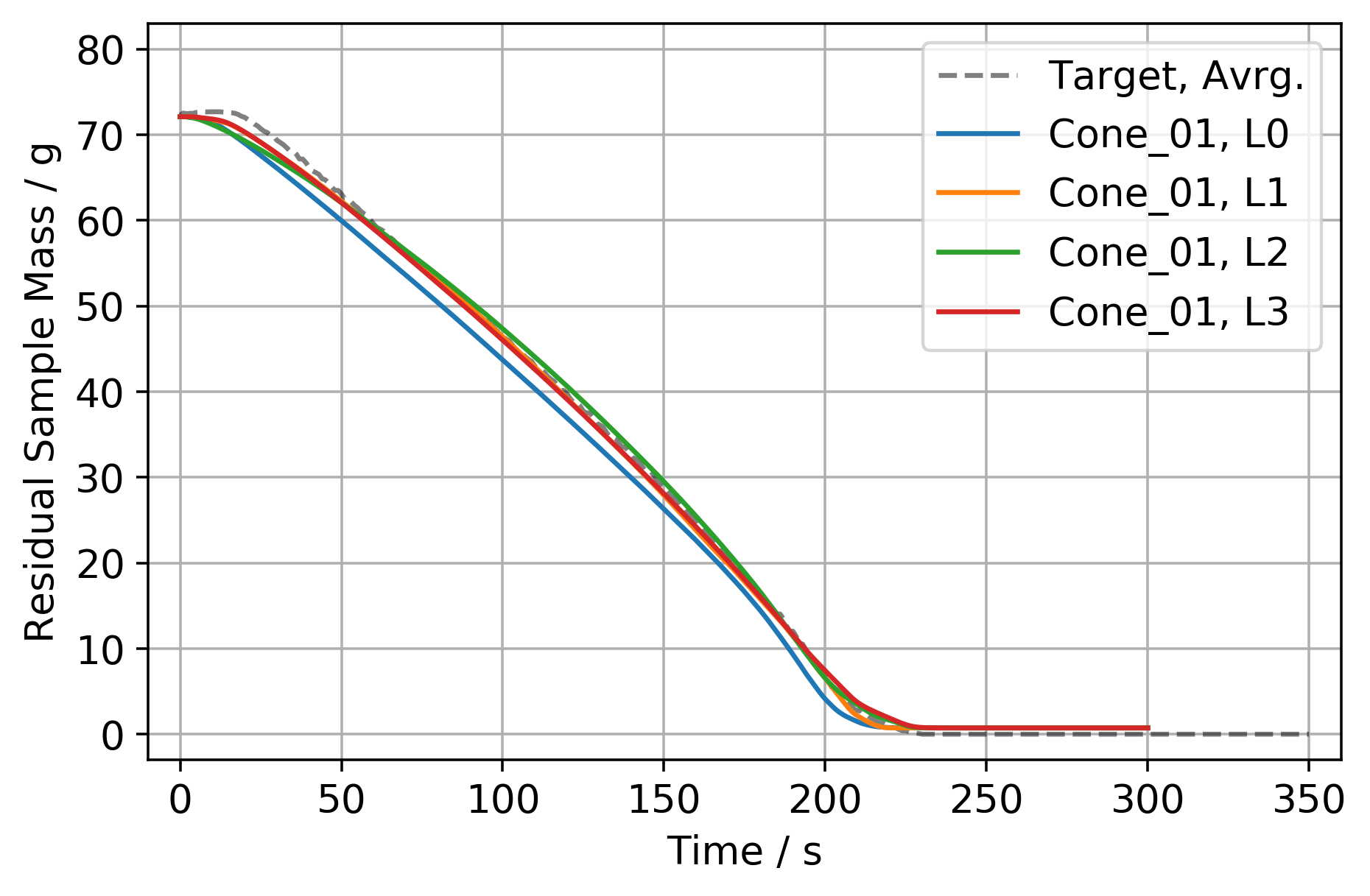} }}%
    \hfill
    \subfloat[\centering IMP Setup Cone\_02.]{{\includegraphics[width=0.395\columnwidth]{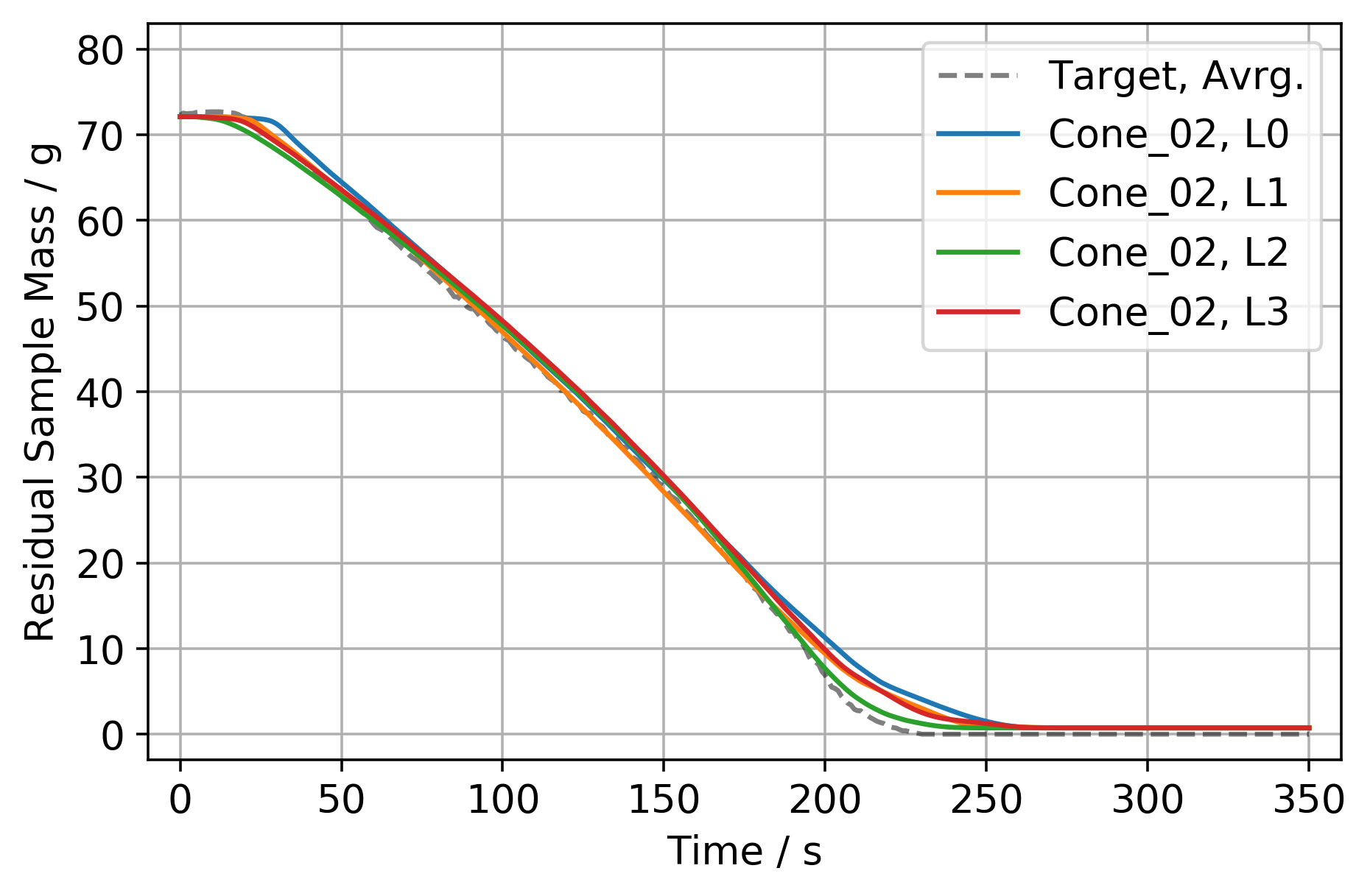} }}%
    \hfill
    \subfloat[\centering IMP Setup Cone\_03.]{{\includegraphics[width=0.395\columnwidth]{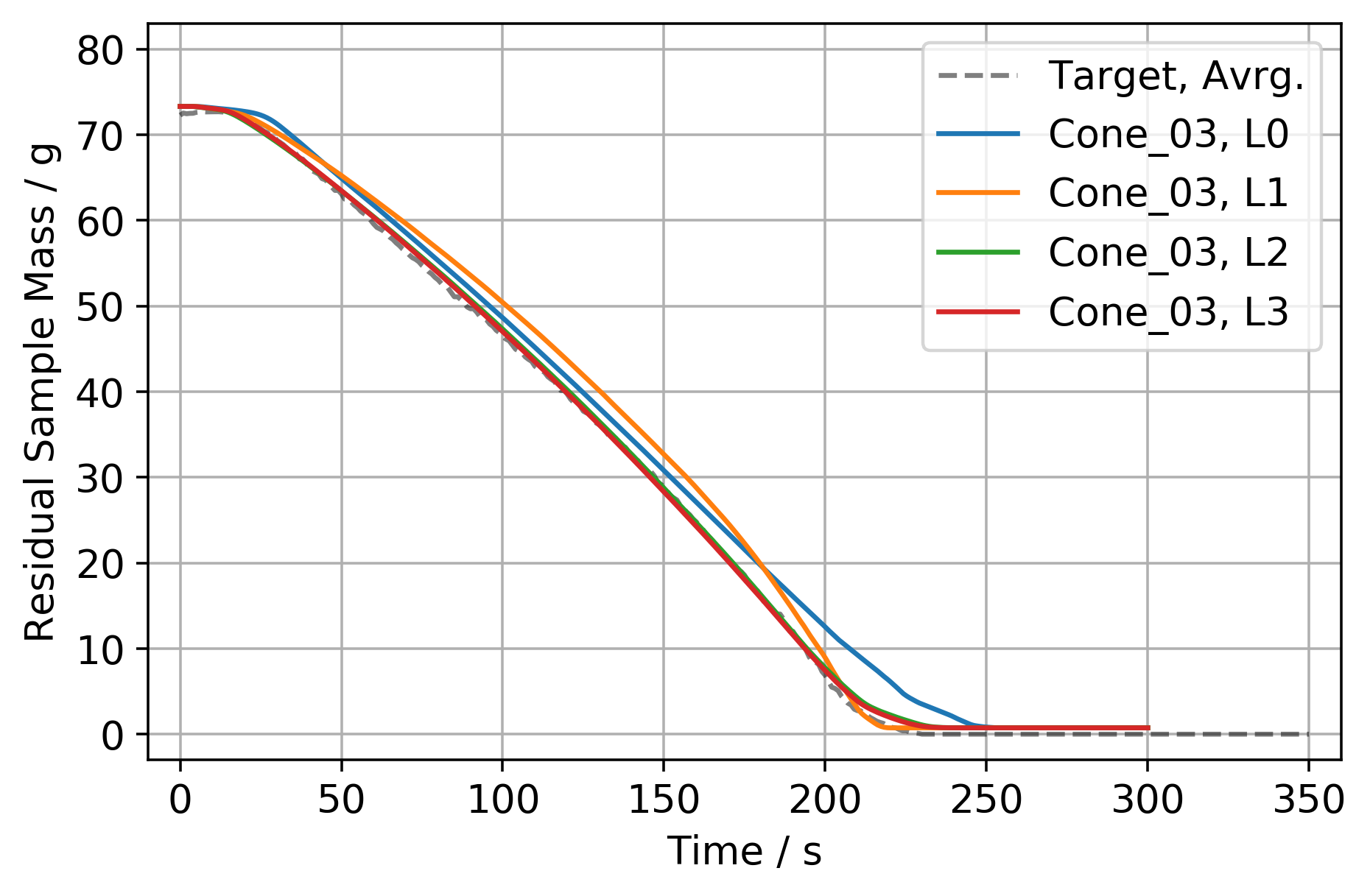} }}%
    \hfill
    \subfloat[\centering IMP Setup Cone\_04.]{{\includegraphics[width=0.395\columnwidth]{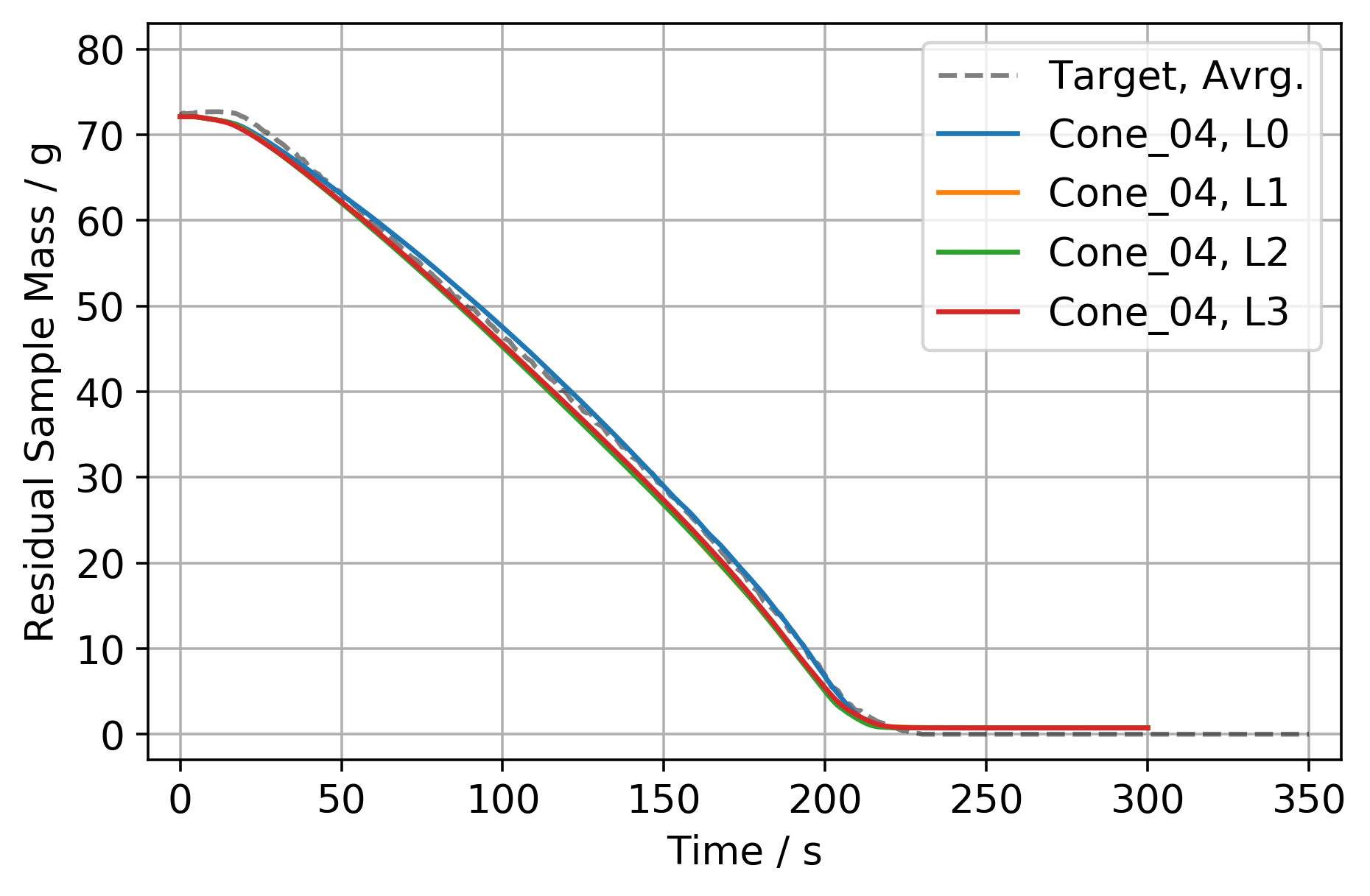} }}%
    \hfill
    \subfloat[\centering IMP Setup Cone\_05.]{{\includegraphics[width=0.395\columnwidth]{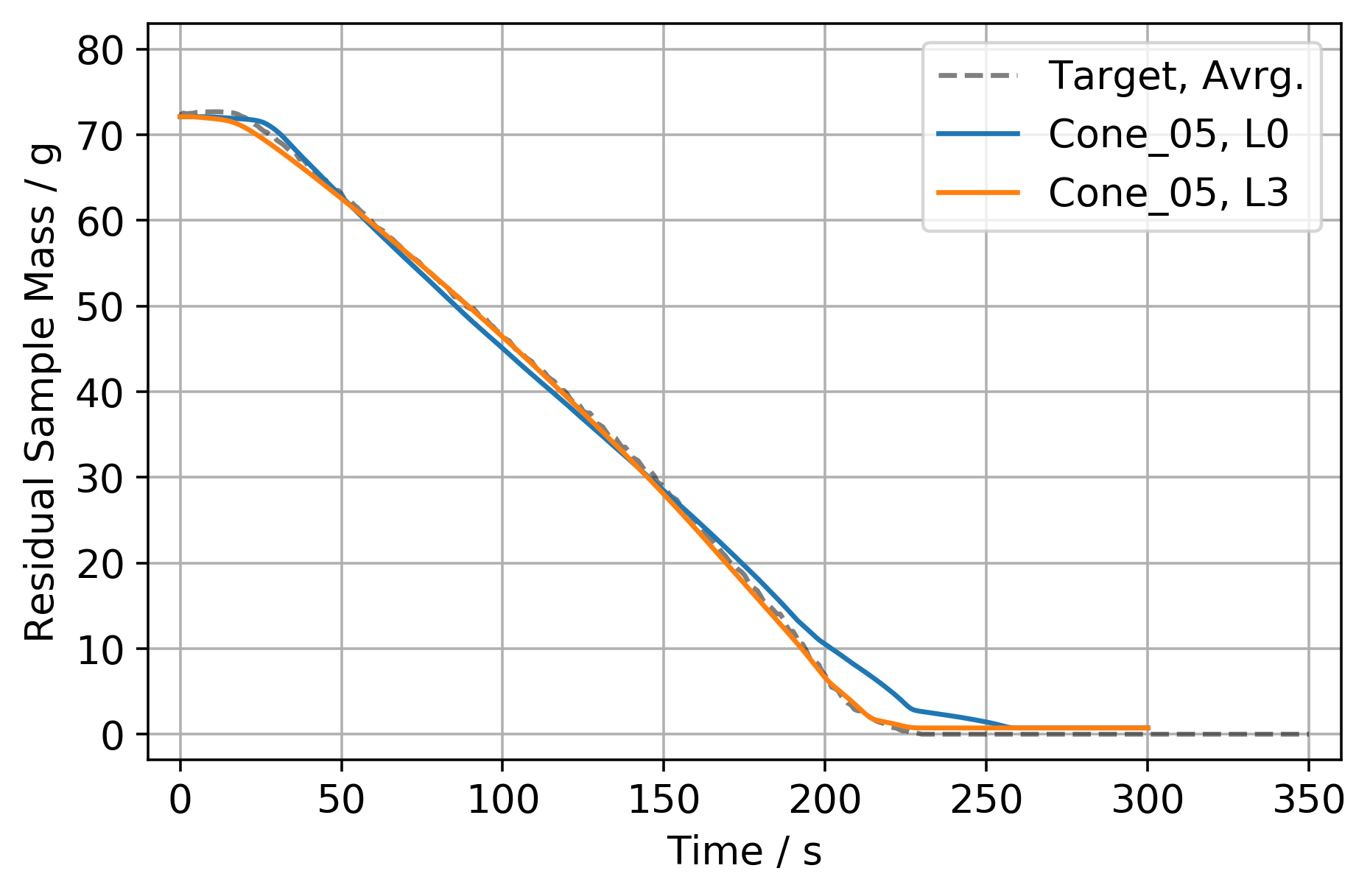} }}%
    \hfill
    \subfloat[\centering IMP Setup Cone\_06.]{{\includegraphics[width=0.395\columnwidth]{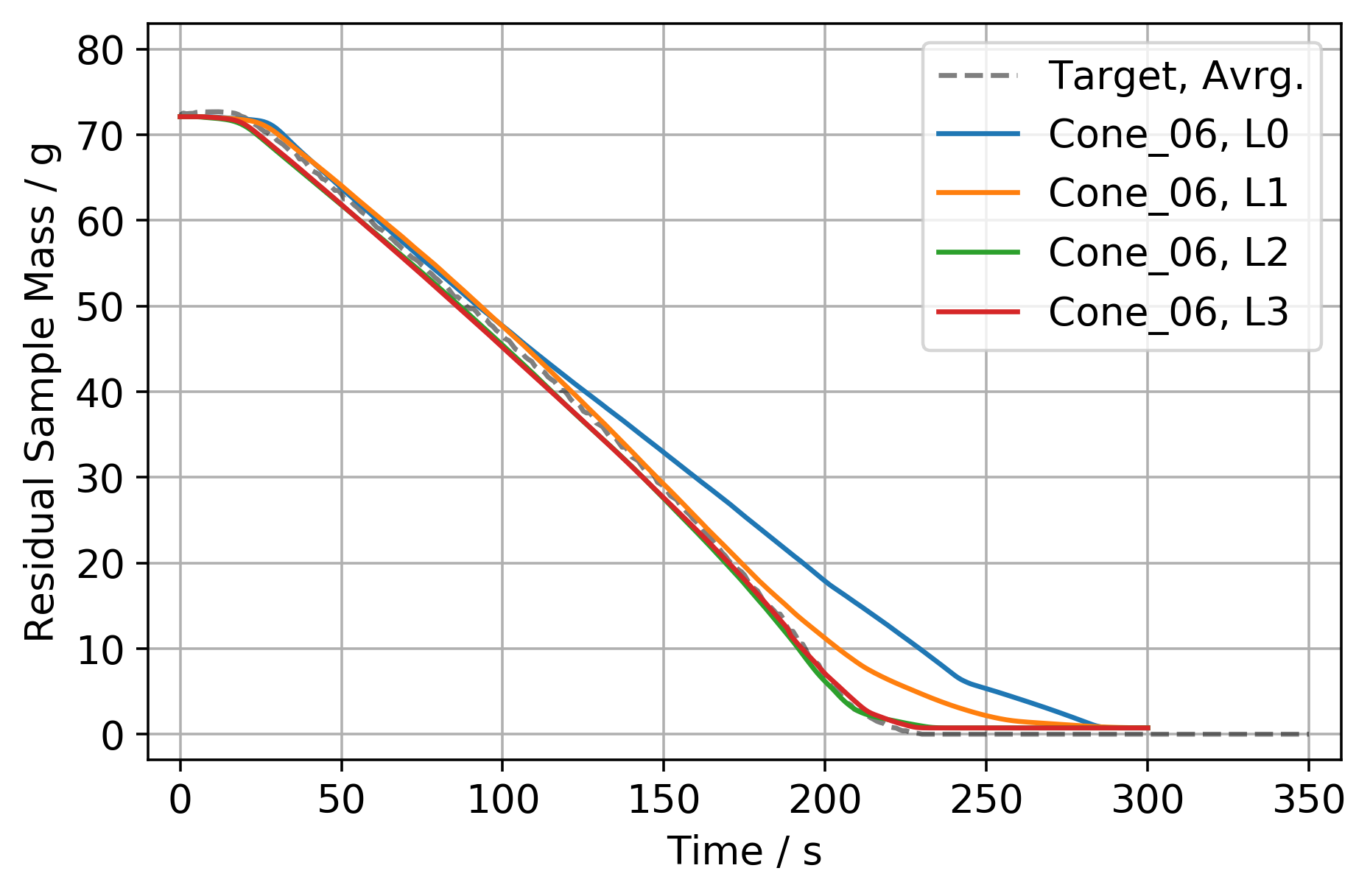} }}%
    \hfill
    \subfloat[\centering IMP Setup Cone\_07.]{{\includegraphics[width=0.395\columnwidth]{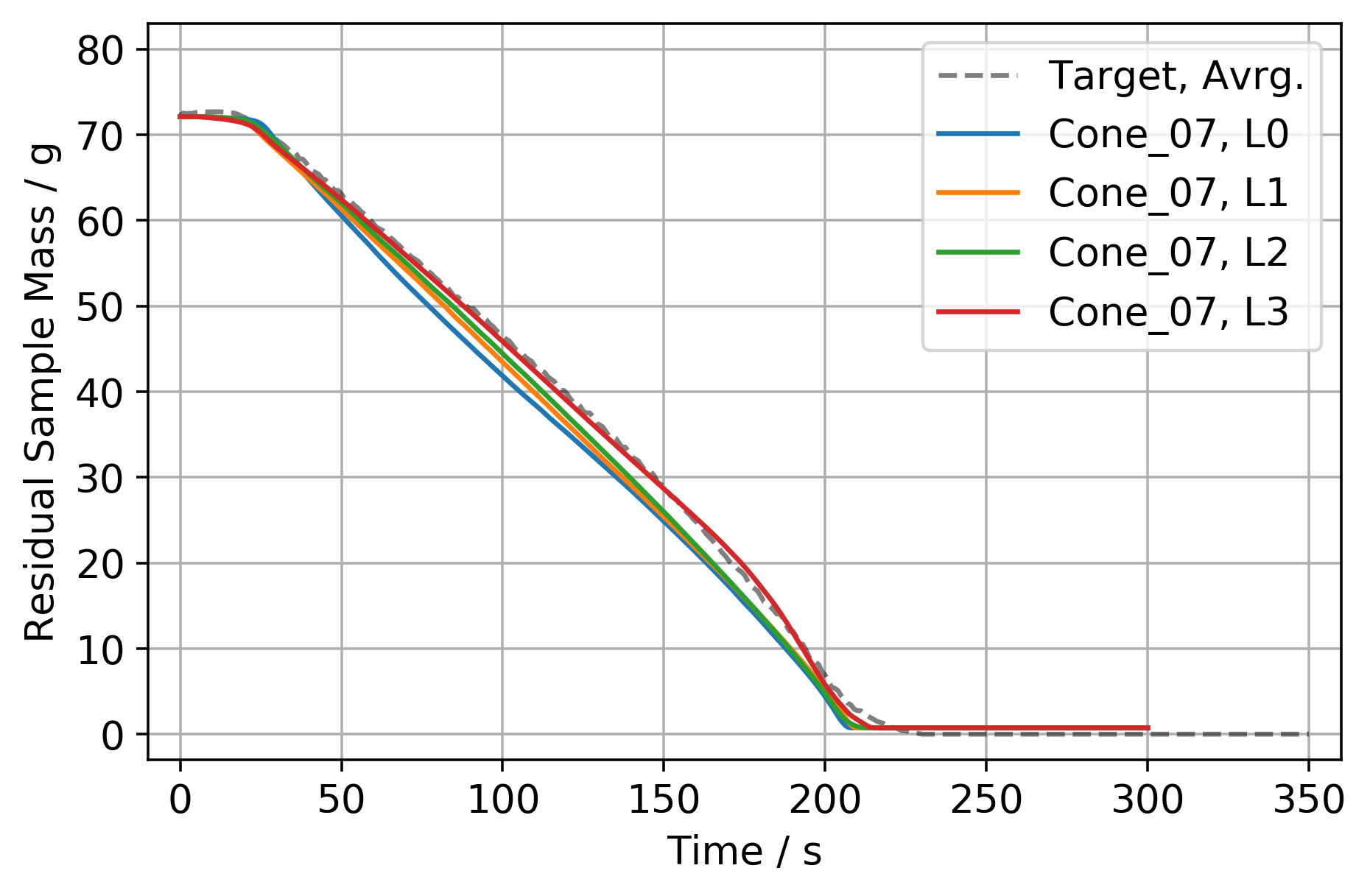} }}%
    \hfill
    \subfloat[\centering IMP Setup Cone\_08.]{{\includegraphics[width=0.395\columnwidth]{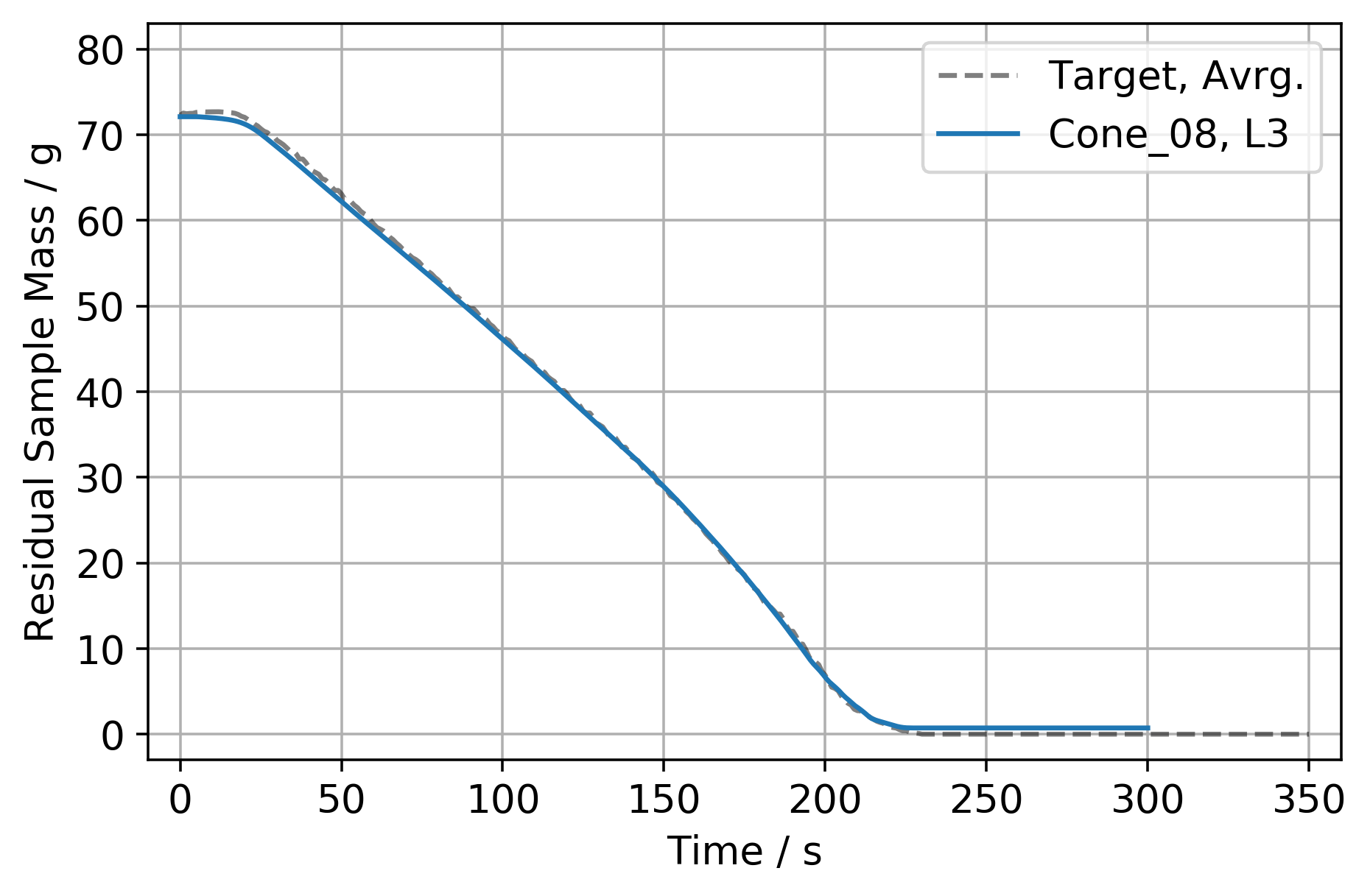} }}%
    
    \caption{Residual sample mass in simplified cone calorimeter simulation at 65 kW/m².}%
    \label{fig:ConeSimBestParaSampleMass_Aalto}%
\end{figure*}

\clearpage
\section{Simple Cone Calorimeter Simulation -- Fluid Cell Convergence}
\label{appendix_SimpleConeComparison}

\begin{figure*}[h]%
    \centering
    \subfloat[\centering Limit 0.]{{\includegraphics[width=0.4\columnwidth]{Figs_Art03/ConeScale/ConeSimCheck_EnergyReleaseRate_Aalto_04_L0.png} }}%
    \qquad
    \subfloat[\centering Limit 1.]{{\includegraphics[width=0.4\columnwidth]{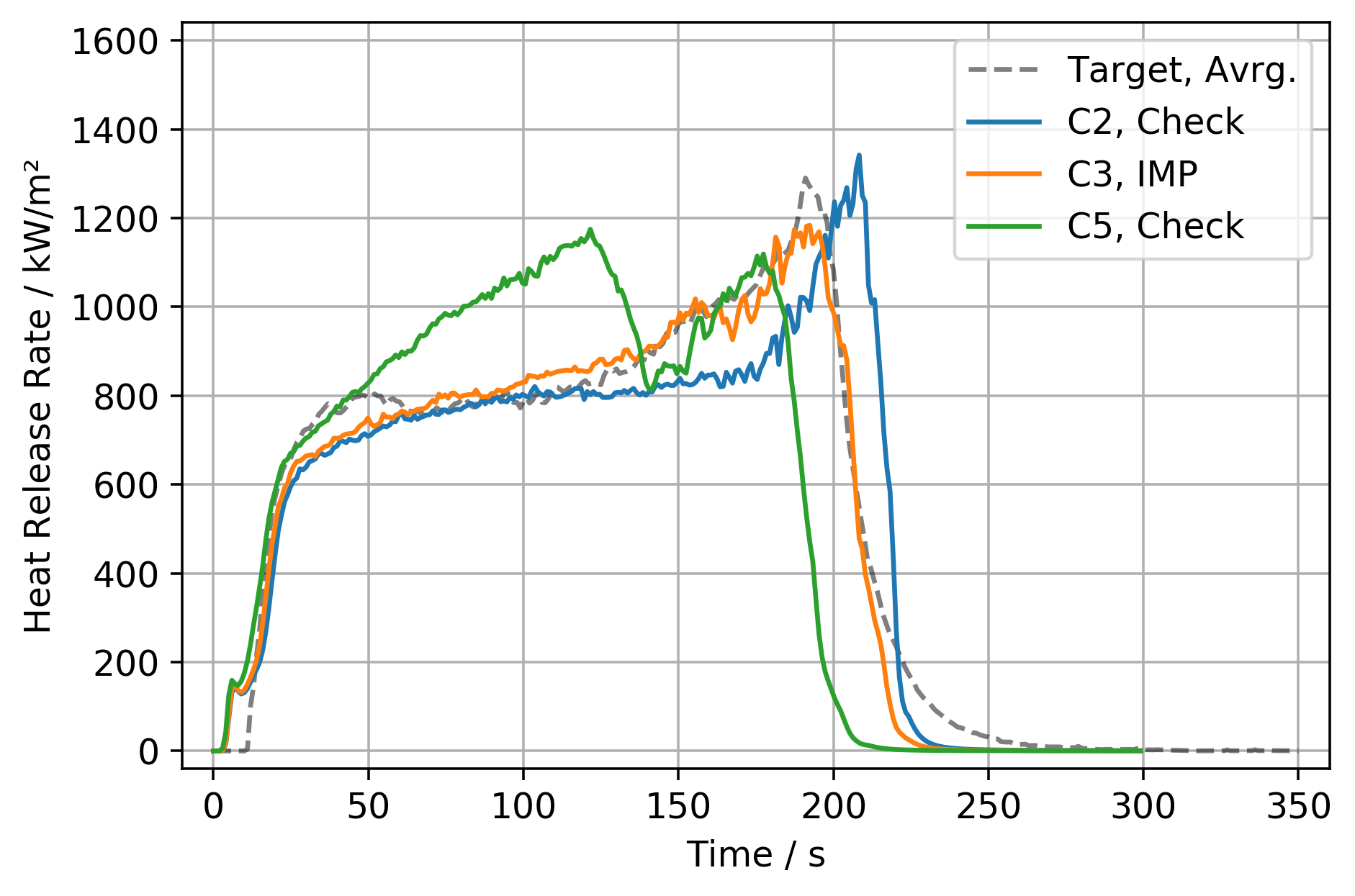} }}%
    
    \subfloat[\centering Limit 2.]{{\includegraphics[width=0.4\columnwidth]{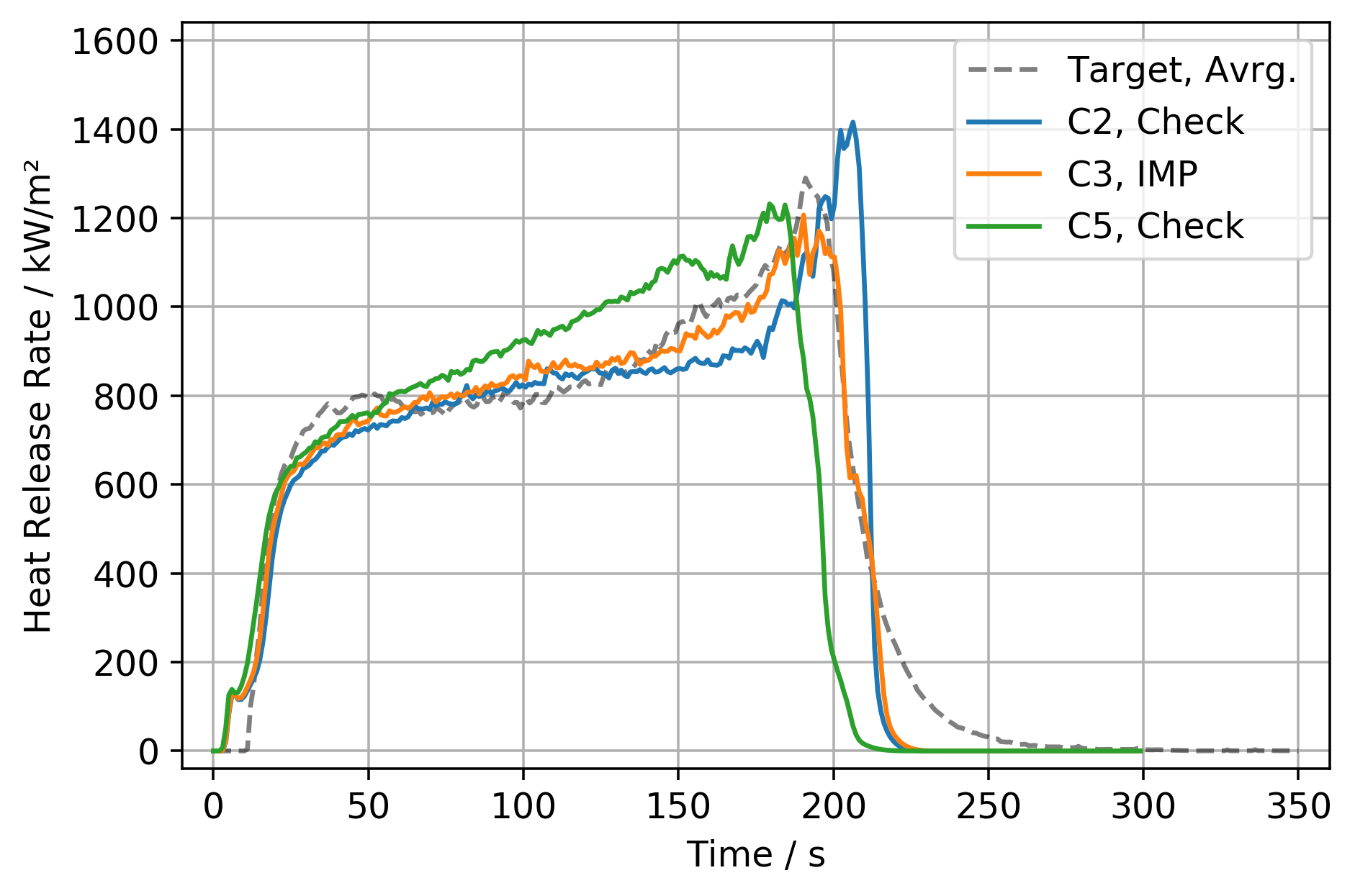} }}%
    \qquad
    \subfloat[\centering Limit 3.]{{\includegraphics[width=0.4\columnwidth]{Figs_Art03/ConeScale/ConeSimCheck_EnergyReleaseRate_Aalto_04_L3.png} }}%
    
    \caption{Comparison of the energy release of Cone\_04 across different fluid cell resolutions for simple cone calorimeter setup. Best parameter set of IMP conducted in 3.3~cm resolution (3C), same parameter set used in 2.0~cm resolution (5C).}%
    \label{fig:SimpleCone_CellSize_full_Aalto04}%
\end{figure*}

\clearpage
\section{Parallel Panel Simulation Results}
\label{App:PP_results}

\begin{figure*}[h]%
    \centering
    \subfloat[\centering Fluid cell size: 5C.]{{\includegraphics[width=0.42\columnwidth]{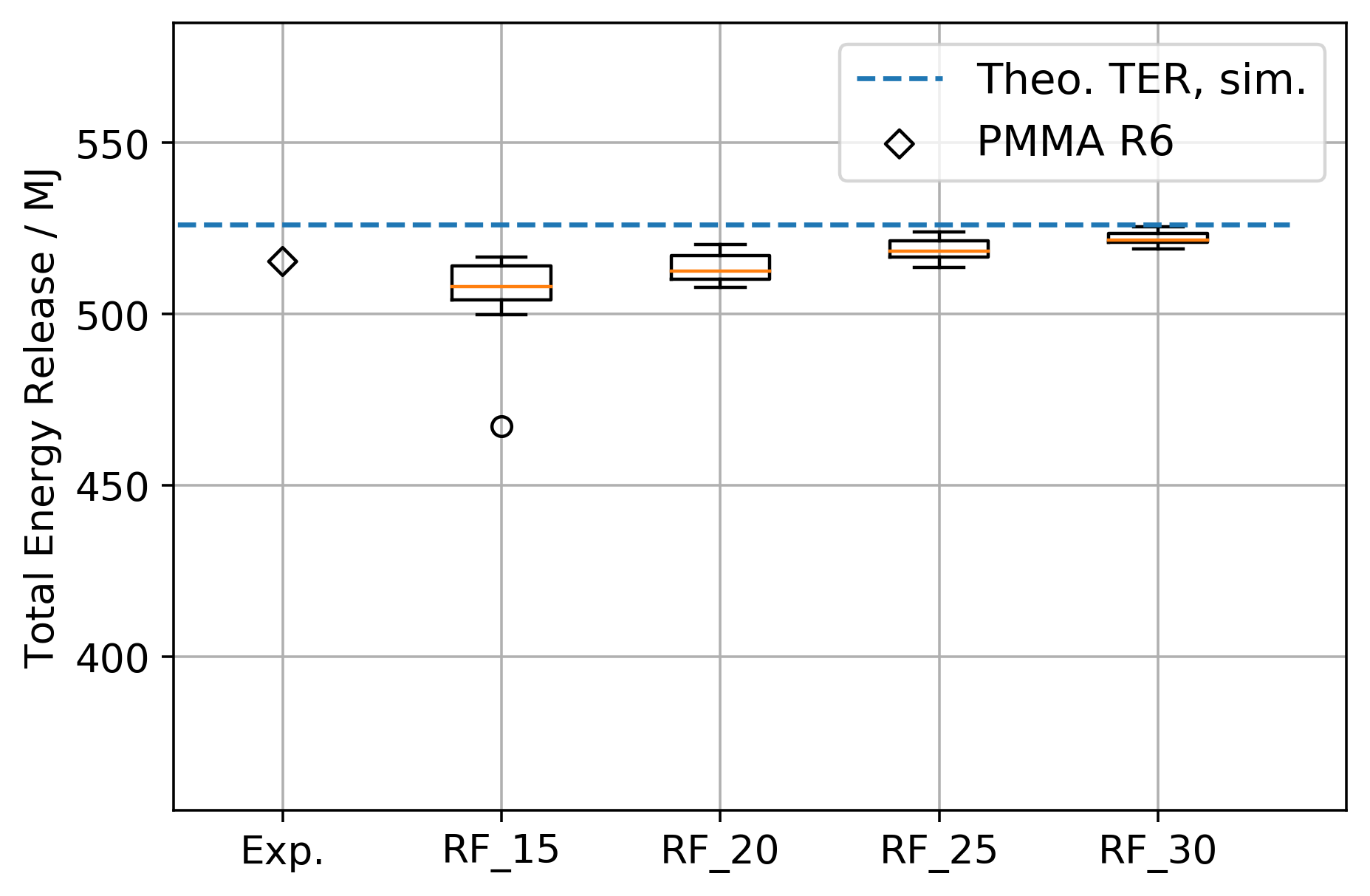} }}%
    \qquad
    \subfloat[\centering Fluid cell size: 3C.]{{\includegraphics[width=0.42\columnwidth]{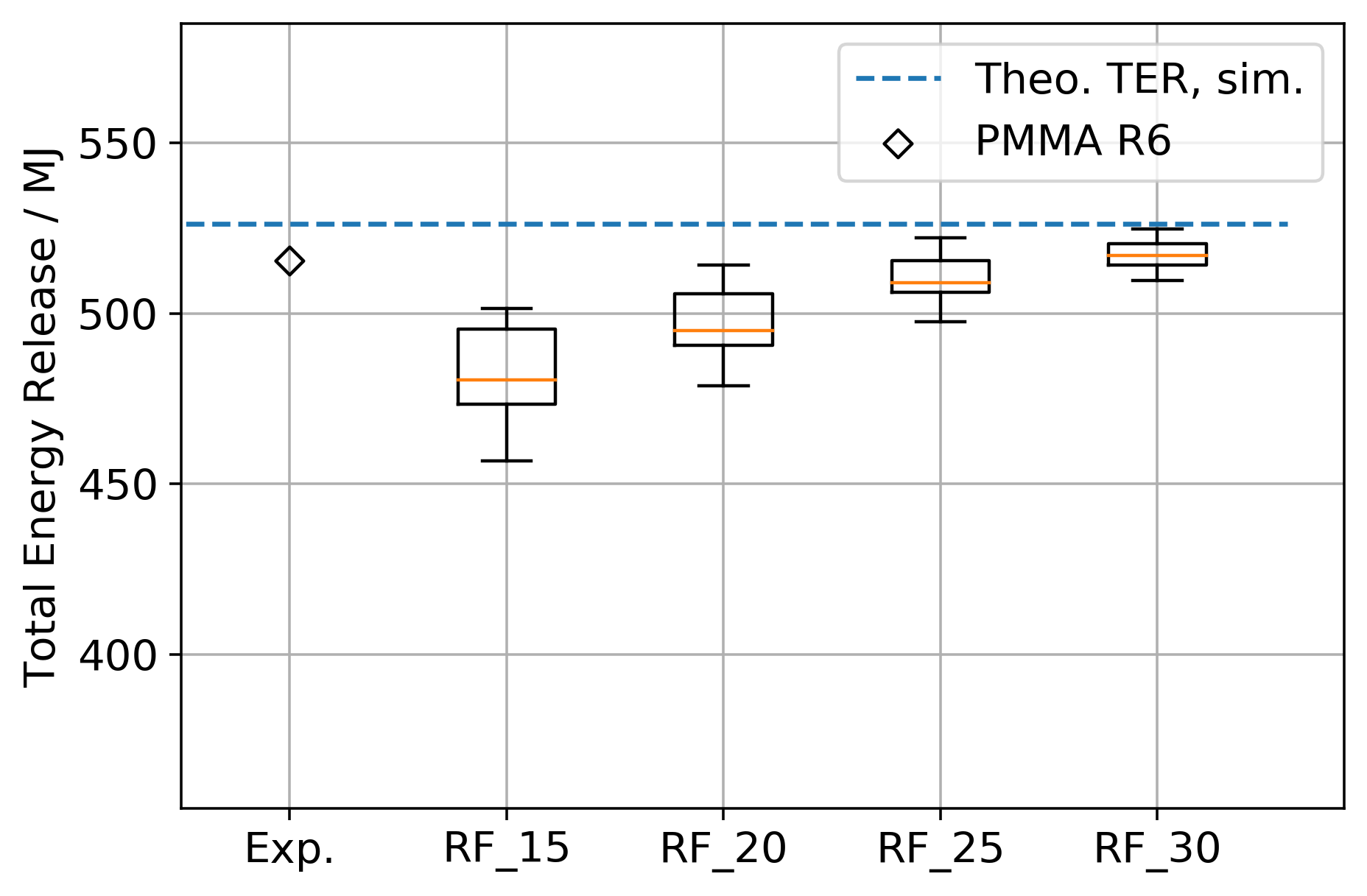} }}%
    \qquad
    \subfloat[\centering Fluid cell size: 2C.]{{\includegraphics[width=0.42\columnwidth]{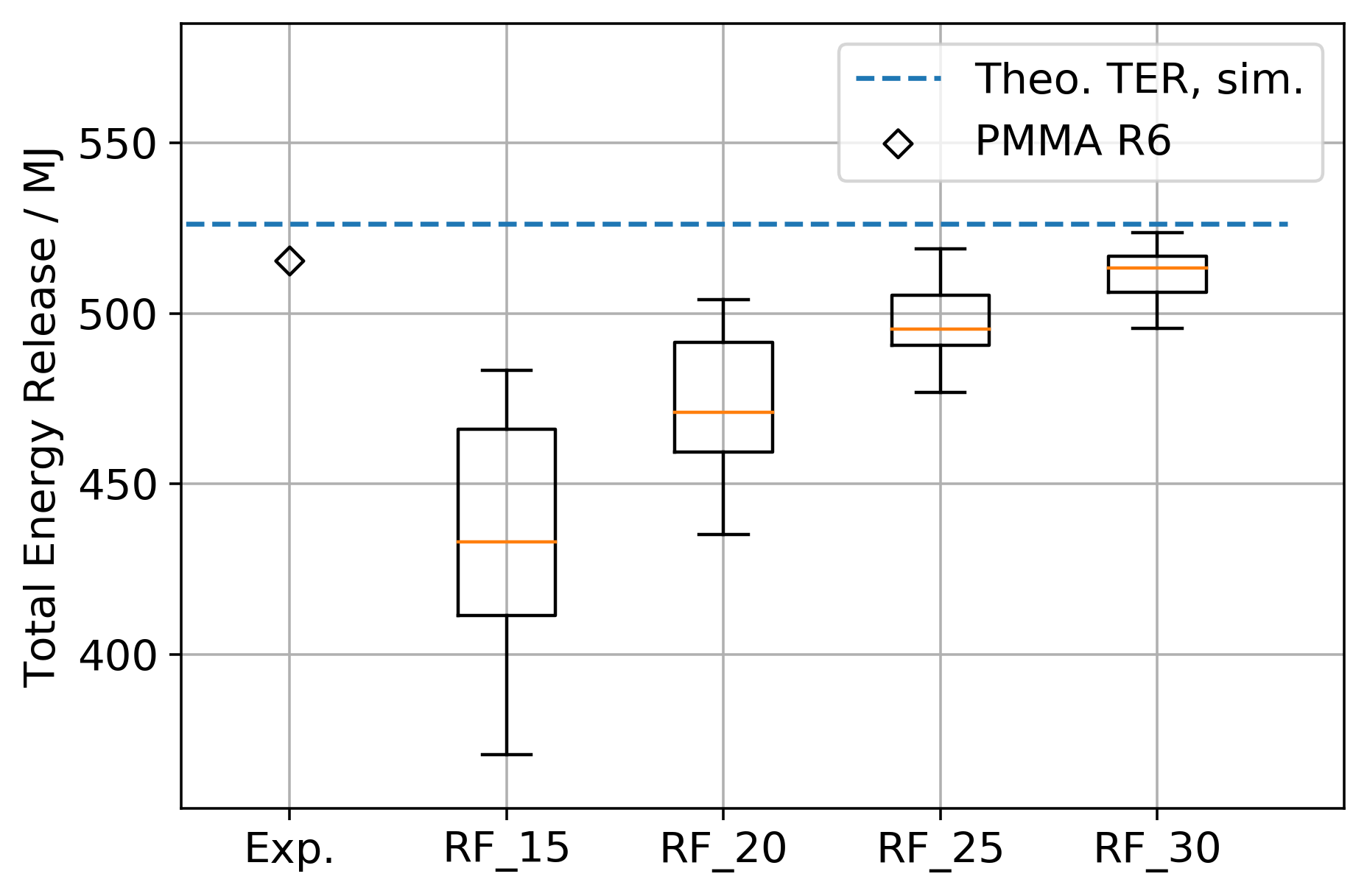} }}%
    
    \caption{Total energy release (TER) of best parameter sets in parallel panel setup. Comparison between different radiative fractions (RF) and fluid cell sizes. Dashed line indicates theoretical total energy release in the simulation.}%
    \label{fig:PP_TER_Complete}%
\end{figure*}

\begin{figure*}[h]%
    \centering
    \subfloat[\centering Burner shut off at 120~s.]{{\includegraphics[width=0.47\columnwidth]{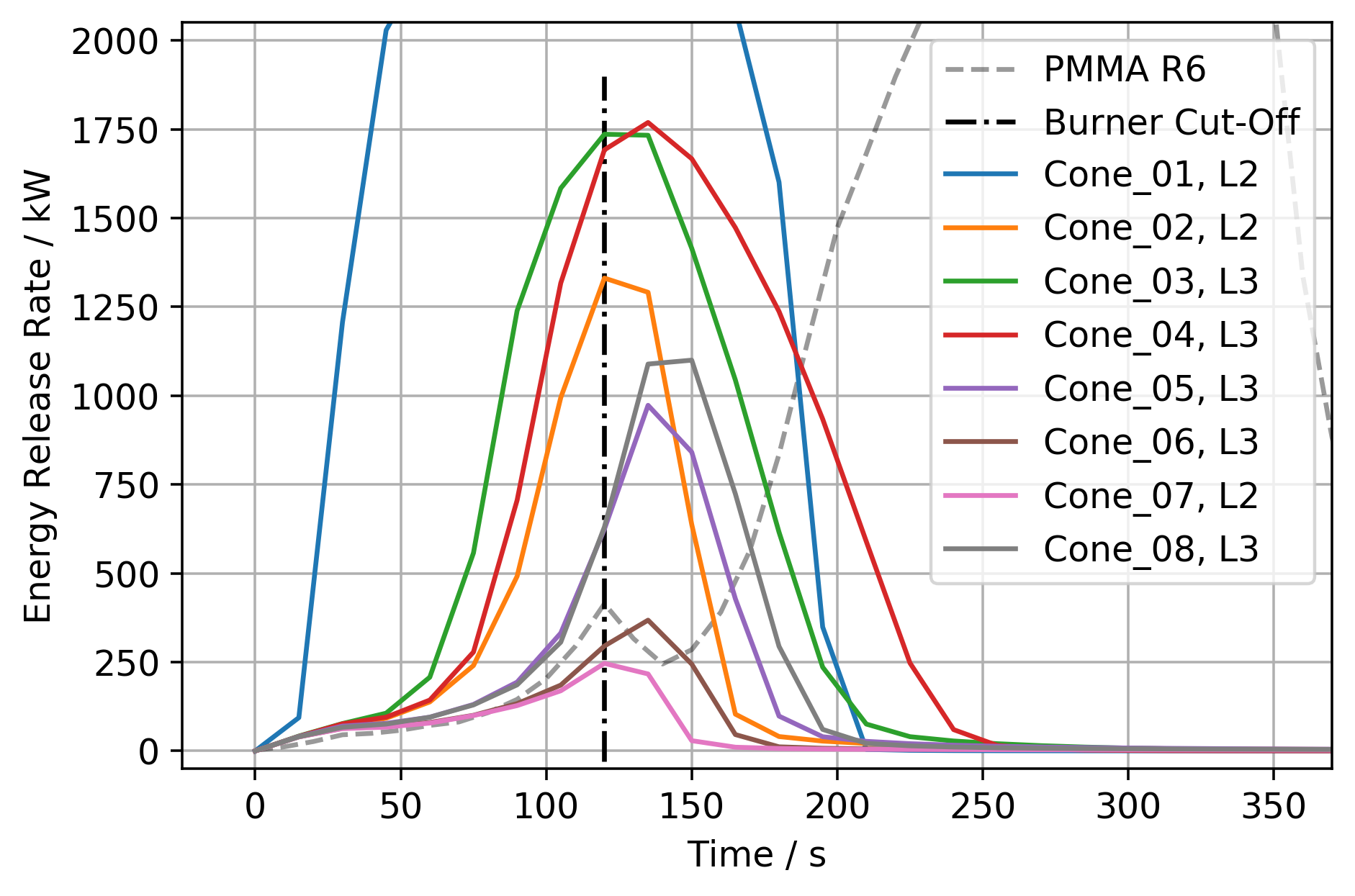} }}%
    \qquad
    \subfloat[\centering Burner shut off at 220~s.]{{\includegraphics[width=0.47\columnwidth]{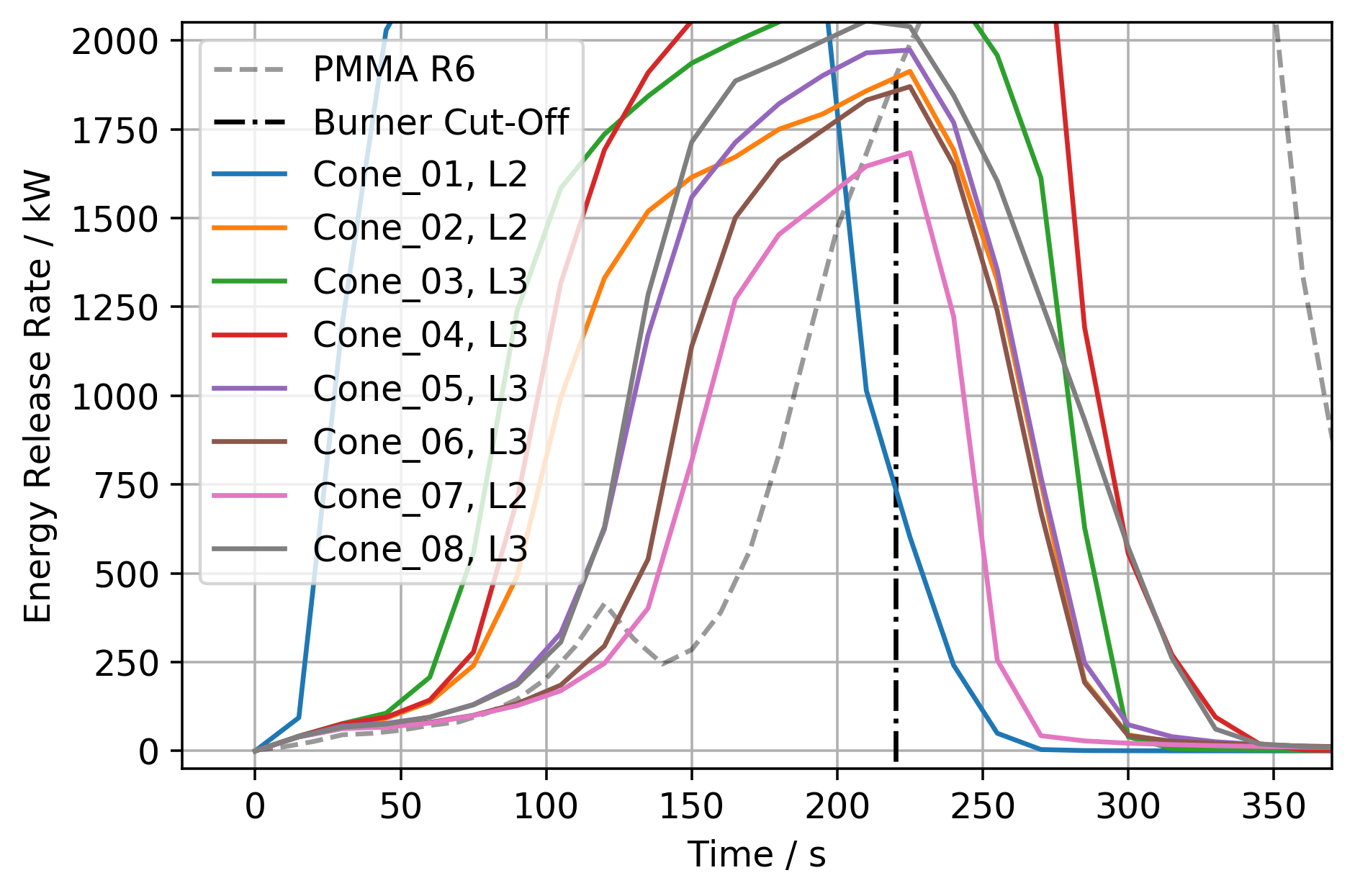} }}%
    
    \caption{Parallel panel simulation with burner cut-off at 120~s and 220~s, ramp down over 6~s. Gas burner fuel is propane.}%
    \label{fig:PPSimBestParaHRR_BurnerOff}%
\end{figure*}

\clearpage
\bibliographystyle{unsrtnat}
\bibliography{literature}

\begin{thebibliography}{52}
\providecommand{\natexlab}[1]{#1}
\providecommand{\url}[1]{\texttt{#1}}
\expandafter\ifx\csname urlstyle\endcsname\relax
  \providecommand{\doi}[1]{doi: #1}\else
  \providecommand{\doi}{doi: \begingroup \urlstyle{rm}\Url}\fi

\bibitem[Matala(2013)]{Matala2013PhdApplicationsOfPyrolysisModelling}
Anna Matala.
\newblock \emph{Methods and applications of pyrolysis modelling for polymeric
  materials}.
\newblock PhD thesis, Aalto University, Department of Mathematics and Systems
  Analysis Laboratory, Finland, 2013.
\newblock URL \url{http://urn.fi/URN:ISBN:978-951-38-8102-3}.

\bibitem[Ding et~al.(2019)Ding, I.~Stoliarov, and
  H.~Kraemer]{Stoliarov2019PyrolysisModelDevelopment}
Yan Ding, Stanislav I.~Stoliarov, and Roland H.~Kraemer.
\newblock Pyrolysis model development for a polymeric material containing
  multiple flame retardants: Relationship between heat release rate and
  material composition.
\newblock \emph{Combustion and Flame}, 202:\penalty0 43--57, 04 2019.
\newblock \doi{10.1016/j.combustflame.2019.01.003}.

\bibitem[Rogaume(2019)]{ThermalDecompositionAndPyrolysisOfSolidFuels_Rogaume}
Thomas Rogaume.
\newblock Thermal decomposition and pyrolysis of solid fuels: Objectives,
  challenges and modelling.
\newblock \emph{Fire Safety Journal}, 106:\penalty0 177 -- 188, 05 2019.
\newblock \doi{10.1016/j.firesaf.2019.04.016}.

\bibitem[Lautenberger et~al.(2006)Lautenberger, Rein, and
  Fernandez-Pello]{LAUTENBERGER2006204}
Chris Lautenberger, Guillermo Rein, and Carlos Fernandez-Pello.
\newblock The application of a genetic algorithm to estimate material
  properties for fire modeling from bench-scale fire test data.
\newblock \emph{Fire Safety Journal}, 41\penalty0 (3):\penalty0 204 -- 214,
  2006.
\newblock ISSN 0379-7112.
\newblock \doi{https://doi.org/10.1016/j.firesaf.2005.12.004}.
\newblock URL
  \url{http://www.sciencedirect.com/science/article/pii/S0379711205001372}.

\bibitem[Alonso et~al.(2019)Alonso, L{\'a}zaro, L{\'a}zaro, L{\'a}zaro, and
  Alvear]{Alonso2019_LLDPEKineticPropertiesEstimation}
Alain Alonso, Mariano L{\'a}zaro, Pedro L{\'a}zaro, David L{\'a}zaro, and
  Daniel Alvear.
\newblock Lldpe kinetic properties estimation combining thermogravimetry and
  differential scanning calorimetry as optimization targets.
\newblock \emph{Journal of Thermal Analysis and Calorimetry}, 138\penalty0
  (4):\penalty0 2703--2713, Nov 2019.
\newblock ISSN 1588-2926.
\newblock \doi{10.1007/s10973-019-08199-4}.
\newblock URL \url{https://doi.org/10.1007/s10973-019-08199-4}.

\bibitem[Lautenberger and Fernandez-Pello(2011)]{lautenberger2011optimization}
Chris Lautenberger and AC~Fernandez-Pello.
\newblock Optimization algorithms for material pyrolysis property estimation.
\newblock \emph{Fire Safety Science}, 10:\penalty0 751--764, 2011.

\bibitem[Batiot et~al.(2022)Batiot, Bruns, Hostikka, Leventon, Nakamura,
  Reszka, Rogaume, and Stoliarov]{macfp_matl_git}
B.~Batiot, M.~Bruns, S.~Hostikka, I.~Leventon, Y.~Nakamura, P.~Reszka,
  T.~Rogaume, and S.~Stoliarov.
\newblock {Measurement and Computation of Fire Phenomena (MaCFP) Condensed
  Phase Material Database.}
\newblock \url{https://github.com/MaCFP/macfp-db}, Commit:
  7f89fd85f75cd2d4999c262f9b39f2f8109e12ef, DOI:
  \url{https://doi.org/10.18434/mds2-2586}, 2022.

\bibitem[Viitanen et~al.(2022)Viitanen, Hostikka, and
  Vaari]{AlexandraViitanen_CableTrays}
Alexandra Viitanen, Simo Hostikka, and Jukka Vaari.
\newblock {CFD} {S}imulations of {F}ire {P}ropagation in {H}orizontal {C}able
  {T}rays {U}sing a {P}yrolysis {M}odel with {S}tochastically {D}etermined
  {G}eometry.
\newblock \emph{Fire Technology}, July 2022.
\newblock \doi{10.1007/s10694-022-01291-6}.

\bibitem[Rinta-Paavola and Hostikka(2022)]{Aleksi_WoodPyrolysis}
Aleksi Rinta-Paavola and Simo Hostikka.
\newblock A model for the pyrolysis of two nordic structural timbers.
\newblock \emph{Fire and Materials}, 46\penalty0 (1):\penalty0 55--68, 2022.
\newblock \doi{10.1002/fam.2947}.
\newblock URL \url{https://onlinelibrary.wiley.com/doi/abs/10.1002/fam.2947}.

\bibitem[Hehnen et~al.(2020)Hehnen, Arnold, and La~Mendola]{Hehnen_fire3030033}
Tristan Hehnen, Lukas Arnold, and Saverio La~Mendola.
\newblock Numerical {F}ire {S}pread {S}imulation {B}ased on {M}aterial
  {P}yrolysis -- {A}n {A}pplication to the {CHRISTIFIRE} {P}hase 1 {H}orizontal
  {C}able {T}ray {T}ests.
\newblock \emph{Fire}, 3\penalty0 (3), 2020.
\newblock ISSN 2571-6255.
\newblock \doi{10.3390/fire3030033}.
\newblock URL \url{https://www.mdpi.com/2571-6255/3/3/33}.

\bibitem[Hostikka and Matala(2017)]{HostikkaMatala2017PyrolysisBirchWood}
Simo Hostikka and Anna Matala.
\newblock Pyrolysis model for predicting the heat release rate of birch wood.
\newblock \emph{Combustion Science and Technology}, 189\penalty0 (8):\penalty0
  1373--1393, 8 2017.
\newblock ISSN 0010-2202.
\newblock \doi{10.1080/00102202.2017.1295959}.

\bibitem[Alonso et~al.(2023)Alonso, Lázaro, Lázaro, and
  Alvear]{Alonso2023_NumericalPredictionofCablesFireBehaviour}
Alain Alonso, David Lázaro, Mariano Lázaro, and Daniel Alvear.
\newblock Numerical prediction of cables fire behaviour using non-metallic
  components in cone calorimeter.
\newblock \emph{Combustion Science and Technology}, 0\penalty0 (0):\penalty0
  1--17, 2023.
\newblock \doi{10.1080/00102202.2023.2182198}.
\newblock URL \url{https://doi.org/10.1080/00102202.2023.2182198}.

\bibitem[Korobeinichev et~al.(2019)Korobeinichev, Paletsky, Gonchikzhapov,
  Glaznev, Gerasimov, Naganovsky, Shundrina, Snegirev, and
  Vinu]{Korobeinichev2019}
O.P. Korobeinichev, A.A. Paletsky, M.B. Gonchikzhapov, R.K. Glaznev, I.E.
  Gerasimov, Y.K. Naganovsky, I.K. Shundrina, A.Yu. Snegirev, and R.~Vinu.
\newblock Kinetics of thermal decomposition of {PMMA} at different heating
  rates and in a wide temperature range.
\newblock \emph{Thermochimica Acta}, 671:\penalty0 17--25, 2019.
\newblock ISSN 0040-6031.
\newblock \doi{10.1016/j.tca.2018.10.019}.
\newblock URL
  \url{https://www.sciencedirect.com/science/article/pii/S0040603118303460}.

\bibitem[Zeng et~al.(2002)Zeng, Li, and Chow]{Zeng_chemreac_burning_PMMA}
W.~R. Zeng, S.~F. Li, and W.~K. Chow.
\newblock Review on chemical reactions of burning poly(methyl methacrylate)
  pmma.
\newblock \emph{Journal of Fire Sciences}, 20\penalty0 (5):\penalty0 401--433,
  September 2002.
\newblock \doi{10.1177/0734904102020005482}.

\bibitem[Fiola et~al.(2021)Fiola, Chaudhari, and
  Stoliarov]{FIOLA2021103083PMMA_Pyrolysis_Model}
Gregory~J. Fiola, Dushyant~M. Chaudhari, and Stanislav~I. Stoliarov.
\newblock Comparison of {P}yrolysis {P}roperties of {E}xtruded and {C}ast
  {P}oly(methyl methacrylate).
\newblock \emph{Fire Safety Journal}, 120, 2021.
\newblock ISSN 0379-7112.
\newblock \doi{https://doi.org/10.1016/j.firesaf.2020.103083}.
\newblock URL
  \url{https://www.sciencedirect.com/science/article/pii/S0379711219307301}.
\newblock Fire Safety Science: Proceedings of the 13th International Symposium.

\bibitem[Karpov et~al.(2018)Karpov, Korobeinichev, Shaklein, Bolkisev, Kumar,
  and Shmakov]{KARPOV2018937}
Alexander~I. Karpov, Oleg~P. Korobeinichev, Artem~A. Shaklein, Andrey~A.
  Bolkisev, Amit Kumar, and Andrey~G. Shmakov.
\newblock {Numerical study of horizontal flame spread over PMMA surface in
  still air}.
\newblock \emph{Applied Thermal Engineering}, 144:\penalty0 937--944, 2018.
\newblock ISSN 1359-4311.
\newblock \doi{https://doi.org/10.1016/j.applthermaleng.2018.08.106}.
\newblock URL
  \url{https://www.sciencedirect.com/science/article/pii/S1359431118335750}.

\bibitem[Karpov et~al.(2015)Karpov, Shaklein, Korepanov, and
  Galat]{KARPOV_UpwardFlameSpread2015}
Alexander Karpov, Artem Shaklein, Mikhail Korepanov, and Artem Galat.
\newblock {Numerical Study of the Radiative and Turbulent Heat Flux Behavior of
  Upward Flame Spread Over PMMA}.
\newblock In Kazunori Harada, Ken Matsuyama, Keisuke Himoto, Yuji Nakamura, and
  Kaoru Wakatsuki, editors, \emph{Fire Science and Technology 2015}, pages
  841--848, Singapore, 2015. Springer Singapore.
\newblock ISBN 978-981-10-0376-9.

\bibitem[Wu et~al.(2003)Wu, Fan, Chen, Liou, and Pan]{WU2003697}
K.K. Wu, W.F. Fan, C.H. Chen, T.M. Liou, and I.J. Pan.
\newblock {Downward flame spread over a thick PMMA slab in an opposed flow
  environment: experiment and modeling}.
\newblock \emph{Combustion and Flame}, 132\penalty0 (4):\penalty0 697--707,
  2003.
\newblock ISSN 0010-2180.
\newblock \doi{https://doi.org/10.1016/S0010-2180(02)00520-5}.
\newblock URL
  \url{https://www.sciencedirect.com/science/article/pii/S0010218002005205}.

\bibitem[Rauwoens et~al.(2010)Rauwoens, Degroote, Wasan, Vierendeels, and
  Merci]{Rauwoens_UpwardFlameSpreadEnthalpy2010}
P.~Rauwoens, J.~Degroote, S.~Wasan, J.~Vierendeels, and B~Merci.
\newblock {Upward Flame Spread Simulations by Coupling an Enthalpy-Based
  Pyrolysis Model with CFD}.
\newblock In \emph{Proceedings of the 6th International Seminar on Fire and
  Explosion Hazards}, pages 241--251, Leeds, UK, 2010.

\bibitem[Kwon(2006)]{KwonMasterThesisFDS4FlameSpreadPMMA}
Jae-Wook Kwon.
\newblock {Evaluation of FDS V.4: Upward Flame Spread}.
\newblock Master's thesis, Worcester Polytechnic Institute, Fire Protection
  Engineering, MA, USA, 2006.
\newblock URL \url{https://web.wpi.edu/Pubs/ETD/Available/etd-090606-112948/}.

\bibitem[Chaudhari et~al.(2021)Chaudhari, Fiola, and
  Stoliarov]{CHAUDHARI2021109433PMMA_RoomCorner_Simulation}
Dushyant~M. Chaudhari, Gregory~J. Fiola, and Stanislav~I. Stoliarov.
\newblock Experimental analysis and modeling of buoyancy-driven flame spread on
  cast poly(methyl methacrylate) in corner configuration.
\newblock \emph{Polymer Degradation and Stability}, 183:\penalty0 109433, 2021.
\newblock ISSN 0141-3910.
\newblock \doi{https://doi.org/10.1016/j.polymdegradstab.2020.109433}.
\newblock URL
  \url{https://www.sciencedirect.com/science/article/pii/S0141391020303621}.

\bibitem[Stoliarov and Lyon(2008)]{StoliarovThermaKin}
Stanislav~I. Stoliarov and Richard~E. Lyon.
\newblock Thermo-kinetic model of burning for pyrolyzing materials.
\newblock In \emph{Fire Safety Science -- Proceedings of the Ninth
  Internatinoal Symposium, Karlsruhe, Germany}, page 1141–1152, 2008.
\newblock \doi{10.3801/iafss.fss.9-1141}.

\bibitem[McGrattan et~al.(2021{\natexlab{a}})McGrattan, Hostikka, Floyd,
  McDermott, and Vanella]{fdsUserGuide676}
Kevin McGrattan, Simo Hostikka, Jason Floyd, Randall McDermott, and Marcos
  Vanella.
\newblock \emph{Fire {D}ynamics {S}imulator {U}ser’s {G}uide}, 05
  2021{\natexlab{a}}.

\bibitem[Moinuddin et~al.(2020)Moinuddin, Razzaque, and
  Thomas]{Moinuddin_PMMAConeModelling}
Khalid Moinuddin, Qazi~Samia Razzaque, and Ananya Thomas.
\newblock Numerical simulation of coupled pyrolysis and combustion reactions
  with directly measured fire properties.
\newblock \emph{Polymers}, 12\penalty0 (9), 2020.
\newblock ISSN 2073-4360.
\newblock \doi{10.3390/polym12092075}.
\newblock URL \url{https://www.mdpi.com/2073-4360/12/9/2075}.

\bibitem[Boulet et~al.(2012)Boulet, Parent, Acem, Rogaume, Fateh, Zaida, and
  Richard]{BOULET201253}
P.~Boulet, G.~Parent, Z.~Acem, T.~Rogaume, T.~Fateh, J.~Zaida, and F.~Richard.
\newblock Characterization of the radiative exchanges when using a cone
  calorimeter for the study of the plywood pyrolysis.
\newblock \emph{Fire Safety Journal}, 51:\penalty0 53--60, 2012.
\newblock ISSN 0379-7112.
\newblock \doi{10.1016/j.firesaf.2012.03.003}.
\newblock URL
  \url{https://www.sciencedirect.com/science/article/pii/S037971121200046X}.

\bibitem[Hostikka and
  Axelsson(2003)]{HostikkaAxelssonRadiativeFeedbackFlamesConeCalorimeter}
Simo Hostikka and Jesper Axelsson.
\newblock {Modelling of the radiative feedback from the flames in cone
  calorimeter}.
\newblock Technical report, NORDTEST, 2003.
\newblock URL
  \url{https://www.nordtest.info/wp/2003/10/02/modelling-of-the-radiative-feedback-from-the-flames-in-cone-calorimeter-nt-tr-540/}.

\bibitem[Leventon et~al.(2022)Leventon, Heck, McGrattan, Bundy, and
  Davis]{Leventon2022ParallelPanel}
I.T. Leventon, M.V. Heck, K.B. McGrattan, M.F. Bundy, and R.D. Davis.
\newblock {Experimental Measurements for Fire Model Validation - Parallel Panel
  Tests on PMMA}.
\newblock
  \url{https://github.com/MaCFP/macfp-db/tree/master/Fire_Growth/NIST_Parallel_Panel},
  Commit: 25614bd527b658fca72265ed2940ea1287e83343, DOI:
  \url{https://doi.org/10.18434/mds2-2812}, 2022.

\bibitem[Hehnen and Arnold(2023)]{zenodo:ArticleDataset}
Tristan Hehnen and Lukas Arnold.
\newblock {PMMA} {P}yrolysis {S}imulation – from {M}icro- to {R}eal-{S}cale
  -- {D}ataset.
\newblock Technical report, Forschungszentrum Jülich and Bergische
  Universität Wuppertal, Germany, January 2023.
\newblock URL \url{https://zenodo.org/record/7065338}.

\bibitem[Hehnen(2022)]{firesimandcoding_playlist}
Tristan Hehnen.
\newblock {PMMA Pyrolysis and Fire Propagation in FDS}.
\newblock \emph{\url{www.youtube.com/@firesimandcoding}; \newline
  \url{www.youtube.com/playlist?list=PLWziITJoPnJKtVuaVudyO2Lz-OWaCE9Co}},
  2022.

\bibitem[McGrattan et~al.(2021{\natexlab{b}})McGrattan, Hostikka, Floyd,
  McDermott, and Vanella]{fdsTechGuide1_676}
Kevin McGrattan, Simo Hostikka, Jason Floyd, Randall McDermott, and Marcos
  Vanella.
\newblock \emph{{Fire Dynamics Simulator Technical Reference Guide Volume 1:
  Mathematical Model}}, 05 2021{\natexlab{b}}.

\bibitem[Duan et~al.(1993)Duan, Gupta, and Sorooshian]{duan1993shuffled}
Qingyun Duan, Vijai~K. Gupta, and Soroosh Sorooshian.
\newblock Shuffled complex evolution approach for effective and efficient
  global minimization.
\newblock \emph{Journal of optimization theory and applications}, 76\penalty0
  (3):\penalty0 501--521, 1993.

\bibitem[Arnold et~al.(2018{\natexlab{a}})Arnold, Hehnen, Lauer, Trettin, and
  Vinayak]{propti_nancy}
Lukas Arnold, Tristan Hehnen, Patrick Lauer, Corinna Trettin, and Ashish
  Vinayak.
\newblock {PROPTI} -- {A} {G}eneralised {I}nverse {M}odelling {F}ramework.
\newblock In \emph{Journal of Physics: Conference Series}, volume 1107, page
  032016. IOP Publishing, 2018{\natexlab{a}}.

\bibitem[Arnold et~al.(2019)Arnold, Hehnen, Lauer, Trettin, and
  Vinayak]{propti:ARNOLD2019102835}
Lukas Arnold, Tristan Hehnen, Patrick Lauer, Corinna Trettin, and Ashish
  Vinayak.
\newblock Application cases of inverse modelling with the {PROPTI} framework.
\newblock \emph{Fire Safety Journal}, page 102835, 2019.
\newblock ISSN 0379-7112.
\newblock \doi{10.1016/j.firesaf.2019.102835}.
\newblock URL
  \url{http://www.sciencedirect.com/science/article/pii/S0379711219300438}.

\bibitem[Arnold et~al.(2018{\natexlab{b}})Arnold, Hehnen, Lauer, Trettin, and
  Vinayak]{zenodo:PROPTI}
Lukas Arnold, Tristan Hehnen, Patrick Lauer, Corinna Trettin, and Ashish
  Vinayak.
\newblock {PROPTI}.
\newblock Technical report, Forschungszentrum Jülich and Bergische
  Universität Wuppertal, Germany, March 2018{\natexlab{b}}.
\newblock URL \url{https://zenodo.org/record/1438349}.

\bibitem[Houska et~al.(2015)Houska, Kraft, Chamorro-Chavez, and
  Breuer]{houska2015spotting}
Tobias Houska, Philipp Kraft, Alejandro Chamorro-Chavez, and Lutz Breuer.
\newblock Spotting model parameters using a ready-made python package.
\newblock \emph{PLoS ONE}, 10:\penalty0 e0145180, 12 2015.
\newblock URL \url{doi:10.1371/journal.pone.0145180}.

\bibitem[McNeill(1994)]{McNeill:ThermalDegradationMMA}
Ian~C. McNeill.
\newblock Products of {T}hermal {D}egradation in {P}olymer {S}ystems {B}ased on
  {M}ethyl {M}ethacrylate.
\newblock \emph{International Journal of Polymeric Materials and Polymeric
  Biomaterials}, 24\penalty0 (1-4):\penalty0 31--45, 1994.
\newblock \doi{10.1080/00914039408028548}.

\bibitem[Bowman(2019)]{BowmanShockInitiatedMethaneOxidation}
Graig~T. Bowman.
\newblock Non-equilibirum radical concentrations in shock-initiated methane
  oxidation.
\newblock \emph{Fire Safety Journal}, page 102835, 2019.
\newblock ISSN 0379-7112.
\newblock \doi{10.1016/j.firesaf.2019.102835}.
\newblock URL
  \url{http://www.sciencedirect.com/science/article/pii/S0379711219300438}.

\bibitem[Zhukov and Kong(2018)]{ZhukovCompactReactionMechanismMethane}
Victor~P. Zhukov and Alan~F. Kong.
\newblock A compact reaction mechanism of methane oxidation at high pressures.
\newblock \emph{Progress in Reaction Kinetics and Mechanism}, 43\penalty0
  (1):\penalty0 62--78, 2018.
\newblock \doi{10.3184/146867818X15066862094914}.
\newblock URL \url{https://doi.org/10.3184/146867818X15066862094914}.

\bibitem[Zhukov et~al.(2005)Zhukov, Sechenov, and
  Starikovskii]{ZhukovAutoignitionLeanPropaneAirMixtureHighPressures}
Victor~P. Zhukov, V.~A. Sechenov, and A.~Yu. Starikovskii.
\newblock Autoignition of a lean propane-air mixture at high pressures.
\newblock \emph{Kinetics and Catalysis}, 46\penalty0 (3):\penalty0 319--327,
  2005.
\newblock \doi{10.1007/s10975-005-0079-7}.
\newblock URL \url{https://doi.org/10.1007/s10975-005-0079-7}.

\bibitem[Westbrook et~al.(2002)Westbrook, Pitz, Boercker, Curran, Griffiths,
  Mohamed, and
  Ribaucour]{WESTBROOKDetailedChemicalKineticReactionMechanismsHeptane}
C.K. Westbrook, W.J. Pitz, J.E. Boercker, H.J. Curran, J.F. Griffiths,
  C.~Mohamed, and M.~Ribaucour.
\newblock Detailed chemical kinetic reaction mechanisms for autoignition of
  isomers of heptane under rapid compression.
\newblock \emph{Proceedings of the Combustion Institute}, 29\penalty0
  (1):\penalty0 1311--1318, 2002.
\newblock ISSN 1540-7489.
\newblock \doi{10.1016/S1540-7489(02)80161-4}.
\newblock URL
  \url{https://www.sciencedirect.com/science/article/pii/S1540748902801614}.
\newblock Proceedings of the Combustion Institute.

\bibitem[Wolfhard and Parker(1949)]{Wolfhard_1949}
H.~G. Wolfhard and W.~G. Parker.
\newblock A new technique for the spectroscopic examination of flames at normal
  pressures.
\newblock \emph{Proceedings of the Physical Society. Section A}, 62\penalty0
  (11):\penalty0 722--730, nov 1949.
\newblock \doi{10.1088/0370-1298/62/11/305}.

\bibitem[Curran(2019)]{CURRANDevelopingDetailedChemicalKineticMechanisms}
Henry~J. Curran.
\newblock Developing detailed chemical kinetic mechanisms for fuel combustion.
\newblock \emph{Proceedings of the Combustion Institute}, 37\penalty0
  (1):\penalty0 57--81, 2019.
\newblock ISSN 1540-7489.
\newblock \doi{10.1016/j.proci.2018.06.054}.
\newblock URL
  \url{https://www.sciencedirect.com/science/article/pii/S1540748918302372}.

\bibitem[Lannoye et~al.(2023)Lannoye, Trettin, Belt, Reinecke, Goertz, and
  Arnold]{KarenCorinna_PMMA}
Karen~De Lannoye, Corinna Trettin, Alexander Belt, E.A. Reinecke, Roland
  Goertz, and Lukas Arnold.
\newblock {The Influence of Experimental Conditions on the Mass Loss for TGA in
  Fire Safety Science}.
\newblock \emph{Submitted to: {F}ire {S}afety {J}ournal}, 2023.

\bibitem[McGrattan et~al.(2021{\natexlab{c}})McGrattan, Hostikka, Floyd,
  McDermott, and Vanella]{fdsValiGuide676}
Kevin McGrattan, Simo Hostikka, Jason Floyd, Randall McDermott, and Marcos
  Vanella.
\newblock \emph{Fire Dynamics Simulator Technical Reference Guide Volume 3:
  Validation}, 05 2021{\natexlab{c}}.

\bibitem[Quintiere(2017)]{Quintire:PrinciplesOfFireBehaviour}
James~G. Quintiere.
\newblock \emph{Principles of Fire Behavior, Second Edition}.
\newblock CRC Press, 2017.
\newblock ISBN 9781498735629.

\bibitem[Babrauskas(1982)]{BabrauskasConeCalorimeter}
Vyto Babrauskas.
\newblock Development of the {C}one {C}alorimeter: {A} {B}ench-{S}cale {H}eat
  {R}elease {R}ate {A}pparatus {B}ased on {O}xygen {C}onsumption.
\newblock Technical report, National Institute of Standards and Technology,
  Gaithersburg, MD, 1982-11-01 1982.

\bibitem[iso(2015)]{iso5660_1}
{ISO} 5660-1:2015({E}) — {P}art 1: {H}eat release rate (cone calorimeter
  method) and smoke production rate.
\newblock \emph{ISO}, 2015.

\bibitem[Brännström(2016)]{braennstroem_cone_radiation_Interflam}
Fabian Brännström.
\newblock Application of fire simulation within the railway industry.
\newblock In \emph{14th International Conference and Exhibition on Fire Science
  and Engineering}. Interflam, July 2016.

\bibitem[De~Lannoye et~al.(2023)De~Lannoye, Belt, Reinecke, Markert, and
  Arnold]{Karen_ConeDeformationASTM2023}
Karen De~Lannoye, Alexander Belt, Ernst-Arndt Reinecke, Frank Markert, and
  Lukas Arnold.
\newblock {Comparison of Black and Transparent PMMA in the Cone Calorimeter}.
\newblock In Morgan~C. Bruns and Marc~L. Janssens, editors, \emph{ASTM SELECTED
  TECHNICAL PAPERS STP1642: Obtaining Data for Fire Growth Models}, page
  150–160. ASTM International, March 2023.
\newblock \doi{10.1520/STP164220210105}.

\bibitem[Elliott et~al.(2014)Elliott, Temple, Maluk, and
  Bisby]{ElliottIntumescentCoatings}
Angus Elliott, Alastair Temple, Cristian Maluk, and Luke Bisby.
\newblock Novel {T}esting to {S}tudy the {P}erformance of {I}ntumescent
  {C}oatings under {N}on-{S}tandard {H}eating {R}egimes.
\newblock In \emph{Fire Safety Science}, volume~11, pages 652--665, 02 2014.
\newblock \doi{10.3801/IAFSS.FSS.11-652}.

\bibitem[Quaresma(2023)]{Tassia_Sensitivity_SFPE2023}
Tássia Quaresma.
\newblock Sensitivity analysis of simulation input parameters, a comparison
  between flame spread and cone calorimeter setups.
\newblock In \emph{European Conference \& Expo on Fire Safety Engineering,
  SFPE23}. SFPE, March 2023.

\bibitem[{J\"{u}lich Supercomputing Centre}(2021)]{JURECA}
{J\"{u}lich Supercomputing Centre}.
\newblock {JURECA: Data Centric and Booster Modules implementing the Modular
  Supercomputing Architecture at J\"{u}lich Supercomputing Centre}.
\newblock \emph{Journal of large-scale research facilities}, 7\penalty0 (A182),
  2021.
\newblock \doi{10.17815/jlsrf-7-182}.
\newblock URL \url{http://dx.doi.org/10.17815/jlsrf-7-182}.

\end{thebibliography}

\end{document}